\documentclass[11pt]{article}
\usepackage{fullpage}
\usepackage{amsthm,amsmath,amssymb}
\usepackage{xspace}
\usepackage{nicefrac}
\usepackage{booktabs}
\usepackage{graphicx}
\usepackage[table,xcdraw]{xcolor}
\usepackage{tikz}
\usetikzlibrary{decorations.pathreplacing}
\usetikzlibrary{patterns}
\usepackage{pgfplots}
\pgfplotsset{compat=1.18}
\usepackage{caption}
\usepackage{subcaption}
\usepackage[authoryear]{natbib}
\usepackage{hyperref}
\usepackage{bbm}
\usepackage{multirow}
\usepackage{colortbl}
\usepackage{enumitem}
\usepackage{tablefootnote}

\newtheorem{theorem}{Theorem}[section]
\newtheorem{proposition}[theorem]{Proposition}
\newtheorem{definition}[theorem]{Definition}
\newtheorem{observation}[theorem]{Observation}
\newtheorem{corollary}[theorem]{Corollary}
\newtheorem{lemma}[theorem]{Lemma}

\theoremstyle{remark}
\newtheorem{remark}[theorem]{Remark}
\theoremstyle{definition}
\newtheorem{example}[theorem]{Example}

\DeclareMathOperator*{\argmax}{arg\,max}

\hypersetup{
    colorlinks,
    linkcolor={red!80!black},
    citecolor={blue!80!black},
    urlcolor={blue!80!black}
}

\newcommand{\rew}[0]{r\xspace}
\newcommand{\Rew}[0]{R\xspace}
\newcommand{\rvec}[0]{\mathbf{r}\xspace}

\newcommand{\cost}[0]{c\xspace}
\newcommand{\costs}[0]{\textbf{c}\xspace}
\newcommand{\alloc}[0]{x\xspace}
\newcommand{\Alloc}[0]{\textbf{x}\xspace}
\newcommand{\feat}[0]{F\xspace}
\newcommand{\Feat}[0]{\textbf{F}\xspace}
\newcommand{\prob}[0]{q\xspace}
\newcommand{\Prob}[0]{\mathbf{q}\xspace}

\newcommand{\Bid}[0]{\mathbf{b}\xspace}
\newcommand{\Wel}[0]{W\xspace}
\newcommand{\pay}[0]{t\xspace}
\newcommand{\ambpay}[0]{\tau\xspace}
\newcommand{\Pay}[0]{T\xspace}
\newcommand{\istar}[0]{i^\star\xspace}
\newcommand{\con}[0]{\mathbf{t}\xspace}
\newcommand{\linear}{\alpha\xspace}
\newcommand{\OPT}[0]{\textsf{OPT}\xspace}
\newcommand{\LP}[0]{\textsf{PAY}\xspace}
\newcommand{\classOfContracts}{\mathcal{T}\xspace}
\newcommand{\indicator}[1]{\mathbbm{1}\left[{#1}\right]}
\newcommand{\reals}{\mathbb{R}}
\newcommand{\agents}{\mathcal{A}}
\newcommand{\round}{s}
\newcommand{\numrounds}{S}

\title{Algorithmic Contract Theory: A Survey%
\thanks{This survey evolved from a tutorial at the 20th ACM Conference on Economics and Computation (EC 2019), and a tutorial at the 54th ACM Symposium on Theory of Computing (STOC 2022)~\citep{DuettingT19,DuettingT22,FeldmanL22}. 

We would like to thank the editors and anonymous reviewers of \emph{Foundations and Trends in Theoretical Computer Science} for inviting this survey, and for their very valuable feedback. We would also like to thank Tal Alon, Matteo Castiglioni, Jose Correa, Shaddin Dughmi, Tomer Ezra, Yoav Gal-Tzur, Vasilis Gkatzelis, Zhiyi Huang, Thomas Kesselheim, Ron Lavi, Yingkai Li, Brendan Lucier, Tomasz Ponitka, Manish Raghavan, Shaul Rosner, Tim Roughgarden, Larry Samuelson, Maya Schlesinger, Nicolas Stier-Moses, L\'{a}szl\'{o} V\'{e}gh, Bo Waggoner, Joshua R.~Wang, Haifeng Xu, and Konstantin Zabarnyi for their comments, which greatly improved the survey. 

This survey received funding from the European Research Council (ERC) under the European Union's Horizon 2020 research and innovation program (grant agreement No.~866132 and grant agreement No.~101077862), by the Israel Science Foundation (grant No.~336/18 and grant No.~3331/24), by the Israel Science Foundation Breakthrough Program (grant No.~2600/24), by the NSF-BSF (grant No.~2020788 and grant No.~2021680), by a Google Research Scholar Award, and by an Amazon Research Award.}
}

\author{Paul D\"utting\thanks{Google Research, Z\"urich, Switzerland. Email: \texttt{duetting@google.com}} \and Michal Feldman\thanks{Tel Aviv University, Tel Aviv, Israel. Email: \texttt{mfeldman@tauex.tau.ac.il}} \and Inbal Talgam-Cohen\thanks{Tel Aviv University, Tel Aviv, Israel. Email: \texttt{italgam@tauex.tau.ac.il}; Technion---Israel Institute of Technology, Haifa, Israel. Email: \texttt{italgam@cs.technion.ac.il}}}

\date{December 2024}

\begin{document}

\maketitle

\begin{abstract}
A contract is an economic tool used by a principal to incentivize one or more agents to exert effort on her behalf, by defining payments based on observable performance measures. A key challenge addressed by contracts --- known in economics as \emph{moral hazard} --- is that, absent a properly set up contract, agents might engage in actions that are not in the principal's best interest.  Another common feature of contracts is \emph{limited liability}, which means that payments can go only from the principal --- who has the deep pocket --- to the agents.

With classic applications of contract theory  moving online, growing in scale, and becoming more data-driven, tools from contract theory become increasingly important for incentive-aware algorithm design. At the same time,  algorithm design offers a whole new toolbox for reasoning about contracts, ranging from additional tools for studying the tradeoff between simple and optimal contracts, through a language for discussing the computational complexity of contracts in combinatorial settings, to a formalism for analyzing data-driven contracts.

This survey aims to provide a computer science-friendly introduction to the basic concepts of contract theory. We give an overview of the emerging field of ``algorithmic contract theory'' and highlight work that showcases the potential for interaction between the two areas. We also discuss avenues for future research.
\end{abstract}

\thispagestyle{empty}
\newpage
\thispagestyle{empty}
\tableofcontents
\thispagestyle{empty}
\newpage

\section{Introduction}
\label{sec:intro}
Imagine you are a website owner employing a website designer through an online freelancing platform. The most straightforward payment scheme for the designer’s work, i.e., \emph{contract}, is offering a fixed (lump sum)  transfer $t$ for completing the website’s design. But is this the best in terms of incentives? Anecdotal evidence and everyday experience suggest this is not the case. 
In the words of an Upwork user: ``Remember, Upwork [...] is more like a box of chocolates, you never know what you are going to get'' \citep{upwork}.
Rigorous empirical studies confirm the problem of low-quality, “careless” online work~\citep{ArugueteHB+19}, even when platforms use rating systems (as ratings are often inflated and thus not very informative)~\citep{GargJ21}.
 
This problem stems from a basic misalignment of incentives: The designer (\emph{agent}, he) is doing the hard work, while the owner (\emph{principal}, she) is reaping the rewards. This misalignment is coupled with an information gap---the principal has no way of knowing how much effort the agent invested in designing her website. 
With misaligned interests and imperfect observability, the principal has to rely on the moral behavior of the agent. 
This effect, known as \emph{moral hazard}, is a fundamental obstacle that any task delegation to human (or AI) agents must overcome. 

Fortunately, studies also show that \emph{pay-for-performance} contracts can have a significant impact on work quality~\citep{MasonW09,DellaVignaPope17,FestKNS20,KaynarS22,WangH22}. In our example, paying for performance means paying the agent based on information the principal can track and that determines her own rewards, such as the increase in the number of visitors to the website, the increase in the number of conversions, or the increase in revenue. Since the details of the payment scheme matter a lot towards the agent’s incentives, this raises important economic design questions such as what should the payments be contingent on, or how high these payments should be. 

The rising design challenge can thus be summarized as: compute an optimal (or near-optimal) pay-for-performance contract, where ``optimal’’ is with respect to welfare and revenue implications of the cooperation. 
Questions like this are studied in economics under the umbrella of \emph{contract theory} \citep{Ross73,Mirrlees99,Holmstrom79,GrossmanH83,Innes90,Carroll15}. 
Contract theory is one of the pillars of microeconomic theory, recognized by the 2016 Nobel Prize awarded to Hart and Holmstr\"om \citep{Nobel}. 
However, unlike other well-established areas of microeconomic theory, such as mechanism design or information design, contract design has not seen much work from computer science until recently.

\paragraph{Motivation: Why Algorithms? Why Now?}

We are motivated by a recent spike of interest from computer scientists in contract theory \citep[e.g.,][]{BabaioffFN06,HoSV16,DuttingRT19}.
This spike of interest is caused by the fact that more and more 
of the classic 
applications of contract theory are moving online, growing in scale, and happening in data-rich environments. {These include online labor platforms \citep[e.g.,][]{KaynarS22}, delegating machine learning tasks \citep[e.g.,][]{CaiDP15}, pay-for-performance healthcare \citep[e.g.,][]{BastaniEtAl16,BastaniGB19}, and blockchain
\citep[e.g.,][]{CongHe19}.}

In addition, tools from contract theory are anticipated to play a crucial role in a world in which we increasingly rely on AI agents to perform complex tasks 
\citep{Hadfield-Menell19a,WangEtAl2023,SaigET24}. This direction comes with a number of challenges, which are not addressed by classic contract theory. For instance, outcome and action spaces might be huge. Or, we may have to select a group of agents from a large pool of available agents. Also, naturally, all sides of the problem will involve (machine) learning.
At the same time, the fact that the agents are programmed, might also open up new opportunities. For instance, it seems reasonable to assume programmed AI agents exhibit ``hyper-rationality" that is harder to attribute to humans.

This naturally calls for a field that combines tools from \emph{contract theory} with tools from \emph{computer science} (specifically algorithm design and machine learning). Contract theory offers a well-established formalism to talk about incentives, and prevent detrimental behavior (such as shirking or free-riding). Computer science, in turn, provides a language to talk about computational complexity, offers tools for studying the tradeoffs between simple and optimal solutions, and has a natural focus on (machine) learning algorithms. 

Indeed, similar to other economic areas where the computational lens has been applied (notably, mechanism and information design), the algorithmic perspective is already providing new structural insights, helping to map out the tractability frontiers, and leading to new tools for data-driven contracts. 
Ultimately, the algorithmic approach to contracts has the potential to inform better designs in practice, especially in computational environments.  

This survey aims to provide an introduction to contract theory that is accessible to computer scientists and give an overview of the emerging field of algorithmic contract theory.%
\footnote{Due to the large volume of recent work that takes an algorithmic approach to contracts, we present only a sample of papers from the current main trajectories of research.} {We also discuss what we see as main directions for future work.}

\paragraph{Disambiguation: Contract Theory vs.~Smart Contracts.} 

We emphasize that the goals of the nascent area of algorithmic contract theory are orthogonal to those behind \emph{smart contracts} \citep{Szabo97}. While algorithmic contract theory, just as classic contract theory, aims to design contracts and provide tools to assess the pros and cons between different designs, smart contracts are a tool to \emph{implement} contracts in an automated way, often relying on blockchain technologies to enable execution, control, and documentation. A shared theme of both is the use of computing technology to enable more efficient contracts.

\begin{figure}
\begin{center}
\begin{tabular}{|l| l l|}
\toprule
& \bf Uninformed party & \bf Informed party\\ 
& \bf moves first: & \bf moves first:\\
\midrule
\bf Private information& Adverse selection & Bayesian persuasion 
\\  
\bf is hidden type:& (Mechanism design)& (Information design)\\
\midrule
\bf Private information & Moral hazard  & Not studied
\\
\bf is hidden action:&  (Contract design) &  \\
\bottomrule
\end{tabular}
\end{center}
\caption{ 
\citet[][Chapter 1.1]{Salanie17} proposes to classify problems where an informed party interacts with an uninformed party, along two dimensions: The first distinction is whether the private information bears on \emph{who} the agent is (``hidden type''), or whether it bears on \emph{what} action the agent takes (``hidden action''). The second distinction concerns the timing of the problem, and asks who moves first: the \emph{uninformed} party or the \emph{informed} party.}
\label{fig:great-families-of-models}
\end{figure}

\paragraph{Digression: Contracts within the Wider Context.} 

In this survey, we follow \cite{Salanie17} in classifying incentive problems along two dimensions, as shown in  Figure~\ref{fig:great-families-of-models}.
This leads to three basic incentive problems (because the fourth combination does not seem to capture relevant applications). 
We adopt a terminology that identifies contract design, mechanism design, and information design with the three basic incentive problems that result from this classification.

The division into three basic incentive problems results from viewing incentive problems as interactions between an uninformed party and an informed party, and classifying these interactions according to two criteria: The first is whether the private information concerns \emph{who} the agent is (``hidden type''), or whether it concerns \emph{what} action the agent takes (``hidden action''). The second is whether the \emph{uninformed} party \emph{moves first} and designs the incentive scheme, or whether it is the \emph{informed} party who moves first.

This classification yields three important families of models:%
\footnote{\label{ftnt:fourth-rubric}The fourth case is where the uniformed party cannot observe the actions of the informed party, and the informed party moves first. \citet[][FN1 on p.4]{Salanie17} argues that: ``It is difficult to imagine a real-world application of such a model, and I do not know of any paper that uses it.'' Of course, it is also possible to consider problems that exhibit features of two or more of the ``pure'' problems, e.g.,~\cite{BernasconiEtAl24}.} 
\begin{enumerate}
\item[(1.)] \emph{Adverse selection} models: The uninformed party is imperfectly informed of the characteristics of the informed party; the uninformed party moves first. A canonical example is a first-price auction, where the auctioneer knows that the bidders' valuations are drawn from certain distributions, but only the bidders know the realized valuations. The auctioneer moves first by announcing the rules of the auction. Afterwards, the bidders submit their bids and based on this an allocation and payments are determined.
\item[(2.)] \emph{Bayesian persuasion} models: The uninformed party is imperfectly informed of the characteristics of the informed party; the informed party moves first. A prototypical example here is one in which there is a hidden state  drawn from a publicly known distribution, whose realization is known by only one of the two parties. For example, in a court case, the attorney representing a client, may know whether the client is guilty or innocent, and may seek to structure her arguments so as to convince the judge to acquit her client.
\item[(3.)] \emph{Moral hazard} models: 
The uninformed party is imperfectly informed of the actions of the informed party; the uninformed party moves first. For example, a brand may seek to hire an influencer on a social media platform to create sponsored content. 
The brand proposes a contract that defines how 
the influencer shall get paid. Payments can only be contingent on the observable but typically stochastic outcome 
of the agent's action (e.g., number of views the content receives). After signing the contract, the influencer creates the sponsored content and is paid according to the contract, based on the observed outcome.
\end{enumerate}

Alternative names that can be found in the literature for (1.) and (2.) are \emph{screening} and \emph{signaling}, respectively. 
The majority of the work in computer science has focused on mechanism design (i.e., (1.)) and information design (i.e., (2.)). The focus of this survey is on (3.).

We note that while the division into three basic incentive problems is fairly standard  and widely agreed upon, not all authors identify the three basic incentive problems with the terms mechanism design, information design, and contract design as we do here. We chose to adopt this terminology because it seems very natural from a computer science perspective (where mechanism design and information design/signaling are well established for (1.) and (2.), respectively), and {because contracts are the main object of study in (3.).}

\paragraph{Organization.} 

This survey is organized as follows. 
In Section~\ref{sec:model}, we introduce the basic principal-agent model. 
In Section~\ref{sec:opt-and-linear}, we present the optimal contract problem, and discuss properties of optimal contracts. 
Section~\ref{sec:linear} introduces linear (a.k.a. commission-based) contracts,  and studies the tradeoffs involved in choosing a simple rather than optimal contract from a worst-case approximation angle and a max-min optimality perspective. 
In Section~\ref{sec:comb-contracts}, we explore the computational complexity of finding optimal and near-optimal contracts in complex scenarios. In Section~\ref{sec:types} we study scenarios where agents have private types, and the goal is to construct contracts that incentivize agents to truthfully reveal their types, in addition to exerting effort.  
A modern algorithmic approach to contracts would not be complete without considering learning algorithms. 
In Section~\ref{sec:data-driven}, we consider data-driven contracts, while in Section~\ref{sec:incentive-aware}, we explore contracts and incentive-aware machine learning. 
Section~\ref{sec:ambiguous} explores incomplete, vague, and ambiguous contracts. In Section~\ref{sec:social-good}, we discuss contract design for social good. 
Afterwards, in Section~\ref{sec:beyond-contracts}, we discuss approaches ``beyond contracts,'' such as \emph{delegation} and \emph{scoring rule design}, that tackle related problems.
We mention several open problems and additional directions throughout the survey, and conclude with a discussion in Section~\ref{sec:discussion}.

\section{Basic Principal-Agent Model}
\label{sec:model}
We introduce the default model that we consider in this survey: the hidden-action principal-agent problem with discrete actions due to \citet{Holmstrom79,Ross73,Mirrlees99,GrossmanH83}, with the friction arising from limited liability rather than risk aversion (as in \citep{Innes90,Carroll15,DuttingRT19}). Our coverage of the basic model and properties of that model loosely follows \cite*{DuttingRT19}.

\paragraph{Setting.} 

In the basic principal-agent model,
a principal interacts with an agent. 
The agent has a set of actions $\mathcal{A}$ of size $n$. 
The action \emph{costs} for the agent are $0 \leq \cost_1 \leq \cdots \leq \cost_n$. 
There is a set of $m$ outcomes, with \emph{rewards} $0 \leq \rew_1\leq...\leq \rew_m$ for the principal.
The agent's action stochastically leads to an outcome based on a probability matrix $\Prob=\{\prob_{ij}\}_{i\in [n], j\in [m]}$, where $\prob_{ij}$ is the probability of getting outcome $j$ under action $i$.
So the $i$th row $\Prob_i$ is the distribution (probability mass function) over the rewards induced by action $i$. 
The matrix $\Prob$ is also known as the agent's \emph{technology}.
We use
\begin{equation}
\Rew_i := \mathbb{E}_{j\sim \Prob_i}[\rew_j] =\sum_{j\in [m]} \prob_{ij} \rew_j
\label{eq:expected-reward}
\end{equation}
to denote the expected reward of action $i$. 
The expected welfare from action $i$ is $\Wel_i := \Rew_i-\cost_i$, and the (overall) expected welfare 
of the contractual setting is $\Wel := \max_{i \in [n]} \Wel_i.$
Importantly, the action $i$ that the agent takes is \emph{hidden} from the principal, who only observes the stochastic outcome~$j$ and reward $r_j$ that result 
from the agent's choice of action. This incomplete information coupled with misalignment of interests (the principal enjoys the action's reward while the agent bears the cost) creates an incentive problem.

\paragraph{Contract.} 

A contract is a payment rule $\con$ that consists of $m$~non-negative payments or \emph{transfers} $(\pay_1,...,\pay_m)$, one for each outcome. Solving the principal-agent problem is by designing the contract~$\con$. 
The transfers are associated with outcomes rather than actions since the actions are hidden from the principal.
For action $i\in[n]$ let
\begin{equation}
\Pay_i := \mathbb{E}_{j \sim \Prob_i}[\pay_j] =\sum_{j\in [m]} \prob_{ij} \pay_j\label{eq:expected-transfer}
\end{equation}
denote the expected payment from principal to agent for taking action $i$.

Both the principal and the agent are assumed to be \emph{risk neutral}. For a fixed contract $\con$, the agent's expected utility under action $i$ is $U_A(i \mid \con) := \Pay_i - c_i$. 
The principal's expected utility (a.k.a.~\emph{revenue}) from action $i$ under contract $\con$ is $U_P(i \mid \con) := \Rew_i - \Pay_i$. 

Notice that the sum of the players' expected utilities is always equal to the expected welfare $\Wel_i = \Rew_i-\cost_i$ of the action $i$ chosen by the agent. 
The contract thus influences the agent's choice of the welfare ``pie'' (through his choice of action), in addition to determining how this pie is divided between the principal and the agent.

In addition to risk neutrality, we assume that all transfers are non-negative. This is a standard assumption, known as \emph{limited liability (LL)} of the agent. It reflects the asymmetric roles of the principal and the agent in contractual relations, and also serves to rule out trivial but unrealistic solutions to the contracting problem (see additional discussion below).

\begin{figure}[t]
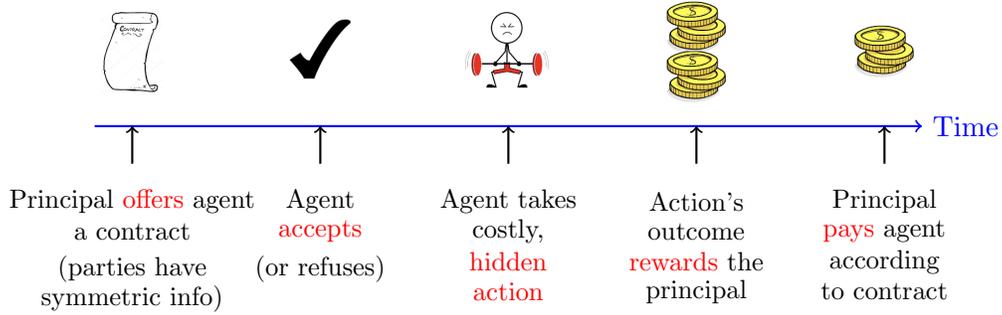

\vspace*{-10pt}
\centering
\include{figures/timing-contracts-classic.tex}
\vspace*{-20pt}
\caption{Timeline.}\label{fig:timeline}
\end{figure}

\paragraph{Best Response.}

Let us now consider the agent's rational behavior. 
When facing contract $\con$, the agent best responds by choosing an action $\istar$
that maximizes his expected utility. Let $\mathcal{A}^\star(\con) := \argmax_{i\in [n]} U_A(i \mid \con) \subseteq [n]$ denote the set of actions that maximize the agent's expected utility. Using this notation, the agent chooses an action
\begin{align}
\istar\in \mathcal{A}^\star(\con) = \argmax_{i\in [n]}\; U_A(i \mid \con) 
\label{eq:istar}
\end{align}
or no action ($\istar = \bot$), 
if the maximum expected utility from any action is negative. In the latter case, both players' utilities are zero.
Any such choice $\istar$ is \emph{incentive compatible} (IC) for the agent, because it is preferred over any other action. 
It is also  \emph{individually rational} (IR) for the agent, namely it ensures his expected utility is non-negative.

Fixing a contract $\con$ and denoting by $i^\star(\con)$ the agent's choice of action under contract $\con$, the agent's and principal's expected utility from contract $\con$ are $U_A(\con):=U_A(i^\star(\con) \mid \con)$ and $U_P(\con):=U_P(i^\star(\con) \mid \con)$, respectively. Note that the principal's expected utility depends on the agent's choice of action $i^\star(\con)$.
It is thus important to specify how the agent breaks ties. 
(An alternative, which we discuss below, is to assume that the principal, in addition to setting up payments, also recommends an action.)

By default, and as is standard in the contracts literature, we adopt the following tie-breaking rule, which is also known as the \emph{canonical tie-breaking rule}:%
\footnote{
This tie-breaking rule is justified by the fact that a small perturbation would make the agent strictly prefer that action (see, e.g., \citep{Carroll15,DuttingRT19} for additional discussion).} %
If there are multiple actions that maximize the agent's expected utility, then the agent breaks ties in favor of the principal by choosing an action that maximizes the principal's expected utility. (For completeness, in the case where there are multiple such actions, we assume that the agent breaks ties in favor of the highest index action.)
As we will argue formally below (in Proposition~\ref{prop:LP}), the canonical tie-breaking rule is without loss when the principal's objective is to maximize revenue.

In summary, we can view the contract design problem as a Stackelberg game, in which the principal moves first by defining the contract $\con$, and the agent responds with a utility maximizing action $i^\star(\con)$ (which, under the canonical tie-breaking rule, maximizes the principal's expected utility among all such actions). See Figure~\ref{fig:timeline}. 

\paragraph{Unifying IC and IR.}

A common approach in the literature, that we will also follow in this survey, is to fold the IR constraint into the IC constraint by assuming that there is a zero-cost action. 
Specifically, we will assume that the first action's cost is $c_1 = 0$, and that the expected reward of that action is $\Rew_1 \geq 0$.

\paragraph{An Example.} 

Consider the following example of a simple principal-agent setting, and the interaction between the principal and the agent in that setting. We will return to this example a few times in the following sections. 

\begin{example}[A simple principal-agent setting]
\label{ex:example-one-new}
Consider a principal-agent setting with three actions $i=1,2,3$  
with costs, rewards, and probabilities as specified in the following table: 
\begin{center}
\begin{tabular}{|l|ccc|c|}
\toprule
& $r_1 = 0$ & $r_2 = 1$ & $r_3 = 7$ & \text{cost}\\
\midrule 
action $1$: & $1$ & $0$ & $0$ & $c_1 = 0$\\
action $2$: & $0$ & $\nicefrac{1}{2}$ & $\nicefrac{1}{2}$ & $\cost_
2 = 1$ \\ 
action $3$: & $0$ & $\nicefrac{1}{6}$ & $\nicefrac{5}{6}$ & $\cost_3 = 2$\\
\bottomrule
\end{tabular}
\end{center}
The expected rewards corresponding to the three actions are $\Rew_1 = 0$, $\Rew_2 =\nicefrac{1}{2} \cdot 1 + \nicefrac{1}{2} \cdot 
7 = 4$, and $\Rew_3= \nicefrac{1}{6} \cdot 1 + \nicefrac{5}{6} \cdot 7 = 6$. Their expected welfares are 
$\Wel_1 = \Rew_1 - \cost_1 = 0$, $\Wel_2 =\Rew_2-\cost_2= 4 - 1 = 3$ and $\Wel_3=\Rew_3-\cost_3= 6 - 2 = 4$.
Consider the contract $\con = (0,1,3)$. 
The expected payment for action~$1$ under this contract is $\Pay_1 = 0$, for action $2$ it is $\Pay_2 =\nicefrac{1}{2}\cdot 1 + \nicefrac{1}{2} \cdot 3 = 2$, and for action $3$ it is 
$\Pay_3 = \nicefrac{1}{6} \cdot 1+\nicefrac{5}{6} \cdot 3=\nicefrac{8}{3}$.
The agent's expected utility is therefore maximimzed by action $2$, which yields an expected utility of $\Pay_2-\cost_2=2-1 = 1$, compared to an expected utility of $\Pay_1 - \cost_1 = 0$ for action $1$ and an expected utility of $\Pay_3-\cost_3 = \nicefrac{8}{3} - 2 = \nicefrac{2}{3}$ for action $3$. 
The principal's expected utility under this contract is $\Rew_2-\Pay_2 = 4 - 2 = 2$.
\end{example}

\subsection{Common Variations and Regularity Assumptions}
\label{sub:regularity}

\paragraph{Tie-Breaking Rule vs.~Recommended Action.} Instead of assuming a certain tie-breaking rule, it is also common to define a contract as a pair consisting of the actual payments $\con$ and a recommended action $i$. We then say that the contract  
$\langle \con,i \rangle$ is IC if action $i$ maximizes the agent's expected utility under $\con$. This approach is sometimes more convenient to work with, and we use it in a few places in this survey. (We will encounter it for a first time below, when we introduce the concept of $\varepsilon$-incentive compatibility.)

\paragraph{Limited Liability vs.~Risk Aversion.} 
The classic hidden-action principal-agent problem comes in two flavors: one models the agent as risk-neutral but adds limited liability (as we do here), the other models the agent as \emph{risk-averse}. 
The principal-agent problem with risk-aversion is well-studied in the economics literature \citep[e.g.,][]{Holmstrom79,Shavell79}. 
Risk-aversion captures the tendency 
to prefer certain outcomes over uncertain ones, and is typically modeled via a concave utility function. 

Both adding limited liability to a risk neutral approach, or adding risk-aversion to a model in which negative transfers are allowed, serve to rule out trivial but unrealistic solutions to the contracting problem, commonly referred to as ``selling the project to the agent.''
In this solution the principal sells the project to the agent, at a price equal to the maximum expected welfare $\Wel = \max_{i \in [n]} \Wel_i = \max_{i \in [n]} (\Rew_i - \cost_i)$, and the agent receives the reward from his actions. A risk-neutral agent would accept the principal's offer since his utility, on top of the negative utility of $-\Wel$ from buying the project, would be the expected reward $R_{i'}$ from any action $i'$ less the action's cost $c_{i'}$, so by choosing the welfare-maximizing action he would exactly break even.\footnote{Note how this solution of ``selling the project to the agent'' can be implemented within the principal-agent model through contract $\con$ that pays $t_j = - \Wel + r_j$ for each outcome $j \in [m]$.} This solution ``solves'' the problem by fully aligning the incentives of the agent and principal. 
However, it overlooks the inherent asymmetry between the two parties, particularly the fact that the principal is typically better suited to bear the risks due to her deeper pockets.

Given the pivotal role that risk-neutral models have played in the economics and computation community, we believe that the risk-neutral model is the natural starting point of an algorithmic theory of contracts.

\paragraph{Discrete Actions vs.~Continuum of Actions.} 

Another dimension in which principal-agent models differ from one another is whether they assume that the agent can choose from a discrete set of actions (as we do here), or from a continuum of actions. We believe that the discrete model is a more natural starting point for computer scientists, and indeed much of the work on algorithmic contract theory has focused on this version of the problem.

A common approach in the continuum model, in combination with the risk-averse agent assumption, is the so-called \emph{first-order approach} \citep{Mirrlees99,Rogerson85}. This approach replaces the requirement that the agent's choice of action is a global maximizer with the requirement that the agent's choice of action is a local optimum. The Mirrlees-Rogerson  condition 
states that MLRP plus CDFP (defined below) ensure that local optimality implies global optimality.

Very recent work of \cite{GeorgiadisRS24} goes one step further, by considering a model in which the agent can freely choose the outcome distribution.

\paragraph{$\varepsilon$-Incentive Compatibility.} 

The following relaxed notion of incentive compatibility mirrors the standard relaxation of IC in algorithmic mechanism design~\cite[e.g.][]{GonczarowskiW21} and equilibrium computation~\cite[e.g.][]{Papadimitriou06,Rubinstein18}. 
Given payments  
$\con$ and a small constant $\varepsilon\ge 0$, an action $i$ is \emph{$\varepsilon$-IC} (a.k.a.~an \emph{$\varepsilon$-best response}) for the agent if it is preferred over any other action up to an additive $\varepsilon$. That is, the agent loses no more
than $\varepsilon$ in expected utility by choosing action $i$:
\begin{equation}
    T_i - c_i \ge T_{i'}-c_{i'}-\varepsilon~~~\forall i'\ne i.\label{eq:approx-IC}
\end{equation}
The contract design problem can be relaxed by assuming the agent is willing to choose an $\varepsilon$-IC action. This assumption is considered especially reasonable in economic settings like ours, where a principal can suggest such an action to the agent. 
An \emph{$\varepsilon$-IC contract} is a pair 
$\langle \con,i \rangle$ such that given contract $\con$, action $i$ is an $\varepsilon$-IC action for the agent. 
Lemma~\ref{lem:approx-IC} in Section~\ref{sub:combi-outcomes} shows how to transform $\varepsilon$-IC to IC contracts while bounding the principal's expected utility loss.
The results of Section~\ref{sub:combi-outcomes} also demonstrate how the relaxation to $\varepsilon$-IC can provably simplify contract design problems and facilitate positive results.

\paragraph{Regularity Assumptions.}

It is quite common in the literature to impose additional structure on the distributions over outcomes, in the form of regularity assumptions. 
Probably the best-known such property is the \emph{monotone likelihood ratio property} (MLRP), which requires that for any two actions $i,i'$ such that $\cost_i < \cost_{i'}$ the likelihood ratio $\prob_{i',j}/\prob_{i,j}$ is increasing in $j$. The MLRP property ensures that the higher the observed outcome, the more likely it is that the agent exerted a higher effort level. A weaker requirement is \emph{first-order stochastic dominance} (FOSD), which requires that for any two actions $i,i'$ such that $\cost_i < \cost_{i'}$ it holds that $\sum_{\ell=j}^{m} \prob_{i',\ell} \geq \sum_{\ell=j}^{m} \prob_{i,\ell}$ for all $j$. That is, for all outcomes $j$, the higher the cost of an action the higher is the probability that the action leads to an outcome that is at least $j$. A proof that shows that MLRP implies FOSD (and that FOSD does \emph{not} imply MLRP) can be found in \citep[][p.104]{TadelisSegal05}.

In addition to MLRP and FOSD, there are other orthogonal (rather strong) regularity assumptions in the literature, for example the
following: An action $i$ satisfies the \emph{concavity of distribution function property} (CDFP) if for every two
actions such that $i$'s cost $c_i$ is a convex combination of their costs, it holds that $i$'s distribution over outcomes first-order stochastically dominates the corresponding convex combination of their distributions. 

\paragraph{Computational Model.}

A focus of this survey is on computational results, and proving or disproving the existence of \emph{efficient} algorithms. Generally speaking, an algorithm is efficient if its running time is upper-bounded by some polynomial function of the input size. 
This requires pinning down how we measure \emph{running time} and \emph{input size}. 
Per default we assume that input numbers are reals represented in binary, and denote by $k$ the maximum number of bits required to represent any number in the input. Hence the contracting problem with $n$ actions and $m$ outcomes can be specified with $O(nmk)$ bits.  In this case, we say that an algorithm is \emph{polynomial time} if it requires  $O(\mathsf{poly}(n,m,k))$ many basic operations.
For notational convenience, we usually omit the dependence on $k$ when talking about the running time of an algorithm. 
An alternative computational model assumes that each real number requires a single memory cell to be stored and that basic operations involving reals take a single step. In this model, an algorithm is polynomial time if it requires $O(\mathsf{poly}(n,m))$ many basic operations, independent of the numbers' magnitude. 
Informally, if the input contains a very large number, such as ${2^{2^n}}$, in the first computational model the algorithm is allowed to run in time $O(\mathsf{poly}(2^n,m))$, whereas in the second model only $O(\mathsf{poly}(n,m))$ time is allowed. 
An algorithm that is polynomial time according to both models is typically called a \emph{strongly} polynomial time algorithm, while an algorithm that is only polynomial time according to the first model is sometimes referred to as a \emph{weakly} polynomial time algorithm. 
As we shall see, some of the algorithms covered in this survey are not only polynomial time but also strongly polynomial time. Naturally, these definitions extend to any computational problem~\citep[see, e.g.,][]{Schrijver03}. 

\section{Optimal Contracts} 
\label{sec:opt-and-linear}
The principal's canonical design problem is to choose a contract $\con$ 
that maximizes her expected utility (a.k.a.~revenue), when the agent takes an action $\istar \in \mathcal{A}^\star(\con)$ 
that maximizes his expected utility. 
This is called the revenue-optimal or simply \emph{optimal} contract. 

While not our focus in this section, other design objectives for contracts exist. For example, the principal may be interested in maximizing welfare (see \citep{BalamcedaEtAl16} and the discussion in Section~\ref{sec:comb-contracts}), maximizing effort subject to a budget constraint (see \citep{SaigTR23,SaigET24}), or maintaining fairness (see \citep{FehrKS07}).

Our plan for this section is as follows. In Section~\ref{sec:lp-formulation}, we discuss a linear programming (LP) approach to (revenue-)optimal contracts. Afterwards, in Section~\ref{sec:implementability}, we present an important implication of the LP formulation, namely a characterization of actions that the principal can \emph{implement} (up to tie breaking) by setting up an appropriate contract. In Section~\ref{sec:optimal-contracts} we identify two special cases---binary action and binary outcome---in which optimal contracts take a simple form. We conclude our discussion of optimal contracts in Section~\ref{sub:shortcomings}, by pointing out some shortcomings of optimal contracts.

\subsection{An LP Approach to Optimal Contracts}
\label{sec:lp-formulation}

Our first result in this section is the following proposition, which is usually credited to \cite{GrossmanH83}. It states that the optimal contract can be found by solving $n$ \emph{linear programs} (LPs), one for each action.

\begin{proposition}[\cite{GrossmanH83}]\label{prop:LP}
    An optimal contract can be found by solving $n$ linear programs, one per action. Each linear program has $m$ variables and $n-1$ constraints. The output is a contract $\con^\star$ along with an action $i^\star \in \mathcal{A}^\star(\con^\star)$ that attains the maximum expected utility the principal can achieve. The choice of action $i^\star \in \mathcal{A}^\star(\con^\star)$ is compatible with the canonical tie-breaking rule.
\end{proposition}

To describe the LP approach, it will be convenient to distinguish between actions that the principal can implement up to tie-breaking, and the action that the agent chooses given a contract under a fixed tie-breaking rule. Formally, we say that an action $i \in [n]$ is \emph{implementable up to tie-breaking}, or simply that it is \emph{implementable}, if there exist a contract $\con$ such that
\[
U_A(i \mid \con) = \sum_{j \in [m]} \prob_{ij} \pay_j - \cost_i \geq U_A(i' \mid \con) = \sum_{j \in [m]} \prob_{i'j} \pay_j - c_{i'} \quad\quad \forall i' \neq i.
\]

The idea is now to formulate an LP for each action $i \in [n]$, that decides whether a given action is implementable (up to tie-breaking), and if it is finds the minimum expected payment required to implement action $i$. We refer to any contract with these properties (implements action $i$, has minimum expected payment), as a \emph{min-pay contract} for action $i$.

The primal LP for finding a min-pay contract for action $i$ and its dual are given in Figure~\ref{fig:primal-dual-lp}.
We refer to these as \textsf{MINPAY-LP($i$)} and \textsf{DUAL-MINPAY-LP($i$)}.
The variables of the primal LP are the payments $\{\pay_{j}\}$, and the constraints ensure that (i) the agent achieves a higher expected utility from action $i$ than from any other action $i' \neq i$ (IC constraints), and that (ii) transfers are non-negative (limited liability). 

\begin{figure}[h!]
\begin{center}
\begin{subfigure}[t]{0.4\textwidth}
\begin{align*}
\min_{\pay_j:\;j \in [m]}\quad &\sum_{j} \prob_{ij} \pay_j\\
\text{s.t.}\quad &\sum_{j} \prob_{ij} \pay_j - \cost_{i} \geq \sum_{j} \prob_{i'j} \pay_j - \cost_{i'} &&\forall i' \neq i  \\
     &t_j \geq 0 &&\forall j 
\end{align*}
\caption{\textsf{MINPAY-LP($i$)}}
\label{fig:primal-lp}
\end{subfigure}
~
~
\begin{subfigure}[t]{0.4\textwidth}
\begin{align*}
\max_{\lambda_{i'}:\; i' \in [n]\setminus\{i\}}\quad &\sum_{i' \neq i} \lambda_{i'}(\cost_i-\cost_{i'})\\
\text{s.t.}\quad &\sum_{i' \neq i} \lambda_{i'} (\prob_{ij} - \prob_{i'j}) \leq \prob_{ij} &&\forall j\\
     &\lambda_{i'} \geq 0 &&\forall i' \neq i
\end{align*}
\caption{\textsf{DUAL-MINPAY-LP($i$)}}
\label{fig:dual-lp}
\end{subfigure}
\end{center}
\vspace*{-8pt}
\caption{The \textsf{MINPAY-LP($i$)} for action $i$ \textbf{(left)} and its dual \textbf{(right)}. }\label{fig:primal-dual-lp}
\end{figure}

\begin{remark} 
Note that the first constraint in \textsf{MINPAY-LP($i$)} assumes that IR is implied by IC.
Without this assumption, we would have
to add an explicit non-negativity constraint. Namely, we would need to add the constraint $\sum_{j} \prob_{ij} \pay_j - \cost_{i} \geq 0$, requiring that the agent's expected utility from action $i$ is non-negative.
\end{remark} 

We are now ready to prove Proposition~\ref{prop:LP}.

\begin{proof}[Proof of Proposition~\ref{prop:LP}]
Consider the algorithm that (1) solves \textsf{MINPAY-LP($i$)} for each action $i \in [n]$ to determine whether action $i$ is implementable (up to tie-breaking), and for each such action determines a min-pay contract $\con^i$, and (2) returns the implementable action $i^\star$ and corresponding min-pay contract $\con^{i^\star}$ that maximizes the principal's expected utility (breaking ties in favor of the highest index if there are multiple such actions and contracts).

Observe that there is at least one action that can be implemented up to tie-breaking (any zero-cost action $i$, of which there is at least one, via $\con^i = (0,\ldots, 0)$); and that the principal's expected utility $U_P(i^\star \mid \con^{i^\star})$ from action $i^\star$ under contract $\con^{i^\star}$ is an upper bound on the principal's expected utility under any tie-breaking rule.

The proof is completed by noting that the choice of action $i^\star \in \mathcal{A}^\star(\con^{i^\star})$ is compatible with the canonical tie-breaking rule. Indeed, suppose by contradiction that under contract $\con^{i^\star}$ the agent would rather choose action $i' \neq i^\star$ because this yields a (strictly) higher principal utility. This would show that action $i'$ can be implemented via contract $t^{i^\star}$, and that $U_P(i' \mid t^{i^\star}) > U_P(i^\star \mid \con^{i^\star}) \geq U_P(i' \mid \con^{i'})$. However, this would imply that $\sum_{j} \prob_{i'j} t^{i^\star}_j < \sum_{j} \prob_{i'j} t^{i'}_j$, in  contradiction to $\con^{i'}$'s definition as a min-pay contract for action $i'$.
\end{proof}

The LP-based approach has a few immediate implications.

\paragraph{Computational Aspects.} 

A first implication is the existence of efficient algorithms for computing an optimal contract.
Specifically, since an optimal contract $\con$ can be found by solving $n$ instances of \textsf{MINPAY-LP($i$)} (one per action $i$) and \textsf{MINPAY-LP($i$)} can be solved in time polynomial in $n,m$ using standard LP algorithms, we obtain: 

\begin{observation}
\label{obs:poly-time-alg}
An optimal contract can be found in time polynomial in the number of actions $n$ and the number of outcomes $m$. 
\end{observation}

Standard LP methods are weakly polynomial time algorithms, and in fact it is a well-known open question whether linear programming in general admits a strongly polynomial time algorithm (this is known to hold for special cases)~\citep[][9th Problem]{Smale98}. Thus, Observation~\ref{obs:poly-time-alg} shows the existence of a weakly polynomial time algorithm for finding an optimal contract.  
Whether the problem of computing an optimal contract admits a strongly polynomial time algorithm (possibly under additional regularity assumptions) is an interesting open question. 

One powerful approach to solving LPs is the ellipsoid method. It can be utilized to solve, in polynomial time, LPs with polynomially-many constraints and exponentially-many variables, whenever there is a computationally-efficient ``separation oracle.'' We will use this approach in Section~\ref{sec:comb-contracts} to deal with large outcome spaces ($m$ is exponential in $n$).

\paragraph{Non-Zero Payments.} 

A second implication of the LP formulation is that there is always an optimal contract $\con$ with at most $n-1$ non-zero payments. This result is of particular importance when the outcome space is huge ($m \gg n$) (see Section~\ref{sub:combi-outcomes}).

\begin{observation}[e.g., \cite*{DuttingRT19}]
\label{obs:bfs}
In a contract setting with $n$ actions, there is an optimal contract with at most $n -1$ non-zero payments.
\end{observation}

The argument is as follows. The dual of \textsf{MINPAY-LP($i$)} is always feasible (e.g., the solution $\lambda_{i'}=0$ $\forall i'\ne i$ is feasible), and so  \textsf{MINPAY-LP($i$)} either has a bounded optimal solution or is infeasible.
Hence, if \textsf{MINPAY-LP($i$)} is feasible, then it is bounded and has an optimal \emph{basic feasible solution} with at most $n -1$ non-zero payments~\cite[e.g.][]{GartnerM06}. Since the optimal contract implements some action $i^\star$ at minimum expected payment, it is without loss of generality an optimal basic feasible solution to \textsf{MINPAY-LP($i^\star$)}.

\subsection{Characterization of Implementable Actions}
\label{sec:implementability}

As another important corollary of the LP formulation in Figure~\ref{fig:primal-dual-lp}, we obtain the following characterization of actions that can be implemented (up to tie-breaking) by the principal. 

\begin{proposition}[\cite{HermalinK91}, Proposition 2]
\label{prop:implementable}
Action $i$ is implementable (up to tie-breaking) if and only if there is no convex combination $\{\gamma_{i'}\}_{i' \neq i}$ of the actions other than $i$ that results in the same distribution over outcomes, i.e., $\sum_{i' \neq i} \gamma_{i'} \prob_{i'j} = \prob_{i,j}$ for all outcomes $j$, with lower weighted cost, i.e., $\sum_{i' \neq i} \gamma_{i'} \cost_{i'} < \cost_i$.
\end{proposition}

The necessity of the condition in Proposition~\ref{prop:implementable} follows by a standard argument. 
Indeed, if the condition is violated, then interpreting the convex combination that yields the same outcome distribution at lower cost as a mixed strategy, it is immediate that this mixed strategy gives the agent a higher expected utility than action $i$ under \emph{any} contract. In particular, for each possible contract, there must be an action $i' \neq i$ in the support of the mixed strategy, that yields a strictly higher expected utility than action $i$. So action $i$ cannot be implemented.

What is less obvious is the sufficiency of the condition in Proposition~\ref{prop:implementable}. 
To prove the second direction, we turn to LP-based considerations. For completeness we provide a full LP-based proof of both directions.

\begin{proof}[Proof of Proposition~\ref{prop:implementable}] (Proof adopted from \cite*{DuettingFP23}.)
Consider the \textsf{MINPAY-LP($i$)} for action $i$ with the objective $\min \sum_j \prob_{ij} \pay_j$ replaced with  $\min 0$ (primal LP, Figure~\ref{fig:mod-primal-lp})  and the dual to this LP (dual LP, Figure~\ref{fig:mod-dual-lp}).
\begin{figure}[ht]
~~
\begin{subfigure}[t]{0.4\textwidth}
\begin{align*}
\min_{\pay_j:\; j \in [m]}\quad &0\\
     &\sum_{j} \prob_{ij} \pay_j - \cost_{i} \geq \sum_{j} \prob_{i'j} \pay_j - \cost_{i'} &&\forall i' \neq i  \\
     &t_j \geq 0 &&\forall j 
\end{align*}
\caption{Primal LP}\label{fig:mod-primal-lp}
\end{subfigure}%
\hspace*{12pt}
\begin{subfigure}[t]{0.4\textwidth}
\begin{align*}
\max_{\lambda_{i'}:\; i' \in [n] \setminus\{i\}}\quad & \sum_{i' \neq i} \lambda_{i'} (\cost_i-\cost_{i'}) \\
& \sum_{i' \neq i} \lambda_{i'} (\prob_{ij} - \prob_{i'j}) \leq 0 && \forall~j\\
&\lambda_{i'} \geq 0 &&\forall~i' \neq i
\end{align*}
\caption{Dual LP}\label{fig:mod-dual-lp}
\end{subfigure}
\caption{\textsf{MINPAY-LP($i$)} for action $i$ with the objective $\min \sum_j \prob_{ij} \pay_j$ replaced with  $\min 0$ \textbf{(left)} and the dual to this LP \textbf{(right)}.}
\end{figure}

Action $i$ is implementable if and only if the primal LP is feasible. 
By strong duality \citep[e.g.,][]{GartnerM06}, for a general primal-dual pair one of the following four cases holds: 
\begin{enumerate}
\item[(1.)] The dual LP and the primal LP are both feasible. 
\item[(2.)] The dual LP is unbounded and the primal LP is infeasible. 
\item[(3.)] The dual LP is infeasible and the primal LP is unbounded. 
\item[(4.)] The dual LP and the primal LP are both infeasible.
\end{enumerate}
In our case the dual LP is always feasible (we can choose $\lambda_{i'} = 0$ for all $i' \neq i$). This rules out cases (3.)~and (4.). So in order to prove the claim it suffices to show that the dual LP is unbounded if and only if there exists a convex combination $\{\lambda_{i'}\}_{i'\neq i}$ of the actions other than $i$ that results in the same distribution over outcomes, i.e., $\sum_{i' \neq i} \gamma_{i'} \prob_{i'j} = \prob_{i,j}$ for all $j$, with lower weighted cost, i.e., $\sum_{i' \neq i} \gamma_{i'} \cost_{i'} < \cost_i$.

$\Longleftarrow$: We first show that if such a convex combination exists, then the dual LP is unbounded. Indeed, if such a convex combination exists, then it corresponds to a feasible solution to the dual LP because, for all $j$, 
\[
\sum_{i' \neq i} \gamma_{i'} \left(\prob_{ij} - \prob_{i'j}\right) = \left(\sum_{i' \neq i} \gamma_{i'} \right) \prob_{ij} - \left(\sum_{i' \neq i} \gamma_{i'} \prob_{i'j}\right) = \prob_{ij} - \left(\sum_{i' \neq i} \gamma_{i'} \prob_{i'j}\right) = 0,
\]
where we used that $\sum_{i' \neq i} \gamma_{i'} =1$ and that 
$\sum_{i' \neq i} \gamma_{i'} \prob_{i'j} = \prob_{ij}$ for all $j$.
Next observe that the objective value achieved by this feasible solution is 
\[\sum_{i' \neq i} \gamma_{i'} (\cost_i-\cost_{i'}) = \left(\sum_{i' \neq i} \gamma_{i'} \right) \cost_i - \left(\sum_{i' \neq i} \gamma_{i'} \cost_{i'}\right) = \cost_i - \left(\sum_{i' \neq i} \gamma_{i'} \cost_{i'}\right) = \delta  
\] 
for some $\delta > 0$. This is because $\sum_{i' \neq i} \gamma_{i'} = 1$, and because $\sum_{i' \neq i} \gamma_{i'} c_{i'} < c_i$. 
But then for any $\kappa \geq 0$ setting the dual variables to 
$\kappa \cdot \gamma_{i'}$ 
for $i' \neq i$ results in a feasible solution whose objective value is equal to 
$\kappa \cdot \delta$.
So the dual LP is unbounded.

$\Longrightarrow$: We next show that if the dual LP is unbounded, then a convex combination with the desired properties must exist. Since the dual LP is unbounded, for any $\delta > 0$ there must be a feasible solution to the dual LP,  $\lambda_{i'}$ for $i' \in [n]\setminus \{i\}$,  such that $\sum_{i' \neq i} \lambda_{i'} (\cost_i - \cost_{i'}) \geq \delta$ and $\sum_{i' \neq i} \lambda_{i'} (\prob_{ij} - \prob_{i'j}) \leq 0$ for all $j$. Note that we must have $\sum_{i' \neq i} \lambda_{i'} > 0$ (since $\delta > 0$). 
Now consider 
$\gamma_{i'} = \lambda_{i'}/(\sum_{i' \neq i} \lambda_{i'})$ 
for all $i' \neq i$. We claim that 
$\{\gamma_{i'}\}_{i' \neq i}$ presents a convex combination with the desired properties. First note that
$\{\gamma_{i'}
\}_{i' \neq i}$ 
is indeed a convex combination, i.e.,
$\gamma_{i'} \in [0,1]$ for all $i' \neq i$ and $\sum_{i' \neq i} \gamma_{i'} = 1$. 
Also note that, 
\[
\sum_{i' \neq i} \gamma_{i'} (\cost_i - \cost_{i'}) = \frac{1}{\sum_{i' \neq i}\lambda_{i'}} \sum_{i' \neq i} \lambda_{i'} (\cost_i - \cost_{i'}) \geq  \frac{1}{\sum_{i' \neq i}\lambda_{i'}} \cdot \delta > 0
\]
and therefore 
$\sum_{i' \neq i} \gamma_{i'} \cost_{i'} < (\sum_{i' \neq i} \gamma_{i'}) \cost_i = \cost_i$. 
Moreover, for all $j$, using the fact that $\sum_{i' \neq i} \lambda_{i'} \prob_{i'j} \geq (\sum_{i' \neq i} \lambda_{i'}) \prob_{ij}$, we must have
\[
\sum_{i' \neq i} \gamma_{i'} \prob_{i' j} = \frac{1}{\sum_{i' \neq i} \lambda_{i'}}\sum_{i' \neq i} \lambda_{i'} \prob_{i' j} \geq \frac{1}{\sum_{i' \neq i} \lambda_{i'}} \left(\sum_{i' \neq i} \lambda_{i'}\right) \prob_{i j} = \prob_{i j}.
\]
So we know that for all $j$, 
$\sum_{i' \neq i} \gamma_{i'} \prob_{i' j} \geq \prob_{i j}$. 
We claim that, for all $j$, this inequality must hold with equality. Indeed, assume for contradiction that for some $j'$ we have a strict inequality. By summing over all $j$, we then have
\begin{align}
\sum_{j} \left(\sum_{i' \neq i} \gamma_{i'} \prob_{i' j}\right) &> \sum_j \prob_{i j} = 1,
\label{eq:more-than-one}
\end{align}
where we used that $\Prob_i$ is a probability distribution over outcomes $j$. 
On the other hand, we have that
\begin{align}
\sum_{j} \left(\sum_{i' \neq i} \gamma_{i'} \prob_{i' j}\right) = \sum_{i' \neq i} \gamma_{i'} \left(\sum_{j}  \prob_{i' j}\right) = \sum_{i' \neq i} \gamma_{i'} = 1,
\label{eq:exactly-one}
\end{align}
where we used that the $\Prob_{i'}$'s are also probability distributions over outcomes $j$ and that
$\{\gamma_{i'}\}_{i'\neq i}$ is a convex combination of the actions other than $i$. Combining \eqref{eq:more-than-one} with \eqref{eq:exactly-one} we get the desired contradiction.
\end{proof}

\begin{remark}[Adopted from \cite*{DuttingRT19}, Proposition 3]
A slight tweak in the characterizing condition appearing in Proposition~\ref{prop:implementable} is that for action $i$ there is no convex combination of the actions other than $i$ with lower weighted cost which, rather than resulting in the exact same distribution over outcomes, results in a distribution that (weakly) \emph{dominates} it (in the first-order
stochastic domination sense). Since this is a stronger condition, less actions will satisfy it. It turns out that this condition precisely characterizes implementability by \emph{monotone} contracts where $\pay_1\le\dots\le \pay_n$, so that the agent is paid more for achieving a higher-reward outcome. We return to monotonicity in Section~\ref{sub:shortcomings}.
\end{remark}

\subsection{Optimal Contracts in Special Cases: Binary Action and Binary Outcome}
\label{sec:optimal-contracts}

We next discuss two important special cases of principal-agent problems in which optimal contracts have nice and interpretable forms. 

\paragraph{Optimal Contract with Binary Action.} 

We first consider the case in which the agent has two non-trivial actions. Formally, in a
\emph{generalized binary-action} principal-agent problem, (1) the first action has zero cost, and leads to a special zero-reward outcome (outcome $1$)  that no other action leads to, and (2) the agent has two additional non-trivial actions, action~$2$ and action~$3$, which correspond to ``low effort'' and ``high effort,'' respectively. We refer to this setting as having \emph{binary action} because the first action plays the role of an explicit ``outside option'' that gives the agent a utility of zero, which could also be an implicit requirement.\footnote{It is also possible to state Proposition~\ref{pro:opt-two-actions} for a setting with two actions, but then per our default assumption one of the two actions would have to have a cost of zero; limiting the generality of the result.}

In a generalized binary-action setting, if the optimal contract incentivizes action $1$ then it is the all-zero contract $\con = (0, \ldots, 0)$. If the optimal contract incentivizes action $i \in \{2,3\}$ then it pays only for one outcome, namely the outcome $j$ that maximizes the \emph{likelihood ratio} $\prob_{i j}/\prob_{i' j}$ where $i' \in \{2,3\} \setminus \{i\}$ is the other nontrivial action. This is cast in the following proposition.

In what follows, we refer to contracts that have a non-zero payment for at most one outcome as \emph{single-outcome payment} (SOP) 
contracts. 

\begin{proposition}[e.g., \citet{LaffontM09}, Chapter 4.5.1%
\footnote{The result in \cite{LaffontM09} is stated for risk-averse agents, but it holds under limited liability too, as the proof here shows. For settings with two actions, there is an alternative LP-based argument that uses Observation~\ref{obs:bfs} \cite[e.g.,][Proposition 5]{DuttingRT19}.}%
]
\label{pro:opt-two-actions} 
Consider a generalized binary-action principal-agent problem. If the optimal contract $\con^\star$ incentivizes a non-trivial action $i \in \{2,3\}$, then without loss it takes the following form.
Let $i'\in\{2,3\}\setminus \{i\}$ be the complementary non-trivial action, and let $j$ be an outcome that maximizes the \emph{likelihood ratio} $\prob_{i j}/\prob_{i' j}$ (interpreting $0/0$ as zero). Then $\pay^\star_{j'} = 0$ for all $j' \neq j$, while $\pay^\star_j \geq 0$ is the smallest payment such that 
\begin{equation}
\prob_{i j} \cdot \pay^\star_j - \cost_i \geq \max\{\prob_{i' j} \cdot \pay^\star_j - \cost_{i'}, 0\}.\label{eq:opt-two-payment}
\end{equation}
\end{proposition}

\begin{proof}
By the same argument as in the proof of Proposition~\ref{prop:LP} it suffices to show that for any non-trivial action $i \in \{2,3\}$ that is implementable (up to tie-breaking), there is a min-pay contract of the claimed form.
Towards this goal, assume that non-trivial action $i$ is implementable, and fix a max-likelihood outcome $j$. Suppose there is a contract $\con$ that implements action $i$, but that pays a non-zero amount for some outcome $j' \neq j$. 
We first show how to zero-out the payment on $j'$ and increase the payment on $j$, while keeping the same expected payment for action $i$ and (weakly) lowering the expected payment for the other actions.

The argument is as follows: By zeroing-out the payment on $j'$, the expected payment for action $i$ loses $t_{j'}\prob_{ij'}$, so we can pay an additional $t_{j'}\prob_{ij'}/\prob_{ij}$ upon outcome $j$ to exactly regain the loss. 
For action $i'$, the loss from this change is $t_{j'}\prob_{i'j'}$ and the gain is $t_{j'}\prob_{ij'}\prob_{i'j}/\prob_{ij}$. To show that the loss is at least the gain, i.e., 
$t_{j'}\prob_{i'j'} \ge t_{j'}\prob_{ij'}\prob_{i'j}/\prob_{ij}$, it suffices to show $\prob_{i'j'}/\prob_{i'j} \ge \prob_{ij'}/\prob_{ij}$, or equivalently by taking the inverse,
$$
\frac{\prob_{ij'}}{\prob_{i'j'}} \le 
\frac{\prob_{ij}}{\prob_{i'j}}.
$$
Since $j$ was chosen to maximize the right-hand side among all $j'$, we conclude that the inequality holds. Thus by shifting payment from $j'$ to $j$ at a rate of $\prob_{ij'}/\prob_{ij}$ as we have done, the expected payment of action $i'$ is (weakly) reduced. For action $1$ that leads deterministically to the first outcome, the claim follows from noting that $j$ cannot be the first outcome, so $\prob_{1j}=0 \leq \prob_{1j'}$. Thus, zeroing out the payment for $j'$ and increasing the payment for $j$ only lowers the agent's expected payment from action $1$. 

We have thus found a contract that implements action $i$ (up to tie-breaking) with weakly lower expected payment, and zero payment for outcome $j'$. By repeating the argument with other outcomes $j' \neq j$ as needed, we conclude that there is a contract that implements action $i$ with expected payment at most $\Pay_i$ that pays a (possibly) non-zero amount for outcome $j$ only. Since this holds for any contract $\con$ that implements action $i$, this shows that there is a min-pay contract $\con^\star$ for action $i$ of this form.

The proof is completed by noting that the minimum payment $\pay^\star_j \geq 0$ for outcome $j$ must satisfy Equation~\eqref{eq:opt-two-payment} in order to satisfy IC.
\end{proof}

Using Proposition~\ref{pro:opt-two-actions}, we can revisit Example~\ref{ex:example-one-new} and find the optimal contract. 

\medskip

\noindent\textbf{Example~\ref{ex:example-one-new}, revisited} (Optimal contract)\textbf{.}
Consider the principal-agent setting from Example~\ref{ex:example-one-new}.  The best (revenue-maximizing) contract for incentivizing action $1$ is $\con = (0,0,0)$. The principal's expected utility under this contract is $0$. 
In order to find the overall optimal contract, we apply Proposition~\ref{pro:opt-two-actions}: If the optimal contract incentivizes action $2$ or $3$, then without loss it pays only for the outcome that maximizes the likelihood ratio. 
Observe that the likelihood ratio is maximized on outcome $2$ for action $2$ (where it is $(\nicefrac{1}{2})/(\nicefrac{1}{6})$), and on outcome $3$ for action $3$ (where it is $(\nicefrac{5}{6})/(\nicefrac{1}{2})$). The candidate contract for action~$2$ can thus be found by letting $\pay_1 = \pay_3 = 0$, and solving for the smallest $\pay_2 \geq 0$ such that $\nicefrac{1}{2} \cdot \pay_2 - 1 \geq \max\{\nicefrac{1}{6} \cdot \pay_2 - 2,0\}$. This yields $\con = (0, 2, 0)$. The principal's expected utility is then $\Rew_2 - T_2 = 4 - \nicefrac{1}{2} \cdot 2 = 3$. Similarly, the candidate contract for incentivizing action $3$ can be found by letting $\pay_1 = \pay_2 = 0$, and solving for the smallest $\pay_3 \geq 0$ such that $\nicefrac{5}{6}\cdot \pay_3 - 2 \geq \max\{\nicefrac{1}{2} \cdot \pay_3 - 1,0\}$.
This yields $\con = (0,0,3)$.
The principal's expected utility is then $\Rew_3 - T_3 = 6 - \nicefrac{5}{6} \cdot 3 = \nicefrac{7}{2}$. 
The overall optimal contract is thus the one that incentivizes action $3$, where the principal's expected utility is $\nicefrac{7}{2}$.

\medskip

\begin{remark}[The connection between contract design and statistical inference]
\label{rem:inference}
The optimal contract in the generalized binary-action case highlights an interesting connection between optimal contract design and statistical inference. As discussed in \citet[Section 5.2.2]{Salanie17}, the intuitive connection is that the maximum likelihood ratio outcome is the ``strongest'' signal that the agent chose the desired action (and not some other action). As such, it makes sense for the principal to concentrate all payment on this outcome. 
Recently, \cite{SaigTR23,SaigET24} further formalize this connection, by showing a transformation from optimal \emph{hypothesis tests} to optimal contracts and vice versa, where a hypothesis test is optimal if it minimizes a combination of its type I and type II errors. 
Beyond the generalized binary-action case, the connection between contract design and statistical inference is diluted, but some of the intuition carries over. 
\end{remark}

\paragraph{Optimal Contract with Binary Outcome.} 

Another special case in which the optimal contract has a nice and interpretable form is the \emph{binary-outcome} case, where there are only two outcomes---``failure'' and ``success''---with rewards $r_1 = 0$ and $r_2 = r \geq 0.$ 
This outcome/reward structure captures important applications such as settings where the principal delegates the execution of a project to an agent, and the project can either succeed or fail. 

The optimal contract for this case turns out to be linear. A contract $\con = (t_1, \ldots, t_r)$ is said to be \emph{linear} (or commission-based) if there is a parameter $\alpha \in [0,1]$ such that $\pay_j = \alpha \cdot \rew_j$ for all $j \in [m]$. In other words, the principal transfers a fixed percentage of the rewards to the agent. Linear contracts are regarded as simple and frequently occur in practice. We will discuss linear contracts in more detail in the following sections. 

\begin{proposition}[e.g., \cite*{DuettingEFK21}]
\label{prop:linear-with-two-outcomes}
    In binary-outcome principal-agent problems, a linear contract is optimal.
\end{proposition}

\begin{proof}
For $r = 0$ the claim is trivially true ($\alpha = 0$ is optimal). So suppose $r > 0$.
Consider an arbitrary contract $\con = (t_1, t_2)$. We claim that then there is a linear contract, which yields a (weakly) higher principal utility than $\con$. To this end we will show that there is a contract $\con' = (0,t'_2)$ with this property. The proof is completed by observing that we can convert any such contract $\con'$ into an equivalent linear contract by letting $\alpha = t'_2/r.$

For our argument, it will be convenient to sort the actions by non-decreasing probability of success $\prob_{i2}$, so that $\prob_{12} \leq \prob_{22} \leq \ldots \prob_{n2}$. Let $i^\star$ be the action chosen under contract $\con$. Let's denote the expected payments of $\con$ and $\con'$ for action $i$ by $\Pay_i$ and $\Pay'_i$.

If $t_1 = 0$, then there is nothing to show ($\con$ already has the properties we want $\con'$ to have). So suppose $t_1 > 0$.   If $\prob_{i^\star 2} = 0$, then $\prob_{i^\star 1} = 1 - \prob_{i^\star 2} = 1$, and the principal's expected utility from $\con$ is $\Rew_{i^\star} - \Pay_{i^\star} = -t_1 < 0$, and we are better off with contract $\con' = (0, 0)$. So suppose $\prob_{i^\star 2} > 0$. Then we can choose $\con' = (0,t'_2)$ such that $t'_2 = \nicefrac{T_{i^\star}}{q_{i^\star 2}}$. We claim that, under contract $\con'$, the agent will choose an action $i' \geq i^\star$. This is because for action $i^\star$, 
\[
T'_{i^\star} = \prob_{i^\star  2} \cdot \frac{T_{i^\star}}{\prob_{i^\star 2}} = T_{i^\star},
\]
while for actions $i < i^\star$,
\[
T'_{i} = \prob_{i 2} \cdot \frac{T_{i^\star}}{\prob_{i^\star 2}} = \prob_{i 2} \cdot \frac{\prob_{i^\star 1} \cdot t_1 + \prob_{i^\star 2} \cdot t_2}{\prob_{i^\star 2}} \leq \prob_{i 1} \cdot t_1 + \prob_{i 2} \cdot t_2 = T_i,
\]
where the inequality holds because $\prob_{i 2} \cdot \prob_{i^\star 1} \leq \prob_{i^\star 2} \cdot \prob_{i 1}$.

This shows that the principal's expected utility under contract $\con'$ is at least 
\[
\prob_{i' 2} \cdot (r - t'_2) \geq \prob_{i^\star 2} \cdot (r - t'_2) = \prob_{i^\star 2} \cdot r - T_{i^\star} = \Rew_{i^\star} - T_{i^\star},
\]
where the inequality holds because $i' \geq i^\star$ and thus $\prob_{i' 2} \geq \prob_{i^\star 2}$, the first equality holds by definition of $t'_2$, and the final equality holds because $r_1 = 0$. Since the expected principal utility under $\con$ is $\Rew_{i^\star} - \Pay_{i^\star}$, this completes the proof. 
\end{proof}

\begin{remark}
\label{rem:zero-reward-not-wlog}
The assumption in Proposition~\ref{prop:linear-with-two-outcomes} that one of the two outcomes has reward zero is important. If there are two outcomes and both outcomes can have positive reward, then linear contracts may be suboptimal (see Example~\ref{ex:equal-rev}).
\end{remark}

\subsection{Shortcomings of Optimal Contracts}
\label{sub:shortcomings}

Beyond special cases, optimal contracts tend to be opaque and typically lack an intuitive interpretation. 
In addition, optimal contracts are known to exhibit a number of properties that run counter to economic intuition. A particularly important one is that optimal contracts generally fail to be \emph{monotone}. That is, it is  possible that in the optimal contract $\con$, a higher principal reward $\rew_j$ may entail a lower payment $\pay_j$. 
The next example gives a concrete setting where the optimal contract exhibits non-monotonicity.\footnote{Note that Example~\ref{ex:non-monotone} satisfies the regularity property of FOSD but not the more demanding property of MLRP (see Section~\ref{sub:regularity} for details). 
Proposition~\ref{pro:opt-two-actions} implies that even MLRP is insufficient to ensure monotonicity, even for just two actions. 
\cite{GrossmanH83}, working in a model with a risk-averse agent, show that MLRP together with CDFP implies monotonicity.}

\begin{example}[Non-monotone optimal contract]\label{ex:non-monotone}
Consider the principal-agent setting depicted in the following table: 
\begin{center}
\begin{tabular}{|l|cccc|c|}
\toprule
& $r_1 = 0$ & $r_2 = 3$ & $r_3 = 9$ & $r_4 = 12$ & \text{cost} \\
\midrule
action $1$ & $1$ & $0$ & $0$ & $0$ & $\cost_1 = 0$\\
action $2$: & $0$ & $\nicefrac{1}{3}$ & $0$ & $\nicefrac{2}{3}$ & $\cost_2 = 1$\\
action $3$: & $0$ & $0$ & $\nicefrac{1}{3}$  & $\nicefrac{2}{3}$ & $\cost_3 = 2$\\
\bottomrule
\end{tabular}
\end{center}

\noindent In this setting the unique optimal contract for action $i \in \{1,2,3\}$ pays just enough for outcome $i$ to cover the action's cost and nothing for the other two outcomes. The optimal contract is the best contract for incentivizing action $3$, which is $\con = (0,0,6,0)$. This contract is non-monotone as $\rew_3 < \rew_4$ but $\pay_3 > \pay_4$. In this example the non-monotonicity is caused by the fact that outcome $4$---the one with the highest reward---doesn't help differentiate between action $2$ and action $3$, and so it doesn't make sense for the principal to pay for that outcome. 
\end{example}

A possible economic interpretation of Example~\ref{ex:non-monotone} is that the agent is a salesperson, and the rewards corresponds to number of units sold. With no effort the agent sells nothing, with some effort the agent sells either $3$ units or $12$ units, and if the agent exerts maximum effort he sells $9$ or $12$ units. The counter-intuitive property of the optimal contract is that the best-possible outcome (i.e., selling $12$ units) does not warrant any payment.

In addition to demonstrating that optimal contracts may fail to be  monotone, Example~\ref{ex:non-monotone} also highlights a general challenge in contracts; namely, that outcomes serve a dual role: On the one hand they determine the principal's reward, on the other they serve as (imperfect) signals of which hidden action the agent chose to take. This creates a tension between incentivizing actions that lead to outcomes with high rewards, versus actions that are ``easy'' to incentivize.%
\footnote{The reader may find it useful to connect this to the \emph{welfare pie} analogy from Section~\ref{sec:model}. Properties of the outcome distribution determine both the size of the welfare pie, and how it can be split between the principal and the agent.} 

Another important critique of optimal contracts is that they require
perfect knowledge of the input, such as the distributions $\Prob$ and the costs $\costs$, and may be sensitive to slight perturbations (see  Section~\ref{sec:linear} for robust optimization approaches, and Section~\ref{sec:data-driven} for learning-based approaches).
Furthermore, the LP formulation can fail to capture structure in the contract setting, and so may be exponential in the natural representation size of the setting (see Section~\ref{sec:comb-contracts} for a range of succinctly-representable contract settings for which this is the case).

\section{Linear Contracts: Simplicity versus Optimality}
\label{sec:linear}
\label{sec:simple}
The complexity and shortcomings of optimal contracts motivate the study of ``simple'' contracts. Arguably the most prominent class of simple contracts are \emph{linear} (or commission-based) contracts. 
A linear contract is fully described by a single real number $\linear \in [0,1]$, with the interpretation that the payment $t_j$ is $\pay_j=\linear \cdot \rew_j \in [m]$ for every outcome $j$; i.e., the principal pays the agent an $\linear$-fraction of the obtained reward (see also Section~\ref{sec:optimal-contracts}). Consequently, the agent's and principal's expected utilities when the agent takes action~$i$ are $\alpha\Rew_i - c_i$ and $(1-\alpha)\Rew_i$, respectively. 
As part of their simplicity, we already note here the intrinsic robustness of linear contracts: The players' expected utilities do not depend on the details of the underlying distributions over outcomes. They just depend on the expected rewards $\{\Rew_i\}$ and the costs $\{\cost_i\}$. 

In Section~\ref{sec:geometric-approach} we present a geometric approach to linear contracts. Afterwards, in Section~\ref{sec:linear-properties}, we discuss some basic properties of linear contracts that follow from this geometric approach. We explore worst-case approximation guarantees of linear contracts in Section~\ref{sub:wc-approx}, and results that establish the robust optimality of linear contracts with (non-Bayesian) uncertainty about the principal-agent setting in Section~\ref{sec:robust}.

\subsection{The Geometry of Linear Contracts}
\label{sec:geometric-approach}%

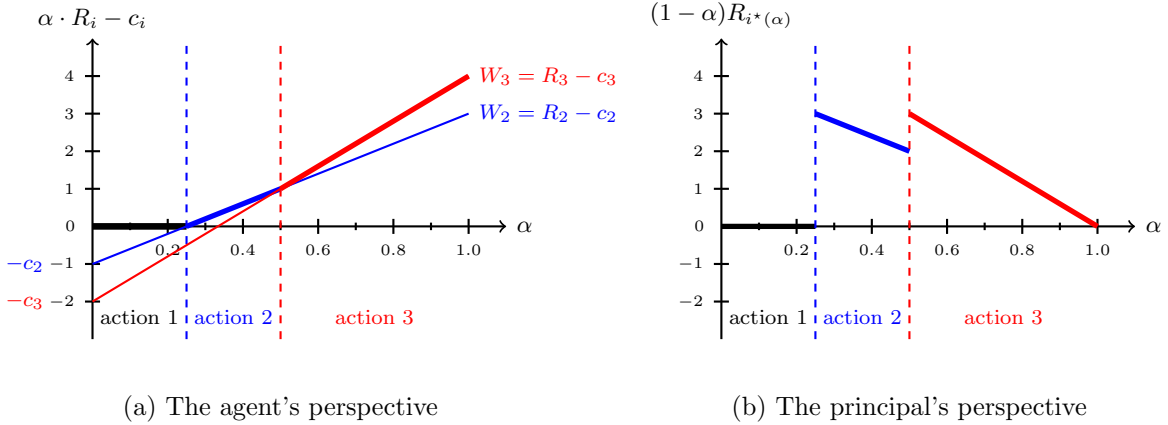
\begin{figure}
    \centering
    \begin{subfigure}[t]{0.4\textwidth}
    \centering
    \begin{tikzpicture}
    \useasboundingbox (0,-2) rectangle (5,3);
    \draw[thick,->] (0,-1.5) -- (0,2.5) node[above] {\footnotesize $\alpha \cdot \Rew_i - \cost_i$};
    \draw[thick,->] (0,0) node[left] {\tiny $0~$} -- (5.5,0) node[right] {\footnotesize $\alpha$};
    \draw[thick,-] (0.5,0.05) -- (0.5,-0.05); 
    \draw[thick,-] (1,0.1) -- (1,-0.1) node[below] {\tiny $0.2$}; 
    \draw[thick,-] (1.5,0.05) -- (1.5,-0.05);
    \draw[thick,-] (2,0.1) -- (2,-0.1) node[below] {\tiny $0.4$};
    \draw[thick,-] (2.5,0.05) -- (2.5,-0.05);
    \draw[thick,-] (3,0.1) -- (3,-0.1) node[below] {\tiny $0.6$};
    \draw[thick,-] (3.5,0.05) -- (3.5,-0.05);
    \draw[thick,-] (4,0.1) -- (4,-0.1) node[below] {\tiny $0.8$};
    \draw[thick,-] (4.5,0.05) -- (4.5,-0.05);
    \draw[thick,-] (5,0.1) -- (5,-0.1) node[below] {\tiny $1.0$};
    \draw[thick,-] (0.1,-1.0) -- (-0.1,-1.0) node[left] {\tiny $-2$};
    \draw[thick,-] (0.1,-0.5) -- (-0.1,-0.5) node[left] {\tiny $-1$};
    \draw[thick,-] (0.1,0.5) -- (-0.1,0.5) node[left] {\tiny $1$};
    \draw[thick,-] (0.1,1.0) -- (-0.1,1.0) node[left] {\tiny $2$};
    \draw[thick,-] (0.1,1.5) -- (-0.1,1.5) node[left] {\tiny $3$};
    \draw[thick,-] (0.1,2.0) -- (-0.1,2.0) node[left] {\tiny $4$};
    \draw[line width=2.5pt,black] (0,0) -- (1.25,0);
    \node at (0.625,-1.2) {\scriptsize \textcolor{black}{action $1$}};
    \draw[thick,blue,dashed] (1.25,-1.5) -- (1.25,2.5);
    \draw[thick,blue] (0,-0.5) -- (5,1.5) node[right] {\scriptsize{$\Wel_2 = \Rew_2 - \cost_2$}};
    \node at (-0.9,-0.5) {\textcolor{blue}{\scriptsize{$-\cost_2$}}};
    \draw[line width=2pt,blue] (1.25,0) -- (2.5,0.5);
    \node at (1.875,-1.2) {\scriptsize \textcolor{blue}{action $2$}};
    \draw[thick,red,dashed] (2.5,-1.5) -- (2.5,2.5);
    \draw[thick,red] (0,-1) -- (5,2) node[right] {\scriptsize{$\Wel_3 = \Rew_3 - \cost_3$}};
    \node at (-0.9,-1) {\textcolor{red}{\scriptsize{$-\cost_3$}}};
    \draw[line width=2pt,red] (2.5,0.5) -- (5,2);
    \node at (3.75,-1.2) {\scriptsize \textcolor{red}{action $3$}};
\end{tikzpicture}
    \caption{The agent's perspective}
    \label{fig:upper-envelope-agent}%
    \end{subfigure}
    \hspace{1.5cm}
    \begin{subfigure}[t]{0.4\textwidth}
    \centering
    \begin{tikzpicture}
    \useasboundingbox (0,-2) rectangle (5,3);
    \draw[thick,->] (0,-1.5) -- (0,2.5) node[above] {\footnotesize $(1-\alpha) \Rew_{i^\star(\alpha)}$};
    \draw[thick,->] (0,0) node[left] {\tiny $0~$} -- (5.5,0) node[right] {\footnotesize $\alpha$};
    \draw[thick,-] (0.5,0.05) -- (0.5,-0.05); 
    \draw[thick,-] (1,0.1) -- (1,-0.1) node[below] {\tiny $0.2$}; 
    \draw[thick,-] (1.5,0.05) -- (1.5,-0.05);
    \draw[thick,-] (2,0.1) -- (2,-0.1) node[below] {\tiny $0.4$};
    \draw[thick,-] (2.5,0.05) -- (2.5,-0.05);
    \draw[thick,-] (3,0.1) -- (3,-0.1) node[below] {\tiny $0.6$};
    \draw[thick,-] (3.5,0.05) -- (3.5,-0.05);
    \draw[thick,-] (4,0.1) -- (4,-0.1) node[below] {\tiny $0.8$};
    \draw[thick,-] (4.5,0.05) -- (4.5,-0.05);
    \draw[thick,-] (5,0.1) -- (5,-0.1) node[below] {\tiny $1.0$};
    \draw[thick,-] (0.1,-1.0) -- (-0.1,-1.0) node[left] {\tiny $-2$};
    \draw[thick,-] (0.1,-0.5) -- (-0.1,-0.5) node[left] {\tiny $-1$};
    \draw[thick,-] (0.1,0.5) -- (-0.1,0.5) node[left] {\tiny $1$};
    \draw[thick,-] (0.1,1.0) -- (-0.1,1.0) node[left] {\tiny $2$};
    \draw[thick,-] (0.1,1.5) -- (-0.1,1.5) node[left] {\tiny $3$};
    \draw[thick,-] (0.1,2.0) -- (-0.1,2.0) node[left] {\tiny $4$};
    \draw[black,line width=2pt] (0,0) -- (1.25,0);
    \node at (0.625,-1.2) {\scriptsize \textcolor{black}{action $1$}};
    \draw[thick,blue,dashed] (1.25,-1.5) -- (1.25,2.5);
    \node at (1.875,-1.2) {\scriptsize \textcolor{blue}{action $2$}};
    \draw[blue,line width=2pt] (1.25,1.5) -- (2.5,1);
    \draw[thick,red,dashed] (2.5,-1.5) -- (2.5,2.5);
    \draw[red,line width = 2pt] (2.5,1.5) -- (5,0);
    \node at (3.75,-1.2) {\scriptsize \textcolor{red}{action $3$}};
\end{tikzpicture}
    \caption{The principal's perspective}
    \label{fig:upper-envelope-principal}%
    \end{subfigure}
    \caption{The agent's expected utility as a function of the linear contract's parameter $\alpha$ {\bf (left)}, and the principal's expected  utility as a function of $\alpha$ {(\bf right)}, for the principal-agent setting in Example~\ref{ex:example-one-new}. 
    }
    \label{fig:upper-envelope}
\end{figure}

We follow \cite*{DuttingRT19}, and describe a geometric approach to linear contracts.  
This approach considers 
the agent's and principal's 
expected utilities as a function of the linear contract's parameter $\alpha$.
We start with the agent's perspective 
(see Figure~\ref{fig:upper-envelope-agent}). 
Intuitively, the more $\alpha$ is raised by the principal, the more the agent's utility is determined by how much reward is generated by his action rather than by his cost for the action. 
Thus, as $\alpha$ increases, the agent shall be more incentivized to take high-reward actions, even if they come at a higher cost.
To make this precise,
for every action $i$, let us plot the agent's expected utility 
$\alpha \Rew_i -\cost_i$ as a function of $\alpha$. 

Then, to figure out which action is incentivized by a linear contract with parameter $\alpha$, we can just check which line is highest at that $\alpha$. 
The agent's best response is thus given by the \emph{upper envelope} of the lines given by $\alpha \Rew_i - \cost_i$ (see thick lines in Figure~\ref{fig:upper-envelope-agent}).
Actions that do not appear on the upper envelope cannot be incentivized by a linear contract. 
Let us denote the number of actions on the upper envelope by $n' \leq n$. Now re-index the actions that appear on the upper envelope, in the order they appear (from left to right). 
Note that, after the re-indexing, the  actions 
will be sorted by increasing expected reward $\Rew_i$ (slope), by increasing cost $\cost_i$ (negative height at $\alpha = 0$), and by increasing expected welfare $\Wel_i = \Rew_i - \cost_i$ (height at $\alpha = 1$).\footnote{In principle, two actions that appear on the upper envelope could also have the same cost or same expected reward (or both). In any such pair of actions, however, one of the actions would yield a weakly lower principal utility for all $\alpha$ and would thus be dominated.}

A significant role is played by the values of $\alpha$ at which the segments of the upper envelope intersect. 
These points, are called \emph{indifference} points (or \emph{breakpoints} or \emph{critical} $\alpha$'s).  
For action $1 \leq i \leq n'$ on the upper envelope, the intersection point with the previous action on the upper envelope (or the $x$-axis for action $1$) is at $\alpha_i$, where
\begin{align}
\alpha_i := \begin{cases}
0 & \text{for $i = 1$,}\\
(\cost_i - \cost_{i-1})/(\Rew_i - \Rew_{i-1}) & \text{for $2 \leq i \leq [n']$,}\\
1 & \text{for $i =  n'+1$}.
\end{cases}
\label{eq:critical-point}
\end{align}

At $\alpha_i$ the agent is indifferent between action $i$ and action $i-1$. The agent's choice of action at these points is thus determined by the tie-breaking rule. Under the canonical tie-breaking rule, in which the agent chooses the action that is better for the principal, the agent favors action $i$ over action $i-1$.
The $[0,1]$ interval of possible $\alpha$'s is thus subdivided into $n'$ intervals, one for each action that can be incentivized. For every $i\in[n']$, action's $i$'s interval $[\alpha_i,\alpha_{i+1})$ includes all values of $\alpha$ for which the agent will choose to take this action. (For notational convenience, we let all intervals be half-open. The last interval is actually closed.)
 
Let us now take the principal's perspective (see Figure~\ref{fig:upper-envelope-principal}).
Intuitively, the principal prefers to lower the agent's share $\alpha$ as much as possible, while still incentivizing the agent to take a rewarding action. 
To make this precise, observe that 
for a given $\alpha \in [0,1]$, the principal's utility is $(1-\alpha)\Rew_{i^\star(\alpha)}$, where $i^\star(\alpha)$ is the agent's best response to $\alpha$.  In the interval $[\alpha_i,\alpha_{i+1})$ in which the agent is incentivized to take action $i$, the principal's expected utility is thus given by a line that starts at height $(1-\alpha_i)\Rew_i$ and that decreases throughout the entire interval with slope $-\Rew_i$. 
The best way for the principal to incentive the agent to take action $i$ is therefore via $\alpha_i$ --- the left endpoint of that action's interval $[\alpha_i,\alpha_{i+1})$. 

\subsection{Basic Properties of Linear Contracts}
\label{sec:linear-properties}

The geometric approach enables the following characterization of actions that are implementable by a linear contract. It also implies that linear contracts are monotone in a number of ways. Namely, as we increase a linear contract's parameter $\alpha$, the agent's best-response action will have a weakly higher cost, expected reward, and expected welfare.

\begin{proposition}[Implementability and monotonicity, \citet*{DuttingRT19}]
\label{prop:implementable-linear}
The actions that can be implemented by a linear contract are precisely those that appear on the upper envelope.
Under the canonical tie-breaking rule, considering the implementable actions in increasing order of the minimum $\alpha$ that implements them, it holds that:
(1) The costs $\{c_i\}$, expected rewards $\{R_i\}$, and expected welfares $\{R_i-c_i\}$ are  increasing in $i$. 
(2) Action $i$ is implemented by linear contracts with $\alpha\in[\alpha_{i},\alpha_{i+1})$ where $\alpha_i$ is defined as in Equation~\eqref{eq:critical-point}. 
\end{proposition}

The geometric analysis also implies that we can compute the optimal linear contract in (strongly) polynomial time. This can be done (na\"ively) by building the upper envelope, enumerating over the critical $\alpha$'s, and finding the one that maximizes the principal's expected revenue. 

The following example illustrates this. 
It also shows that linear contracts might be \emph{suboptimal}; their suboptimality is further quantified in Section~\ref{sub:wc-approx}. 

\medskip

\noindent\textbf{Example~\ref{ex:example-one-new}, revisited} (Optimal linear contract and suboptimality)\textbf{.} 
Consider the principal-agent setting from Example~\ref{ex:example-one-new}. The three actions correspond to lines $\alpha \Rew_1 - \cost_1= 0$, $\alpha \Rew_2 - \cost_2 = \alpha \cdot 4 - 1$, and $\alpha \Rew_3 - \cost_3 = \alpha \cdot 6 - 2$ for $\alpha \in [0,1]$. See Figure~\ref{fig:upper-envelope}. For $\alpha \in [0,\nicefrac{1}{4})$ the agent prefers action $1$, for $\alpha \in [\nicefrac{1}{4},\nicefrac{1}{2})$ the agent prefers action $2$, and for $\alpha \in [\nicefrac{1}{2},1]$ the agent prefers action $3$. The critical $\alpha$'s are $\alpha_1 = 0$ for action $1$, $\alpha_2 = \nicefrac{1}{4}$ for action $2$, and $\alpha_3 = \nicefrac{1}{2}$ for action $3$. The resulting principal's expected utility for action $1$ is $0$, while for action $2$ it is $(1-\nicefrac{1}{4}) \cdot 4 = 3$, and for action $3$ it is $(1-\nicefrac{1}{2}) \cdot 6 = 3$. Recall that the best-possible expected utility that the principal can achieve in this setting with a general contract is $\nicefrac{7}{2} > 3$. 

\begin{remark}
Recall that Proposition~\ref{prop:linear-with-two-outcomes} identified the binary-outcome case as an important special case in which linear contracts are optimal. We will discuss combinatorial versions of the binary-outcome case in Section~\ref{sec:comb-contracts}.
\end{remark}
\subsection{Worst-Case Approximation Guarantees}
\label{sub:wc-approx}

A major contribution of computer science to economics is the study of worst-case approximation guarantees of simple mechanisms relative to the optimal mechanism (e.g., \citep{HartlineR09}). Denote by $\mathsf{ALG}(I)$ the performance of a simple mechanism on instance $I$, and by $\mathsf{OPT}(I)$ the performance of the optimal mechanism on the same instance. For a maximization problem, the goal is a guarantee of the form 
$\rho \cdot \mathsf{ALG}(I) \geq \mathsf{OPT}(I)$ for all $I$. Here $\rho 
\geq 1$ is the approximation guarantee, and the closer it is to $1$ the better. 

In the context of contracts, a natural performance measure is the principal's expected utility (a.k.a.~revenue). 
\citet{DuttingRT19} explore the worst-case gap between the revenue achievable with a linear contract and that achievable with an optimal contract, and give (asymptotically) tight approximation guarantees in all natural parameters of the problem (see Theorem~\ref{thm:wc-approx} and Table~\ref{table:approx-guarantees}).\footnote{Below we discuss
work by \citet{BalamcedaEtAl16}, which bounds the gap between the optimal welfare and the welfare achievable with a revenue-maximizing contract and/or a revenue-maximizing linear contract.}
These bounds show that the gap $\rho$ \emph{can} be large, but also that the gap \emph{is} indeed large only when the instance is rather pathological.

\begin{table}[t]
\begin{center}
\begin{tabular}{|l|c|c|c|c|}
\toprule
& \# actions & spread of rewards & spread of costs & \# outcomes\\
\midrule
approx.~ratio: & $n$ & $\Theta(\log H)$ & $\Theta(\log C)$ & unbounded ($m \geq 3$)\\
\bottomrule
\end{tabular}
\end{center}
\caption{Approximation guarantees of linear contracts (as shown in \citep{DuttingRT19}), comparing the revenue achievable with a linear contract to that of an optimal contract. 
Here $H$ denotes the ratio between the highest and lowest expected reward of an action, while $C$ is the ratio between the highest and lowest cost of an action.
The upper bounds that appear in this table apply to the potentially wider gap between the optimal welfare and the revenue achievable with a linear contract. The lower bounds that appear in this table compare the revenue achievable with any contract to that achievable with a linear contract, and apply even when the contract setting is required to satisfy MLRP. 
The worst-case approximation ratio for $m = 2$ outcomes is still (partially) open: 
If one of the outcomes has reward zero then a linear contract is optimal, i.e., the approximation ratio is 1 (Proposition~\ref{prop:linear-with-two-outcomes}). If both outcomes have positive rewards then Example~\ref{ex:equal-rev} with $n=m=2$ actions and outcomes shows that the approximation ratio is at least $2$; it is unknown whether the approximation ratio of~$2$ is tight.
}
\label{table:approx-guarantees}
\end{table}

\begin{theorem}[\citet*{DuttingRT19}] \label{thm:wc-approx} 
Let $\rho$ denote the worst-case ratio between the principal's expected utility under the optimal contract, and the principal's expected utility under the optimal linear contract. Then: 
(1) Among all principal-agent settings with $n$ actions, $\rho = n$. 
(2) Among all principal-agent settings where the ratio between highest and lowest expected reward is $H$, $\rho = \Theta(\log H)$. 
(3) Among all principal-agent settings where the ratio between highest and lowest cost is $C$, $\rho = \Theta(\log C)$. 
(4) Among all principal-agent settings with $m \geq 3$ outcomes, $\rho$ can be arbitrarily large relative to $m$. 
\end{theorem}

The upper bounds in the above theorem
hold even against the strongest-possible benchmark, the optimal welfare, rather than merely the optimal principal utility. Moreover, the lower bounds apply even if one insists on the setting satisfying the regularity assumption of MLRP (as defined in Section~\ref{sec:model}).
Let us take a closer look at the proof of this result for the parameter $n$---the number of actions.  

\paragraph{Proof of Upper Bound.}

We first sketch how to establish the upper bound on the approximation guarantee $\rho$ in terms of the number of actions $n$. A common approach for showing an upper bound on the approximation guarantee is to show an upper bound on $\mathsf{OPT}$ and a lower bound on $\mathsf{ALG}$. In our case, $\mathsf{OPT}$ and $\mathsf{ALG}$ correspond to the maximum expected utility the principal can achieve with a general contract and a linear contract, respectively. 

First observe that because of IR, the payment that the principal needs to make to incentivize the agent to take any action $i$ is at least $\cost_i$. So the maximum expected utility the principal can extract from any action $i$ is at most $\Wel_i = \Rew_i - \cost_i$. This shows that $\mathsf{OPT} \leq \Wel := \max_{i \in [n]}(\Rew_i - \cost_i)$. 
To show a lower bound on $\mathsf{ALG}$, we will rely on the geometric approach to linear contracts developed in Section~\ref{sec:geometric-approach}. Following this approach, let us re-index the actions in the order in which they appear on the upper envelope from left to right by $1, \ldots, n'$ for $n' \leq n$ (see Figure~\ref{fig:upper-envelope-agent}). Recall that then the actions $i \in [n']$ are sorted by increasing cost $\cost_i$, expected reward $\Rew_i$, and expected welfare $\Rew_i - \cost_i$ (Proposition~\ref{prop:implementable-linear}).
In particular, we have $\Wel = \Rew_{n'}-\cost_{n'}$.  On the other hand, for any action $i$ that appears on the upper envelope, the best way to incentivize it with a linear contract is to choose the smallest $\alpha$ for which this action is on the upper envelope (see Figure~\ref{fig:upper-envelope-principal}). Recall that we denote this value of $\alpha$ by $\alpha_{i}$, and that $\alpha_i$ is determined by solving $\alpha \Rew_i - \cost_i = \alpha \Rew_{i-1} - \cost_{i-1}$, where we let $R_0 = 0$ and $c_0 = 0$ (see Equation~\eqref{eq:critical-point}).  Hence, the principal's expected utility from using a linear contract is $\mathsf{ALG} = \max_{1 \leq i \leq n'} (1-\alpha_{i})\Rew_i$. 

The key observation of the upper bound proof of \cite{DuttingRT19} 
is now that for any action~$i$ that appears on the upper envelope, \begin{align}
\Rew_i - \cost_i \leq \sum_{i'=1}^{i} (1-\alpha_{i'}) \Rew_{i'}.\label{eq:telescope}
\end{align}
This observation follows from the inequality $(\Rew_i - \cost_i) - (\Rew_{i-1} - \cost_{i-1}) \leq (1-\alpha_{i}) \Rew_i$, summed up telescopically. The inequality reflects that since the agent is indifferent among actions $i-1$ and $i$ at $\alpha_{i}$, then the increase in expected welfare by switching from $i-1$ to $i$ (the left-hand side of the inequality) should go entirely as revenue to the principal (the right-hand side of the inequality). To see that this inequality holds note that
\[
(1-\alpha_i) R_i = \left(1-\frac{c_i-c_{i-1}}{R_i-R_{i-1}}\right)R_i = \frac{(R_i-R_{i-1})-(c_i-c_{i-1})}{R_i-R_{i-1}} R_i \geq (R_i-c_{i})-(R_{i-1}-c_{i-1}),
\]
where we used Equation~\eqref{eq:critical-point} and that $R_i/(R_i-R_{i-1}) \geq 1$.

So, in particular, by applying Equation~\eqref{eq:telescope}  to action $i=n'$, we get
$$\mathsf{OPT} \leq \Wel = \Rew_{n'} - \cost_{n'} \leq \sum_{i'=1}^{n'} (1-\alpha_{i'})\Rew_{i'} \leq n' \cdot \max_{1 \leq i \leq n'} (1-\alpha_{i'})\Rew_{i'} = n' \cdot \textsf{ALG}.$$ 
Since $n' \leq n$ we conclude that the gap between the revenue achieved by the best contract and the revenue achieved by the best linear contract---and in fact the potentially larger gap between the revenue of the best linear contract and the optimal welfare,
typically referred to as \emph{first best} in economics---is at most $n$.

\paragraph{Proof of Lower Bound.}

Complementing the upper bound $\rho\le n$, the gap between the best linear contract and the best overall contract is shown to be at least $n$ in the worst case. \citet{DuttingRT19} show this by introducing the following worst-case instance, in which no matter what action the principal incentivizes the agent to take through a linear contract, her expected revenue remains the same, namely $1$. At the same time, the example is such that the expected welfare of each action $i$ is $\Wel_i = \Rew_i - \cost_i \approx i$, and, with a general contract, it is possible to incentivize each action $i$ with an expected payment of $\Pay_i =  c_i$. So the maximum expected utility the principal can achieve with a general contract is $\Wel_n \approx n$.   
Since introduced, this ``equal revenue'' principal-agent setting has proven useful as a benchmark instance for contract design, similarly to the equal revenue distribution being a benchmark instance for Bayesian mechanism design \citep[e.g.,][]{HartlineR09}.

\begin{example}[Equal revenue contract setting, \citet*{DuttingRT19}]
\label{ex:equal-rev}
Consider a setting with $n$ actions. 
The important features of the setting are summarized by the actions' costs and expected rewards. For concreteness and to allow the optimal contract to extract the full welfare as the principal's revenue, let there be $m=n$ outcomes, and let action $i$ lead to outcome $i$ with certainty for every $i\in[n]$ (i.e., $\prob_{i,i} = 1$).%
\footnote{The setting is ``full information'' in the sense that upon observing the outcome, the principal has full knowledge of the agent's action, and can pay directly for action $i$ by paying for outcome $i$. She can thus simply tell the agent to take her preferred action $i$ by paying its cost $c_i$, while paying zero for all other actions.} 
Let $\epsilon > 0$. The rewards and costs are defined as follows:
\begin{eqnarray}
\Rew_i = 1/\epsilon^{i-1}; & \cost_i = \Rew_i - i + \epsilon(i-1); & \Wel_i = \Rew_i - \cost_i = i - \epsilon(i-1),
\end{eqnarray}
where $\Wel_i$ is the welfare from action $i$.%
\footnote{In this example, because action $i$ leads to outcome $i$ with certainty then $r_i=R_i$. In particular $r_1=1$; the example can be easily ``normalized'' by adding an outcome with zero reward.}
\end{example}

The equal revenue contract setting has actions with exponentially growing expected rewards and costs. The optimal contract can incentivize action $n$ with an expected payment equal to the agent's cost, achieving the principal expected utility of $\Wel_n = n - \epsilon(n-1)\approx n$.
On the other hand, by analyzing the upper envelope as in Section~\ref{sec:geometric-approach}, one can show that with a linear contract the principal can achieve expected utility at most $1$ from any of the actions. To see this, note that actions are already sorted as required, and that each action $i$ appears on the upper envelope. Moreover, for each action $i$, the smallest $\alpha$ that incentivizes action $i$ is $\alpha_1 = 0$ for action $1$ and
\[
\alpha_i = \frac{\cost_i-\cost_{i-1}}{\Rew_{i}-\Rew_{i-1}}  
= 1 - \frac{1}{\Rew_i}.
\]
for action $i > 1$. So, for all actions $i \in [n]$, the maximum utility the principal can extract from action $i$ through a linear contract is $(1-\alpha_i) \Rew_i = 1$.
We conclude that, for $\epsilon \rightarrow 0$, the gap between best contract and the best linear contract goes to $n$.

\begin{table}[t]
\begin{center}
\begin{tabular}{|l|cccc|}
\toprule
& Welfare & Opt.~contract & Monotone & Linear\\
\midrule
Opt.~contract & $n$ & $1$ & -- & --\\
Monotone & $n$ & $\Theta(n)$ & 1 & -- \\
Linear & $n$ & $n$ & $n$ & $1$ \\ 
\bottomrule
\end{tabular}
\end{center}
\caption{Gaps between the revenue achieved by different classes of contracts (rows), with respect to different benchmarks (columns). All upper bounds of $n$ follow from the approximation guarantee of linear contracts shown with respect to welfare shown in \citep{DuttingRT19}. The lower bounds of $n$ for linear contracts follow from the lower bound construction of \cite{DuttingRT19} discussed in the main text, because in that example the optimal welfare and the revenue of the best contract coincide, and the best contract is monotone. The lower bound that compares optimal contract and monotone contract also appears in \citep{DuttingRT19}. The lower bound of $n$ on the gap between welfare and best contract (and hence monotone and linear contract) is shown in \citep{DuttingRT21}. For the lower bounds that apply to optimal contracts and linear contracts also see \citep{BalamcedaEtAl16}.}
\label{table:classes-of-contracts}
\end{table}

\paragraph{Additional Gaps.} 

The previous analysis implies that the more general class of monotone contracts provides at least a factor $n$ approximation to the optimal contract, and that the best contract provides at least a factor $n$ approximation to the optimal welfare. 

One might wonder if either of these gaps could be improved by using a more sophisticated (monotone) contract. That is, does the class of monotone contracts provide a better approximation guarantee than the class of linear contracts? Are there really instances where the gap between the revenue of any contract and the welfare is of order $\Omega(n)$?

An answer to the first question can be found in \citep{DuttingRT19}, which provides a construction in which the gap between the revenue of the optimal contract and the best monotone contract is at least $n-1$. For the latter question, consider Example~\ref{ex:first-betst-second-best} 
which appears in \citep{DuttingRT21}. In this contract setting the gap between the optimal welfare and the revenue of \emph{any} 
contract is at least $n$, thus showing a gap between revenue and welfare in contract design. 
See Table~\ref{table:classes-of-contracts} for a summary of the gaps between different classes of contracts and different benchmarks.

\begin{example}[First best vs.~second best, \citet*{DuttingRT21}]\label{ex:first-betst-second-best}
Consider the following instance with $n$ actions, two outcomes and $\gamma \in (0,1), \gamma\to 0$: \begin{table}[h!]
    \centering
    \begin{tabular}{|l|c|c|c|c|}
    \toprule
        \rule{0pt}{2ex}  \hspace{1.0mm} \hspace{1.0mm}  & \hspace{3.0mm} $r_1 = 0$ \hspace{3.0mm} & \hspace{3.0mm} $r_2 = \frac{1}{\gamma^{n-1}}$ \hspace{3.0mm} & \hspace{3.0mm} cost \hspace{3.0mm} \\ \hline
        \rule{0pt}{2ex} action $i \in [n]$: & $(1-\gamma^{n-i}$) & $\gamma^{n-i}$ & $c_i = \frac{1}{\gamma^{i-1}}-i+(i-1)\gamma$ \\ 
    \bottomrule
    \end{tabular}
\end{table}

\noindent In this instance, the action with the highest welfare is action $n$, with welfare $\approx n$. Indeed, the welfare of action $i$ is, $\Wel_i = \gamma^{n-i} \frac{1}{\gamma^{n-1}}-(\frac{1}{\gamma^{i-1}}-i+(i-1)\gamma) = i - (i-1)\gamma \approx i$ for $\gamma \rightarrow 0$.
We next show that the best revenue the principal can get is $\approx 1$. First note that $\Rew_1=1$.
For $i \geq 2$, we obtain a lower bound on the expected payment required to incentivize action $i$, by only considering the incentive constraint that compares action $i$ to action $i-1$. That is, we want to find $\con = (\pay_1,\pay_2)$ that minimizes $\Pay_i=\prob_{i,1} \pay_1 + \prob_{i,2} \pay_2$ subject to
\[
\Pay_i-\cost_i=\prob_{i,1} \pay_1 + \prob_{i,2} \pay_2 - \cost_i \geq \prob_{i-1,1} \pay_1 + \prob_{i-1,2} \pay_2 - \cost_{i-1}=\Pay_{i-1}-\cost_{i-1}.
\]
First note that the likelihood ratio of outcome 2 exceeds that of outcome 1, namely: $\frac{\prob_{i,2}}{\prob_{i-1,2}} \geq \frac{\prob_{i,1}}{\prob_{i-1,1}}$. 
This follows simply by plugging in the $q_{i,j}$'s and using $\gamma \leq 1$. 
It follows that in order to minimize $T_i$, we can set $t_1 = 0$, and find the smallest $t_2$ such that $\prob_{i,2} \pay_2 - \cost_i \geq \prob_{i-1,2} \pay_2 - \cost_{i-1}$. Plugging in the probabilities and the costs, we obtain 
\[
\gamma^{n-i} t_2 - \left(\frac{1}{\gamma^{i-1}} - i +(i-1)\gamma\right) \geq \gamma^{n-i+1} t_2 - \left(\frac{1}{\gamma^{i-2}} - (i-1) +(i-2)\gamma\right).
\]
Rearranging, we get
\[
\gamma^{n-i} t_2 \geq \frac{1}{\gamma^{i-1}}-1,
\]
which shows that the revenue from action $i$ is at most
\[
\Rew_i - \gamma^{n-i} \pay_2 \leq  \frac{1}{\gamma^{i-1}} - \left(\frac{1}{\gamma^{i-1}} -1\right)= 1.
\]
\end{example}

\begin{remark}
In Example~\ref{ex:first-betst-second-best} there are two outcomes, and one of the two outcomes has a reward of zero. So, by Proposition~\ref{prop:linear-with-two-outcomes}, a linear contract is optimal. The example thus presents another setting, in which the gap between the optimal welfare and the utility from the best linear contract is at least $n$. The added value of  Example~\ref{ex:equal-rev} is that it shows that the same worst-case gap of $n$ occurs when the benchmark is the (potentially smaller) optimal revenue.
\end{remark}

\paragraph{Further Work.}
\citet*{BalamcedaEtAl16} also take a worst-case approximation approach to contracts, but focus on welfare rather than revenue. They bound the worst-case gap between optimal welfare (``first best'' welfare), and the welfare achieved by a revenue-maximizing contract (``second best'' welfare). 
This quantifies the loss in welfare from the agency relation, caused by the principal choosing a utility-maximizing contract.
They also bound the gap between the optimal welfare, and the best welfare achieved by a revenue-maximizing linear contract. Their bounds hold under the assumptions of MLRP (for the first gap) and FOSD (for the second gap), in combination with additional assumptions (see their paper for details). In the worst case, both gaps are of order $\Theta(n)$ where $n$ is the number of actions.

\subsection{Robust (Max-Min) Optimality}
\label{sec:robust}

A central challenge in the economic literature on contracts is to find formal justifications for simple contracts \cite[e.g.,][]{milgrom-holmstrom87}.
An important recent line of work has established that simple---in particular linear---contracts are robustly optimal under (non-Bayesian) uncertainty. The high-level approach in this line of work is to assume that certain aspects of the problem instance are uncertain (while other aspects remain known to the principal). 
It is then shown that linear contracts maximize the principal's minimum expected utility, where the minimum is taken over all instances that are compatible with the principal's restricted knowledge about the setting. 
We present two results of this form: A canonical result by \citet{Carroll15} on robustness to action uncertainty (Section~\ref{sub:robustness-carroll}), and a recent contribution by \citet{DuttingRT19} on robustness to distributional uncertainty (Section~\ref{sub:robustness-moments}).
Additional results in this direction include \citep{Diamond98,DaiT22,DuettingEFK21,YuK20,Kambhampati23,AnticG23,PengT24}.

\subsubsection{Robustness to Uncertainty about the Action Set} 
\label{sub:robustness-carroll}

We first explore the result of \cite{Carroll15}. In his model, the principal is aware of some of the actions the agent may take, but the actual set of actions the agent can choose from can be any superset of this known action set.

\paragraph{Model.}

There is a known set of possible rewards $\mathcal{R}$, assumed to be a compact subset of $\reals$ normalized such that $\underline{r}:=\min(\mathcal{R}) = 0$ (compactness is used so that limits are attained). Denote $\overline{r}:=\max(\mathcal{R})$. 
Since the proofs will involve changing the distribution and cost of an action, it will be convenient to define an action as 
a pair  $(\Prob_i,\cost_i) \in \Delta(\mathcal{R}) \times \reals_{\geq 0}$, where $\Prob_i$ is a distribution over rewards and $\cost_i$ is the cost of the action. 
A \emph{technology} (set of actions) is a compact subset of $\Delta(\mathcal{R}) \times \reals_{\geq 0}$. The agent has a technology $\mathcal{A}$ which is unknown to the principal. The principal only knows a subset of the available actions $\mathcal{A}_0 \subseteq \mathcal{A}$. 

\begin{figure}[t]
\vspace{-10pt}
\begin{center}
\begin{tikzpicture}
\draw[black, thick,->] (-0.25,0) -- (5.5,0) node[right] {\small{$r$}};
\draw[black, thick,->] (0,-0.25) -- (0,3.5) node[above] {\small{$t$}};
\node[] at (0,-0.6) {\small{$\underbar{$r$} = 0$}};
\draw[thick,-] (4.5,-0.1) -- (4.5,0.1);
\node[] at (4.5,-0.6) {\small{$\bar{r}$}};
\draw[thick,-] (0.1,0.6) -- (-0.1,0.6) node[left] {\small{$t(\underbar{$r$})$}};
\draw[thick,-] (0.1,3) -- (-0.1,3) node[left] {\small{$t(\bar{r})$}};
\draw[-,thick] (0,0.6) --node[below,rotate=30,xshift=+1.3cm]{\small{$\alpha_1 r + \alpha_0$}} (4.5,3);
\draw[thick,-,color=red] (0,0) --node[below,rotate=23,xshift=1.5cm]{\small{$\alpha r$}} (4.5,1.9);
\filldraw[black] (0,0.6) circle (2pt) {};
\filldraw[black] (4.5,3) circle (2pt) {};
\draw[very thick,-,dashed] plot [smooth, tension=0.6] coordinates { (0,0.6) (1,1.5) (2,1) (3,3) (4.5,3)};
\node at (3.5,3.5) {\small{$t(r)$}};
\draw[color=red,dashed,thick] (2.7,1.1) -- (2.7,0) node[below,yshift=-0.1cm] {\small{\textcolor{red}{$R_{i^\star}$}}};
\draw[color=red,dashed,thick] (2.7,1.1) -- (0,1.1) node[left,xshift=-0.05cm] {\small{\textcolor{red}{$T_{i^\star}$}}};
\end{tikzpicture}
\end{center}
\vspace*{-35pt}
\caption{Illustration of the linear and affine contracts constructed in Sections~\ref{sub:robustness-carroll} and \ref{sub:robustness-moments}, respectively, to prove Theorem~\ref{thm:carroll} and Theorem~\ref{thm:max-min-moments}. 
In both these results, a general contract $t$, depicted here as a non-affine function from reward $r$ to transfer $t$ (heavy dashed line), is compared to a contract $t'$, where $t'$ is linear (lower solid line, in red) or affine (upper solid line, in black). An adversarial choice of action sets (in Theorem~\ref{thm:carroll}) or distributions (in Theorem~\ref{thm:max-min-moments}) shows that $t'$ beats $t$ in terms of robust revenue guarantees. The linear contract (lower red) is obtained by finding the action $i^\star$ chosen by the agent given $t$, and using its expected reward $R_{i^\star}$ and payment $T_{i^\star}$ to find the slope. The affine contract (upper black) is obtained by ``linearizing'' $t$ between its endpoints.
 }\label{fig:simple-vs-opt}
\end{figure}
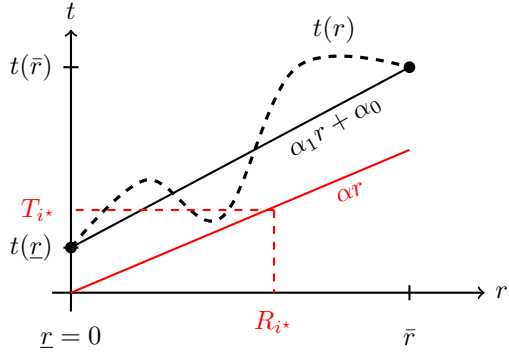

A contract in Carroll's model is any continuous function $t: \mathcal{R} \rightarrow \reals_{\geq 0}$ mapping rewards to transfers---see Figure~\ref{fig:simple-vs-opt} (in the rest of this section we use $t$ instead of $\con$ to denote the contract since we treat it as a mapping rather than as a vector of transfers).
As in the vanilla model, the agent's expected utility from action $(\Prob_i,\cost_i)$ under contract~$t$ is the expected payment minus cost. We introduce the following notation for this utility: $U_A( (\Prob_i,\cost_i) \mid t ) := \mathbb{E}_{r\sim \Prob_i}[t(r)] - \cost_i$. Similarly, the principal's expected utility for action $(\Prob_i,c_i)$ under contract $t$ is defined as the expected reward minus payment. We introduce the following notation: $U_P((\Prob_i,\cost_i) \mid t) :=\mathbb{E}_{r \sim \Prob_i}[r-t(r)]$. 
If the distribution and cost of an action $(\Prob_i,\cost_i)$ as well as the contract $t$ are clear from the context, we use the shorthands $\Rew_i := \mathbb{E}_{r \sim \Prob_i}[r]$ and $\Pay_i := \mathbb{E}_{r \sim \Prob_i}[t(r)]$ (recovering the notation from Section~\ref{sec:model}). 
As usual, the agent is assumed to choose a utility-maximizing action from the set of all available actions $\mathcal{A}$, while breaking ties in favor of the principal. 

In what follows we write $U_A(\mathcal{A} \mid t) := \max_{(\Prob_i,\cost_i) \in \mathcal{A}} U_A((\Prob_i,\cost_i) \mid t)$ 
for the agent's maximum expected utility from the set of actions $\mathcal{A}$ under contract $t$. 
We further use $\mathcal{A}^\star( \mathcal{A} \mid t) := \arg\max_{(\Prob_i,\cost_i) \in \mathcal{A}} U_A( (\Prob_i,\cost_i) \mid t)$ 
to denote all actions $(\Prob_i,\cost_i) \in \mathcal{A}$ that maximize the agent's expected utility under contract $t$. 
Using this notation, the principal's expected utility for a given set of actions $\mathcal{A}$ under contract $t$ is
$U_P(\mathcal{A} \mid t) := \max_{(\Prob_i,\cost_i) \in \mathcal{A}^\star(\mathcal{A} \mid t)} U_P((\Prob_i,\cost_i) \mid t)$. Finally, given a known technology $\mathcal{A}_0$, we denote 
by $U_P(t) = \min_{\mathcal{A} \supseteq \mathcal{A}_0} U_P(\mathcal{A} \mid t)$ the principal's minimum expected utility over all sets of actions $\mathcal{A} \supseteq \mathcal{A}_0$ under contract $t$.\footnote{We write minimum here, but technically it is an infimum as the lowest principal's utility under a given contract $t$ may not be attained.}
Notice that for any $\mathcal{A}_0$ and $t$ it holds that $U_P(t) \le U_P(\mathcal{A}_0 \mid t)$, while for any $\mathcal{A} \supseteq \mathcal{A}_0$ and $t$ it holds that $U_A(\mathcal{A} \mid t) \ge U_A(\mathcal{A}_0 \mid t)$.

\paragraph{Carroll's Main Result.} The question now is: knowing the set of actions $\mathcal{A}_0$, which contract $t$ maximizes the principal's minimum expected utility $U_P(t)$, where the minimum is taken over all technologies $\mathcal{A} \supseteq \mathcal{A}_0$? That is, the principal seeks to solve: 
\[
\max_t \min_{\mathcal{A}\supseteq \mathcal{A}_0} U_P(\mathcal{A} \mid t). 
\]

Carroll shows that the max-min principal's utility can always be achieved by a \emph{linear} contract. The main take-away is that linear contracts are robustly optimal to uncertainty about the agent's technology. Informally, even if the agent's capabilities are unknown to  the principal, she can align incentives with the agent by transferring to him a cut of the rewards; and there is nothing better she can hope to guarantee (in the worst case) when facing such uncertainty.

\begin{theorem}[\cite{Carroll15}]\label{thm:carroll}
For any known technology $\mathcal{A}_0$ and set of possible rewards $\mathcal{R}$ such that $\underline{r} = \min (\mathcal{R}) = 0$, a \emph{linear} contract maximizes 
$U_P(t) = \min_{\mathcal{A} \supseteq \mathcal{A}_0} U_P(\mathcal{A} \mid t)$ over all contracts $t$.
\end{theorem}

\begin{remark}\label{rem:zero_carroll}
Without the assumption that $\underline{r} = \min (\mathcal{R}) = 0$, a max-min optimal contract is affine rather than linear \cite[][Footnote 2, p.~546]{Carroll15}.
\end{remark}

We present a proof of this result suggested by Lucas Maestri \citep[see][Appendix C]{Carroll15}. The high-level approach is to begin with an arbitrary contract $t$ and technology $\mathcal{A}_0$, and to show that $t$ is outperformed by some linear contract $t'$ whose parameter $\alpha$ is derived from the agent's choice of action from $\mathcal{A}_0$ under contract $t$. 

\begin{proof}[Proof of Theorem~\ref{thm:carroll}]
Let $\mathcal{A}_0$ be any technology and let $t$ be an arbitrary contract. 
We construct a linear contract $t'$ such that $U_P(t')\ge U_P(t)$.
Let $(\Prob_{i^\star},\cost_{i^\star})  \in \mathcal{A}_0$
be the action chosen by the agent under contract $t$ when the set of actions is $\mathcal{A}_0$.
If $U_P(\mathcal{A}_0 \mid t) \leq 0$, then we are done: since $\mathcal{A}_0$ is a valid instantiation of $\mathcal{A}$, it holds that
$\min_{\mathcal{A} \supseteq \mathcal{A}_0} U_P(\mathcal{A} \mid t)\le 0$, which is (weakly) 
outperformed by the linear contract $t'$ with $\alpha = 1$ that gives $0$ utility to the principal.

Assume from now on that $U_P(\mathcal{A}_0 \mid t) > 0$, and let  $\alpha := \Pay_{i^\star}/\Rew_{i^\star}$ (see Figure~\ref{fig:simple-vs-opt}). Note that $\Rew_{i^\star} - \Pay_{i^\star} = U_P(\mathcal{A}_0 \mid t)>0$. Therefore, in the definition of $\alpha$ the denominator must be positive, and the ratio must be $< 1$. Consider the linear contract $t'(r) = \alpha \cdot r$. First observe that, under contract $t'$, for any set of actions $\mathcal{A} \supseteq \mathcal{A}_0$, the agent may take action $(\Prob_{i^\star},\cost_{i^\star}) \in \mathcal{A}_0$ to earn an expected utility of
\begin{align}
U_A((\Prob_{i^\star},\cost_{i^\star}) \mid t') =  
\Pay'_{i^\star} - \cost_{i^\star}
= \alpha \cdot \Rew_{i^\star} - \cost_{i^\star} = \Pay_{i^\star} - \cost_{i^\star} = U_A(\mathcal{A}_0 \mid t),\label{eq:agent-utility}
\end{align}
where the third equality holds by definition of $\alpha$.
Next observe that the principal's expected utility for action $(\Prob_{i^\star},\cost_{i^\star})$ under contract $t'$ satisfies
\begin{align}
U_P((\Prob_{i^\star},\cost_{i^\star}) \mid t') 
= \Rew_{i^\star} - \Pay'_{i^\star} 
= (1-\alpha) \cdot \Rew_{i^\star} = \Rew_{i^\star} - \Pay_{i^\star} = U_P(\mathcal{A}_0 \mid t) \geq U_P(t),\label{eq:principal-utility}
\end{align}
where the third equality again holds by definition of $\alpha$.

Now consider an arbitrary set of actions $\mathcal{A}$. 
Let $(\Prob_i,\cost_i) \in \mathcal{A}$ be the action chosen by the agent under the linear contract $t'$ when the set of actions is $\mathcal{A}$. 
We will prove that $U_P(\mathcal{A} \mid t') \geq U_P(t)$. Since this will hold for any $\mathcal{A}$, this will imply that $U_P(t') \geq U_P(t)$, as desired.

First consider the case where $\Rew_i \geq \Rew_{i^\star}$. In this case, it holds that
\[
U_P(\mathcal{A} \mid t') = (1-\alpha) \cdot \Rew_i \geq (1-\alpha)\cdot \Rew_{i^\star} 
\geq U_P(t),
\]
as required, where the last inequality follows by Equation~\eqref{eq:principal-utility}.

So consider the case where 
$\Rew_{i} < \Rew_{i^\star}$. 
Note that by the definition of $(\Prob_i,\cost_i) \in \mathcal{A}$ as the agent's best response to contract $t'$ when the set of actions is $\mathcal{A}$, it must hold that
\begin{align}
\Pay'_{i} - \cost_i  = U_A(\mathcal{A} \mid t') \geq U_A(\mathcal{A}_0 \mid t') \geq \Pay_{i^\star} - \cost_{i^\star} = U_A(\mathcal{A}_0 \mid t ),\label{eq:sandwich}
\end{align}
where the first inequality holds because $\mathcal{A} \supseteq \mathcal{A}_0$ and thus $U_A(\mathcal{A} \mid t') \geq U_A(\mathcal{A}_0 \mid t')$ and the second inequality holds because $U_A(\mathcal{A}_0 \mid t') \geq U_A((\Prob_{i^\star},\cost_{i^\star}) \mid t')$ and $U_A((\Prob_{i^\star},\cost_{i^\star}) \mid t') \geq T_{i^\star} - \cost_{i^\star}$ by Equation~\eqref{eq:agent-utility}.

In the case where $\Pay'_{i} - \cost_i = U_A(\mathcal{A}_0 \mid t)$ 
we are good, because then by Equation~\eqref{eq:sandwich} the agent facing actions $\mathcal{A}$ and contract $t'$ would also be willing to choose action $(\Prob_{i^\star},\cost_{i^\star})$, resulting in a principal utility of at least $U_P(t)$ (by Equation~\ref{eq:principal-utility}). 

So we can assume that 
$\Pay'_{i} - \cost_i > U_A(\mathcal{A}_0 \mid t)$.
For this case, consider the following construction. Let 
$\lambda := \Rew_i/\Rew_{i^\star}$. 
(Note that $\lambda \in [0,1)$ because we are in the case where 
$\Rew_{i} < \Rew_{i^\star}$.) 
Consider the action $(\Prob_{i'},\cost_{i'})$ whose distribution over outcomes is given by $\lambda \Prob_0 + (1-\lambda) \delta_0$ where $\delta_0$ puts probability $1$ on reward $0$, and whose cost is $\cost_{i'}=\cost_i$. 
Consider contract $t$ when the set of actions is $\mathcal{A}' = \mathcal{A}_0 \cup \{(\Prob_{i'},\cost_{i'})\}$. We will show that $U_P(\mathcal{A}' \mid t) \leq U_P(\mathcal{A} \mid t')$. For this, first observe that the agent's expected utility from action $(\Prob_{i'},\cost_{i'})$ under contract $t$ is
\begin{align*}
U_A((\Prob_{i'},\cost_{i'}) \mid t) = \Pay_{i'} - \cost_{i'}
&= \lambda \cdot \Pay_{i^\star} + (1-\lambda) \cdot t(0) - \cost_{i}\\
&\geq \lambda \cdot \Pay_{i^\star} - \cost_{i}\\
&= \lambda \cdot \alpha \cdot \Rew_{i^\star} - \cost_{i}\\
&= \alpha \cdot \Rew_i - \cost_{i}
= \Pay'_{i} - \cost_{i}
> U_A(\mathcal{A}_0 \mid t),
\end{align*}
where the fourth and fifth step hold by the definitions of $\alpha$ and $\lambda$, respectively.
This shows that under contract $t$, the agent prefers action $(\Prob_{i'},\cost_{i'})$ over any action in $\mathcal{A}_0$. Using this, we can then conclude that
\begin{align*}
U_P(\mathcal{A'} \mid t) = \Rew_{i'} - \Pay_{i'} 
&= \lambda \cdot (\Rew_{i^\star} - \Pay_{i^\star}) - (1-\lambda) \cdot t(0)\\
&\leq \lambda \cdot (\Rew_{i^\star} - \Pay_{i^\star}) \\
&=\lambda \cdot (1-\alpha) \cdot  \Rew_{i^\star}\\
&= (1-\alpha) \cdot \Rew_{i}
= \Rew_{i} - \Pay'_{i} 
= U_P(\mathcal{A} \mid t'),
\end{align*}
where the fourth and fifth step hold by the definitions of $\alpha$ and $\lambda$, respectively. We have thus shown that $U_P(\mathcal{A}' \mid t) \leq U_P(\mathcal{A} \mid t')$. Since $U_P(t) \leq U_P(\mathcal{A}' \mid t)$, this shows that $U_P(t) \leq U_P(\mathcal{A} \mid t')$ also in this case.
\end{proof}

\subsubsection{Robustness to Uncertainty about the Distributions} 
\label{sub:robustness-moments}

We next explore the result of \citet{DuttingRT19}, which uses a different notion of uncertainty:
instead of assuming there are completely unknown actions available to the agent alongside fully-known actions, \citet{DuttingRT19} assume that each available action is partially known, in the sense that the principal knows all actions, costs and rewards, but has partial knowledge of the distributions over the rewards that are associated with these actions. This partial knowledge consists of the expectation of each distribution. 

\paragraph{Model.} 

In more detail, consider the following variant of the vanilla model in Section~\ref{sec:model}: There are $n$ actions, with (known) costs $\cost_i \geq 0$ for every $i \in [n]$ (where $c_1=0$, as usual). There are $m$ (known) rewards $r_j \geq 0$ for every $j \in [m]$. Recall, that per default, we index actions and outcomes so that $\cost_1 \le \cost_2 \le \ldots \le \cost_n$ and $\rew_1 \le \rew_2 \le \ldots \le \rew_m.$ As in Section~\ref{sub:robustness-carroll}, assume that rewards are normalized so that $\underline{r}:= \rew_1 = 0$, and let $\overline{r}:=\max_j \rew_j = \rew_m$. 
 
Each action $i \in [n]$ is associated with a distribution $\Prob_i$ over outcomes $j \in [m]$. The principal does not know the actions' exact distributions $\Prob_1, \ldots, \Prob_n$.
Rather, she is only given the expected reward (first moment) $\Rew_i \geq 0$ of each action $i \in [n]$. 
We say that a distribution $\Prob_i$ is \emph{compatible} (with $\Rew_i$) 
if its expected reward $\sum_{j} \prob_{ij} \rew_j$ is equal to $\Rew_i$. Denote the set of compatible distribution profiles $(\Prob_1,\dots,\Prob_n)$ by $\mathcal{D}=\mathcal{D}(R_1,\dots,R_n)$.
In the proofs we will vary the distributions $(\Prob_1,\dots,\Prob_n)$ associated with the actions. It will therefore be convenient to sometimes refer to an action through its distribution $\Prob_i$ (rather than its index $i$).

A contract $\con \in \mathbb{R}_+^m$ is a vector of (non-negative) payments. 
We define expected utilities of the agent and the principal as in Section~\ref{sec:model}, and introduce notation for the principal's worst-case expected utility across compatible distributions.
Adopting notation similar to the one in Section~\ref{sub:robustness-carroll}, we write 
$U_A(\Prob_i \mid \con) := \mathbb{E}_{j \sim \Prob_i}[\pay_j] - \cost_i$ 
for the agent's expected utility from action $\Prob_i$ under contract $\con$,
and $U_A((\Prob_1,\dots,\Prob_n) \mid \con) := \max_{i \in [n]} U_A(\Prob_i \mid \con)$ 
for the agent's maximum expected utility from actions $\Prob_1,\dots,\Prob_n$ under contract $\con$.
We let $\mathcal{Q}^\star((\Prob_1,\dots,\Prob_n) \mid \con) \subseteq \{\Prob_1, \ldots, \Prob_n\}$ denote the set of actions that maximize the agent's expected utility under contract $\con$. 
The principal's expected utility for action $\Prob_i$ under contract $\con$ is 
$U_P(\Prob_i \mid \con) := \mathbb{E}_{j \sim \Prob_i} [\rew_j - \pay_j]$, 
and the principal's expected utility for a set of actions $\Prob_1, \ldots, \Prob_n$ under contract $\con$ is $U_P((\Prob_1, \ldots, \Prob_n) \mid \con) := \max_{\Prob_i \in \mathcal{Q}^\star((\Prob_1, \ldots,  \Prob_n) \mid \con)} U_P(\Prob_i \mid \con)$. Finally, we use $U_P(\con) :=  \min_{(\Prob_1, \ldots, \Prob_n) \in \mathcal{D}} U_P((\Prob_1, \ldots, \Prob_n) \mid \con)$ to denote the principal's minimum utility from contract $\con$ over all compatible distribution profiles. 

For ease of presentation and consistency with Section~\ref{sub:robustness-carroll}, in what follows we will make the following simplifying assumption: We assume that the rewards $\rew_1, \ldots, \rew_m$ are all distinct. We can thus interpret $\Prob_i$ as a distribution over rewards (rather than outcomes), and a contract $\con$ as a mapping $t: \{r_j\}_{j \in [m]} \rightarrow \mathbb{R}_{+}$ from rewards to payments (rather than as a vector of payments, one for each outcome). 

\paragraph{The Result.}

The goal is to design a contract $t$ that maximizes the principal's minimum utility over all distribution profiles $(\Prob_1, \ldots, \Prob_n)$ that are compatible with the expected rewards $\Rew_1, \ldots, \Rew_n$.
That is, the principal seeks to solve 
\[
\max_t \min_{(\Prob_1, \ldots, \Prob_n) \in \mathcal{D}(\Rew_1,\ldots,\Rew_n)} U_P((\Prob_1, \ldots, \Prob_n) \mid t). 
\]

Assuming moment information has a computational flavor and is
standard in robust optimization and in particular robust mechanism design (see, e.g., Scarf’s seminal paper on distributionally-robust stochastic programming \citep{Scarf58} and works like \citep{AzarDM13,BandiB14,CarrascoEtAl17} on prior-independent mechanism design).

The main result of \citeauthor{DuttingRT19}~is that linear contracts are max-min optimal in the above model, where only the first moment of each distribution is known. This result offers an alternative formulation of the inherent robustness of linear contracts, in a natural model of moment
information that is easy to interpret. The following theorem summarizes this result:

\begin{theorem}[\citet*{DuttingRT19}]\label{thm:max-min-moments}
Consider a contract setting with known costs $c_1 \leq \dots \leq c_n$ and rewards $0 =r_1 \leq \dots \leq r_m$. 
For any expected rewards $\Rew_1, \ldots, \Rew_n$, a \emph{linear} contract maximizes
$U_P(t) = \min_{(\Prob_1, \ldots, \Prob_n) \in \mathcal{D}(\Rew_1,\ldots,\Rew_n)} U_P((\Prob_1, \ldots, \Prob_n) \mid t)$ over all contracts $t$.
\end{theorem}

\begin{remark}
\label{rem:zero_r1}
As in Section~\ref{sub:robustness-carroll}, 
without the assumption that $r_1=0$, affine (rather than linear) contracts are max-min optimal (see Remark~\ref{rem:zero_carroll}). 
To see the necessity of $r_1=0$ for robust optimality of linear contracts, consider Example~\ref{ex:equal-rev} with $n=2$ actions and outcomes, 
and rewards $r_1=1, r_2=1/\epsilon$. Recall that the expected rewards are $R_1 = 1, R_2= 1/\epsilon$, and that costs are $c_1 = 0, c_2 = 1/\epsilon - 2 + \epsilon$. 
In this setting, the set of compatible distributions $\mathcal{D}(\Rew_1,\Rew_2)$ is a singleton, since it must be the case that $\Prob_1=(1,0)$ and $\Prob_2=(0,1)$. The analysis of Example~\ref{ex:equal-rev} above shows that no linear contract can provide revenue $>1$. 
However, the optimal contract $\con=(0,c_2)$ gives a revenue of $W_2\approx 2$, and is max-min optimal since $\mathcal{D}$ is a singleton. Thus no linear contract is max-min optimal.
\end{remark}

The high-level proof idea for Theorem~\ref{thm:max-min-moments} is as follows. We first observe that linear contracts, and in fact the larger class of (positive) affine contracts (defined formally below), is agnostic to distributional details. That is, the agent's and principal's utilities only depend on the actions' expected rewards and are thus the same across all compatible distributions (Observation~\ref{obs:agnostic}). We then prove that for every positive affine contract $t'$ there is always a linear contract $t''$, which guarantees the principal at least the same utility (Observation~\ref{obs:linear_dominates_affine}). The proof is completed by showing that for any general contract $t$ there is a (positive) affine contract $t'$ such that $U_P(t') \geq U_P(t)$ (Lemma~\ref{lem:key-proof-by-pic}).

\paragraph{Proof of Theorem~\ref{thm:max-min-moments}.}

Consider the class of (positive) \emph{affine} contracts, where $t'(\rew_j) =\alpha_0 + \alpha_1 \rew_j$ for some parameters $\alpha_0,\alpha_1 \in \mathbb{R}$ such that $t'(\rew_j) \geq 0$ for all $j \in [m]$ to ensure limited liability. Note that, since $\underline{r} := \min_j \rew_j = 0$, the latter implies that $\alpha_0 \geq 0$.
A linear contract is an affine contract with $\alpha_0=0$ and $\alpha_1 \in [0,1]$.

The non-negative parameter $\alpha_0\ge 0$ in an affine contract plays the role of a minimum wage.
Intuitively, since $\alpha_0\ge 0$ is paid for every outcome, it does not affect the incentives. Removing it only helps the principal. We formalize this below.

First, we make a simple but crucial observation regarding affine contracts---that they are agnostic to the details of the distributions beyond their first moments. 

\begin{observation}[Affine contracts are agnostic to distributional details]
\label{obs:agnostic}
The agent's and principal's utilities $U_A(\Prob_i \mid t')$ and $U_P(\Prob_i \mid t')$ from each action $\Prob_i$ under an affine contract $t'(\rew_j) = \alpha_0 + \alpha_1 \cdot \rew_j$ depend only on the costs $c_1,\dots,c_n$ and expected rewards $\Rew_1, \Rew_2, \ldots, \Rew_n$; they are thus the same across all compatible distributions. 
\end{observation}

Next, we show that any affine contract $t'$ can be switched to a linear contract $t''$, without lowering the principal's expected utility.

\begin{observation}
\label{obs:linear_dominates_affine}
For every costs $c_1,\dots,c_n$, distribution profile $(\Prob_1,\ldots,\Prob_n)$ with expected rewards $\Rew_1, \Rew_2, \ldots, \Rew_n$, and  affine contract $t'(r_j) = \alpha_0 + \alpha_1 \rew_j$, there is a linear contract $t''(\rew_j) = \alpha \cdot \rew_j$ such that $U_P((\Prob_1, \ldots, \Prob_n) \mid t'') \ge U_P( (\Prob_1, \ldots, \Prob_n) \mid t')$.
\end{observation}

\begin{proof} 
If $\alpha_1<0$, denote the agent's chosen action under affine contract $t'$ by $i$, and its expected payment by $T'_i=\alpha_0+\alpha_1 R_i$. Consider the IC constraint for action $i$ compared to the zero-cost action (action $1$): $\alpha_0+\alpha_1 R_i-c_i\ge\alpha_0+\alpha_1 R_1$. This implies $\alpha_1(R_i-R_1)\ge c_i$, thus $(R_i-R_1)$ must be non-positive. 
By limited liability, $T'_i\ge 0$, and the principal's expected utility is $R_i - T'_i \le R_i \le R_1$. The principal is thus better off with the zero-pay linear contract $\alpha=0$, given which the agent chooses action 1 (or some other zero-cost action with higher expected reward), so the revenue is at least $R_1$. 
If $\alpha_1>1$, the principal's revenue is negative and again $\alpha=0$ is better.
So assume $\alpha_1\in [0,1]$. Since the minimum wage $\alpha_0\ge 0$ is paid regardless of the outcome, it has no effect on the agent's choice of action $i^\star$. In particular, removing the minimum wage does not make the expected utility from $i^\star$ negative: The agent can always choose the zero-cost action (action $1$) for expected utility $\alpha_0+\alpha_1 R_1$ where $\alpha_1 R_1\ge 0$. The agent's expected utility $\alpha_0+\alpha_1 R_{i^\star}-c_{i^\star}$ from action $i^\star$ is only higher, and so necessarily it holds that $\alpha_1 R_{i^\star} - c_{i^\star} \ge 0$.
It is therefore better for the principal to set $\alpha_0=0$, resulting in a linear contract $\alpha=\alpha_1$. 
\end{proof}

As a corollary of Observation~\ref{obs:linear_dominates_affine}, to prove Theorem~\ref{thm:max-min-moments} it now suffices to show that for every contract $t$ there is an affine contract $t'$ such that $U_P(t') \geq U_P(t).$

\begin{lemma}
    \label{lem:key-proof-by-pic}
    Consider a contract setting with known costs $c_1,\dots,c_n$ and rewards $r_1,\dots,r_m$. For any set of expected rewards $R_1, \dots, R_n$ and any contract $t$, there exists an affine contract $t'$ such that $U_P(t')\ge U_P(t)$.
\end{lemma}

\begin{proof}
The proof is by showing that an adversarially-chosen distribution profile from $\mathcal{D}$ can cause the expected revenue of contract $t$ to drop below that of an affine contract $t'$ (while the latter remains unaffected).
We construct $t'$ from $t$ as follows: 
Treat $t$ as a general function, mapping rewards to transfers (see Figure~\ref{fig:simple-vs-opt} for a visualization).
Consider the points $(\rew_1,\pay(\rew_1))$ and $(\rew_m,\pay(\rew_m))$, i.e., the lowest reward $\underline{r}=r_1=0$ and the highest one $\overline{r} =r_m$, with their respective transfers. These points can be connected by a line graph $\ell$, and we denote its function by $\ell(r)=\alpha_0 + \alpha_1 r$. By definition of $\ell$, $\ell(\rew_1)=t(\rew_1)\ge 0$ and $\ell(\rew_m) = t(\rew_m) \geq 0$ (in both cases, non-negativity is by LL), and hence also $\ell(r_j) \geq 0$ for all $j \in [m]$ (by linearity). 
In particular, $\alpha_0 = \ell(0) = \ell(r_1) \geq 0$. 
Thus line $\ell$ defines a (positive) affine contract $t'(\rew_j) = \alpha_0 + \alpha_1 \rew_j$.
 
For affine contract $t'$, Observation~\ref{obs:agnostic} applies, so only the expectation $R_i$ of action $i$ matters, and we can write $U_A(\Rew_i \mid t')$ (resp.,~$U_P(\Rew_i \mid t')$) for the agent's (resp., principal's) expected utility for action $i$ under contract $t'$.
Let $i^\star$ be the agent's utility-maximizing action when facing affine contract $t'$.
Note that neither the choice of $i^\star$, nor the utilities $U_A(\Rew_i \mid t')$ and $U_P(\Rew_i \mid t')$ for any $i \in [n]$ depend on the details of the distributions. In particular, under $t'$, for any compatible profile of distributions, the agent will choose action $i^\star$, and we have $U_P(t') = U_P( R_{i^\star} \mid t')$.

Below we show that for every action $i$ there is a compatible distribution $\bar{\Prob}_i$ s.t.:  
\begin{equation}
U_A(\bar{\Prob}_i \mid t)=U_A( R_i \mid t');~~~U_P(\bar{\Prob}_i \mid t)=U_P(R_i \mid t').
\label{eq:above-below}
\end{equation}
By~Equation~\eqref{eq:above-below}, if the profile of compatible distributions $(\bar{\Prob}_1,\ldots, \bar{\Prob}_n)$ is chosen by the adversary, the agent's utility-maximizing action when facing contract $t$ (after tie-breaking in favor of the principal) is $i^\star$. 
Thus, the principal's expected revenue from contract~$t$ given $(\bar{\Prob}_1,\ldots,\bar{\Prob}_n)$ equals her expected revenue $U_P(t')$ from $t'$. Since $U_P(t)$ is only lower, $U_P(t)\le U_P(t')$ as required.

It remains to show that for every action $i \in [n]$ with expected reward $R_i$ there exists a compatible distribution $\bar{\Prob}_i$ such that Equation~\eqref{eq:above-below} holds. 
We define a compatible distribution $\bar{\Prob}_i$ with expectation $\Rew_i\in [\rew_1,\rew_m]$ over a support consisting of the endpoints of the interval, by decomposing $\Rew_i$: placing probability $\bar{q}_{i,1}=\Rew_i/\rew_m$ on $\rew_m$, and the remaining probability $\bar{q}_{i,m}=(\rew_m-\Rew_i)/\rew_m$ on $\rew_1=0$. Note that then the expected reward is indeed $\Rew_i$.
Under contract $t$, the agent's expected payment for action $i$ when the distribution is $\bar{\Prob}_i$ is $\bar{q}_{i,1}t(\rew_1) + \bar{q}_{i,m}t(\rew_m)$.
Since we defined $t'$ such that $t'(\rew_1)=t(\rew_1)$ and $t'(\rew_m)=t(\rew_m)$, this is also the agent's expected payment for action $i$ under contract $t'$. The agent's expected utility is the expected payment minus cost $\cost_i$, and the principal's expected utility is $\Rew_i$ minus the expected payment, thus~Equation~\eqref{eq:above-below} holds.
\end{proof}

\begin{remark}[Agnostic vs.~Non-Agnostic Designs]
    There is an interesting difference between the two robustness results---robustness to uncertainty about the action set and robustness to uncertainty about the distributions. In the latter case, 
    the given information is all one needs to derive the optimal linear contract, and the agent's and principal's expected utilities do not depend at all on the adversarial choice of the distributions. 
    In the former case, while a linear contract is max-min optimal, given a linear contract, the agent's and principal's expected utility typically do depend on the adversarial choice of the action set.
    This difference is related to the notion of agnostic robust design, see 
\citep[Sections 1, 2.2]{BabaioffFGLT20} and \citep[Section 1.1]{BachrachT22} for more details.
\end{remark}

\paragraph{Discussion and Open Problems.}

Linear contracts are not the only class of simple contracts. Other simple contracts include the aforementioned \emph{single-outcome payment} contracts (Section~\ref{sec:optimal-contracts}), \emph{step} and \emph{binary-pay} contracts~\citep[e.g.,][]{GeorgiadisS20,DuettingFP23}, \emph{debt} contracts~\cite[e.g.,][]{GaleH85,Hebert17}, and \emph{bounded} contracts~\citep[e.g.,][]{ChenCDH24}.
An interesting open problem is whether there exists a class of simple contracts, which provides a constant-factor approximation to optimal contracts. (Note that Theorem 6.1~in \cite{DuttingRT19} already rules out any monotone class of contracts.)
Another interesting direction is to study additional models with (non-Bayesian) uncertainty, and to explore which types of contracts are max-min optimal under different assumptions. We refer the reader to Section~\ref{sec:ambiguous} for one such model, which drives the max-min optimal contracts to other forms of simple contracts. 

\section{Combinatorial Contracts}  
\label{sec:comb-contracts}
In this section we turn to another natural focal point of an algorithmic approach to contracts: the computational complexity of contract design. 

In the standard model of Section~\ref{sec:model}, there is a single principal, and a single agent. The principal-agent setting is represented by $n$ action costs, $m$ outcome rewards, and an $n\times m$ matrix of distributions. The welfare-optimal contract is trivial (a linear contract with $\alpha=1$ incentivizes the agent to choose the welfare-maximizing action). The revenue-optimal contract can be found in time polynomial in $n,m$ by solving LPs (Section~\ref{sec:lp-formulation}). In either binary-outcome or generalized binary-action settings, the optimal contract even has a simple, practical form---linear contracts in the former case and single-outcome payment contracts in the latter (Section~\ref{sec:optimal-contracts}).

But this is not the end of the story: It is easy to imagine contractual settings that are more complex than this basic setting.  Instead of a single principal-agent pair, the principal may contract with \emph{multiple agents}, or \emph{multiple principals} may share the same agent. Instead of choosing a single action, the agent may choose a \emph{combination of actions}, and instead of measuring performance with a single outcome, the contract may rely on a \emph{combination of outcomes}. 

These complexities often arise in practice. Returning to our introductory example of social media influencers (Section~\ref{sec:intro}), a brand may contract with multiple influencers (agents), and a single influencer may promote multiple brands (principals). The influencer may choose to combine activity on several social media platforms (combination of actions), and the campaign's performance may be measured through different metrics (combinations of outcomes).

In all of these cases, the contract design problem introduces new computational challenges, making it a natural focal point of the computational study of contracts. 
Pioneering studies include the work of \cite*{BabaioffFN06}, who introduced a combinatorial multi-agent contract model, and the work of \cite*{DuttingRT21}, who initiate the study of single-agent combinatorial contracts.
Since then the literature on combinatorial contracts has rapidly grown.

In this section, we present computationally-efficient algorithms that perform well despite the additional complexities, and provide (near-)optimal solutions for the emerging combinatorial settings, as well as impossibility results.
As is often the case, the computational lens offers more than just polynomial-time algorithms; it also reveals valuable structural insights in the process.
We start with preliminaries in Section~\ref{sec:set-funct-oracles}, covering several concepts that may be familiar to readers with a background in combinatorial optimization, especially those with expertise in combinatorial auctions.
We then organize the rest of the material around which aspect of the problem is combinatorial: the set of actions (in Section~\ref{sec:multiple-actions}), the set of agents (in Section~\ref{sec:teams}), the set of outcomes (in Section~\ref{sub:combi-outcomes}), and finally, the set of principals (in Section~\ref{sub:multi-principals}).

\subsection{Combinatorial Contracts Preliminaries}
\label{sec:set-funct-oracles}

In this section, the contract settings we consider are assumed to have \emph{bounded} rewards (thus w.l.o.g.~also bounded costs), normalized such that the highest reward $\max_j \{r_j\}$ is equal to $1$ (unless stated otherwise).\footnote{Boundedness is an important assumption and generally \emph{not} without loss; it is also commonly assumed in (algorithmic) mechanism design~\citep[e.g.][]{Myerson81}.} 
We are interested in algorithms that optimize the principal's expected revenue, or alternatively, approximate it. 
An algorithm is said to provide a \emph{$\rho$-approximation} (where our convention will be that $\rho\ge 1$) if the expected revenue of the contract it finds is at least $\frac{1}{\rho}\OPT$, where $\OPT$ is the optimal expected revenue.
A \emph{fully polynomial-time approximation scheme (FPTAS)} is an algorithm that provides a multiplicative $(1+\varepsilon)$-approximation, in time polynomial in the input size and $1/\varepsilon$.
A \emph{polynomial-time approximation scheme (PTAS)}  is an algorithm that provides a multiplicative $(1+\varepsilon)$-approximation, in time polynomial in the input size for any fixed $\varepsilon$.

\paragraph{Set Functions.}

Given a set $U$ of $n$ \emph{elements}, a set function 
$f: 2^U \rightarrow \reals$ 
assigns a real \emph{value} to every subset of $U$, where $f(S)$ denotes the value of $S \subseteq U$. 
Below, $U$ will represent different sets, such as the set of actions (Section~\ref{sec:multiple-actions}), the set of agents (Section~\ref{sec:teams}), or the set of outcomes (Section~\ref{sub:combi-outcomes}).
We focus on 
\emph{normalized} set functions, for which $f(\emptyset) = 0$, that are \emph{monotone}, 
i.e., for every $S\subseteq T \subseteq U$, it holds that $f(S)\le f(T)$.
The \emph{marginal} value of a set $S$ given a set $T$ is denoted by $f(S \mid T)$, and defined as $f(S \mid T) = f(S \cup T) - f(T)$. When $S$ is a singleton, we sometimes abuse notation and omit the brackets, i.e., for the marginal value of $S=\{j\}$ given $T$, we write $f(j \mid T)$.
We use mainly the classes of set functions within the hierarchy of complement-free set functions of \citep*{LehmannLN06}, introduced in the context of combinatorial auctions (see also~\cite{BlumrosenN06}), but also set functions that exhibit complementarities.

\begin{definition}
Let $U$ be a set of size $n$. 
A set function 
$f: 2^U \rightarrow \reals$
is said to be: 
\begin{itemize}
    \item \emph{Additive} if there exist $f_1,\ldots,f_n$
    such that $f(S)=\sum_{i\in S}f_i$ for every set $S \subseteq U$.
    \item \emph{Gross substitutes (GS)} if it is submodular (see below) and it satisfies the following triplet condition: for any set $S \subseteq U$, and any three elements $i,j,k \not\in S$, it holds that 
    $$
    f(i \mid S) + f(\{j,k\} \mid S) \leq \max\left(f(j \mid S) + f(\{i,k\} \mid S), f(k \mid S) + f(\{i,j\} \mid S)\right).
    $$
    \item \emph{Submodular} if for any two sets $S \subseteq T \subseteq U$, and any element $j \not\in T$, $f(j \mid T) \leq f(j \mid S)$.
    \item \emph{XOS} if it is a maximum over additive functions. 
    That is, there exists a set of additive functions $f_1,\ldots, f_{\ell}$ such that for every set $S \subseteq U$, 
    $
    f(S)=\max_{i \in [\ell]} \left( f_i(S)\right)
    $.
    \item \emph{Subadditive} if for any two sets $S, T \subseteq U$, it holds that $f(S)+f(T) \geq f(S \cup T)$.
    \item \emph{Supermodular} if for any two sets $S \subseteq T \subseteq U$, and any action $j \not\in T$, $f(j \mid T) \geq f(j \mid S)$.
\end{itemize}
\label{def:classes}
\end{definition}

All classes above are \emph{complement free (CF)} except for the supermodular class.
It is well known that $Additive \subset GS \subset Submodular \subset XOS \subset Subadditive$, with strict containment relations \citep{LehmannLN06}. 

\paragraph{Oracle Access.}
Since $f$ is typically of exponential size, it is standard to consider two primitives by which we can access $f$, defined by the following types of queries: 
\begin{itemize}
    \item A {\em value} query receives a set $S \subseteq U$ and returns $f(S)$.
    \item A {\em demand} query receives a vector of prices $p = (p_1, \dots, p_n) \in \reals^n_{\geq 0}$, and returns a set $S\subseteq U$ that maximizes $f(S) - \sum_{i \in S} p_i$.
\end{itemize}

Computationally, assuming demand oracle access is generally a stronger assumption than assuming value oracle access. For most classes of set functions, a demand query cannot be answered with a polynomial number of value queries under standard complexity assumptions, while demand queries do induce value queries~\citep{BlumrosenN09}. 
Two exceptions are the GS class and the supermodular class. 
It is well-known that solving a demand query for GS functions can be done in polyonmial time using a greedy algorithm---this is, in fact, a characterization of GS functions \citep{Bertelsen05,PaesLeme17}. 
For supermodular functions, it is also the case that a demand query requires only polynomially-many value queries. 
This is because, for supermodular $f$, $f(S) - \sum_{i \in S} p_i$ is supermodular, and maximizing a supermodular function is equivalent to minimizing a submodular function, known to admit a polynomial algorithm~\citep{IwataFF01}.

\paragraph{Multilinear Extension.}
Given a set function $f$ over a set $U$ of $n$ elements, its \emph{multilinear extension} $F:[0,1]^n\to\reals^+$ is defined as follows~\citep{CalinescuCPV11}: Treat ${\bf x}\in [0,1]^n$ as a vector of probabilities for selecting each element in $U$ independently at random, and denote by 
\begin{equation}
    q_S({\bf x}):=\prod_{i\in S} x_i \prod_{j\in U\setminus S} (1-x_j)\label{eq:multilinear} 
\end{equation}
the probability of selecting the set $S\subseteq U$. Then
$F({\bf x}):=\sum_S{q_S({\bf x}) f(S)}$, i.e., the expectation of $f(S)$ where the elements of $S$ are selected independently according to vector ${\bf x}$.
For additive~$f$, the multilinear extension $F$ can be computed in time polynomial in $n$, since
$F({\bf x}) := \mathbb{E}_{S}[f(S)] = \mathbb{E}_{S}[\sum_{j=1}^n \mathbbm{1}_{j\in S} f_j] = \sum_{j=1}^n {x_j f_j}$, where the second equality is by additivity and the third is by linearity of expectation.
For general (bounded) $f$, random sampling evaluates $F({\bf x})$ up to an arbitrary precision with high probability using a polynomial number of value queries~\citep{Vondrak10}. 
We thus follow~\citep{Shioura09} and assume oracle access to $F$. 

\paragraph{Linear Programming and the Ellipsoid Method.} 

Recall that a standard approach to computing the optimal contract is by solving $n$ LPs (one per action), each with $n-1$ constraints, and as many payment variables as there are outcomes (Section~\ref{sec:lp-formulation}). 
Below we will discuss natural scenarios where the number of outcomes is $\mu = 2^m$ 
(see Section~\ref{sub:combi-outcomes}). While solving such LPs na\"ively requires exponential time in $m$, the well-known \emph{ellipsoid method} can be used to improve upon this if a \textsf{poly}$(n,m)$-time \emph{separation oracle} is given for the dual program. 
A separation oracle is an algorithm that, given a candidate solution to the program, either decides that the candidate is feasible or returns a violated constraint.

\begin{observation}\label{obs:ellipsoid}
Consider a principal-agent setting with $n$ actions and $\mu = 2^m$ outcomes. Given a $\mathsf{poly}(n,m)$-time separation oracle to \textsf{DUAL-MINPAY-LP($i$)} for each action $i \in [n]$, there exists a $\mathsf{poly}(n,m)$-time algorithm that finds the optimal contract.
\end{observation}

For completeness, we provide more details about this procedure below. 
These details are not necessary for comprehending the majority
of this section (we revisit them only in the proof of Theorem~\ref{thm:multi-outcome-pos}). 

\begin{proof}[Proof sketch for Observation~\ref{obs:ellipsoid}]
Consider the dual program \textsf{DUAL-MINPAY-LP($i$)} for action~$i$ (see Figure~\ref{fig:dual-lp}). This program has $n-1$ variables and as many constraints as there are outcomes (under our assumptions, $\mu = 2^m$ many constraints). 
Recall that \textsf{DUAL-MINPAY-LP($i$)} is always feasible, but may be unbounded. Moreover, if the dual is bounded, then the primal is feasible; otherwise, the primal is infeasible.
As shown in~\citep{Grotschel81}, using $\mathsf{poly}(n)$-many queries to the separation oracle, the ellipsoid method finds the optimal value $\OPT(i)$ of \textsf{DUAL-MINPAY-LP($i$)}, or decides that \textsf{DUAL-MINPAY-LP($i$)} is unbounded. 
In the former case, by duality,
$\OPT(i)$ is also the optimal value of the primal \textsf{MINPAY-LP($i$)}. We can thus decide, for every action $i \in [n]$, whether it is implementable, and, if it is, determine the minimum expected payment required for implementing it. Thus, by enumerating over all actions $i \in [n]$, we can derive the optimal contract's expected revenue in \textsf{poly}($n,m$) time, 
given access to a \textsf{poly}($n,m$)-time 
separation oracle.

The missing piece of the puzzle is how to find the optimal contract itself; we give a high-level description for completeness: Let $i \in [n]$ be the action implemented by the optimal contract.
Reduce optimizing \textsf{DUAL-MINPAY-LP($i$)} to determining feasibility of the same program, with the 
additional constraint that $\sum_{i' \neq i} \lambda_{i'}(\cost_i-\cost_{i'})\ge \OPT(i)$.
Run the ellipsoid method on the new program---this will result in a feasible dual solution. The crux of the argument is that every one of the polynomially-many calls to the separation oracle (except for the last one in which a feasible solution is found) identifies a violated dual constraint. Construct a new dual $\textsf{DUAL-MINPAY-LP($i$)}'$ with only these \textsf{poly}($n$)-many constraints; because of the ellipsoid method's correctness, $\textsf{DUAL-MINPAY-LP($i$)}'$ is equivalent to the original. We can now take the dual of this program to obtain a new primal program, $\textsf{MINPAY-LP($i$)}'$, with \textsf{poly}($n$)-many variables and constraints. Solving this primal results in the optimal contract.
\end{proof}

\subsection{Combinatorial Actions}
\label{sec:multiple-actions}

The classic principal-agent model fails to capture an important aspect of complex task performance, which is a widely recognized phenomenon in economics. 
This aspect is the idea that performing a complex task often involves choosing a set of actions out of a given pool of available actions. 
This concept has been extensively explored in economics in the influential paper on \emph{multi-tasking} by \cite{HolmstromMilgrom91}. To explore this aspect computationally, in this subsection 
we present and discuss results for the
principal-agent model introduced in \citet*{DuettingEFK21}. 

\paragraph{Model.}

In the model of \cite{DuettingEFK21} the principal seeks to delegate a project to an agent. The project can either succeed or fail. The (normalized) rewards for success and failure, are $1$ and $0$, respectively. So we are in a binary-outcome setting (see Section~\ref{sec:optimal-contracts}). The agent has a set $\mathcal{A} = [n]$ of $n$ actions from which he can choose any {\em subset}.

The combinatorial structure is captured by a \emph{success probability} function, $f:2^{\mathcal{A}}\rightarrow [0,1]$, a set function which assigns a (not-necessarily additive) success probability $f(S)$ to every set of actions $S\subseteq \mathcal{A}$. 
Note that since the reward for success is normalized to $1$, $f(S)$ is also the expected reward for the set of actions $S \subseteq \mathcal{A}$, so we sometimes refer to $f$ as the \emph{(expected) reward} function. 
The cost function is additive, so for each action $i \in [n]$ there is a cost $c_i \geq 0$, and the cost of a set of actions $S \subseteq \mathcal{A}$ is $c(S) := \sum_{i \in S} c_i$.\footnote{More general cost structures have been considered in, e.g., \cite{VuongDPP23} and \cite{DuettingFG23}, see discussion below.} 

The optimal contract in the binary-outcome case is linear (see Proposition~\ref{prop:linear-with-two-outcomes}), i.e., it pays $\alpha$ for
success and $0$ for
failure. Given a (linear) contract $\alpha \in [0,1]$, the agent chooses the set $S$ that maximizes his expected utility $U_A(S \mid \alpha) := \alpha f(S)-c(S)$. As before, we assume the agent breaks ties in favor of the principal (alternatively, the principal recommends a best response $S$, and the agent follows that recommendation). The principal's goal is to find a contract $\alpha$, that maximzies her expected utility $U_P(S \mid \alpha) := (1-\alpha)f(S)$, where $S$ is the agent's response to $\alpha$. 

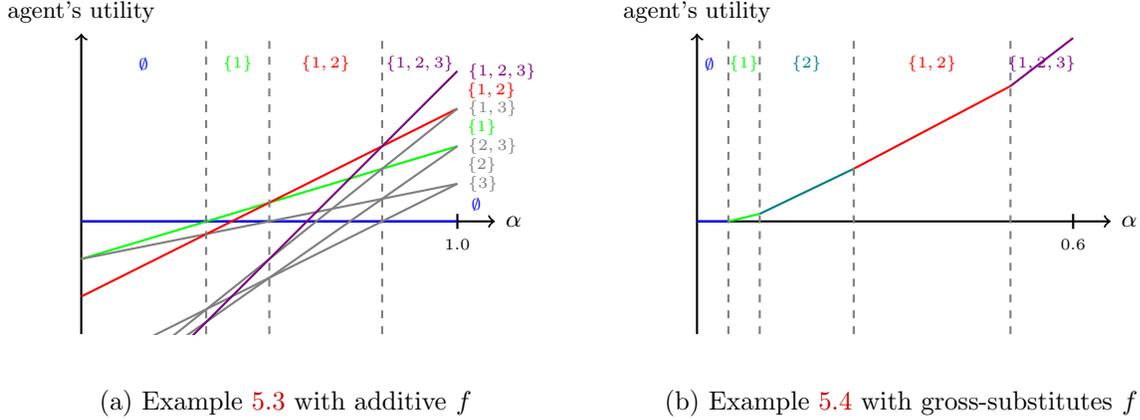
\begin{figure}
\begin{subfigure}[t]{0.45\textwidth}
\begin{tikzpicture}
    \useasboundingbox (-1,-2) rectangle (6,3);
    \clip (-1,-1.5) rectangle (6,3);
    \draw[thick,->] (0,-1.5) -- (0,2.5) node[above] {\footnotesize agent's utility};
    \draw[thick,->] (0,0) -- (5.5,0) node[right] {\footnotesize $\alpha$};
    \draw[thick,-] (5,0.1) -- (5,-0.1) node[below] {\tiny $1.0$};
    \def\z{5}
    \draw[-,blue,thick] (0,0) -- (5,0) node[above,xshift=0.25cm,blue] {\tiny{$\emptyset$}};
    \draw[-,green,thick] (0,-0.1*\z) -- (5,0.2*\z) node[above,xshift=0.325cm,green]{\tiny $\{1\}$};
    \draw[-,gray,thick] (0,-0.1*\z) -- (5,0.1*\z) node[above,xshift=0.325cm,gray]{\tiny $\{2\}$};
    \draw[-,gray,thick] (0,-0.4*\z) -- (5,0.1*\z) node[right]{\tiny $\{3\}$};;
    \draw[-,red,thick] (0,-0.2*\z) -- (5,0.3*\z) node[above,xshift=0.45cm,red]{\tiny $\{1,2\}$};
    \draw[-,gray,thick] (0,-0.5*\z) -- (5,0.2*\z) node[right]{\tiny $\{2,3\}$};;
    \draw[-,gray,thick] (0,-0.5*\z) -- (5,0.3*\z) node[right]{\tiny $\{1,3\}$};
    \draw[-,violet,thick] (0,-0.6*\z) -- (5,0.4*\z) node[right,violet]{\tiny$\{1,2,3\}$};    
    \draw[gray,thick,dashed] (1.66,-1.5) -- (1.66,2.5);
    \draw[gray,thick,dashed] (2.5,-1.5) -- (2.5,2.5);
    \draw[gray,thick,dashed] (4,-1.5) -- (4,2.5);
    \node[blue] at (0.83,2.1) {\tiny $\emptyset$};
    \node[green] at (2.08,2.1) {\tiny $\{1\}$};
    \node[red] at (3.25,2.1) {\tiny $\{1,2\}$};
    \node[violet] at (4.5,2.1) {\tiny $\{1,2,3\}$};
\end{tikzpicture}
\caption{Example~\ref{ex:additive} with additive $f$}
\label{fig:additive}
\end{subfigure}
\hspace*{0.5cm}
\begin{subfigure}[t]{0.45\textwidth}
\begin{tikzpicture}
    \useasboundingbox (-1,-2) rectangle (6,3);
    \clip (-1,-1.5) rectangle (6,3);
    \def\z{2}
    \def\x{0.6}
    \draw[thick,->] (0,-1.5) -- (0,2.5) node[above] {\footnotesize agent's utility};
    \draw[thick,->] (0,0) -- (5.5,0) node[right] {\footnotesize $\alpha$};
    \draw[thick,-] (5,0.1) -- (5,-0.1) node[below] {\tiny {\x}}; 
    \draw[-,blue,thick] (0,0) -- (0.05*5/\x,0);
    \draw[-,green,thick] (0.05*5/\x,0.05*1*\z-0.05*\z) -- (0.1*5/\x,0.1*1*\z-0.05*\z);
    \draw[-,teal,thick] (0.1*5/\x,0.1*2*\z-0.15*\z) -- (0.25*5/\x,0.25*2*\z-0.15*\z);
    \draw[-,red,thick] (0.25*5/\x,0.25*2.2*\z-0.2*\z) -- (0.5*5/\x,0.5*2.2*\z-0.2*\z);
    \draw[-,violet,thick] (0.5*5/\x,0.5*3.2*\z-0.7*\z) -- (\x*5/\x,\x*3.2*\z-0.7*\z);
    \draw[gray,thick,dashed] (0.05*5/\x,-1.5) -- (0.05*5/\x,2.5);
    \draw[gray,thick,dashed] (0.1*5/\x,-1.5) -- (0.1*5/\x,2.5);
    \draw[gray,thick,dashed] (0.25*5/\x,-1.5) -- (0.25*5/\x,2.5);
    \draw[gray,thick,dashed] (0.5*5/\x,-1.5) -- (0.5*5/\x,2.5);
    \node[blue] at (0.02*5/\x,2.1) {\tiny $\emptyset$};
    \node[green] at (0.075*5/\x,2.1) {\tiny $\{1\}$};
    \node[teal] at (0.175*5/\x,2.1) {\tiny $\{2\}$};
    \node[red] at (0.375*5/\x,2.1) {\tiny $\{1,2\}$};
    \node[violet] at (0.55*5/\x,2.1) {\tiny $\{1,2,3\}$};
\end{tikzpicture}
\caption{Example~\ref{ex:gs} with gross-substitutes $f$}
\label{fig:gs}
\end{subfigure}
\caption{Upper envelopes of the agent's utility.}
\label{fig:multi-action}
\end{figure}

The following examples give 3-action instances with additive and  gross-substitutes success probability functions $f$ (see Definition~\ref{def:classes}), respectively. Their corresponding upper envelopes are given in Figures~\ref{fig:additive} and~\ref{fig:gs}.

\begin{example}[Additive $f$] 
\label{ex:additive} 
There are three actions $\{1,2,3\}$. The success probability function $f$ is additive, with $f(\{1\}) = 0.3, f(\{2\}) = 0.2$, and $f(\{3\})=0.5$. The action costs are $c_1=c_2=0.1$, and $c_3=0.4$.
Consider, for example, the contract $\alpha=0.5$. 
The agent's utility for taking action $1$ is $\alpha f(\{1\})-c_1=0.5\cdot 0.3-0.1=0.05$, for action $2$ it is $\alpha f(\{2\})-c_2=0.5\cdot 0.2-0.1=0$, and for action $3$ it is $\alpha f(\{3\})-c_3=0.5\cdot 0.5-0.4 = -0.15$. 
Therefore, among all singletons, 
action $1$ 
is best. However, the agent may be better off selecting more than a single action. The agent's utility for the set $\{1,2\}$ is $\alpha f(\{1,2\})-(c_1+c_2)=0.5\cdot 0.5-0.2=0.05$, for the set $\{1,3\}$ it is $\alpha f(\{1,3\})-(c_1+c_3)=0.5\cdot 0.8-0.5=-0.1$, for the set $\{2,3\}$ it is 
$\alpha f(\{2,3\})-(c_2+c_3)=0.5\cdot 0.7-0.5=-0.15$,
and for the set $\{1,2,3\}$ it is $\alpha f(\{1,2,3\})-(c_1+c_2+c_3)=0.5\cdot 1-0.6=-0.1$. 
Therefore at $\alpha = 0.5$ the agent is indifferent between $\{1\}$ and $\{1,2\}$ and tie breaks in favor of the set $\{1,2\}$ (this point is the intersection of the green and red curves in Figure~\ref{fig:additive}).
Below we provide more details about how the agent's best response changes as a function of $\alpha$, and how that affects the principal's choice of $\alpha$.
\end{example}

\begin{example}[Gross-substitutes $f$] 
\label{ex:gs} 
There are three actions $\{1,2,3\}$.
The success probability function $f$ is as follows:
	$f(\emptyset)=0,~f(\{1\})=0.25,~f(\{2\}) = 0.5, f(\{3\}) = 0.25, f(\{1,2\})= 0.55,~f(\{1,3\}) = 0.5, f(\{2,3\}) = 0.75,$ and $f(\{1,2,3\}) = 0.8$.
The action costs are $c_1 =0.0125, c_2 =0.0375,$ and $c_3=0.125$.
Consider, for example, the contract $\alpha=0.5$ (this point is the intersection of the red and violet curves in in Figure~\ref{fig:gs}).
Given this contract, the agent's utility is maximized by set $\{1,2\}$ and set $\{1,2,3\}$.
Since the set $\{1,2,3\}$ yields a higher principal utility, 
the agent 
breaks the tie in favor of
this set. 
Below we provide more details about the transitions in the agent's best response, and how they differ from those in the additive case.
\end{example}

\paragraph{Challenges.} 

The combinatorial action model of \cite{DuettingEFK21} fits within the classic model 
by defining a meta-action for each of the $2^n$ possible subsets of actions. 
The linear programming approach can then be applied (see Section~\ref{sec:lp-formulation}).
However, 
this na\"ive approach disregards the inherent structure of the problem and specifically, computing the optimal contract through this blueprint would entail an exponential running time.

Since the optimal contract is linear, it is also possible to tackle the problem of computing an optimal contract via  the geometric approach in Section~\ref{sec:geometric-approach}, specifically the upper envelope diagram (Figure~\ref{fig:upper-envelope}).
Recall that this diagram traces the 
agent's expected utility for each action as a function of $\alpha \in [0,1]$ (including the empty set, represented by the $x$-axis).
The issue is that now there are potentially exponentially-many $\alpha f(S)-c(S)$ curves, one for each \emph{set} of actions $S \subseteq \mathcal{A}$, so the $[0,1]$ interval may be subdivided into up to $2^n$ intervals. 

Figure~\ref{fig:additive} demonstrates the upper envelope diagram for the setting given in Example~\ref{ex:additive}, where $f$ is additive. 
By inspecting the upper envelope, one can verify that the agent's best response is to engage in no action for small values of $\alpha$, then engage in action 1, then in the action set $\{1,2\}$, and finally in the action set $\{1,2,3\}$ for sufficiently large $\alpha$.
This is no coincidence: for every scenario with an additive~$f$, every action $i$ belongs to the agent's best response if and only if 
$\alpha \geq \cost_i/f(\{i\})$, independent of the other actions. This is the point $\alpha$ satisfying 
$\alpha f(S \cup \{i\}) - c(S \cup \{i\}) = \alpha f(S) - c(S)$, independent of the set $S$.  
Thus, there are at most $n$ {\em indifference points} (a.k.a.~\emph{breakpoints} or \emph{critical $\alpha$'s})---values of $\alpha$ for which the agent's best response changes. 

Figure~\ref{fig:gs} demonstrates the upper envelope diagram 
for the setting given in Example~\ref{ex:gs}, where $f$ is gross substitutes. 
By inspecting the upper envelope, one can verify that the agent's best response is to engage in no action for small values of $\alpha$, then engage in action 1 (obtaining the set $\{1\}$), then replace action 1 by action 2 (obtaining the set $\{2\}$), then add action 1 again (obtaining the set $\{1,2\}$), and finally add action 3 as well (obtaining the set $\{1,2,3\}$).
We immediately observe that the nice structure in Example~\ref{ex:additive} no longer holds. In particular, action 1 is included for some $\alpha$, later abandoned for a larger $\alpha$, and then reselected for an even larger $\alpha$.  
Unlike the case of additive $f$, this means that we cannot bound the number of indifference points without further exploration.

\paragraph{A Positive Result for Gross Substitutes Rewards.} 
While the geometric approach does not yield a poly-time algorithm \emph{per se}, it does suggest 
a natural algorithm for the optimal contract problem: 
iterate over all critical $\alpha$'s, for each one compute the agent's best response $S_{\alpha}$, and choose an $\alpha$ that yields the maximal principal's expected utility $(1-\alpha)f(S_{\alpha})$.

For the natural algorithm to run in polynomial time, one needs: (i)  a poly-time algorithm that given an $\alpha$, finds the agent's best response $S_{\alpha}$, (ii) a polynomial number of critical $\alpha$'s, and (iii) a poly-time algorithm for iterating over the critical $\alpha$'s (for example, a poly-time algorithm that given a critical $\alpha$, returns the next higher critical $\alpha$).

As explained above, all three requirements are satisfied for scenarios with an additive $f$, thus the natural algorithm solves the best contract problem in polynomial time. 
The situation, however, becomes more challenging for 
more complex $f$ functions, as demonstrated by the gross-substitutes function $f$ given in Example~\ref{ex:gs} and Figure~\ref{fig:gs}.

The class of gross-substitutes functions plays a major role both in economics, where it is the frontier for the existence of market equilibrium \citep{kelso1982job,gul1999walrasian}, and in computer science, where it admits a poly-time algorithm for social welfare maximization in combinatorial auctions \citep{NisanSegal06}.%
\footnote{We refer the interested reader to the work of \cite{RoughgardenT15}, which explores connections between the two roles.} 

The main positive result of \cite{DuettingEFK21}
is that for the case where the success probability function $f$ is gross substitutes, the optimal contract can be computed in polynomial time (with value oracles).  
They also show that for the larger class of submodular success probability functions $f$ (see Definition~\ref{def:classes}) computing an optimal contract is \textsf{NP}-hard, and thus gross-substitutes is a ``frontier" for exact optimization. 

\begin{theorem}[\citet*{DuettingEFK21}] 
In binary-outcome settings where the agent can take any combination of $n$ actions, for gross substitutes success probability functions, the optimal contract can be computed in time polynomial in $n$, given access to a value oracle.
\end{theorem}

The theorem is established by showing that for GS functions, all three of the aforementioned requirements are satisfied, and as a result, the optimal contract can be computed in polynomial time, using the natural algorithm suggested above.

Let us consider requirement (i) first. 
Namely, an algorithm that, given $\alpha$, and using only value queries, finds a set $S$ that maximizes $\alpha f(S) - c(S)$. 
Note that, this is equivalent to finding a set $S$ that maximizes $f(S)- \frac{1}{\alpha}\sum_{i \in S}\cost_i$. 
This problem is precisely solving a {\em demand query} at prices $\frac{1}{\alpha}\cost_i$ in the framework of combinatorial auctions (see Section \ref{sec:set-funct-oracles}). 
It is well-known that solving a demand query for GS functions can be done greedily in polynomial time (see Section~\ref{sec:set-funct-oracles}). 
It follows that requirement (i) is satisfied 
for GS functions. 
Moreover, it is not too difficult to show that the greedy algorithm for answering a demand query can be utilized to satisfy requirement (iii) as well (details omitted).

It remains to show that requirement (ii) is satisfied for gross-substitutes functions. 
Figure~\ref{fig:gs} demonstrates that, unlike additive functions, the number of breakpoints may be larger than $n$, and more complex transitions may happen along the $\alpha$ axis. 
For example, we observe that action $1$ is added at some point, then replaced with action $2$, then added back to action $2$. 
Nevertheless, the key lemma in \cite{DuettingEFK21} shows that only one of two things can happen at a breakpoint: either an action joins the best response set (as in additive $f$), or an action in the best-response set is replaced with a more costly action (as in the transition from action 1 to action 2 in Figure~\ref{fig:gs}). Using this key lemma, a simple potential function argument shows that the number of transitions is at most $O(n^2)$.
The potential function assigns every action its rank according to the cost function (where the lowest-cost action is ranked 1, and the highest-cost action is ranked $n$), and the potential of a set of actions is the sum of its actions' potentials. Thus, the potential function is upper bounded by $\sum_{i=1}^{n}i=O(n^2)$. Observing that the potential of the best response is an integer that monotonically increases in $\alpha$ completes the argument.
Interestingly, this bound is tight, i.e., there exists a GS function with $\Omega(n^2)$ breakpoints.

\begin{table}[t]
\fontsize{10pt}{12pt}
\selectfont
\scalebox{0.95}{
\begin{tabular}{|c|cc|cc|}
\hline
\rowcolor[HTML]{C0C0C0} 
\textbf{\begin{tabular}[c]{@{}c@{}}Multiple\\ actions\end{tabular}} & \multicolumn{2}{c|}{\cellcolor[HTML]{C0C0C0}\textbf{Value Oracle}}                                                                                                                                                                       & \multicolumn{2}{c|}{\cellcolor[HTML]{C0C0C0}\textbf{Value and Demand Oracle}}                                                                                                              \\ \hline
\rowcolor[HTML]{C0C0C0} 
\multicolumn{1}{|l|}{\cellcolor[HTML]{C0C0C0}}                      & \multicolumn{1}{c|}{\cellcolor[HTML]{C0C0C0}\textbf{\begin{tabular}[c]{@{}c@{}}Upper bound \\ (pos)\end{tabular}}} & \textbf{\begin{tabular}[c]{@{}c@{}}Lower bound\\ (neg)\end{tabular}}                                                & \multicolumn{1}{c|}{\cellcolor[HTML]{C0C0C0}\textbf{\begin{tabular}[c]{@{}c@{}}Upper bound \\ (pos)\end{tabular}}}  & \textbf{\begin{tabular}[c]{@{}c@{}}Lower bound\\ (neg)\end{tabular}} \\ \hline
\cellcolor[HTML]{C0C0C0}\textbf{GS}
               
& \multicolumn{1}{c|}{\cellcolor[HTML]{FFFFC7}\begin{tabular}[c]{@{}c@{}}1\\ \cite{DuettingEFK21}\end{tabular}}                    &  {\cellcolor[HTML]{EFEFEF} 1}                                                                                                                   & \multicolumn{1}{c|}{\cellcolor[HTML]{EFEFEF}1}                                                                      &  {\cellcolor[HTML]{EFEFEF} 1}                                                                     \\ \hline
\cellcolor[HTML]{C0C0C0}\begin{tabular}[c]{@{}c@{}}\textbf{Sub-}\\\textbf{modular}\end{tabular}                          & \multicolumn{1}{c|}{\cellcolor[HTML]{FFFFFF}}                                                                      & \cellcolor[HTML]{FFFFC7}\begin{tabular}[c]{@{}c@{}}No constant\\ approx\\ (if \textsf{P}$\neq$\textsf{NP})\\ \footnotesize{\cite{EzraFS24}}\end{tabular}       & \multicolumn{1}{c|}{\cellcolor[HTML]{EFEFEF}
FPTAS}
& 
\multicolumn{1}{c|}{\cellcolor[HTML]{FFFFC7}
\begin{tabular}[c]{@{}c@{}}$>1$\\ 
\footnotesize{\cite{DuettingEFK21}}\\
\footnotesize{\cite{DuettingFGR24}}
\end{tabular}}
\\ 
\hline
\cellcolor[HTML]{C0C0C0}\textbf{XOS}                                & \multicolumn{1}{c|}{\cellcolor[HTML]{FFFFFF}}                                                                      & \cellcolor[HTML]{FFFFC7}\begin{tabular}[c]{@{}c@{}}No better\\ than $\Omega(n^{1/2})$\\ (if \textsf{P}$\neq$\textsf{NP})\\ \footnotesize{\cite{EzraFS24}}\end{tabular} & \multicolumn{1}{c|}{\cellcolor[HTML]{EFEFEF}
FPTAS}
& 
\multicolumn{1}{c|}{\cellcolor[HTML]{EFEFEF}
$>1$}\\ \hline
\cellcolor[HTML]{C0C0C0} 
\begin{tabular}[c]{@{}c@{}}\textbf{Sub-}\\\textbf{additive}\end{tabular}  
& \multicolumn{1}{c|}{\cellcolor[HTML]{FFFFFF}\textbf{}}                                                             & \cellcolor[HTML]{EFEFEF}\begin{tabular}[c]{@{}c@{}}No better\\ than $\Omega(n^{1/2})$\end{tabular}                       & 
\multicolumn{1}{c|}{\cellcolor[HTML]{FFFFC7}
\begin{tabular}[c]{@{}c@{}}FPTAS\\ \footnotesize{\cite{DuettingEFK21}}\\
\footnotesize{\cite{DuettingEFK24}}
\end{tabular}} &
\multicolumn{1}{c|}{\cellcolor[HTML]{EFEFEF}
$>1$}\\ \hline \hline
\cellcolor[HTML]{C0C0C0}\cellcolor[HTML]{C0C0C0}\begin{tabular}[c]{@{}c@{}}\textbf{Super-}\\\textbf{modular}\end{tabular}                       & 
\multicolumn{1}{c|}{\cellcolor[HTML]{FFFFC7}\begin{tabular}[c]{@{}c@{}}1\\ \footnotesize{\citet{DuettingFG23}}\\ \footnotesize{\citet{VuongDPP23}}\end{tabular}}
& {\cellcolor[HTML]{EFEFEF} 1}                                                                                                                   & \multicolumn{1}{c|}{\cellcolor[HTML]{EFEFEF}1}                                                                      &  {\cellcolor[HTML]{EFEFEF} 1}                                                                      \\ \hline
\end{tabular}
}
\caption{This table presents approximation results for the combinatorial multi-action binary-outcome model.  
The left part presents results under access to value oracle, and the right part presents results under access to both value and demand oracles. For each one we present both upper bounds (positive results) and lower bounds (negative results) on the achievable approximation.
The rows represent different reward function classes.
Yellow cells give the results, whereas gray cells represent results derived from other cells (where positive results carry over north (to sub-classes) and east (from value oracle to value and demand oracle), and negative results carry over south and west).   
For example, the FPTAS for subadditive rewards implies  the same result for all subclasses of subadditive rewards.
}
\label{tab:multi-action}
\end{table}

\paragraph{Complement-Free Rewards, Beyond Gross Substitutes.}

The \textsf{NP}-hardness result for submodular success probability functions $f$ (under value queries) \citep{DuettingEFK21} is shown for a family of instances, in which the optimal contract $\alpha^\star$ takes one of two values, and the difficulty stems from answering a demand query with value queries. \cite{DuettingEFK21} also show a structural result, namely that there exist instances with submodular $f$ that admit  exponentially-many critical $\alpha$'s.
Follow-up work by \citet*{EzraFS24} strengthens the hardness result for submodular $f$, by showing that no polynomial-time algorithm with value oracle access can approximate the optimal contract to within any constant factor, assuming $\mathsf{P} \neq \mathsf{NP}$.
In addition, \cite{EzraFS24} show an impossibility of $\Omega(n^{1/2})$ for XOS $f$ that applies to any polynomial-time algorithm with value oracle access, 
again assuming $\mathsf{P \neq NP}$. Together these negative results show a sharp transition in the computational tractability of the optimal contract problem with value oracle access, when going from gross substitutes to more general complement-free settings.

On the positive side, \cite{DuettingEFK21} devise a weakly-polynomial FPTAS for any (monotone) reward function, given access to value and demand oracles. The FPTAS of \cite{DuettingEFK21} is only weakly poly-time as its running time is 
polynomial in $k$, where $k$ denotes the number of bits required to represent $f$ and $c$. Recent work by \cite*{DuettingEFK24} strengthens this result, by giving a strongly-polynomial FPTAS for any (monotone) reward function, with value and demand oracles.

Recall that the NP-hardness for submodular reward functions in \cite{DuettingEFK21} arises from the hardness of answering a demand query using only value queries. Could it be that with access to a demand oracle, the FPTAS developed in \cite{DuettingEFK24} could be improved to a polynomial-time algorithm that finds the optimal contract? This question motivated the work of \cite*{DuettingFGR24}, who proved that finding the optimal contract for submodular rewards requires exponentially many queries, even when both value and demand oracles are available.
In addition to showing tightness of the FPTAS, this demonstrates
that the hardness of the optimal contract problem is is inherently rooted in the nature of the optimal contract problem itself, not only in the hardness of solving a demand query.

\paragraph{A Positive Result for Supermodular Rewards.}

\citet*{VuongDPP23} and \citet*{DuettingFG23}, in a pair of recent papers, give an algorithm for finding all critical $\alpha$'s for a general (monotone) reward function $f$ with access to value and 
demand oracles, whose running time is polynomial in the number of critical values.

The algorithm for enumerating all critical values operates recursively. 
It identifies critical values by querying the agent's demand oracle at the endpoints of a segment $[\alpha, \alpha'] \subseteq [0, 1]$. If the agent's best response for the two contracts is the same, i.e., $S_\alpha = S_{\alpha'}$, then the segment $(\alpha, \alpha']$ admits no critical values. Otherwise, the procedure is recursively applied to the sub-segments $[\alpha, \gamma]$ and $[\gamma, \alpha']$, where $\gamma$ is the contract at which the agent is indifferent between $S_\alpha$ and $S_{\alpha'}$. 

An important implication of this algorithm is that in order to obtain a polynomial-time algorithm for finding an optimal contract with value oracle access only, it suffices to establish the aforementioned properties (i) and (ii), i.e., that it is possible to efficiently answer a demand query with value queries and that there is a polynomial-number of critical values.

By arguing that both these conditions are satisfied for supermodular $f$ and additive $c$ (and more generally submodular $c$), \cite{VuongDPP23} and \cite{DuettingFG23} obtain a polynomial-time algorithm for such settings (in the value oracle model). 

Condition (i) is satisfied because the agent's utility is a supermodular function (as the difference of a supermodular and submodular functions), 
and maximizing a supermodular function is equivalent to minimizing a submodular function, which admits a polynomial time algorithm \citep{IwataFF01}.
For condition (ii), the following lemma shows that at every critical point the best response is a superset of the previous best response, implying an upper bound of $n$ on the number of breakpoints. We state and prove the lemma for additive costs, but note that the lemma extends to submodular cost functions, using the same proof.

\begin{lemma}
[\citet*{DuettingFG23,VuongDPP23}]
\label{lem:supmod-critical}
    For any supermodular reward function $f$ and additive cost function $c$, and any two contracts $\alpha < \alpha'$ and corresponding agent's best response sets $S_\alpha$, $S_{\alpha'}$, it holds that $S_\alpha \subseteq S_{\alpha'}$.
\end{lemma}

\begin{proof}
    If $S_\alpha = S_{\alpha'}$ the lemma obviously hold.
    Otherwise, let $S_{\alpha'}$ be a maximal best-response for contract $\alpha'$ (in line with our tie-breaking assumption), and let $R=S_\alpha \setminus S_{\alpha'}$. Suppose towards contradiction that $R \ne \emptyset$.
    We show that $\alpha' f(R \mid S_{\alpha'}) - c(R) \geq 0$, contradicting the maximality of $S_{\alpha'}$. Indeed, 
    $$\alpha' f(R \mid S_{\alpha'}) - c(R)
        \ge
        \alpha' f(R \mid S_\alpha \cap S_{\alpha'}) - c(R) 
        \ge
        \alpha f(R \mid S_\alpha \cap S_{\alpha'}) - c(R) \geq 0,
    $$  
    where the first inequality follows from the supermodularity of $f$, the second inequality follows by the monotonicity of $f$ combined with $\alpha'>\alpha$, and the last inequality follows by the optimality of $S_\alpha$ at $\alpha$.
\end{proof}

\begin{remark}[Connection to sensitivity analysis]
The recursive algorithm for finding all breakpoints of the agent's best response with access to value and demand oracles has been previously discovered in a variety of contexts, including in the field of \emph{sensitivity analysis} of combinatorial optimization problems \citep{gusfield1980sensitivity}, where it is known as the Eisner-Severance technique \citep{eisner1976mathematical}. 
\end{remark}

\paragraph{Summary and Open Problems.} 

We summarize the state-of-the-art for the combinatorial multi-action binary-outcome model in Table~\ref{tab:multi-action}.
An interesting direction for future work is to explore the best-possible approximation guarantees that can be given for submodular, XOS, and subadditive success probabilities with value oracle access. Another direction is to explore the problem beyond binary outcome. \cite{DuettingEFK21} show that linear contracts remain max-min optimal when only the expected reward of each set of actions is known, but they are suboptimal when it comes to worst-case approximation.
\subsection{Multiple Agents}
\label{sec:teams}

Another very natural extension of the contracting problem concerns situations where the principal seeks to incentivize a \emph{team} of agents. The seminal work on moral hazard in teams in economics is by \citet*{Holmstrom82}.
Clearly, how ``effective'' a team is depends on the composition of the team. We discuss the algorithmic aspects of identifying the optimal (or a near-optimal) contract for a team of agents. This problem is interesting already in the basic (but fundamental) case, in which each agent can either exert effort or not. In this case, the problem boils down to identifying the best set of agents to contract with. 

This direction was pioneered by \cite*{BabaioffFN06} and \citet*{BabaioffFNW12}, who referred to the problem as \emph{combinatorial agency}.
In their model,  
every agent either succeeds or fails in their individual task, and there exists a Boolean function mapping individual outcomes to success or failure of the project. Two natural examples are the OR Boolean function, where the project succeeds iff at least one of the agent succeeds, and the AND Boolean function, where the project succeeds iff all agents succeed.
A computational analysis of these two extreme cases reveals that the optimal contract problem admits a polynomial-time algorithm under the AND Boolean function \citep{BabaioffFN06}, whereas it is NP-hard under the OR Boolean function but admits an FPTAS \citep{EmekF12}.
\citet*{DuettingEFK23} generalize this model by considering a general (monotone) set function $f$ that maps every set of agents who exert effort to a success probability of the project.

We first explore results obtained in the model of \cite{DuettingEFK23}. Afterwards, in Section~\ref{sec:additional-results}, we discuss additional models and results by \cite{CastiglioniEtAl23,CacciamaniEtAl24}, and \cite{DuettingEFK24}.

\paragraph{Model.}

Consider a setting in which a single principal seeks to hire a team of agents from a set of agents $\agents = [n]$ to work on a project.
Every agent has a binary choice of action. He can either 
exert effort or not (be part of the team or not). Agent $i$ incurs a cost $\cost_i \in \reals_{\geq 0}$ for exerting effort. 
We focus here on the binary outcome case, where the project either succeeds, with a principal's reward of $r$, or fails (with 0 reward).
A success probability function, 
$f: 2^\agents \rightarrow [0,1]$, maps every set of agents that exert effort to a success probability.

For this special case, it is without loss of generality to restrict attention to linear contracts (this follows by a slight generalization of Proposition~\ref{prop:linear-with-two-outcomes}).
A linear contract for this setting is given by a vector $\alpha=(\alpha_1, \ldots, \alpha_n)$, where $\alpha_i$ denotes the fraction of the reward that goes to agent $i$ in case the project succeeds. 
In addition, it is again without loss of generality to assume that the principal's reward for success is normalized to $r = 1$.
Fix a contract $\alpha$ and let $S$ be the set of agents that exert effort.
Then,
the principal's utility is given by 
$U_P(S \mid \alpha) := (1-\sum_{i \in \agents} \alpha_i) f(S)$, and agent $i$'s utility is $U_{i}(S \mid \alpha) :=  \alpha_i f(S) - \indicator{i \in S} \cdot c_i$, where $\indicator{i \in S} = 1$ if $i \in S$ and $\indicator{i \in S} = 0$ otherwise. Note that agent $i$ may be paid a non-zero amount even if he does not exert effort.

Every  
contract thus induces a game among the agents; we analyze the (pure) Nash equilibria (possibly more than one) of the game---an action profile from which no agent wishes to deviate.
Namely, a linear contract $\alpha$ incentivizes a set of agents $S$ to exert effort in equilibrium if (i) for every $i \in S$, $\alpha_i f(S) - c_i \geq \alpha_i f(S \setminus \{i\})$ (thus, an agent exerting effort cannot benefit from shirking), and (ii)  for every $i \not\in S$, $\alpha_i f(S) \geq \alpha_i f(S \cup \{i\}) - c_i$ (thus, an agent which currently does not exert effort does not benefit from exerting effort). 
Our benchmark is the best principal utility (a.k.a.~revenue) in any (pure Nash) equilibrium.

\paragraph{Approach/Challenges.}
We pursue an approach in which the principal computes both a contract $\alpha$ and a set of agents $S$ that should exert effort. The interpretation is then that the principal recommends each agent $i \in S$ to exert effort and each agent $i \not \in S$ to \emph{not} exert effort, and following the recommendation 
should be a (pure) Nash equilibrium.

Towards this goal, observe that for a
given a set of agents $S$ it is easy to check whether it can be incentivized, and it is also clear what the $\alpha_i$'s should be in that case. Namely: When $f(i \mid S \setminus\{i\})>0$ then we can incentivize agent $i$ to exert effort, and the  optimal choice of $\alpha_i$ is $\alpha_i = c_i / f(i \mid S \setminus\{i\})$. When $f(i \mid S \setminus\{i\}) = 0$ and $c_i =0$ we can also incentivize agent $i$ to exert effort, and the optimal choice of $\alpha_i$ is $\alpha_i = 0$. Finally, the only case where we can't incentivize agent $i$ to exert effort is when $f(i \mid S \setminus\{i\}) = 0$ and $c_i  > 0$. If we define $0/0 = 0$ and $c/0 = \infty$ 
when $c > 0$ we get that the optimal contract for a set of agents $S$ is 
\[
\alpha_i = \frac{c_i}{f(i \mid S\setminus\{i\})} \quad \text{for $i \in S$} \quad \text{and} \quad \alpha_i = 0 \quad\text{for $i \not\in S$}.
\]

This way the problem of finding the optimal contract reduces to finding the set of agents $S^\star$ that maximizes the function 
$g: 2^{\agents} \to \mathbb{R} \cup \{-\infty\}$ defined by
\[
g(S) := \left( 1 - \sum_{i \in S} \frac{c_i}{f(i \mid S \setminus \{i\})} \right) f(S).
\]

\begin{figure}
\vspace*{-10pt}
\begin{center}
\begin{tikzpicture}
\draw[black, thick,->] (-0.25,0) -- (8,0) node[right] {size of $S$};
\draw[black, thick,->] (0,-0.25) -- (0,4); \node at (-0.8,4) {value};
\def\y{1.3}
\draw[blue,thick,-] (0,0*\y) -- (3,1*\y);
\draw[blue,thick,-] (3,1*\y) --node[above] {\small{\textcolor{blue}{$f$}}} (3.8,1.1*\y);
\draw[blue,thick,-] (3.8,1.1*\y) -- (7,2.25*\y);
\draw[red,thick,-] (0,0*\y) -- (1.5,0.35*\y) -- (3,0.5*\y);
\draw[red,thick,-] (3,0.5*\y) -- (3.5,-0.25*\y);
\draw[red,thick,-] (3.5,-0.25*\y) -- (4,0.4*\y);
\draw[red,thick,-] (4,0.4*\y) --node[above]{\small{\textcolor{red}{$g$}}} (5.5,0.15*\y);
\draw[red,thick,-] (5.5,0.15*\y) -- (7,-0.3*\y);
\end{tikzpicture}
\vspace*{-30pt}
\end{center}
\caption{
An example of an XOS
success probability $f$ that only depends on the size of $S$, and the corresponding expected revenue $g$ under the best contract incentivizing $S$.}
\label{fig:f-vs-g}
\end{figure}
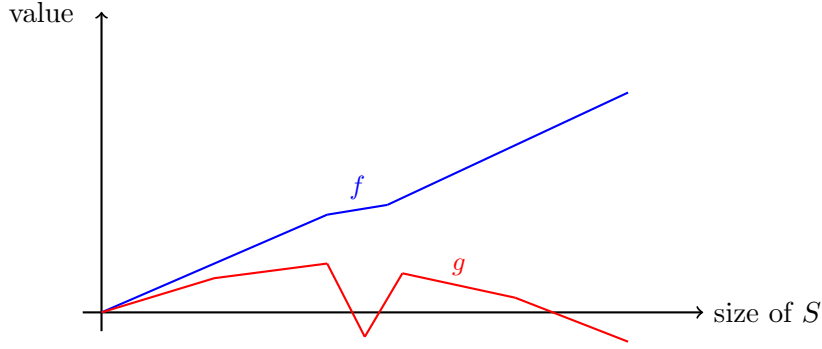

The challenge is now that, even in cases where $f$ is highly structured this structure does \emph{not} necessarily carry over to $g$. 
For example, even in cases where $f$ is non-negative, monotone,  
and submodular (see Definition~\ref{def:classes}), the induced $g$
will usually not be monotone and take negative values. If $f$ is only XOS (an important super-class of submodular valuations, see Definition~\ref{def:classes}), 
$g$ may not even be subadditive. This issue arises even when $f$ depends only on the size of $S$; see Figure~\ref{fig:f-vs-g} for an illustration.

The following example presents a scenario with two identical agents and a submodular function~$f$. It demonstrates that even for a submodular function $f$ over two identical agents, the principal's utility $g$ may be non-monotone and negative.

\begin{example}[Multiple agents with submodular $f$]
Consider a setting with two agents $\mathcal{A} = \{1,2\}$, with costs $c_1=c_2=0.25$, and  with the following submodular success probability function $f$: $f(\{1\})=f(\{2\})=0.5, f(\{1,2\})=0.75$. 
For implementing an equilibrium in which no agent exerts effort, the best contract is $\alpha_1=\alpha_2=0$, for a principal's utility of $0$. For implementing an equilibrium where only agent 1 (resp., agent 2) exerts effort, the optimal contract is $\alpha_1=c_1/f(\{1\})=0.5$ and $\alpha_2=0$ (resp., $\alpha_1=0, \alpha_2=c_2/f(\{2\})=0.5$), for a principal's utility of $(1-\alpha_1)f(\{1\})=(1-\alpha_2)f(\{2\}) = 0.25$. Finally, for implementing an equilibrium in which both agents exert effort, the best contract is $\alpha_1=c_1/(f(\{1,2\})-f(\{2\}))=0.25/(0.75-0.5)=1$ and similarly $\alpha_2=1$, for a principal's utility of $(1-2)f(\{1,2\})<0$. Thus, the optimal contract is either $\alpha_1=0.5, \alpha_2=0$ or $\alpha_1=0, \alpha_2=0.5$, implementing an equilibrium where a single agent exerts effort.
\end{example}

\begin{remark}[Pure vs.~Mixed Nash equilibria]
Focusing on pure Nash equilibria is very natural, but it is \emph{not} without loss. 
For a concrete example in which the principal can achieve a strictly higher utility by inducing a mixed rather than a pure Nash equilibrium, see Example 3.1~in \cite{BabaioffFN10}.
In this example, the success of the project is an OR Boolean function of the agent individual outcomes, which is a submodular success probability function. 
While it is not normalized, it can be easily normalized to yield a submodular function $f$ adhering to our model.
\end{remark}

\paragraph{Positive Results for Complement-Free Rewards.}
\citet{DuettingEFK23} study the problem of computing (near-optimal) contracts under (pure) Nash equilibrium, for the hierarchy of complement-free set functions $f$. They show that, even in the case where $f$ is additive, the optimal contract problem is \textsf{NP}-hard (via a reduction from PARTITION), but admits an FPTAS.
The main result in \citet{DuettingEFK23} is a constant-factor approximation for submodular and XOS functions under suitable oracle access models.

\begin{theorem}[\citet*{DuettingEFK23}]\label{thm:multi-agents}
For both submodular and XOS success probability functions $f$ it is possible to compute a $O(1)$-approximation to the optimal contract with (a) polynomially-many-value queries in the case of submodular $f$ and with (b) polynomially-many-value and demand oracle queries in the case of XOS $f$.
\end{theorem}

Below, we provide a proof sketch for this result, which reveals another (perhaps surprising) connection between contract design 
and prices / demand queries. 

\begin{proof}[Proof sketch for Theorem~\ref{thm:multi-agents}] Let $S^\star$ be a set that maximizes $g$. 
Our goal is to find a set $S$ such that $g(S) \geq O(1) \cdot g(S^\star)$. 
For the purpose of conveying the intuition behind the proof, 
assume in the following that $f(S^\star)$ is known to the algorithm
(but not the set $S^\star$ itself)
and the contribution of a single agent is negligible. The actual proof in \cite{DuettingEFK23} does not need these assumptions.
Also assume that we have access to both value and demand oracles. The actual proof 
shows that, for submodular $f$, value queries suffice.

A key ingredient in the proof is a pair of lemmas.
The first lemma, let's call it Lemma~A, shows that $\sum_{i \in S^\star} \sqrt{c_i f(S^\star)} \leq f(S^\star)$. The other lemma, let's call it Lemma B, 
shows that if for a set $S$ it holds that $f(i \mid S \setminus \{i\}) \geq \sqrt{2 c_i f(S)}$ for every $i \in S$, then $g(S) \geq \frac{1}{2} f(S)$. The first lemma shows that the costs for the optimal set $S^\star$ are not too high. The second lemma shows that if the marginals for each agent $i \in S$ are sufficiently high, then in the optimal contract for set $S$ the principal's expected utility is at least half of $f(S)$. Moreover, the ``not too high'' and ``sufficiently high''  in the two lemmas is in terms of a similar-looking quantity, which involves the square root of an agent's cost times the reward associated with a set of agents.

These observations motivate an approach for finding a ``good'' set $S$ by defining a ``price'' for each agent. 
Namely, imagine that we let $p_i = \frac{1}{2} \sqrt{c_i f(S^\star)}$ for each agent $i$ and consider the \emph{demand set} $T$, which is defined to maximize $f(T) - \sum_{i \in T}p_i$. We now have $f(T) \geq f(T) - \sum_{i \in T} p_i \geq f(S^\star) - \sum_{i \in S^\star} p_i \geq \frac{1}{2} f(S^\star)$ by the definition of a demand set and Lemma A.
By definition, the marginal contribution of every agent in the demand set must exceed its price, that is $f(i \mid T \setminus \{i\}) \geq p_i = \frac{1}{2} \sqrt{c_i f(S^\star)}$.  This condition looks almost like the one that is necessary to invoke Lemma B.
However, note that we only have a lower bound on $f(T)$, no upper bound. Therefore it is possible that $f(T)$ is much larger than $f(S^\star)$. 

To deal with this, \cite{DuettingEFK23} establish a novel scaling property of XOS functions, showing that one can scale down the value of any set $T$ to essentially any level, by removing some of its elements, while keeping the marginals of the remaining elements sufficiently high with respect to their original marginals. Namely, for every set $T$ and every $\Psi < f(T)$, one can compute a subset $U \subseteq T$ such that 
$\frac{1}{2} \Psi \leq f(U) \le \Psi$ and $f(i \mid U \setminus \{ i \}) \geq \frac{1}{2} f(i \mid T \setminus \{ i \})$ for every $i \in U$.
While this property is not too surprising for submodular functions, for XOS functions this is far from obvious, given the apparent lack of control over marginals, and may be of independent interest. 

Let's set $\Psi = \frac{1}{32}  f(S^\star)$. If $\Psi = \frac{1}{32}  f(S^\star) \geq f(T)$, then we know that $f(i \mid T \setminus\{i\}) \geq p_i = \frac{1}{2} \sqrt{c_i f(S^\star)} \geq \sqrt{2 c_i f(T)}$. So Lemma B applied to $T$ shows that $g(T) \geq \frac{1}{2} f(T)$ and therefore $g(T) \geq \frac{1}{2} f(T) \geq \frac{1}{4} f(S^\star)$. 
Otherwise, $\Psi = \frac{1}{32}  f(S^\star) < f(T)$ and we can apply the scaling property to obtain set $U$. It then holds that 
$f(i \mid U \setminus\{i\}) \geq \frac{1}{2} f(i \mid T \setminus \{i\}) \geq \frac{1}{2} p_i = \frac{1}{4} \sqrt{c_i f(S^\star)} \geq \sqrt{2 c_i f(U)}$.
So we can apply Lemma B to $U$ and conclude that 
$g(U) \geq \frac{1}{2} f(U) \geq \frac{1}{128} f(S^\star) \geq \frac{1}{128} g(S^\star)$. We conclude that by either incentivizing $T$ or $U$ we obtain a constant-factor approximation to $g(S^\star)$. 
\end{proof}

Beyond submodular and XOS success probability, the following observation yields a factor-$n$ approximation for subadditive success probability, with polynomially many value queries.

\begin{observation}[Approximation for subadditive]\label{rem:subaddtive}
For subadditive success probability functions $f$, it is possible to compute a $O(n)$-approximation to the optimal contract with polynomially many value queries. 
\end{observation}

\begin{proof}[Proof sketch.]
Observe that for subadditive $f$, for any set of agents $S \subseteq \agents$ such that $g(S)\ge 0$, it holds that $g(\{i\}) \geq 0$ for all $i \in S$ and $g(S) \leq \sum_{i \in S} g(\{i\}) \leq n \cdot \max_{i \in S}g(\{i\})$. The approximation can thus be obtained by (i) computing the best single-agent contract $g(\{i\})$ for each agent $i \in \agents$ and (ii) returning the best such contract.
\end{proof}

\paragraph{Impossibility Results for Complement-Free Rewards.}

\cite{DuettingEFK23}
show two lower bounds that apply to any algorithm that uses polynomially-many demand or value queries. The first result is a constant-factor lower bound for XOS $f$, and the second result is a $\Omega(\sqrt{n})$ lower bound for subadditive $f$. 
More recent work by \citet*{EzraFS24} 
shows that for submodular success
probability functions, there exists a constant $c > 1$ such that no polynomial-time algorithm with value oracle access can approximate
the optimal contract to within a factor
better than $c$, assuming \textsf{P}$\neq$\textsf{NP}.
In addition, for XOS functions, \citet{EzraFS24} show that no
algorithm that makes poly-many value queries can approximate the optimal contract (with high probability) to within a factor 
$\Omega(n^{1/6})$.
More recently, \cite*{DuettingEFK24}, showed that even with both value and demand oracle access to the submodular function, there exists a constant $\eta > 1$, such that any algorithm that uses a sub-exponential number of queries returns an $\eta$-approximation with probability exponentially-small in~$n$ (see Theorem~\ref{thm:submod-no-ptas}).

Together, these impossibility results show that both the positive result of \cite{DuettingEFK23} for submodular $f$ with value oracle access (Theorem~\ref{thm:multi-agents}, Part (a)), as well as the positive result of \cite{DuettingEFK23} for XOS $f$ with value and demand oracle access (Theorem~\ref{thm:multi-agents}, Part (b)), are best possible (up to constant factors). In addition, they identify submodular $f$ and XOS $f$ as the frontier for constant-factor approximation, with value oracle access and with value \emph{and} demand oracle access, respectively.


\begin{table}[t]
\fontsize{10pt}{12pt}
\selectfont
\scalebox{0.9}{
\begin{tabular}{|
>{\columncolor[HTML]{C0C0C0}}c |c
>{\columncolor[HTML]{EFEFEF}}c |
>{\columncolor[HTML]{EFEFEF}}c 
>{\columncolor[HTML]{FFFFC7}}c |}
\hline
\textbf{\begin{tabular}[c]{@{}c@{}}Multiple\\ agents\end{tabular}}    & \multicolumn{2}{c|}{\cellcolor[HTML]{C0C0C0}\textbf{Value Oracle}}                                                                                                                                                          & \multicolumn{2}{c|}{\cellcolor[HTML]{C0C0C0}\textbf{Value and Demand Oracle}}                                                                                                                                     \\ \hline
\multicolumn{1}{|l|}{\cellcolor[HTML]{C0C0C0}}                        & \multicolumn{1}{c|}{\cellcolor[HTML]{C0C0C0}\textbf{\begin{tabular}[c]{@{}c@{}}Upper bound \\ (pos)\end{tabular}}} & \cellcolor[HTML]{C0C0C0}\textbf{\begin{tabular}[c]{@{}c@{}}Lower bound\\ (neg)\end{tabular}}           & \multicolumn{1}{c|}{\cellcolor[HTML]{C0C0C0}\textbf{\begin{tabular}[c]{@{}c@{}}Upper bound \\ (pos)\end{tabular}}} & \cellcolor[HTML]{C0C0C0}\textbf{\begin{tabular}[c]{@{}c@{}}Lower bound\\ (neg)\end{tabular}} \\ \hline
\textbf{Additive}                                                     & \multicolumn{1}{c|}{\cellcolor[HTML]{FFFFC7}\begin{tabular}[c]{@{}c@{}}FPTAS\\ \cite{DuettingEFK23}\end{tabular}}                & \begin{tabular}[c]{@{}c@{}}OPT is\\ NP-hard\end{tabular}                                               & \multicolumn{1}{c|}{\cellcolor[HTML]{EFEFEF}FPTAS}                                                                 & \begin{tabular}[c]{@{}c@{}}OPT is\\ NP-hard\\ \cite{DuettingEFK23}\end{tabular}                            \\ \hline
\textbf{GS}
& \multicolumn{1}{c|}{\cellcolor[HTML]{EFEFEF}\begin{tabular}[c]{@{}c@{}}Constant \\ approx\end{tabular}}            & \begin{tabular}[c]{@{}c@{}}OPT is\\ NP-hard\end{tabular}                                               & \multicolumn{1}{c|}{\cellcolor[HTML]{EFEFEF}\begin{tabular}[c]{@{}c@{}}Constant\\ approx\end{tabular}}             & \cellcolor[HTML]{EFEFEF}\begin{tabular}[c]{@{}c@{}}OPT is\\ NP-hard\end{tabular}             \\ \hline
\textbf{\begin{tabular}[c]{@{}c@{}}Sub-\\ modular\end{tabular}}& \multicolumn{1}{c|}{\cellcolor[HTML]{FFFFC7}\begin{tabular}[c]{@{}c@{}}Constant\\ approx\\ \cite{DuettingEFK23}\end{tabular}}    & 
\multicolumn{1}{c|}{\cellcolor[HTML]{EFEFEF}\begin{tabular}[c]{@{}c@{}}No PTAS\\ 
\cite{EzraFS24}\\
\cite{DuettingEFK24}
\end{tabular}}  
& \multicolumn{1}{c|}{\cellcolor[HTML]{EFEFEF}\begin{tabular}[c]{@{}c@{}}Constant\\ approx\end{tabular}}             & 
\cellcolor[HTML]{FFFFC7}\begin{tabular}[c]{@{}c@{}}No PTAS\\ 
\cite{DuettingEFK24}\end{tabular}  
\\ \hline
\textbf{XOS}                                                          & \multicolumn{1}{c|}{\cellcolor[HTML]{EFEFEF} $O(n)$-approx}                                                                      & \cellcolor[HTML]{FFFFC7}\begin{tabular}[c]{@{}c@{}}No better\\ than $\Omega(n^{1/6})$\\ \cite{EzraFS24}\end{tabular} & \multicolumn{1}{c|}{\cellcolor[HTML]{FFFFC7}\begin{tabular}[c]{@{}c@{}}Constant\\ approx\\ \cite{DuettingEFK23}\end{tabular}}    & \begin{tabular}[c]{@{}c@{}}No PTAS\\ \cite{DuettingEFK23}\end{tabular}                                     \\ \hline
\textbf{\begin{tabular}[c]{@{}c@{}}Sub-\\ additive\end{tabular}}& 
\multicolumn{1}{c|}{\cellcolor[HTML]{FFFFC7}\begin{tabular}[c]{@{}c@{}}$O(n)$-approx\\(Obs.~\ref{rem:subaddtive}) \end{tabular}}
& 
\begin{tabular}[c]{@{}c@{}}No better\\ than $\Omega(n^{1/6})$\end{tabular}                                  & \multicolumn{1}{c|}{\cellcolor[HTML]{EFEFEF} $O(n)$-approx}                                                                      & \begin{tabular}[c]{@{}c@{}}No better\\ than $\Omega(n^{1/2})$\\ \cite{DuettingEFK23}\end{tabular}               \\ \hline \hline
\textbf{\begin{tabular}[c]{@{}c@{}}Super-\\ modular\end{tabular}}     & 
\multicolumn{1}{c|}{\cellcolor[HTML]{FFFFFF}}                                                                     & \begin{tabular}[c]{@{}c@{}}No constant\\ approx\end{tabular}                                           & \multicolumn{1}{c|}{\cellcolor[HTML]{FFFFFF}}                                                                              & \begin{tabular}[c]{@{}c@{}}No constant\\ approx\\(if \textsf{P}$\neq$\textsf{NP})\\ {\footnotesize\citet{VuongDPP23}}\end{tabular}                      \\ \hline
\end{tabular}
}
\caption{
This table presents approximation results for multi-agent contracts {with binary actions and binary outcome.} The left part presents results under access to value oracle, and the right part presents results under access to both value and demand oracles. For each one we present both upper bounds (positive results) and lower bounds (negative results) on the achievable approximation.
The rows represent different reward function classes.
Yellow cells give the results, whereas gray cells represent results derived from other cells (where positive results carry over north (to sub-classes) and east (from value oracle to value and demand oracle), and negative results carry over south and west). 
}
\label{tab:multi-agent}
\end{table}

\paragraph{The Supermodular Case.} Work by \citet*{VuongDPP23} shows additional results for the multi-agent contracting problem when $f$ is supermodular. 

They show that this problem admits no polynomial-time constant-factor multiplicative approximation algorithm nor an additive fully-polynomial time approximation scheme (additive FPTAS\footnote{An additive FPTAS guarantees a solution with value at least $\mathsf{OPT} - \epsilon$, where $\textsf{OPT}$ is the value of the optimal solution, in time polynomial in the input size and $1/\epsilon$. An additive PTAS (see Theorem~\ref{thm:additive-ptas} below) provides the same approximation guarantee, but its running time is only required to be polynomial in the input size.}), assuming $\mathsf{P} \neq \mathsf{NP}$. The hardness applies also with respect to a special case that they call the uniform-cost graph-supermodular contract problem (U-GSC), described next.

The U-GSC problem: The input to this problem is an undirected graph $G = (V,E)$ on $|V| = n$ vertices, and a cost $c > 0$. Each vertex corresponds to an agent. The reward function $f(S)$ for a set of agents $S \subseteq V$ is given by
\[
f(S) := \frac{|E(S)|}{E_{max}},
\]
where $E(S)$ is the set of edges for which both endpoints are contained in $S$ and 
$E_{\max} = \binom{n}{2}$. 
The cost for including a vertex $v \in V$ in $S$ is $c \geq 0$, irrespective of the identity of the vertex, so that the cost of a set of vertices $S$ is $c(S) = |S| \cdot c$. 

\begin{theorem}[\citet*{VuongDPP23}]
For supermodular success probability functions $f$, even when restricting to U-GSC instances, there can be no constant-factor multiplicative approximation algorithm nor an additive FPTAS, assuming $\mathsf{P} \neq \mathsf{NP}$.
\end{theorem}

On the positive side, they show that the U-GSC special case admits an additive polynomial-time approximation scheme (additive PTAS). 
The proof of this result 
establishes a connection to the $k$-densest subgraph problem, and exploits tools developed for that problem.

\begin{theorem}[\citet*{VuongDPP23}]\label{thm:additive-ptas}
The U-GSC problem admits an additive PTAS.
\end{theorem}

A special case of supermodular success probability functions (up to normalization) was previously studied by \cite{BabaioffFN06,BabaioffFNW12}, who presented 
the Boolean AND function, where the project succeeds if and only if all agents succeed in their individual tasks. Note that, in this non-normalized version, an agent may succeed in their individual task even if the agent doesn't exert effort. Among other scenarios, they consider the case of identical agents, where each agent has a binary action and achieves a higher probability of success in their individual task when exerting effort. 
For this scenario, they identify an interesting phase transition: the optimal contract either induces an equilibrium in which no agent exerts effort or one in which all agents do.

\paragraph{Summary and Open Problems.}

We summarize the known results for the multi-agent binary-action model in Table~\ref{tab:multi-agent}. As can be seen from the table, several gaps remain between upper and lower bounds. A particular interesting one is the gap between upper and lower bounds for gross-substitutes (GS) $f$. Here it would be interesting to determine whether the problem of computing an optimal contract admits a PTAS/FPTAS. 
Another interesting direction is to explore the design of contracts that approximately maximize welfare rather than revenue. 
Finally, it would be insightful to explore 
the design of contracts under budget constraints, with respect to both welfare and revenue maximization.

\subsubsection{Additional Directions}
\label{sec:additional-results}

We conclude our discussion of multi-agent contracts with a brief overview of 
additional directions that have been explored.
We first discuss work by \citet*{DuettingEFK24}, who study a joint generalization of the model from this section and the previous section (still with binary outcome). We then discuss work by \citet*{CacciamaniEtAl24}, which explores a multi-agent multi-action model with general $m$-dimensional outcome space. Finally, we discuss work by \cite*{CastiglioniEtAl23}, who study a multi-agent multi-action model, in which each agent's action leads to an individual outcome that is observable by the principal.

\paragraph{Multiple Agents and Combinatorial Actions.}

In the model of \cite*{DuettingEFK24} a principal delegates a project (that can succeed or fail) to a team of agents, each capable of performing any subset of a given set of actions (without loss, the action spaces of the agents can be assumed to be disjoint). A success probability function maps each set of actions to a success probability.
This scenario extends both the single-agent combinatorial-actions setting (of Section~\ref{sec:multiple-actions}) and the multi-agent binary-action setting of this section. 
The main result of \cite{DuettingEFK24} is a constant-factor approximation for submodular success probability, with access to value and demand oracles. 
We note that, since the action spaces of the agents are disjoint, submodularity over actions is well defined.

\begin{theorem}[\cite*{DuettingEFK24}]
\label{thm:pos-multi-multi}
For any submodular success probability function $f$, given access to value and demand oracles, one can compute a contract $\alpha$ such that any equilibrium of $\alpha$ gives a constant approximation to the optimal principal's utility, measured by the optimal equilibrium of any contract.
\label{thm:multi-agent-submod}
\end{theorem}

Note that, this result is quite strong: it compares the {\em worst} equilibrium of the computed contract to the {\em best} equilibrium of any contract.
Also note that, since for gross substitutes $f$, a demand query can be resolved with poly-many value queries, this result implies a constant-factor approximation for  
instances with gross substitutes $f$, with value oracle access only. 

The proof reduces the problem to one of two cases: Either no agent is ``large", or only a single agent is incentivized.
In the former case, they give a constant-factor approximation with access to a value oracle.
In the latter case, they first devise an FPTAS for the single-agent case, with access to value and demand oracles, then extend this to multiple agents losing only a constant factor. 
Notably, the combined problem lacks certain monotonicity properties that are essential for analyzing the previous special cases, and so novel machinery and tools are needed for both cases.  
 
In addition, as mentioned earlier, \cite{DuettingEFK24} 
show that the positive result for submodular success probability functions is best possible (up to constant factors), even in the special case of binary actions and even when considering the best equilibrium under a contract.

\begin{theorem}[\cite*{DuettingEFK24}]
\label{thm:submod-no-ptas}
There exists a constant $\eta > 1$ such that any algorithm that achieves an $\eta$-approximation for the multi-agent combinatorial-action problem with submodular success probability function $f$ must issue exponentially many (value or demand) queries, even in the special case of binary actions and even when considering the best equilibrium under a contract.
\end{theorem}

A natural open problem is whether the constant-factor approximation for submodular functions can be extended to XOS functions, with value and demand oracle access (as in the binary-action case). 
One can show that Theorem~\ref{thm:multi-agent-submod}, in its current form, cannot apply to XOS functions.
In particular, there exists an example with XOS function $f$ such that for any contract the worst equilibrium is a factor $\Omega(n)$ worse than the best equilibrium of the best contract.
Nevertheless, one could still hope for a constant-factor approximation for XOS functions, in a weaker sense: Either aim for a pair of a contract and a recommended equilibrium for that contract that approximates the best equilibrium under any contract (as in Theorem~\ref{thm:multi-agents}), or
change the benchmark to be the best contract measured in terms of worst-case equilibrium.

\paragraph{Free-Riding and Free-Labor in Multi-Agent Contracts.}
\cite*{BabaioffFN09} explore scenarios in which the principal can achieve greater utility by foregoing effort that is freely available.
Obviously, such scenarios strike us as counterintuitive because there is unutilized ``free-labor''---the principal prefers that some agents will not participate despite the fact that their labor increases the probability of success with no additional cost. Yet, free labor increases \emph{free riding}, resulting in a lower utility for the principal overall, since increased effort of some agents may significantly increase the cost of incentivizing others to work.
This is demonstrated in the following example.

\begin{example}[Free labor decreases principal's utility]
    \label{ex:free-ride}
    Consider a setting with two agents, where each agent can exert effort or not, and suppose that, when exerting effort, agent $1$ succeeds in his own task with probability $p_1$, agent $2$ succeeds in his own task with probability $p_2$, and the project succeeds iff at least one of the agents succeeded. The induced success probability function is $f(\emptyset)=0, f(\{1\})=p_1, f(\{2\})=p_2$, and $f(\{1,2\})=1-(1-p_1)(1-p_2)$. 
    Suppose further that the costs of effort are $c$ for agent $1$ and $0$ for agent $2$, and normalize the principal's reward from the project's success to $1$. Since agent $2$'s cost is $0$, he has no reason to shirk.
    
    Given that agent $2$ exerts effort, in order to incentivize agent $1$ to exert effort, the payment to agent $1$ upon success of the project, denoted $\alpha$, should satisfy
    $\alpha (1-(1-p_1)(1-p_2))-c \geq \alpha p_2$.
    Thus, $\alpha = \frac{c}{p_1(1-p_2)}$ is the best way to incentivize agent $1$ to exert effort in this case.
    The principal's utility is then $(1-(1-p_1)(1-p_2))(1-\frac{c}{p_1(1-p_2)})$.

    Now suppose that agent $2$ does not exert effort. Then, in order to incentivize agent $1$ to exert effort, it should hold that $\alpha p_1 - c \geq 0$. That is, the best way to incentivize agent $1$ to exert effort is via $\alpha=\frac{c}{p_1}$. The principal's utility is then $p_1(1-\frac{c}{p_1})$.

    Consider the case where $p_1=0.6, p_2=0.3$, and $c=0.2$. Then, in the former case, where agent $2$ works for free, the principal's utility is $\approx 0.377$, while in the latter case, where agent $2$ does not work, the principal's utility is $0.4$. Thus, the principal gains utility by foregoing free labor by agent 2. Note also that the principal's utility in the latter case is greater than enjoying agent $2$'s effort for free, which would yield a principal's utility of $p_2=0.3$.
    


\end{example}

Such scenarios raise the question of which success probability functions may give rise to this phenomenon, where free labor is effectively wasted; namely, situations in which the principal prefers that some agents refrain from participating, even when their work increases the probability of success with no additional cost.

\cite{BabaioffFN09} find that for success probability functions that exhibit ``increasing returns to scale" (essentially, super-modularity, where the marginal contribution of any action is non-decreasing in the effort of the other agents), there exists an optimal contract that does not waste free labor.
Moreover, for the special case of Boolean-function induced success probability functions (where every agent succeeds or fails in his own task and the success of the entire project is obtained from a Boolean function that maps individual success and failures into a success of the project), they show that the AND Boolean function (where the project succeeds iff all agents succeed in their individual tasks) is, in some technical sense, a maximal class that does not waste free labor. In particular, for any other Boolean function (that is not constant), there exist parameters where any optimal contract wastes free labor. 

This model and results raise intriguing algorithmic and computational problems for future research, including determining the optimal level of free labor, analyzing its potential impact, and addressing fairness concerns related to the free-riding behavior that may emerge.

\paragraph{Multiple Agents and Randomized Contracts.} 
\citet*{CacciamaniEtAl24} consider a very general (explicitly represented) 
multi-agent multi-action contracting problem with non-binary outcome. 
In their model, there are $n$ agents, each of which can take one of $\ell$ actions. Each agent has a cost for each action. Each action profile (of which there are up to $\ell^n$ many) 
induces a probability distribution over $m$ outcomes. The principal has a reward for each outcome. It is assumed that the rewards, costs, and probability matrices are given explicitly. So overall, the input consist of $O(\ell^n m)$ numbers. 

A main innovation of \cite{CacciamaniEtAl24} is that they introduce a natural class of randomized contracts, and an associated equilibrium concept for this class of contracts. Informally, they define a randomized contract to consist of (1) a joint distribution over recommended action profiles, and (2) for each action profile one classic contract 
for each agent. They then look for an equilibrium, in which no agent can benefit from deviations of the form \emph{whenever being recommended action $a_i$, play some other action $a'_i$}.
Note how randomized contracts encompass deterministic contracts as a special case. Also note how in the deterministic case the equilibrium notion coincides with that of a pure Nash equilibrium, while in the randomized case it is similar to a  \emph{correlated equilibrium} \cite[e.g.,][Chapter 13.1.4]{Roughgarden16}.

Since \cite{CacciamaniEtAl24} work with an explicitly represented model, an optimal deterministic contract can be found in time polynomial in the input size (via linear programming, by finding the optimal classic contract for each action profile). Their study thus focuses on two questions: (1) How much better are randomized contracts as opposed to deterministic contracts? (2) Can (near-)optimal randomized contracts be found in polynomial time?

\medskip 
\noindent \emph{The Model.} More formally, in their model a single principal interacts with $n$ agents. Each agent $i$ has a finite set of (unobservable) actions $\mathcal{A}_i$, with $\ell := \max_i |\mathcal{A}_i|$. There is a finite set of outcomes $\Omega$, with $|\Omega| = m$. Each action profile $\mathbf{a} \in \mathcal{A} := \mathcal{A}_1 \times \ldots \times \mathcal{A}_n$ is associated with a probability distribution $\Prob_{\mathbf{a}}$ over outcomes $\omega \in \Omega$, with $\prob_{\mathbf{a},\omega}$ denoting the probability of outcome $\omega$ under action profile $\mathbf{a}$. Each action $a \in \mathcal{A}_i$ comes with a cost of $c^i_{a} \in [0,1]$ to agent $i$. Each outcome $\omega \in \Omega$ is associated with a reward $r_\omega \in [0,1]$, which goes to the principal.

A \emph{randomized} contract is a tuple $(\mu,\pi)$, where $\mu$ is a probability distribution over action profiles $\mathbf{a} \in \mathcal{A}$ (i.e., recommendations) and $\pi = (\pi^i_{\mathbf{a}})_{i \in [n], \mathbf{a} \in \mathcal{A}}$ is a tuple of payment functions $\pi^i_{\mathbf{a}}: \Omega \rightarrow \reals_{\ge 0}$. 
The interpretation is that, when the principal recommends action profile $\mathbf{a}$, then $\pi^i_{\mathbf{a}}(\omega)$ is the payment to agent $i$ when outcome $\omega$ is realized. A \emph{deterministic} contract is a randomized contract, which  puts probability $\mu(\mathbf{a}) = 1$ on a single action profile $\mathbf{a} \in \mathcal{A}$.

We can now define $U_i(a_i \rightarrow a'_i \mid (\mu,\pi)) := \sum_{\mathbf{a}_{-i} \in \mathcal{A}_{-i}} \mu(a_i,\mathbf{a}_{-i}) \sum_{\omega \in \Omega} \prob_{(a'_i,\mathbf{a}_{-i}),\omega} \pi^i_{(a_i,\mathbf{a}_{-i})}(\omega) - c^i_{a'_i}$ as the (unnormalized) utility of agent $i$ for choosing action $a'_i$ when being recommended action $a_i$. 
Note that the choice of action $a'_i$ impacts the probability distribution over outcomes, but not the payment function. The equilibrium requirement is that $U_i(a_i \rightarrow a_i \mid (\mu,\pi)) \geq U_i(a_i \rightarrow a'_i \mid (\mu,\pi))$ for all $i \in [n]$ and all $a_i, a'_i \in \mathcal{A}_i$. 

The principal's goal is to design a contract $(\mu,\pi)$ that is an equilibrium, and maximizes the principal's expected payoff $U_P(\mu,\pi) = \sum_{\mathbf{a} \in \mathcal{A}} \mu(\mathbf{a}) \sum_{\omega \in \Omega} \prob_{\mathbf{a},\omega} \left(r_\omega - \sum_{i \in [n]} \pi^i_\mathbf{a}(\omega)\right)$. 

\medskip
\noindent \emph{Key Results.}
An important qualitative insight of \cite{CacciamaniEtAl24} is that the gap between randomized and deterministic contracts can be unbounded.
For a fixed instance of the contracting problem, denote by $\OPT_R$ and $\OPT_D$ the optimal principal utility under randomized and deterministic contracts, respectively. Then there is an instance with two agents, two outcomes, and two actions per agent such that no deterministic contract can achieve a positive utility, while there is a randomized contract with strictly positive utility. The instance has success/failure structure (so one outcome has reward zero, while the other has a positive reward), and the success probability is supermodular. 
We thus have:

\begin{proposition}[\citet*{CacciamaniEtAl24}]
There is an instance of the multi-agent contract problem with two agents, two outcomes, and two actions per agent  
such that $\OPT_R/\OPT_D = \infty$.
\end{proposition}

Motivated by this result, \cite{CacciamaniEtAl24} explore whether it's possible to compute optimal randomized contracts in polynomial time. 
They first observe that the problem of finding an optimal randomized contract can be cast as a quadratic program; and that, in general, this program (and hence the problem) only admits a supremum and not a maximum. 
They then present an algorithm, which for any $\varepsilon > 0$ returns a $(1+\varepsilon)$-approximate randomized contract.
This algorithm solves a linear relaxation of the quadratic program that defines the optimal contract, and converts the solution of the relaxed problem into an arbitrarily close-to-optimal solution of the original problem.

\begin{theorem}[\citet*{CacciamaniEtAl24}]
For any fixed $\varepsilon > 0$, there is an algorithm that runs 
in time polynomial in 
$\ell^n$, $m$, and $\log(\nicefrac{1}{\varepsilon})$, and finds a $(1+\varepsilon)$-approximate randomized contract.
\end{theorem}

\noindent\emph{Additional Results.} The paper of \cite{CacciamaniEtAl24} contains a number of additional results, including extensions of the aforementioned results to Bayesian settings, where agents have private types that determine the agents' cost functions and probability distributions over outcomes. We refer the reader to the paper for details, and return to typed contract settings in Section~\ref{sec:types}.

\paragraph{Multiple Agents with Observable Individual Outcomes.}
\citet*{CastiglioniEtAl23} explore
multi-agent contracts in a different, incomparable setting.
In their model, the agents' actions lead to an individual outcome, \emph{observable} by the principal, and the principal's reward is a combinatorial function of the agents' individual outcomes. 
The ability to observe an agent's individual outcome 
gives the principal additional contracting power, as contracts can now depend on the individual outcomes rather than just the joint outcome of the agents' actions. \citet{CastiglioniEtAl23} explore the computational aspects of exercising this additional power, in settings which exhibit economies of scale or diseconomies of scale, corresponding to IR-supermodular and DR-submodular rewards (see Definition~\ref{def:scale}).

\medskip
\noindent\emph{The Model.} 
The problem is as follows. There is a single principal, which interacts with $n$ agents. Each agent can take any action from an action set $\mathcal{A}$ of size $|\mathcal{A}| = \ell$. In addition, there is an (individual) outcome space $\Omega$,  assumed to be a subset of $\reals^q_{\ge 0}$ of dimension $q \in \mathbb{N}_{>0}$, with $|\Omega| = m$. The interpretation is that each agent $i$ takes an (unobservable) action $a_i \in \mathcal{A}$, and for each agent $i$ this stochastically leads to an observable outcome $\omega_i \in \Omega$.
Formally, each agent-action pair $(i,a) \in [n] \times \mathcal{A}$ is associated with a distribution $\Prob^i_{a}$ over individual outcomes $\omega \in \Omega$. If agent $i$ takes action $a$, then outcome $\omega$ is realized independently with probability $\prob^i_{a,\omega}$. Each agent-action pair $(i,a)$ is associated with a cost $c^i_{a} \in [0,1]$. The principal has a reward function $r: \Omega^n \rightarrow [0,1]$, which maps vectors of individual outcomes $\boldsymbol\omega = (\omega_1, \ldots, \omega_n)$ to a reward. The expected reward of an action profile $\mathbf{a} \in \mathcal{A}^n$ is given by $R_{\mathbf{a}} = \sum_{\boldsymbol\omega} r(\boldsymbol\omega) \prod_{i \in [n]} \prob^i_{a_i,\omega_i}$.

A contract $(\con^1, \ldots, \con^n)$ consists of $n$ classic contracts $\con^i$, one for each agent $i$, specifying a payment $\pay^i_{\omega}$ for each (individual) outcome $\omega \in \Omega$. 
Agent $i$'s expected payment under classic contract $\con^i$ for action $a \in \mathcal{A}$ is $\Pay^i_{a} = \sum_{\omega} \prob^i_{a,\omega} \pay^i_{\omega}$, and his utility is $U_i(a \mid \con_i) := \Pay^i_{a} - c^i_{a}.$ An action $a \in \mathcal{A}$ is a \emph{best response} of agent $i$ to contract $\con_i$ if it maximizes the agent's utility among all actions.
Given a contract $(\con^1, \ldots, \con^n)$ and an action profile $\mathbf{a} \in \mathcal{A}^n$, denote the overall expected payment of the principal to the agents by $\Pay_{\mathbf{a}} := \sum_{i \in [n]} \Pay^i_{a_i}$.  The principal's expected utility for contract $(\con^1, \ldots, \con^n)$ and action profile $\mathbf{a} \in \mathcal{A}^n$ is $U_P(\mathbf{a} \mid (\con^1, \ldots, \con^n)) := R_{\mathbf{a}} - T_{\mathbf{a}}$. The principal's goal is to find a contract $(\con^1, \ldots, \con^n)$ and a recommended action profile $\mathbf{a} \in \mathcal{A}^n$ such that for each agent $i$ action $a_i$ is a best response to $\con_i$, that maximizes the principal's expected utility among all such contract and action profile pairs. 

\cite{CastiglioniEtAl23} explore this problem, focusing on monotone reward functions $r$, such that $r(\boldsymbol\omega) \geq r(\boldsymbol\omega')$ whenever $\boldsymbol\omega \geq \boldsymbol\omega'$, under decreasing or increasing marginal returns, as captured by the following definition.

\begin{definition}[DR-submodular and IR-supermodular]\label{def:scale}
A reward function $r: \Omega^n \rightarrow [0,1]$ is \emph{decreasing-return submodular} (DR-submodular) if for all $\boldsymbol\omega, \boldsymbol\omega', \boldsymbol\omega'' \in \Omega^n$ with $\boldsymbol\omega \leq \boldsymbol\omega'$ it holds that
\[
r(\boldsymbol\omega + \boldsymbol\omega'') - r(\boldsymbol\omega) \ge  r(\boldsymbol\omega' + \boldsymbol\omega'') - r(\boldsymbol\omega').
\]
It is \emph{increasing-return supermodular} (IR-supermodular) if the inequality is reversed.
\end{definition}

\noindent\emph{Key Results.} \cite{CastiglioniEtAl23} present results for both settings, the IR-supermodular case and the IR-submodular case. For the IR supermodular case, they show an interesting separation between instances that satisfy (a suitable generalization of) first-order stochastic dominance (FOSD) (see Definition 6 of their paper), and those that don't. 

\begin{theorem}[\citet*{CastiglioniEtAl23}]
    For IR-supermodular rewards: 
    \begin{itemize}
    \item For any constant $\rho > 1$, it is $\mathsf{NP}$-hard to compute a $\rho$-approximation to the optimal principal utility with value oracle access to the reward function, even when both the number of outcomes $m$ and the number of actions $\ell$ are fixed.
        \item For instances satisfying FOSD, there is a poly-time algorithm for computing an optimal contract with value oracle access to the reward function.
    \end{itemize}
\end{theorem}

The negative result for IR-supermodular rewards is obtained via a reduction from the \textsf{LABEL-COVER} problem. The positive result is obtained through a reduction to an optimization problem over matroids, and by showing that FOSD implies a particular structure on the objective function (called ordered-supermodularity) which is known to admit a poly-time algorithm.

For DR-submodular rewards, \cite{CastiglioniEtAl23} show a strong impossibility result, ruling out any sublinear (in the number of agents $n$) approximation. They complement this negative result with a positive result
that holds with respect to a weaker benchmark.

\begin{theorem}[\citet*{CastiglioniEtAl23}]
For DR-submodular rewards:
\begin{itemize}
    \item For any constant $\gamma > 0$, it is $\mathsf{NP}$-hard to compute a $n^{1-\gamma}$ approximation with value oracle access to the reward function, even when both the number of actions $\ell$ and the dimension $q$ of the outcome space are fixed.
    \item There is a poly-time algorithm, that, for any $\varepsilon > 0$, with value-oracle access to the reward function, computes a contract with principal utility at least 
$
(1-\nicefrac{1}{e}) R^\star - T^\star - \varepsilon,
$ 
where $R^\star$ and $T^\star$ are the expected reward and payment of an optimal contract.
\end{itemize}
\end{theorem}

The negative result for DR-submodular rewards is shown by reduction from \textsf{INDEPENDENT-SET}. The positive result is again established through a reduction to an optimization problem over matroids. In this case the objective function is submodular, but neither monotone nor non-negative. To circumvent these challenges, the authors show how the objective function can be decomposed into the sum of a (monotone, non-negative) submodular function and a linear one, and apply algorithms for such objective functions.

\medskip
\noindent{{\emph{Additional Results.}}} In addition, \cite{CastiglioniEtAl23} provide results for the multi-agent problem with observable individual outcomes in a Bayesian setting, where each agent has a private type that determines their cost and probability matrix. We discuss contracting problems with hidden types in Section~\ref{sec:types}.
\subsection{Combinatorial Outcomes}
\label{sub:combi-outcomes}

In this section, we explore 
the model introduced by \citet*{DuttingRT21}, in which the classic principal-agent problem is allowed to have a complex \emph{outcome space}---with exponentially-many outcomes, and combinatorial structure that enables its succinct description. The computational challenge is to compute an optimal or near-optimal contract. The algorithmic results mirror a recurring theme in 
the emerging literature on combinatorial contracts.
In settings where the optimal contract is (in some sense) simple, it is tractable to compute or closely approximate it, while in general, even approximation is computationally hard. 

\paragraph{Motivation.} 

There is a well-known principle in classic contract theory called the \emph{informativeness principle}~\citep{Holmstrom79,Shavell79}, stating that any informative signal (even if noisy) is valuable, in the sense that it allows the principal to write a better contract. According to this principle, it is worth incorporating fine-grained outcomes into the contract whenever these are available. A main advantage of modern computerized contracts is that they make it increasingly feasible to pay for performance where performance is measured on \emph{multiple dimensions}. 
For example, imagine a machine-learned assessment of an agent's quality of work---this will naturally depend on a \emph{combination} of multiple features.\footnote{We elaborate on the machine learning connection in Section~\ref{sec:incentive-aware}.}
Further motivation comes from real-world applications. For example, in revenue-sharing contracts between platforms like YouTube and content creators,
``the amount of money YouTube pays depends on a variety of factors'' including views, clicks, audience features, ad quality measures, etc.~\citep[e.g.,][]{Dunn24}.

The best contract for incentivizing high-quality work in such settings will pay the agent based on a combination of performance measures. 
The model of \citet{DuttingRT21} formalizes this idea by using a combination of performance measures as the contract's outcome. 
While introducing more complexity, this generalization of the classic model can lead to significantly better incentives for the agent, motivating its study. 

\paragraph{Model and Examples.}

The multi-outcome model is based on the standard contract design model from Section~\ref{sec:model}, but with an outcome space of size $\mu=2^m$, structured as follows: There are $m$ \emph{dimensions} on which the agent can succeed or not. An \emph{outcome} $S\in 2^{[m]}$ is given by the set $S \subseteq [m]$ of dimensions on which the agent succeeds. 
We refer to such contract settings as having an $m$-dimensional outcome space (where $m$ is logarithmic in the actual outcome space size $\mu$). For concreteness we provide two examples:

\begin{enumerate}
    \item Each dimension $j\in[m]$ represents an \emph{item}, which could either be sold or not by a sales agent selling the principal's products. Outcome $S$ is then the bundle of items successfully sold by the agent.%
    \footnote{This example resembles multi-item/combinatorial \emph{auctions}, but here the agent exerts effort to sell items on behalf of the principal.} See also Example~\ref{ex:comb-outcome}.
    \item Each dimension $j\in[m]$ represents a desired \emph{skill}, which could either be possessed or not by a job candidate found for the principal by a headhunter agent. Outcome $S$ is then the skill set of the candidate found by the agent. 
\end{enumerate}

The rest of the notation is as usual, but with $S$ replacing $j$ as an outcome: Action $i\in [n]$ has cost $c_i$ and induces a distribution $\Prob_i$ over the $\mu$ outcomes, where $q_{i,S} \in [0,1]$ for every $S$, and $\sum_S q_{i,S} = 1$. The principal derives a reward $r_S\le 1$ when the realized outcome is $S\in 2^{[m]}$, and the contract determines a payment $t_S$. The goal is to design a contract that maximizes the principal's expected utility (a.k.a.~expected revenue), 
namely $U_P(i \mid \con) :=\; R_i-T_i=\mathbb{E}_{S \sim \Prob_i}[r_S]-\mathbb{E}_{S \sim \Prob_i}[t_S]$, where $i$ is the action chosen by the agent to maximize his expected utility 
$U_A(i \mid \con) =\; T_i-c_i$. As usual, we assume tie-breaking in favor of the principal. 
Returning to our examples: 

\begin{enumerate}
    \item In the sales agent example,  
actions represent marketing
strategies of the agent. Each marketing
strategy leads to a distribution over the set of items sold. The payment depends on the bundle of sold items.

\item In the headhunter agent example, actions represent search strategies for finding candidates.
Each search strategy leads to a distribution over the skill set of the candidate. The payment depends on the skill set. 
\end{enumerate}

\paragraph{Computational Problem.}

Consider the computational complexity of finding the optimal contract. 
In full generality, the rewards $\{r_S\}$ and probabilities $\{q_{i,S}\}$ of every distribution $\Prob_i$ have an exponential-in-$m$ description size, since they require one value for each set 
$S\in 2^{[m]}$. Thus the explicit/na\"ive problem description is of size polynomial in $n$ and exponential in $m$. There can also be exponentially-many contractual payments $\{t_S\}$, making the solution size exponential as well. 

A crucial observation is that since the number of actions is $n$, there is an optimal contract with at most $n-1$ non-zero payments---this is an immediate implication of Observation~\ref{obs:bfs}. 
Thus, if the rewards and distributions have combinatorial structure that allows for a succinct description of size $\textsf{poly}(n,m)$ (logarithmic in the size $\mu$ of the outcome space),  
the computational problem becomes:
\emph{Is it possible to compute an optimal or near-optimal contract in time polynomial in $n,m$?} 

\paragraph{Succinct Representation of Rewards and Distributions.}

Consider first the reward function $r(\cdot)$ mapping outcomes $S$ to their rewards $r_S$, and impose a natural combinatorial structure like additive, GS, or submodular on $r$ (see Section~\ref{sec:set-funct-oracles} for definitions). 

The combinatorial structure allows for succinct representation:
For example, if the reward function $r$ is additive, this means that for every dimension~$j$ the principal gets reward $r_j$ whenever the agent succeeds in this dimension, and the total reward for success in $S$ is $r(S) = \sum_{j \in S} r_j$. Thus, $r$ has a succinct representation by $m$ values $r_1,\dots,r_m$.  If $r$ is GS or submodular, we assume standard oracle representation (see Section~\ref{sec:set-funct-oracles}). 

For the distributions $\{\Prob_i\}$,
we assume that for every action $i$ there is an independent probability $q_{i,j}$ for succeeding in dimension $j$. 
Thus $\Prob_i$ is a product distribution with a succinct representation by its marginals ${\bf x}_i:=(q_{i,1},\dots,q_{i,m})$.
Since we are working with product distributions, it will be convenient to use the following notation: The probability $q_{i,S}$ of succeeding in a set of dimensions $S$ is the product of $q_{i,j}$ for every $j\in S$ and of $(1-q_{i,j})$ for every $j\notin S$; using the notation in Equation~\eqref{eq:multilinear},
\begin{equation}
q_{i,S}=q_{i,S}({\bf x}_i).\label{eq:prob-multi-outcome}
\end{equation} 
The next observation follows directly from the definition of multilinear extensions (Section~\ref{sec:set-funct-oracles}):

\begin{observation}
\label{obs:R-by-multilinear}
    Given a reward function $r$ with a multilinear extension $R$, and a product distribution $\Prob_i$ over $2^{[m]}$ with marginals ${\bf x}_i$, 
    the expected reward $\Rew_i$ is equal to $R({\bf x}_i)$. 
\end{observation}

In the remainder of this section, when we refer to a contract setting with an $m$-dimensional outcome space, we mean one with succinctly-represented reward function and product distributions. 

Below, we give an example of a succinctly representable principal-agent setting with combinatorial outcomes. In this example there is a sales agent, which tries to sell $m = 2$ items on behalf of the principal. The principal has additive rewards over the four possible outcomes (no item is sold, only item $1$ is sold, only item $2$ is sold, both items are sold). The selling probabilities depend on the agent's level of effort, but are independent across items.

\begin{example}[Succinctly representable setting with $2$-dimensional outcome space]
\label{ex:comb-outcome}
In the following example, there are $n=3$ actions, $m=2$ dimensions, and $\mu=2^m=4$ outcomes. These may represent, e.g., $3$ possible effort levels of a sales agent for selling $2$ items, with the $4$ outcomes being $\emptyset$ (no item sold), $\{1\}$ (item $1$ sold), $\{2\}$ (item $2$ sold), and $\{1,2\}$ (both items sold):

\begin{center}
\begin{tabular}{|l|cccc|c|}
\toprule
& outcome $1$ & outcome $2$ & outcome $3$ & outcome $4$ & \text{cost}\\
& $r_{\emptyset} = 0$ & $r_{\{1\}} = 3$ & $r_{\{2\}} = 7$ & $r_{\{1,2\}}=10$ & \\
\midrule 
action $1$: & $0.72$ & $0.18$ & $0.08$ & $0.02$ & $\cost_1 = 0$\\
action $2$: & $0.12$ & $0.48$ & $0.08$ & $0.32$ & $\cost_2 = 1$\\ 
action $3$: & $0$ & $0.4$ & $0$ & $0.6$ & $\cost_3 = 2$\\
\bottomrule
\end{tabular}
\end{center}

\noindent Since the reward function mapping every outcome $S$ to reward $r_S$ is additive, and since the rows contain product distributions, there is an equivalent succinct representation:

\begin{center}
\begin{tabular}{|l|cc|c|}
\toprule
& success in dim $1$ & success in dim $2$ & \text{cost}\\
& $r_1 = 3$ & $r_2 = 7$ & \\
\midrule 
action $1$: & $0.2$ & $0.1$ & $\cost_1 = 0$\\
action $2$: & $0.8$ & $0.4$ & $\cost_2 = 1$\\ 
action $3$: & $1$ & $0.6$ & $\cost_3 = 2$\\
\bottomrule
\end{tabular}
\end{center}

\noindent In the lower table, each column corresponds to success in one of the dimensions (selling one of the items), and contains the reward and probability for such success given each action. 
Notice that the representations are indeed equivalent: For every outcome $S\subseteq \{1,2\}$ (bundle of items sold), reward $r_S$ in the upper table is equal to $\sum_{j\in S} r_j$ where $r_1,r_2$ appear in the lower table. Also, the probability $\prob_{i,S}$ in the upper table is equal to the product of $\prob_{i,j}$ for every $j\in S$ and $(1-\prob_{i,j})$ for every $j\notin S$. 
\end{example}

\paragraph{Linear Contracts: An Impossibility.} 

While linear contracts are known to be optimal for the binary-outcome setting (Proposition~\ref{prop:linear-with-two-outcomes}), this is the limit of their optimality: they are no longer optimal even for generalized binary-outcome settings, in which there are two outcomes with non-zero rewards (not even if there are only two actions---see Example~\ref{ex:equal-rev}). There is also no hope that linear contracts will perform \emph{near}-optimally for our settings of interest in this section which have $m$-dimensional outcomes. The reason is that the construction in Example~\ref{ex:equal-rev} can be interpreted as the succinct representation of a setting with an $m$-dimensional outcome space (in which each action leads deterministically to success in exactly one dimension). It is known that for this setting, the best linear contract can achieve no better than an $\Omega(n)$-approximation to the optimal expected revenue (Theorem~\ref{thm:wc-approx}). 
    
It is worth noting that if the principal     restricts attention to the (sub-optimal) class of linear contracts (say, to gain robustness to distributional details---see Theorem~\ref{thm:max-min-moments}), then she can compute the optimal linear contract in \textsf{poly}$(n)$-time even with $m$-dimensional outcomes. The reason is that towards finding the optimal linear contract, the number of outcomes doesn't matter---only the expected reward $R_i$ of each action $i$ plays a role (Observation~\ref{obs:agnostic}). 
Thus, with either additive $r$ or oracle access to the multilinear extension of  $r$, the poly-time algorithm in Section~\ref{sec:linear-properties} for optimal linear contracts can be applied to $m$-dimensional outcomes.

\paragraph{Warm-up: A Positive Result for Generalized Binary-Action.} 

The optimal contract is known to have a simple form in the generalized binary-action case (see Section~\ref{sec:optimal-contracts}), where recall there are two non-trivial actions (actions $2$ and $3$) in addition to a trivial ``opt out'' action (action $1$). Assuming one of the non-trivial actions $i\in\{2,3\}$ is implementable, the optimal contract that implements $i$ has a single non-zero payment for an outcome with maximum likelihood-ratio, and the payment is straightforward to compute (Proposition~\ref{pro:opt-two-actions}). Thus, computing the optimal contract implementing $i$ in the generalized binary-action case with $m$-dimensional outcomes boils down to finding a set $S^\star$ that maximizes $\frac{\prob_{i,S}}{\prob_{i',S}}$ for $i'\in\{2,3\} \setminus\{i\}$. 
\citet{DuttingRT21} observe that while there are exponentially-many possibilities, 
it is possible to find $S^\star$ in polynomial time.  
This follows since with product distributions, the likelihood ratio~is
\begin{equation}
\frac{\prob_{i,S}}{\prob_{i',S}} = \frac{q_S({\bf x}_i)}{q_S({\bf x}_{i'})} = \prod_{j\in S}{\frac{\prob_{i,j}}{\prob_{i',j}}} \prod_{j\notin S}{\frac{1-\prob_{i,j}}{1-\prob_{i',j}}},\label{eq:prod-of-ratios}
\end{equation}
where the first equality is by Equation~\eqref{eq:prob-multi-outcome} above, and the second is by Equation~\eqref{eq:multilinear} in Section~\ref{sec:set-funct-oracles}.
By Equation~\eqref{eq:prod-of-ratios}, the likelihood ratio is maximized by taking $S$ to be the collection of every item $j$ such that $\prob_{i,j}\ge \prob_{i',j}$ (equivalently, \emph{not} taking every item $j$ such that $1-\prob_{i,j}> 1-\prob_{i',j}$). 

There is one subtle point: As we have now seen, finding the contract $\con$ that incentivizes action $i\in\{2,3\}$ with minimum expected payment can be done in polynomial time. However, computing the expected revenue $R_i-T_i$ from this contract (to identify the best overall contract) requires evaluating the expected reward $R_i$. By Observation~\ref{obs:R-by-multilinear}, this is equivalent to evaluating the multilinear extension of the reward function $r$. 
With polynomially-many value queries, 
an exact evaluation is achievable for additive~$r$, while for general $r$ the evaluation is up to arbitrary precision (see Section~\ref{sec:set-funct-oracles}). For simplicity, we ignore the small evaluation error by assuming oracle access to the multilinear extension as in~\citep{Shioura09} (without oracle access we would get an arbitrarily-close approximation to the optimal contract, with high probability). The next proposition summarizes the generalized binary-action case:

\begin{proposition}[\citet*{DuttingRT21}]
\label{pro:multi-outcome-binary}
Consider a (generalized) binary-action contract setting with an $m$-dimensional outcome space. If the reward function $r$ is additive, the optimal contract can be found in time polynomial in 
$m$. 
The same holds for general $r$ assuming oracle access to its multilinear extension.
\end{proposition}

\paragraph{A Positive Result for Constantly-Many Actions.}

\begin{figure}[t]
\begin{center}
\begin{subfigure}[t]{0.45\textwidth}
\begin{align*}
\min_{t_S:\;S\subseteq[m]}\quad &\sum_{S} \prob_{i,S} \pay_S\\
\text{s.t.}\quad &\sum_{S} \prob_{i,S} \pay_S - \cost_{i} \geq \sum_{S} \prob_{i',S} \pay_S - \cost_{i'} &&\forall i' \neq i  \\
&t_S \geq 0 &&\forall S 
\end{align*}
\caption{\textsf{MINPAY-LP($i$)}}
\end{subfigure}
\begin{subfigure}[t]{0.45\textwidth}
\begin{align*}
\max_{\substack{\lambda_{i'}:\\ i' \in [n]\setminus\{i\}}}\quad &\sum_{i' \neq i} \lambda_{i'}(\cost_i-\cost_{i'})\\
\text{s.t.}\quad &\sum_{i' \neq i} \lambda_{i'} (\prob_{i,S} - \prob_{i',S}) \leq \prob_{i,S} &&\forall S\\
     &\lambda_{i'} \geq 0 &&\forall i' \neq i
\end{align*}
\caption{\textsf{DUAL-MINPAY-LP($i$)}}
\end{subfigure}
\end{center}
\caption{The \textsf{MINPAY-LP($i$)} for action $i$ {\bf (left)} and its dual {\bf (right)} for the multi-outcome model.} 
\label{fig:lp-pair}
\end{figure}

Beyond the generalized binary-action case, the optimal contract is no longer necessarily simple. Perhaps surprisingly, it is still possible to get a positive result for computing the (near-)optimal contract (see Theorem~\ref{thm:multi-outcome-pos} below). This will be achieved by using a more sophisticated algorithm and slightly relaxing the constraints from IC to $\varepsilon$-IC (see Section~\ref{sub:regularity}).

As a first attempt, consider applying the standard approach from Section~\ref{sec:lp-formulation}, of finding the optimal contract by solving \textsf{MINPAY-LP($i$)} for each of the $n$ actions $i \in [n]$ (see Figure~\ref{fig:lp-pair}). However, the primal LP now has exponentially-many variables (but polynomially-many constraints), hence this no longer yields a polynomial-time algorithm.

An alternative approach in this case is to turn to the dual, which has polynomially-many variables (namely $n-1$ many) and exponentially-many constraints (namely $2^m$ many), in 
hope that it can be solved via the ellipsoid method (see discussion in Section~\ref{sec:set-funct-oracles}). 
This approach hinges on the existence of
a polynomial-time algorithm that implements 
the separation oracle. So the question is, given dual variables $\lambda_{i'}$ for all $i' \neq i$, can we tractably decide whether there exists a set $S$ that violates the dual constraint $\sum_{i' \neq i} \lambda_{i'}(\prob_{i,S} - \prob_{i',S}) \leq \prob_{i,S}$.
By rearranging, this question is equivalent to asking whether
\[
\sum_{i' \neq i} \lambda_{i'} - 1 \leq \sum_{i' \neq i} \lambda_{i'} \frac{\prob_{i',S}}{\prob_{i,S}} \quad\text{for all $S$, $\prob_{i,S} > 0$.}
\]

For a fixed set of dual variables $\{\lambda_{i'}\}_{i' \neq i}$, the left-hand side of this inequality is a constant, so 
in order to answer this question we need to minimize the right-hand side over $S, \prob_{i,S} > 0$. This problem amounts to finding a set $S$ with minimum likelihood-ratio, where the ratio is between the mixture distribution $\sum_{i'\ne i}\lambda_{i'} \Prob_{i'}$, 
and the product distribution $\Prob_i$.%
\footnote{The fact that the separation oracle boils down to minimizing the likelihood ratio reinforces the connection between optimal contracts and statistical inference---see Remark~\ref{rem:inference}.}
If there is a single (non-trivial) action $i'\ne i$ in addition to $i$, this conclusion coincides with our analysis of the generalized binary-action case, which turns out to be tractable.

Unfortunately, solving the separation oracle exactly turns out to be \textsf{NP}-hard for more than two actions.
The difference from the generalized binary-action case is that a mixture $\sum_{i'\ne i}\lambda_{i'} \Prob_{i'}$ of two or more product distributions is, in general, no longer a product distribution. Thus, minimizing its likelihood ratio with respect to $\Prob_i$ can no longer be done greedily.  
In fact, \cite{DuttingRT21} show that the problem of finding the optimal contract itself (whether by ellipsoid or some other method) is NP-hard beyond the generalized binary-action case. This is by reduction from the min-max product partition problem of~\cite{KovalyovP10}.

One source of hope is that a mixture of \emph{constantly-many} product distributions is known to be ``well-behaved'' algorithmically in some contexts. This turns out to be the case for minimizing the likelihood ratio, and~\cite{DuttingRT21} give a dynamic programming based FPTAS for the separation oracle when $n$ is constant. 
The approximation factor in the separation oracle then translates, via an ellipsoid-like algorithm, to a contract that maximizes the principal's expected utility and approximately maximizes the agent's.
The next theorem summarizes this result.
To state the theorem, recall that an $\varepsilon$-IC contract is a pair of payment vector $\con$ and action~$i$, where $i$ is the agent's $\varepsilon$-best response action given $\con$ (maximizing his expected utility up to $\varepsilon$---see Equation~\eqref{eq:approx-IC} in Section~\ref{sub:regularity}). 
Let $\textsf{OPT}$ be the optimal expected utility the principal can obtain with a fully IC contract. 
Then: 

\begin{theorem}[\citet*{DuttingRT21}] 
\label{thm:multi-outcome-pos}
There is a poly-time algorithm that takes a parameter $\varepsilon > 0$, as well as a contract setting with constantly-many actions and an $m$-dimensional outcome space, and returns an $\varepsilon$-IC contract with expected principal utility $\geq \OPT$. The running time is $\textsf{poly}(m, 1/\varepsilon)$. 
\end{theorem}

Theorem~\ref{thm:multi-outcome-pos} holds for additive rewards with no assumptions, and for general rewards assuming oracle access to the reward function's multilinear extension.
Before sketching the proof, a natural question is: what does Theorem~\ref{thm:multi-outcome-pos} imply for fully IC contracts? By the following lemma of \cite{DuttingRT21}, the implication is a poly-time algorithm for approximating the optimal IC contract, up to vanishing multiplicative and additive losses in the principal's expected utility. The lemma is formulated here as it appears in \cite[Lemma 14]{Zuo24}:

\begin{lemma}[From $\varepsilon$-IC to IC contracts, \cite*{DuttingRT21,Zuo24}]
\label{lem:approx-IC}
Consider a contract setting with reward vector $\rvec$. Let $(\con,i)$ be an $\varepsilon$-IC contract with expected revenue $R_i-T_i$. Then IC contract $\con'=(1-\sqrt{\varepsilon})\con + \sqrt{\varepsilon} \rvec$ 
achieves expected revenue $\ge (1-\sqrt{\varepsilon})(R_i-T_i) - (\sqrt{\varepsilon}-\varepsilon)$.  
\end{lemma}

We now turn to sketch the proof of Theorem~\ref{thm:multi-outcome-pos}. 

\begin{proof}[Proof Sketch for Theorem~\ref{thm:multi-outcome-pos}.]

Consider an algorithm that iterates over the actions. For every action $i$, 
let $\LP_i$ denote the lowest expected payment required from the principal to incentivize action $i$ if $i$ is implementable. That is, $\LP_i$ is the optimal objective value of \textsf{MINPAY-LP($i$)} and its dual \textsf{DUAL-MINPAY-LP($i$)} (see Figure~\ref{fig:lp-pair}).
We show below how, if action $i$ is implementable (by a fully IC contract), the algorithm can find an $\varepsilon$-IC contract $\con^i$ implementing $i$ 
with expected payment $\le \LP_i$.
The running time of the algorithm is $\textsf{poly}(m, 1/\varepsilon)$.
By returning the revenue-maximizing contract among all $\{\con^i\}_{i\in[n]}$, the expected principal utility of the returned contract is guaranteed to be $\ge \OPT$. 

As a preliminary step, we transform \textsf{DUAL-MINPAY-LP($i$)} to an equivalent form as follows: For every set $S$ such that $\prob_{i,S}>0$, we divide both sides of the dual constraint corresponding to $S$ by $\prob_{i,S}$; rearranging we get
\begin{equation}
    \sum_{i' \neq i} (\lambda_{i'}) - 1 \leq \sum_{i' \neq i} \lambda_{i'}\frac{\prob_{i',S}}{\prob_{i,S}}.\label{eq:massaged-dual-constraint}
\end{equation}
For any remaining set $S$ such that $\prob_{i,S}=0$, we simply remove the corresponding dual constraint since it is guaranteed to hold for any dual solution. We thus have a new, equivalent version of \textsf{DUAL-MINPAY-LP($i$)} with constraints as in Equation~\eqref{eq:massaged-dual-constraint}. In the remainder of the proof sketch, \textsf{DUAL-MINPAY-LP($i$)} refers to this version of the dual. 

As discussed above, \cite{DuttingRT21} give an FPTAS for the problem of finding a set~$S$ that minimizes the likelihood ratio on the right-hand side of Equation~\eqref{eq:massaged-dual-constraint}. 
A first attempt is then to run the ellipsoid method described in Section~\ref{sec:set-funct-oracles} on the dual, using the FPTAS as an \emph{approximate} separation oracle~\cite[see, e.g.,][]{FleischerGMS06}. 
For a constant number of actions, the approximate separation oracle runs in \textsf{poly}$(m,1/\varepsilon)$-time for $\varepsilon>0$, 
and returns a set $S$ with likelihood ratio that is at most $(1+\varepsilon)$ times the minimum. 
Running the ellipsoid method with the approximate separation oracle either finds the dual is unbounded or returns a dual solution that possibly violates the constraints, but only slightly so. The main question is now: by how much can the objective value of this solution exceed $\LP_i$?
One approach to establishing that it cannot exceed $\LP_i$ by too much is to regain feasibility by scaling the solution. 
However, this approach fails in our case, as we now show. Suppose we have a solution $\{\lambda_{i'}\}_{i'\ne i}$ that slightly violates the constraints, i.e., for every $S$ it holds that
$$
\sum_{i' \neq i} (\lambda_{i'}) - 1 \le  (1+\varepsilon) \sum_{i' \neq i} \lambda_{i'}\frac{\prob_{i',S}}{\prob_{i,S}}.
$$
Observe that multiplying the dual solution $\{\lambda_{i'}\}_{i'\ne i}$ by $1/(1+\varepsilon)$ does \emph{not} regain feasibility.
So we must take a different approach.

Instead of scaling, we define a new dual LP, \textsf{DUAL-SCALED-LP($i$)}, by multiplying the left-hand side of each constraint of \textsf{DUAL-MINPAY-LP($i$)} by $(1+\varepsilon)$:
\begin{equation}
(1+\varepsilon) \left(\sum_{i' \neq i} (\lambda_{i'}) - 1 \right) \leq \sum_{i' \neq i} \lambda_{i'}\frac{\prob_{i',S}}{\prob_{i,S}}.\label{eq:scaled-constraint}
\end{equation}
We solve \textsf{DUAL-SCALED-LP($i$)} by running the ellipsoid method with the FPTAS as an approximate separation oracle. If this returns a dual solution then it is guaranteed to only slightly violate the constraints in Equation~\eqref{eq:scaled-constraint}. I.e., for every $S$,
\begin{equation}
(1+\varepsilon) \left(\sum_{i' \neq i} (\lambda_{i'}) - 1 \right) \le  (1+\varepsilon) \sum_{i' \neq i} \lambda_{i'}\frac{\prob_{i',S}}{\prob_{i,S}}.\label{eq:back-to-original}
\end{equation}
Note that the constraints in Equation~\eqref{eq:back-to-original} are equivalent to those in Equation~\eqref{eq:massaged-dual-constraint}. This in particular implies that for an action $i$ that is implementable by a fully IC contract, the algorithm cannot return that the solution is unbounded.

Moreover, since \textsf{DUAL-SCALED-LP($i$)} is always feasible (by setting $\lambda_{i'} = 0$ for all $i' \neq i$), it is also feasible in the approximate sense of Equation~\eqref{eq:back-to-original}. The algorithm thus finds an approximately feasible solution (in the sense of Equation~\eqref{eq:back-to-original}), which is fully feasible for the original dual \textsf{DUAL-MINPAY-LP($i$)}. So for the objective value $\gamma$ of the solution found by the algorithm, it holds that
\begin{equation}
    \gamma\le \LP_i.\label{eq:infeasible-is-bounded} 
\end{equation}
Thus we have shown that the approximately-feasible solution has value upper-bounded by $\LP_i$ despite the use of only an approximate separation oracle. 

We now sketch how to compute $\con^i$. Since clear from context, we omit $i$ from the notation and refer to $\con$ for simplicity. Because the ellipsoid method applied to \textsf{DUAL-SCALED-LP($i$)} returns a solution with value $\gamma$, then for any higher value $\gamma^+$ (where the notation $\gamma^+$ means any number higher than $\gamma$), the following holds: The program \textsf{DUAL-SCALED-LP($i$)} with an additional constraint $\sum_{i'\ne i} \lambda_{i'}(c_i-c_{i'})\ge \gamma^+$ (requiring the objective to reach at least $\gamma^+$) is identified as infeasible by the ellipsoid method in polynomial time, while calling the approximate separation oracle. Note that the approximate separation oracle's errors are one-sided, that is, if it identifies a constraint as violated then there is indeed a violating set $S$. 
Once we have polynomially-many violated dual constraints that prove infeasibility of $\gamma^+$, then using similar arguments to Section~\ref{sec:set-funct-oracles}, one can construct in polynomial time a contract $\con$ that is a feasible solution to the dual of \textsf{DUAL-SCALED-LP($i$)}, and has objective value $<\gamma^+$. We name this primal program \textsf{SCALED-LP($i$)}, and by duality it takes the following form: 
\begin{align}
\min_{t_S:S\subseteq[m]}\quad & (1+\varepsilon)\sum_{S} \prob_{i,S} \pay_S \nonumber\\
     &(1+\varepsilon)\sum_{S} \prob_{i,S} \pay_S - \cost_{i} \geq \sum_{S} \prob_{i',S} \pay_S - \cost_{i'} &&\forall i' \neq i \label{eq:eps-IC-constraint} \\
     &t_S \geq 0 & & \forall S \nonumber
\end{align}

It remains to show that contract $\con$ is $\varepsilon$-IC and has expected payment $\le \LP_i$. The $\varepsilon$-IC guarantee follows from the set of constraints in Equation~\eqref{eq:eps-IC-constraint}. The expected payment of $\con$ is the objective value of \textsf{SCALED-LP($i$)} divided by $(1+\epsilon)$, i.e., 
$\sum_S q_{i,S}t_S < \gamma^+/(1+\epsilon)$, and so $\sum_S q_{i,S}t_S \le \gamma/(1+\epsilon)\le \LP_i$, where we transitioned from strict inequality and $\gamma^+$ to weak inequality and $\gamma$, and the final inequality 
is by Equation~\eqref{eq:infeasible-is-bounded}. This completes the proof sketch.  
\end{proof}

\paragraph{A Negative Result for the General Case.}

Interestingly, the problem becomes much harder for a general (non-constant) number of actions. For this version of the problem, \cite{DuttingRT21} establish a hardness-of-approximation result that rules out any constant approximation factor (under standard complexity assumptions). 
In addition, they show a hardness-of-approximation result that applies to $\varepsilon$-IC contracts, provided that $\varepsilon$ is sufficiently small. 
The hardness is established via an intricate gap reduction from a classic problem shown to be hard by~\cite{Hastad01}: Distinguishing between satisfiable \textsf{MAX-3SAT} instances, and instances of \textsf{MAX-3SAT} in which no assignment satisfies more than $7/8+\eta$ of the clauses, where $\eta$ is an arbitrarily-small constant.

\begin{theorem}[\citet*{DuttingRT21}]
For contract settings with arbitrarily-many actions and an $m$-dimensional outcome space, 
for any constant $\rho \geq 1$, it
it is NP-hard to approximate $\OPT$ (the optimal expected principal utility 
achievable by an IC contract) to within a
multiplicative factor of $\rho$, even when rewards are additive.
\end{theorem}

\paragraph{Summary and Open Problems.} 

Overall, the results for combinatorially-many outcomes exhibit a rich computational landscape, with a sharp dichotomy in terms of approximability between a constant and non-constant number of actions.
These results also emphasize the usefulness of approximate incentive compatibility for algorithmic contract design.
An interesting direction for future work would be to go beyond product distributions, and explore (succinct) distributions over outcomes that are correlated.
\subsection{Multiple Principals}
\label{sub:multi-principals}

In this section we focus on another combinatorial aspect of contracts: a multiplicity of principals contracting with a single agent. The problem was first explored by \citet{BernheimW86a}, who refer to this setting as \emph{common agency}.
Their paper led to a substantial amount of follow-up work in the economic literature;
for an entry point to this literature see, e.g., the work of \cite{epstein1999revelation}. In this section we take a computational approach following the work of \citet*{AlonLST23}. 

The section is organized as follows: After introducing the common agency model and the design objective of welfare maximization (as opposed to revenue maximization), we 
zero in on so-called ``VCG contracts'' as candidates for maximizing welfare. We show that while welfare cannot always be maximized with multiple principals, it is algorithmically tractable to identify settings in which it can, and to compute the corresponding welfare-maximizing VCG contracts. The algorithmic approach thus extends the reach of classic contract design.

\paragraph{Common Agency Scenarios: Classic and Modern.} 

In common agency, a single agent operates under more than one system of incentives. 
The competing incentive systems are laid out for the agent by different principals, e.g., by a direct supervisor versus more senior management in an organization~\citep{CarrollW22}.
Additional classic examples include a manager 
acting on behalf of multiple shareholders,
a freelancer working for several employers, or a regulator representing multiple stakeholders. 
In all of these examples, the action of the common agent (e.g., manager) affects the entire group of principals (e.g., shareholders), while interests \emph{within} the principal group diverge. 
The design challenge in common agency is to 
choose which combination of principals the agent's actions will cater to, while
aligning interests not just between the principals and the agent, but also internally among the principals.

Common agency also has many recent applications, especially in online markets. For example, major platforms like Airbnb or Amazon represent multiple sellers, whose listings they promote; a marketing agency bids for online ads on behalf of multiple advertisers; a professional host on websites like Booking.com manages multiple properties for different owners; and a company like OpenAI deploys an LLM model to which multiple principals delegate text-generation tasks. 
Naturally, these applications come with new challenges such as scale and volatility, but also with new opportunities. These motivate the computational approach of \cite{AlonLST23}.

\paragraph{Model.}

Common agency extends the classic contract setting in Section~\ref{sec:model} to multiple principals. 
There are, as usual, $n$ actions from which the agent can choose, where action $i \in [n]$ has cost $c_i$ and induces a distribution $\Prob_i$ over the $m$ outcomes. There are now $k$ principals, where each principal~$\ell\in[k]$ has a reward vector $\rvec^{\ell}$ belonging to a known convex domain $\mathcal{V}^{\ell}\subseteq \mathbb{R}_{\ge 0}^m$.
Reward $\rew^{\ell}_j$ is the value that principal~$\ell$ derives from outcome~$j \in [m]$. 
Each principal~$\ell$ contracts separately with the agent via a contract $\con^{\ell}$, with payment $\pay^{\ell}_j$ for outcome $j\in[m]$. The agent's total payment if outcome~$j$ is realized following his action is the sum of payments $\sum_{\ell=1}^k t^{\ell}_j$. 
The agent chooses an action after observing the full profile of contracts $\con=(\con^1,\dots,\con^k)$.
The expected payment for taking action $i$ is 
$$
T^{\con}_i:=\mathbb{E}_{j \sim \Prob_{i}} \left[\sum_{\ell=1}^k t^{\ell}_j\right].
$$
In words, $T^{\con}_i$ is the expected sum of the principals' payments, where the expectation is taken over action $i$'s stochastic outcome $j$.
The agent picks an action $i\in[n]$ that maximizes his expected utility given by $T^{\con}_i-c_i$, with tie-breaking in favor of the design objective (e.g., revenue or welfare). 

\paragraph{Welfare Maximization.}
In settings with one principal-agent pair, welfare maximization is not hard to achieve.%
\footnote{A linear contract with parameter $\alpha = 1$ (or, equivalently, a contract $\con$ with payments equal to the rewards $\con=\rvec$) transfers the full reward from the principal to the agent. The agent internalizes both reward and cost and thus chooses the welfare-maximizing action.} 
In settings with more than one principal (or, alternatively, more than one agent), both revenue maximization and welfare maximization are natural and non-trivial goals. In this section we depart from previous sections and focus on welfare maximization. 

In common agency with $k$ principals, given a profile of rewards $\rvec=(\rvec^1,\dots,\rvec^k)$, denote the expected total reward when the agent takes action $i$ by 
$$
R^{\rvec}_i:=\mathbb{E}_{j \sim \Prob_{i}}\left[\sum_{\ell=1}^k r^{\ell}_{j}\right].
$$
The expected welfare from action $i$ is then $\Wel^{\rvec}_i:= R^{\rvec}_i - c_{i}$.
We denote the \emph{welfare-maximizing action} by $i^{\star}=i^{\star}(\rvec)=\arg\max_{i\in[n]}\{ \Wel^{\rvec}_i \}$ (where we ignore tie-breaking for simplicity). 
We denote the overall welfare by $\Wel^{\rvec}:=\max_{i\in[n]}\{ \Wel^{\rvec}_i \}= \Wel^{\rvec}_{i^{\star}}$.
We say that $\con=(\con^1,\dots,\con^k)$ is a \emph{welfare-maximizing contract profile} if it incentivizes $i^\star$, i.e., the welfare-maximizing action $i^{\star}$ is a best response of the agent, maximizing his expected utility among all actions: 
$$
i^{\star}\in \arg\max_{i\in[n]} \left\{ T^{\con}_i-c_i\right\}.
$$

\paragraph{Two-Stage Game vs.~Coordinating Platform.}

In a common agency problem, how does a profile of contracts emerge, and when is it considered a stable solution?
The literature considers two main variants~\cite[see, e.g.,][]{BernheimW86a,PratR03,LaviS22,AlonLST23}. 
In the first variant, the principals simultaneously offer contracts $\con^1,\dots,\con^k$ to the agents (Stage 1), and then the agent chooses an action (Stage 2) and utilities are realized. The pure subgame perfect equilibria (SPE) of this game are studied. 
In the second variant, in Stage 1 the principals simultaneously report their rewards to a \emph{coordinating platform}, which outputs a profile of contracts (Stage 2 remains unchanged). We focus here on the second variant, and are most interested in platforms that elicit \emph{truthful} reports by inducing dominant strategy incentive compatibility (DSIC) among the principals. 
Throughout we denote the reported rewards of principal $\ell\in[k]$ by $\mathbf{b}^\ell \in\mathcal{V}^\ell$.

\paragraph{First-Price Contracts.}

For principal $\ell\in[k]$, we say her contract is \emph{first-price} if $\con^\ell = \con^\ell(\mathbf{b}^\ell)=\mathbf{b}^\ell$. Such contracts are quite natural---the principal submits a ``bid'' to the platform stating how much she values each outcome $j\in[m]$, and if this outcome is obtained she pays her bid $b^\ell_j$. 
Unfortunately, the power of first-price contracts to maximize social welfare turns out to be limited in the following ways: 
A first observation is that first-price contracts fail to be truthful. Indeed, under a first-price contract in which principal $\ell$ bids $\mathbf{b}^\ell = \rvec^\ell$, the principal retains no utility at all, and is typically better off shading her bid (i.e., bidding less than her true rewards). 
Moreover,
\cite{AlonLST23} show that with first-price contracts, there can exist an equilibrium $(\mathbf{b}^1,\dots,\mathbf{b}^k)$ that is highly inefficient in terms of social welfare. They do so by considering the worst-case ratio 
between the optimal welfare and the equilibrium welfare, known as the \emph{price of anarchy}~\citep{KoutsoupiasP09,RoughgardenST59}:

\begin{proposition}[\cite*{AlonLST23}]
In common agent settings, the price of anarchy of first-price contracts can be as large as linear in $k$, the number of principals.%
\footnote{Even the \emph{price of stability}---the ratio between the optimal welfare and the welfare of the \emph{best} equilibrium---can be bounded away from 1~\citep{AlonLST23}.}
\end{proposition}

\paragraph{VCG Contracts.}

The weak welfare guarantees of first-price contracts motivate the study of more sophisticated contracts of the form $\con^\ell=\con^\ell(\mathbf{b}^1,\ldots,\mathbf{b}^k)$, which depend not only on the principal's own bid $\mathbf{b}^\ell$, but also on other principals' bids. 
In the domain of resource allocation, 
it is well-known that welfare maximization is achieved by the \emph{VCG auction}~\citep{Vickrey61,Clarke71,Groves73},
which elicits truthful valuations from the buyers and outputs their payments. 
Our goal is to design \emph{VCG contracts}, defined as follows:

\begin{definition}[VCG contracts]
\label{def:VCG-contracts}
    VCG contracts are a profile of contracts, which take the form $\con^\ell(\mathbf{b}^1,\ldots,\mathbf{b}^k)$ for every principal $\ell\in[k]$, and satisfy the following two conditions: (1)~Truthfulness (DSIC), i.e., reporting $\mathbf{b}^\ell=\rvec^\ell$ is a dominant strategy for every principal $\ell$ and results in non-negative expected utility for the principal; (2)~Welfare maximization, i.e., the agent has a best response action that maximizes welfare.
\end{definition}

\citet{LaviS22} are the first to introduce the concept of VCG contracts, which they develop in the context of multi-principal, multi-agent settings with full information (no hidden actions). In this context, VCG contracts are defined via an explicit payment formula. We focus here (as throughout the survey) on the hidden action model, for which VCG contracts were first considered by~\cite{AlonLST23}. 

The following example demonstrates the necessity of VCG contracts' dependence on the entire bid profile $(\Bid^1,\ldots,\Bid^k)$ to achieve truthful welfare maximization.

\begin{example}[A simple multi-principal setting that reduces to resource allocation]
\label{ex:contractible}
Consider an agent with two actions and two outcomes. The distributions associated with the actions are~$\Prob_{1}=(1,0),\Prob_{2}=(0,1)$, and their costs are $\cost_1=\cost_2=0$. 
\begin{center}
\begin{tabular}{|l|cc|c|}
\toprule
& $r^{\ell}_1$ & $r^{\ell}_2$ & \text{cost}\\
\midrule 
action $1$: & $1$ & $0$ & $c_1=0$ \\
action $2$: & $0$ & $1$ & $\cost_
2 = 0$ \\ 
\bottomrule
\end{tabular}
\end{center}
There are $k=2$ principals with reward vectors belonging to domains~$(\mathbb{R}_{\geq 0},0)$ and $(0,\mathbb{R}_{\geq 0})$, respectively. 
Principal $1$'s reward vector is $\rvec^{1} =(\beta,0)$ for some value $\beta$, i.e., principal $1$ wants the agent to take action $1$ to receive the reward from outcome $1$. Principal $2$'s reward vector is $\rvec^{2} = (0, \gamma)$ for some value $\gamma$, i.e., principal $2$ wants the agent to take action $2$ to receive the reward from outcome~$2$. In this example, the expected welfare $\Wel^{\rvec}_1$ of action $1$ is $q_{11}(r^1_1+r^2_1)+q_{12}(r^1_2+r^2_2)=q_{11}r^1_1=\beta$. Similarly, the expected welfare $\Wel^{\rvec}_2$ of action $2$ is $\gamma$. Thus to maximize welfare, the agent must be incentivized to choose action $2$ if and only if $\gamma\ge \beta$ (up to tie-breaking, which we ignore here). 
The principals submit bids $\Bid^1 =(\tilde\beta,0)$ and $\Bid^2 = (0, \tilde\gamma)$, and the coordinating platform returns contracts $\con^1,\con^2$. The agent's expected payment for choosing action $1$ is $T^\con_1=t^1_1$ and for choosing action $2$ it is $T^\con_2=t^2_2$.

Due to the contract setting's simplicity (every action leads deterministically to a unique outcome and the action costs are zero), it is equivalent to a resource allocation setting: A seller (the agent in the contract setting) allocates an indivisible resource among two buyers (the two principals). Notice that the agent's choice of action in the contract setting corresponds to a choice of which principal/buyer wins the resource. The resource is worth $\beta$ to principal $1$ and $\gamma$ to principal $2$; let us denote their reported values by $\tilde\beta,\tilde\gamma$.
By Myerson's theory of truthful mechanisms~\citep{Myerson81}, to elicit truthful value reports from the principals and maximize welfare, the allocation rule chooses principal $2$ as the winner if and only if $\tilde\gamma\ge \tilde\beta$, and otherwise principal $1$ wins; the payment rule charges the winning principal her ``critical bid'', i.e., $\tilde\gamma$ if principal $1$ wins and $\tilde\beta$ if principal $2$ wins.

By the connection between the contract setting and the resource allocation setting, we conclude that if $\tilde\gamma \ge \tilde\beta$ then principal 2's payment $t^2_2$ for outcome $2$ must be set to $\tilde\beta$, and all other payments are set to zero. Similarly if $\tilde\gamma < \tilde\beta$ then principal 1's payment  $t^1_1$ for outcome $1$ must be set to $\tilde\gamma$, and all other payments are set to zero. 
This incentivizes the principals to report truthfully and the agent to take the welfare-maximizing action.
This example thus shows that to maximize welfare truthfully, each principal's contract must sometimes depend on the other's bid. 
As an aside, the example also demonstrates how common agency can encompass resource allocation settings: the principals can be viewed as buyers, rewards play the role of values, the agent is the seller, and what is sold is the outcome of the agent's effort. 
\end{example}

\paragraph{Relation to Contractible Contracts.} 

VCG contracts are an example of \emph{contractible contracts}~\citep{peters2012definable}. 
In such contracts, one principal's
payments to the agent are allowed to depend on the other principals' bids.
This approach is familiar from pricing in auctions---for example, the winner of a second-price auction pays the highest competing offer. 
A simple example from procurement contracts is \emph{price-matching} guarantees, where a principal commits to paying the agent at least as much as the best competing offer.
Contractible contracts are increasingly implementable these days using technology like smart contracts on the blockchain.

\paragraph{Designing VCG Contracts.}

Our goal is to design VCG contracts, defined as 
truthful and welfare-maximizing contracts (Definition~\ref{def:VCG-contracts}). Technically, this means to design a mapping from any bid profile $\Bid\in\mathcal{V}^1\times\dots\times \mathcal{V}^k$ to contracts $\con^1(\Bid),\dots,\con^k(\Bid)$, where the mapping can depend on the common agency instance (including the agent's costs and distributions and the domains of principal rewards $\mathcal{V}^1,\dots,\mathcal{V}^k$).
Ideally, we seek a \emph{universal} design that ``works'' for every instance, i.e., results in a profile of DSIC and welfare-maximizing contracts. 
In what follows, we give an impossibility result in the spirit of \cite{MyersonS83} that rules out the existence of universal VCG contracts. The next best result one could hope for is \emph{instance-specific} VCG contracts, i.e., a mapping $\con^1(\Bid),\dots,\con^k(\Bid)$ for every common agency instance that admits such contracts. We show a polynomial-time construction of instance-specific VCG contracts.

\paragraph{Characterization of Expected Payments.}

We begin by characterizing the expected payments in VCG contracts: for every bid profile $\Bid$, we characterize the expected payment $T^\ell_{i^\star}$ from principal~$\ell$ to the agent, assuming the agent chooses the welfare-maximizing action $i^\star$. 

The characterization is inspired by the expected payments in VCG auctions (see, e.g., \cite[Chapter 7.2]{Roughgarden16}). In a VCG auction, buyer $\ell$'s payment is composed of a term that does not rely on $\ell$'s report, and a term that relies on her report exclusively to determine the welfare-maximizing allocation. The first term is the welfare if buyer~$\ell$ were \emph{not} participating in the allocation, and the second term is the other buyers' welfare assuming buyer~$\ell$ \emph{is} participating. The first term is set according to \emph{Clarke's pivot rule}, to ensure that the buyers not only wish to report accurately to the VCG auction, but also wish to participate in the first place (this is the individual rationality (IR) property for the buyers, which is required as part of a DSIC auction).
The economic intuition for the payment formula is that the two terms together capture buyer~$\ell$'s \emph{externality} on the other buyers from participating, and internalizing this externality makes buyer~$\ell$'s incentives aligned with those of society.

\cite{AlonLST23} use similar intuition to characterize the payments of VCG contracts. Let $T^\ell_{i^\star(\Bid)}$ denote principal~$\ell$'s expected payment to the agent for choosing action~$i^{\star}$.
Let $\Wel^{\Bid}_i$ denote the expected welfare from action $i$ assuming the rewards are $\Bid$, and let $i^{\star}(\Bid)$ denote the welfare-maximizing action under the same assumption.
Let $f^\ell$ denote a function that does not rely on principal $\ell$'s report, to be determined later.
\cite{AlonLST23} show that $T^\ell_{i^\star(\Bid)}$ must be of the following form, which is composed of two terms (similar to VCG auctions):
\begin{equation}
T^\ell_{i^\star(\Bid)}
= \underbrace{f^\ell(\Bid^{-\ell})}_{\text{no dependence on }\Bid^{\ell}}
-
\underbrace{\Wel^{(\mathbf{0},\Bid^{-\ell})}_{i^{\star}(\Bid)}}_{\substack{\Bid^{\ell}\text{ determines only the welfare-} \\ \text{maximizing action }i^{\star}(\Bid)}}.
\label{eq:pivot-contract}
\end{equation}
The second term is the other principals' welfare assuming $i^\star(\Bid)$ is chosen.
Notice that $T^\ell_{i^\star(\Bid)}$ depends on the entire profile of bids $\Bid$, so VCG contracts are indeed contractible.

\paragraph{Impossibility of Universal VCG Contracts.}

The characterization of expected payments in Equation~\eqref{eq:pivot-contract} does not fully specify VCG contracts: First, function $f$ remains to be determined. 
Second, it remains to determine the payment \emph{for every outcome}. To achieve universal VCG contracts, we need a way to break down the required expected payments to per-outcome payments for all possible instances.
It turns out that finding such per-outcome payments ---that are also non-negative to maintain limited liability---is not always possible.%
\footnote{The impossibility holds even if we require only that the payments to the agent are non-negative \emph{in aggregate} over the principals.} 
The following impossibility is related to, but not subsumed by, the impossibility result of~\citet{MyersonS83}:

\begin{theorem}[Impossibility result~\citep*{AlonLST23}]
\label{thm:multi-principal-impossibility}
For any number of principals $k$, there exists a common agency setting for which no contracts $\con^1,\dots,\con^{k}$ satisfy truthfulness for the $k$ principals and limited liability for the agent, while incentivizing the agent to choose the welfare-maximizing action.
\end{theorem}

The proof utilizes the next example.

\begin{example}[Common agency setting for Theorem~\ref{thm:multi-principal-impossibility}]
\label{ex:trade-off} 
Consider an agent with two actions and two outcomes. The distributions associated with the actions are~$\Prob_{1}=(\frac{1}{2},\frac{1}{2}),\Prob_{2}=(0,1)$, and their costs are $\cost_1=0,\cost_2=\epsilon>0$. 
\begin{center}
\begin{tabular}{|l|cc|c|}
\toprule
& $r^{\ell}_1$ & $r^{\ell}_2$ & \text{cost}\\
\midrule 
action $1$: & $\nicefrac{1}{2}$ & $\nicefrac{1}{2}$ & $c_1=0$ \\
action $2$: & $0$ & $1$ & $\cost_
2 = \epsilon$ \\ 
\bottomrule
\end{tabular}
\end{center}
There are $k\ge 1$ principals with reward vectors belonging to domain~$\mathbb{R}^2_{\geq 0}$. All principals but the first have all-zero rewards: $\rvec^{\ell} =(0,0)$ $\forall \ell \in [k]\setminus \{1\}$. The reward vector $\rvec^{1}$ is determined in the proof below.
\end{example}

\begin{proof}[Proof of Theorem~\ref{thm:multi-principal-impossibility}] 
Suppose towards a contradiction that there always exist contracts $\con^1,\dots,\con^{k}$ for the setting in Example~\ref{ex:trade-off}, which satisfy truthfulness for the principals and limited liability for the agent while incentivizing the agent to take the welfare-maximizing action. The contradiction is obtained by applying the characterization of expected payments in Equation~\eqref{eq:pivot-contract} while varying the reward vector of the first principal. Since the first term $f^1(\Bid^{-1})$ does not depend on principal~1's bid, under truthfulness it should remain fixed, but we show that under limited liability there is no suitable fixed term. 

Consider first reward vector $\rvec^{1}=(0,3\epsilon)$. Observe that the welfare from actions 1 and 2 is, respectively, $\Wel_{1}^{\rvec}=\prob_{1,1}\cdot 0 + \prob_{1,2}\cdot 3\epsilon - c_1 = 1.5\epsilon$ and $\Wel_{2}^{\rvec}=\prob_{2,1}\cdot 0 + \prob_{2,2}\cdot 3\epsilon - c_2 = 2\epsilon$ (only the first principal contributes to the welfare). 
Thus, the socially efficient action $i^\star(\rvec)$ is~$2$. 
For the agent to maximize welfare, his utility from action~2 must weakly dominate action~1, which yields the constraint $\sum_{\ell \in [k]} (t^{\ell}_2- t^{\ell}_1) \geq 2\epsilon$. Since no principal but the first pays the agent, it must hold that $t^{1}_2- t^{1}_1 \geq 2\epsilon$. By non-negativity of the payments we conclude
$t^{{1}}_2 \geq 2\epsilon$. 
We now apply the expected payment characterization. Using that $\prob_{2,1}=0,\prob_{2,2}=1$, and applying Equation~\eqref{eq:pivot-contract} where $\Bid=\rvec$ by truthfulness, we get: $t^{1}_2 = \prob_{2,1} \cdot t^{1}_1 + \prob_{2,2} \cdot t^{1}_2 = f^1(\Bid^{-1}) - \Wel^{\Bid^{-1}}_{2}$. Since the welfare $\Wel^{\Bid^{-1}}_{2}$ of action 2 without principal 1 is $-\epsilon$ (minus the agent's cost), we can apply $t^{{1}}_2 \geq 2\epsilon$ to conclude that $f^1(\Bid^{-1}) \geq \epsilon$. 
Consider now a different reward vector in the domain, $\rvec^{1}=(0,0)$. The welfare-maximizing action is now~$1$ and all payments are zero. In particular, principal 1's expected payment must be zero, and by applying Equation~\eqref{eq:pivot-contract} we have $f^1(\Bid^{-1}) = 0$, a contradiction.
\end{proof}

\paragraph{Instance-Specific VCG Contracts.}

To alleviate the impossibility of universal VCG contracts in Theorem~\ref{thm:multi-principal-impossibility} via
an algorithmic approach, \cite{AlonLST23} show how
to compute a welfare-maximizing and incentive compatible contract profile for every common agency setting that admits one. This is in line with the approach of \emph{automated mechanism design}~\citep{ConitzerS03}, which computationally designs mechanisms for given problem instances to circumvent general impossibility results.
Call a common agency setting \emph{applicable}
if truthful, welfare-maximizing VCG contracts exist for it. \cite{AlonLST23} design two polynomial-time algorithms, which can be described as the \emph{detection} algorithm and the \emph{on-the-fly payment} algorithm. The algorithms guarantee the following: 

\begin{theorem}[Applicable common agency settings~\citep*{AlonLST23}]
\label{thm:on-the-fly}
For every common agency setting, the \emph{detection} algorithm determines whether or not it is applicable.
For every applicable common agency setting with $k$ principals, there exist
truthful welfare-maximizing VCG contracts such that given any reports $\Bid$ and outcome~$j$, the \emph{on-the-fly payment} algorithm returns payments $\pay^1_j,\dots,\pay^k_j$ consistent with these contracts.
Both algorithms run in polynomial time.
\end{theorem}

\cite{AlonLST23} give several examples of applicable common agency settings (for which VCG contracts are guaranteed to exist), e.g., 
partially-symmetric settings in which principals share the same expected reward from each action, or settings in which rewards are between $[a, b]$ where
$b \le 2a$. 

\paragraph{Summary and Open Problems.}

The study of VCG contracts demonstrates the challenges of designing welfare-maximizing (rather than revenue-maximizing) contracts in combinatorial contract settings. 
It also highlights the role of contractible contracts in coordination among multiple principals. 
One main take-away from the computational research of common agency so far is that the algorithmic approach, coupled with on-the-fly payments, offers flexibility that can expand the reach of classic contract design. 
An interesting open direction is to explore 
\emph{approximately} welfare-optimal contracts as another avenue for extending the classic theory and circumventing impossibilities.
Since contractible contracts are reminiscent of smart contracts, a more formal exploration of the connections between the two suggests itself as a future research direction.
Contractible contracts also potentially allow principals to \emph{collude} \citep{CalvanoCDP20}, thus harming competition and driving down the agent's payments. It is interesting to study when coordination among the principals is desirable (e.g., for maximizing welfare), versus when it becomes unwanted collusion. 

\section{Contracts for Typed Agents}
\label{sec:types}
In the contract settings we have seen so far, agents implicitly have \emph{types}---for example, the skill set of a CEO (the agent) hired by a company owner (the principal). The agent's type affects the design of the contract---intuitively, the CEO's contract is personalized to his skill set. In full generality, an agent's type is defined as his ability to transform costly actions into outcomes, captured mathematically by his $n\times m$ distribution matrix and vector of $n$ costs, where $n$ is the number of actions and $m$ is the number of outcomes. 
As we have seen, tailoring the contract to the distribution matrix and cost vector (the agent's type) is necessary to get the optimal contract.
However, in many practical contract settings, the distribution matrix and/or related costs are not fully known to the principal; they may be partially or entirely unknown. In this case we say that the agent's type is \emph{hidden}.

In Section~\ref{sec:robust}, we already explored settings where some of the information about the agent is hidden from the principal, and we took a worst-case approach, seeking a design that maximizes the principal's minimum utility over the (non-Bayesian) uncertainty that the principal has. 
In contrast, here we consider a \emph{Bayesian} approach to hidden types, where we assume that types are drawn from a \emph{known} distribution and aim to maximize the principal's expected utility.\footnote{In Section~\ref{sec:data-driven}, we explore an additional approach to hidden types---through learning.}

This approach combines hidden action with private types, and thus generalizes both pure contract theory and pure mechanism design. Several classic papers in economics explore models that combine the two challenges (e.g., \cite{Myerson82}), and problems that exhibit both remain an active field of research (e.g., \cite{GottliebM15,ChadeS21}). Here we focus on recent work that takes an algorithmic approach.

After introducing the model and design goals (in Section~\ref{sec:typed-model}), we consider typed contract settings in which the agent either has a multi-dimensional private type or a single-dimensional private type (in Section~\ref{sub:multi-dim-types} and Section~\ref{sub:single-dim-types}, respectively). Section~\ref{sec:typed-reduction} establishes a link between the two settings. We conclude with a variation of the basic model, in which the agent proposes the contract to the principal, who has a private type (Section~\ref{sub:agent-designed}). 

\subsection{Typed Agents: Model and Design Goals}
\label{sec:typed-model}

We consider a \emph{Bayesian approach}, by which the  private type is drawn from a known population of agents.
The agent population is described by a distribution $G$ over a type space $\theta \in \mathcal{T}$.
The design challenge is then as follows: Given the type distribution $G$ over $\mathcal{T}$, compute a contract that maximizes the expected revenue, where the expectation is over both the agent's type, and (as usual in contract design) over the random outcome of the agent's action. 

In addition to standard contracts (where a contract is a vector of outcome-contingent payments), with private types it will generally be beneficial for the principal to offer contracts that are type-dependent. There are two (equivalent) interpretations of type-dependent contracts.\footnote{This follows from the \emph{Revelation Principle} \citep{Myerson79,Myerson81}, in combination with the \emph{Taxation Principle} \citep{Hammond79,Guesnerie81}.} The first interpretation is to treat them as \emph{menus of contracts}, and the second is to view them as (incentive compatible) \emph{type-soliciting contracts}. In addition, both interpretations come in two flavors---they can either be \emph{deterministic} or \emph{randomized}.

Let's first consider \emph{menus of contracts}. In the deterministic case, a menu of contracts is a collection of (classic) contracts. For each type, the agent chooses a contract and an action, that together maximize his utility.   
In the randomized case, the menu items are lotteries over contracts. Here the agent first chooses a lottery. Then a contract is drawn from the lottery. After learning about the realized contract, the agent chooses an action. For each type, the agent chooses a lottery and subsequent actions that maximize his expected utility.

Next consider \emph{type-soliciting contracts.} In such a contract, the agent is asked to report his type. The agent may report his type truthfully, but may also pretend to be of a different type.
In the deterministic case, the reported type is mapped to a contract and a recommended action. After learning about the contract and recommended action, the agent takes an action---possibly different from the recommended one. 
In the randomized case,  
each reported type is mapped to a distribution over (contract, recommended action) pairs. After learning about the realized contract and recommended action, the agent chooses an action. As in the deterministic case, the agent is free to choose an action that is different from the recommended one.

A type-soliciting contract is \emph{incentive compatible (IC)} if it is in the agent's best interest to report his type truthfully and to follow the recommended action. In other words, the deviations that we need to protect against are (a) the agent might report a type that is different from his truthful one, and/or (b) he might take an action different from the recommended one. The contract design problem is then to design an (incentive compatible) type-soliciting contract---or equivalently a menu of contracts---that maximizes the principal's expected utility. 

A useful observation is that for deterministic menus of contracts it is without loss of generality to consider menus that have at most one contract per type, while for randomized menus of contracts it is without loss of generality to consider menus that have at most one lottery per type and where each lottery has at most one contract per recommended action.\footnote{See, for example, Lemma 5 in the arXiv version of \citep{CastiglioniM022}.} An analogous observation applies to type-soliciting contracts.

The following example illustrates the concept of a (deterministic) menu of contracts, and how it can be interpreted as an incentive-compatible type-soliciting contract.

\begin{figure}[t]
\begin{subfigure}[t]{0.45\textwidth}
\begin{tikzpicture}[scale=0.85]
\draw[->,thick] (-0.7,0) -- (5.7,0);
\node at (5.7,-0.42) {\footnotesize $\Pr[r_2]$};
\draw[->,thick] (0,-1.5) -- (0,4.5);
\draw[-,thick] (5,-1.5) -- (5,4.5);
\node at (-0.35,0.3) {\footnotesize $0$};
\node at (5.35,0.3) {\footnotesize $1$};
\draw[-,thick] (0.1,1) -- (-0.1,1) node[left] {\footnotesize $1$};
\draw[-,thick] (0.1,2) -- (-0.1,2) node[left] {\footnotesize $2$};
\draw[-,thick] (0.1,3) -- (-0.1,3) node[left] {\footnotesize $3$};
\draw[-,thick] (0.1,4) -- (-0.1,4) node[left] {\footnotesize $4$};
\draw[-,thick] (5.1,1) -- (4.9,1);
\draw[-,thick] (5.1,2) -- (4.9,2);
\draw[-,thick] (5.1,3) -- (4.9,3);
\draw[-,thick] (5.1,4) -- (4.9,4);
\node at (-1.4,4.3) {\footnotesize agent's};
\node at (-1.5,3.8) {\footnotesize utility};
\draw[-,thick,red] (0,2) -- (5,1);
\node at (0.6,2.25) {\footnotesize \textcolor{red}{$t^1,a_1$}};
\draw[-,thick,red] (0,0) -- (5,4) node[right] {\footnotesize $t^2, a_1$};
\draw[-,thick,blue] (0,1) -- (5,0);
\node at (0.6,1.25) {\footnotesize \textcolor{blue}{$t^1,a_2$}};
\draw[-,thick,blue] (0,-1) -- (5,3) node[right]{\footnotesize $t^2,a_2$};
\draw[-,thick,red] (0.25*5,0.1) -- (0.25*5,-0.1) node[below] {\footnotesize $\nicefrac{1}{4}$};
\node[label={},circle,fill=red,inner sep=2pt,draw=red] at (0.25*5,1.75) {};
\node[label={},circle,inner sep=2pt,draw=red] at (0.25*5,1) {};
\draw[-,thick,blue] (0.5*5,0.1) -- (0.5*5,-0.1) node[below] {\footnotesize $\nicefrac{1}{2}$};
\node[label={},circle,fill=blue,inner sep=2pt,draw=blue] at (0.5*5,1) {};
\node[label={},circle,inner sep=2pt,draw=blue] at (0.5*5,0.5) {};
\end{tikzpicture}
\caption{Type: $\theta_1$}
\end{subfigure}
\hspace*{12pt}
\begin{subfigure}[t]{0.45\textwidth}
\begin{tikzpicture}[scale=0.85]
\draw[->,thick] (-0.7,0) -- (5.7,0);
\node at (5.7,-0.42) {\footnotesize $\Pr[r_2]$};
\draw[->,thick] (0,-1.5) -- (0,4.5);
\draw[-,thick] (5,-1.5) -- (5,4.5);
\node at (-0.35,0.3) {\footnotesize $0$};
\node at (5.35,0.3) {\footnotesize $1$};
\draw[-,thick] (0.1,1) -- (-0.1,1) node[left] {\footnotesize $1$};
\draw[-,thick] (0.1,2) -- (-0.1,2) node[left] {\footnotesize $2$};
\draw[-,thick] (0.1,3) -- (-0.1,3) node[left] {\footnotesize $3$};
\draw[-,thick] (0.1,4) -- (-0.1,4) node[left] {\footnotesize $4$};
\draw[-,thick] (5.1,1) -- (4.9,1);
\draw[-,thick] (5.1,2) -- (4.9,2);
\draw[-,thick] (5.1,3) -- (4.9,3);
\draw[-,thick] (5.1,4) -- (4.9,4);
\node at (-1.4,4.3) {\footnotesize agent's};
\node at (-1.5,3.8) {\footnotesize utility};
\draw[-,thick,red] (0,2) -- (5,1);
\node at (0.6,2.25) {\textcolor{red}{\footnotesize $t^1,a_1$}};
\draw[-,thick,red] (0,0) -- (5,4) node[right] {$t^2, a_1$};
\draw[-,thick,blue] (0,1) -- (5,0);
\node at (0.6,1.25) {\footnotesize\textcolor{blue}{\footnotesize $t^1,a_2$}};
\draw[-,thick,blue] (0,-1) -- (5,3) node[right]{\footnotesize $t^2,a_2$};
\draw[-,thick,red] (0.5*5,0.1) -- (0.5*5,-0.1) node[below] {\footnotesize $\nicefrac{1}{2}$};
\node[label={},circle,fill=red,inner sep=2pt,draw=red] at (0.5*5,2) {};
\node[label={},circle,inner sep=2pt,draw=red] at (0.5*5,1.5) {};
\draw[-,thick,blue] (0.875*5,0.1) -- (0.875*5,-0.1) node[below] {\footnotesize $\nicefrac{7}{8}$};
\node[label={},circle,fill=blue,inner sep=2pt,draw=blue] at (0.875*5,2.5) {};
\node[label={},circle,inner sep=2pt,draw=blue] at (0.875*5,0.125) {};
\end{tikzpicture}
\caption{Type: $\theta_2$}
\end{subfigure}
\caption{
Visualization of the menu of contracts in Example~\ref{ex:contracts-for-typed-agents}. The left tableau is for type $\theta_1$ and the right tableau is for type $\theta_2$. The lines plot the expected agent utilities for contracts $t^1$ and $t^2$ as a function of $\Pr[\rew_2]$ with cost zero (action $a_1$, red lines) and with cost $1$ (action $a_2$, blue lines). 
The agent's choice of action determines the probability of outcome $r_2$ and thus a point on the $x$-axis.  These points are $\Pr[r_2] = 1/4$ for action $1$ and $\Pr[r_2] = 1/2$ for action $2$ under type $\theta_1$ (tableau on the left), and $\Pr[r_2] = 1/2$ for action $1$ and $\Pr[r_2] = 7/8$ for action $2$ under type $\theta_2$ (tableau on the right).
Consequently, for each type (over which the agent has no control), by choosing the action and the contract, the agent can choose from any of the four dots in the respective plot, and will pick the one with the highest utility. 
For example, under type $\theta_1$, the agent prefers $t^1$ over $t^2$ for action $a_1$, as the former gives a utility of $1.75$ (red dot, filled) while the latter gives a utility of $1$ (red dot, empty). At the same time, the agent prefers $t^2$ over $t^1$ for action $a_2$, as the respective utilities are $1$ (blue dot, filled) and $0.5$ (blue dot, empty). So overall, the agent will choose $t^1$ and action $a_1$ under this type.} 
\label{fig:contracts-for-typed-agents}
\end{figure}
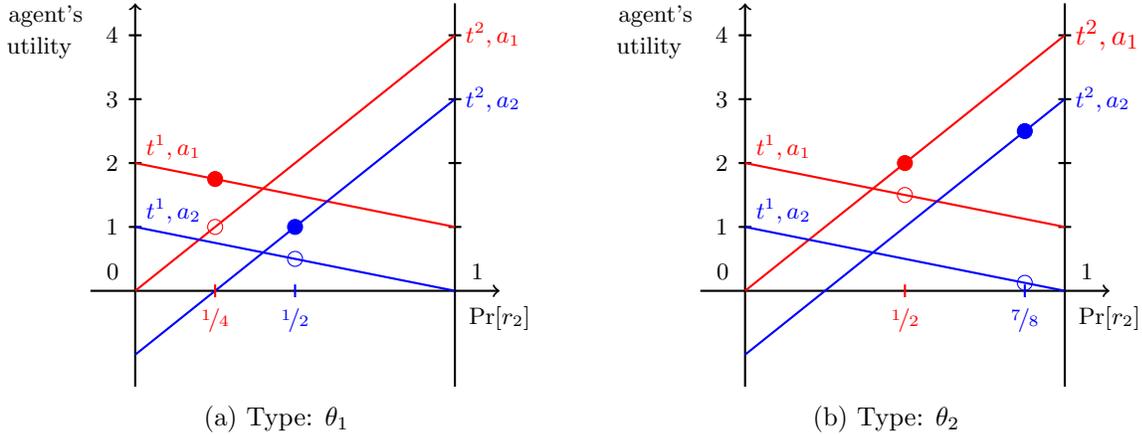

\begin{example}[Contracts for typed agents]
\label{ex:contracts-for-typed-agents}
Consider the following contracting problem, with two actions, two outcomes, and two types.
The rewards are intentionally left unspecified, as they are not relevant to the analysis.

\parbox{.45\linewidth}{
\begin{center}
\begin{tabular}{|l|cc|c|}
\toprule
& $r_1$ & $r_2$ & \text{cost} \\
\midrule
action $1$: & $\nicefrac{3}{4}$ & $\nicefrac{1}{4}$ & $c_1 = 0$\\
action $2$: & $\nicefrac{1}{2}$ & $\nicefrac{1}{2}$ & $c_2 = 1$\\
\bottomrule
\end{tabular}\\[6pt]
Type: $\theta_1$
\end{center}
}
\parbox{.45\linewidth}{
\begin{center}
\begin{tabular}{|l|cc|c|}
\toprule
& $r_1$ & $r_2$ & \text{cost} \\
\midrule
action $1$: & $\nicefrac{1}{2}$ & $\nicefrac{1}{2}$ & $c_1 = 0$\\
action $2$: & $\nicefrac{1}{8}$ & $\nicefrac{7}{8}$ & $c_2 = 1$\\
\bottomrule
\end{tabular}\\[6pt]
Type: $\theta_2$
\end{center}
}

Consider the menu of contracts consisting of two contracts: $\con^1 = (2,1)$ and $\con^2 = (0,4)$. See Figure~\ref{fig:contracts-for-typed-agents} for a visualization of the possible choices of the agent, and the corresponding utilities.
Given this menu of contracts, an agent with type $\theta_1$ chooses contract $\con^1$ and action $a_1$, while an agent with type $\theta_2$ chooses contract $\con^2$ and action $a_2$. 
We can also view this as an IC type-soliciting contract, which maps $\theta_1$ to $\langle \con^1,a_1 \rangle$ and $\theta_2$ to $\langle \con^2,a_2 \rangle$, respectively.
\end{example}

\paragraph{Deterministic vs.~Randomized.} Before we dive into the discussion of what's known about typed contracts, we demonstrate that in settings with typed agents randomized contracts are strictly more powerful than deterministic ones.
Similar separations are known from multi-dimensional mechanism design for the revenue objective, see, e.g., \citet{BriestEtAl2015}.

\begin{example}[Deterministic vs.~randomized contracts, Proposition C.4 in \cite*{AlonDT21}]\label{ex:det-vs-rand} Consider the following typed contract setting, in which the agent's type only affects the cost of the actions. Action $i$ takes $\gamma_i$ units of effort. The agent has a private cost per unit-of-effort, denoted by $c$. The cost of action $i$ is $\gamma_i \cdot c$. 
The agent is of one of two possible types, $\theta_L$ and $\theta_H$, which are equally likely. Type $\theta_L$ correspond to $c = 1$, and type $\theta_H$ corresponds to $c = 3.$

\begin{center}
\begin{tabular}{|l|ccc|c|c|}
\toprule
& $r_1 = 0$ & $r_2 = 20$ & $r_3 = 35$ & \text{units of effort} & cost\\
\midrule
action $1$: & $1$ & $0$ & $0$ & $\gamma_1 = 0$& $0$\\
action $2$: & $0$ & $1$ & $0$ & $\gamma_2 = 1$ & $c$\\
action $3$: & $0$ & $\nicefrac{1}{2}$ & $\nicefrac{1}{2}$& $\gamma_3 = 3$ & $3c$\\
action $4$: & $0$ & $0$ & $1$ & $\gamma_4 = 10$ & $10c$\\
\bottomrule
\end{tabular}
\end{center}

Let's first suppose we knew the agent's type. In this case, for each of the two types, $\theta_L$ and $\theta_H$,
the best way for the principal to incentivize the four actions is by using the respective contracts $(0,0,0)$, $(0,c,0)$, $(0,0,6c)$ and $(0,0,14c)$, with respective expected payments $0$, $c$, $3c$ and $14c$. 
Thus, the principal's utility for incentivizing the four actions are $0, 20-c, 27.5-3c$, and $35-14c$, respectively.
Consequently, for any $7.5/17 \leq c \leq 3.75$, and in particular, under both types, the best contract is $(0,0,6c)$, incentivizing action $3$.
The issue is that if we would post the menu of contracts consisting of the two contracts $(0,0,6)$ and $(0,0,18)$, 
then both types would choose $(0,0,18)$ and action $3$. The principal's expected utility from this menu of contracts would be $18.5$. 

It turns out that the optimal deterministic IC type-soliciting contract maps $\theta_L$ 
to $\langle (0,0,10), 3 \rangle$ and 
$\theta_H$ 
to $\langle (0,3,0), 2\rangle$, for an expected principal utility of $19.75$ (see \cite{AlonDT21}). Intuitively, this contract ``downgrades'' the high-cost type from action $3$ to action $2$, offering the contract that would be optimal for that type and action in the absence of other types. The existence of this option, however, allows the agent to extract a utility of $2$ when he's of the low-cost type (by choosing $(0,3,0)$ and action $2$). To incentivize the low-cost type to take action $3$ we thus need to increase the expected payment for action $3$ by $2$ (leading to the contract $(0,0,10)$ rather than $(0,0,6)$).   

Next consider the randomized type-soliciting contract that maps $\theta_L$ 
to either $\langle (0,1,5), 3 \rangle$ or $\langle (0,0,14), 4 \rangle$ with equal probability, and 
$\theta_H$ 
to $\langle (0,3,0), 2\rangle$. Note that this contract achieves an expected principal utility of $\nicefrac{1}{4}\cdot (27.5-3) + \nicefrac{1}{4}\cdot (35-14) + \nicefrac{1}{2} \cdot (20-3) = 19.875$, which is strictly more than the maximum utility of $ 19.75$ from a deterministic contract, provided that the agent truthfully reveals his type and follows the recommended action.

It remains to verify that this contract is IC. First consider the case where the agent's type is $\theta_L$. 
In this case, the agent's expected utility under truthful reporting and the recommended actions is $\nicefrac{1}{2} \cdot (3-3) + \nicefrac{1}{2} \cdot (14-10) = 2$. First note that when the agent truthfully reports his type, he has no incentive to choose a different action. Indeed, under contract $(0,1,5)$ the agent's utility is maximized by action $3$, while under contract $(0,0,14)$ the agent's utility is maximized by action $4$. The agent could also misreport his type to choose contract $(0,3,0)$. However, under this contract, the agent's best action is action $1$, which yields a utility of $2$ (which is exactly what he already gets). 
Next consider the case, where the agent's type is 
$\theta_H$.
In this case, the agent's utility for reporting truthfully and following the recommendation is $0$. The agent could report his type truthfully and choose a different action, but it is readily verified that under contract $(0,3,0)$ the recommended action (action $2$) indeed maximizes the agent's expected utility. The agent could also misreport his type to choose the lottery of contracts intended for the low-cost type, but then his maximum utility is $0$ (which is exactly what he already gets). This is because under any contract in the support of the lottery, the agent's (unique) utility-maximizing action is action $1$ which yields a utility of $0$ (all other actions yield negative utility). 
\end{example}

\subsection{Multi-Dimensional Types: Private Distributions and Costs}
\label{sub:multi-dim-types}

We first discuss results of \cite*{GuruganeshSW21,GuruganeshSW23,CastiglioniM021,CastiglioniM022}, and \cite*{GanHWX22} for the general case with multi-dimensional types, with $n$ actions and $m$ outcomes, where the agent's type $\theta \in \mathcal{T}$ determines both the probabilities $q^\theta_{i,j}$ with which action $i$ leads to outcome $j$ as well as the cost $c^\theta_i$ of each action $i$.   
In all of these papers, the type space $\mathcal{T}$ is assumed to be given as an explicit list of finitely-many agent types. In the remainder, we will use $T = |\mathcal{T}|$ to denote the number of types.

\paragraph{Computation: A Dichotomy.}

One highlight of the algorithmic study of contracts with multi-dimensional types---established in a sequence of papers \cite{GuruganeshSW21,CastiglioniM021,CastiglioniM022,GanHWX22}---is a stark computational separation between deterministic menus of contracts and randomized ones. It turns out that, while deterministic menus of contracts are intractable (and, in fact, hard to approximate to within any constant), (near-)optimal randomized menus of contracts can be computed efficiently.
It is worth noting that comparable separations have been established in multi-dimensional mechanism design, for example, the problem of designing a revenue-maximizing auction for a single unit-demand buyer \citep{BriestEtAl2015}.
A similar phenomenon also arises in the context of signaling schemes for revenue maximization in auction design. While the problem of determining the optimal deterministic signaling scheme is (strongly) \textsf{NP}-hard, the optimal randomized signaling scheme can be computed in polynomial time using linear programming (see \cite{EmekFGLT14,GhoshNS07}). 

Let's start with the negative results for deterministic menus of contracts. The studies of \cite{GuruganeshSW21,CastiglioniM021,CastiglioniM022} establish a series of negative results, culminating with a proof that the optimal deterministic menu of contracts is hard to approximate to within any multiplicative factor in time polynomial in $n,m,$ and $T$. In fact, the problem remains hard even when the number of actions~$n$ and the number of outcomes~$m$ are both constants. 

\begin{theorem}[%
\citet*{CastiglioniM022}]
\label{thm:APX-hardness-with-types}
	Given a contract setting with $n$ actions, $m$ outcomes, and $T$ multi-dimensional types, it is $\mathsf{NP}$-hard 
    to approximate the principal's expected utility obtainable with a deterministic menu of contracts to within any constant multiplicative factor, even when $n$ and $m$ are both constants.
\end{theorem}

The proof is by reduction from a promise problem called $\text{\textsf{GAP-BOUNDED-IS}}_{\beta,k}$, which is related to the \textsf{INDEPENDENT-SET} problem on undirected graphs with bounded-degree vertices. 
Let $\beta \in [0,1]$ and let $k$ be an integer. The input to the problem is an undirected graph $G = (V,E)$, in which each vertex has degree at most $k$ and a parameter $\eta \in [\frac{1}{k},1]$ such that one of the following is true: Either there exists an independent set of size $\eta|V|$; or all the independent sets have size at most $\beta\eta|V|$. The goal is to determine which of the two conditions apply to the given instance.
The proof exploits that for every $\beta > 0$ there exists a constant $k = k(\beta)$ such that the promise problem $\text{\textsf{GAP-BOUNDED-IS}}_{\beta,k}$ is $\mathsf{NP}$-hard \citep{AlonFWZ95,Trevisan01}.

The same reduction shows that there cannot be an FPTAS for computing an additive approximation to the revenue of the optimal deterministic menu of contracts.

In sharp contrast to these negative results for deterministic menus of contracts, \citet{CastiglioniM022} and \citet{GanHWX22} show that the problem of computing the optimal randomized menu of contracts admits an additive FPTAS; 
namely, it is possible to efficiently compute a randomized menu of contracts that approximates the optimal randomized menu of contracts up to an additive error term $\varepsilon > 0$.
The error term is needed as the problem may only admit a supremum, rather than a maximum.

\begin{theorem}[\citet*{CastiglioniM022} and \citet*{GanHWX22}]
Consider a precision parameter $\varepsilon>0$, and a contract setting with $n$ actions, $m$ outcomes, and $T$ multi-dimensional types. Then, a menu of randomized contracts with revenue at most an additive $\epsilon$ away from the supremum over all such menus can be computed in time $\mathsf{poly}(n,m,T,\nicefrac{1}{\varepsilon})$.\footnote{{The algorithms given by \cite{CastiglioniM022} and \cite{GanHWX22} are based on linear/convex programming, and hence only weakly polynomial-time.}} 
\end{theorem}

The two papers by \cite{CastiglioniM022} and \cite{GanHWX22} differ in how they establish this result. The joint proof strategy of both papers is to first reduce the design space by arguing that one can restrict attention to certain succinct randomized contracts. In \cite{CastiglioniM022} they arrive at a linear program that has exponentially many variables but only polynomially many constraints. They then turn to the Ellipsoid method, and provide an efficient separation oracle for the dual. In \cite{GanHWX22}, in contrast, they arrive at a succinct convex program, which can be solved directly.

\paragraph{Approximation Bounds.}

Another important direction in this line of work quantifies the worst-case (multiplicative) loss in the principal's utility and welfare between different types of contracts and benchmarks \citep{GuruganeshSW21,CastiglioniM021,GuruganeshSW23}. 
From least to most general, the contracts and benchmarks that have been considered include: the principal's utility under linear contracts, single contracts, deterministic menus of contracts, and randomized menus of contracts, as well as social welfare.

\citet{CastiglioniM021} show that the worst-case loss of any linear contract against the best single contract is at least $\Omega(T)$, even when there are only two actions. \citet{GuruganeshSW21} show that this gap is at least $\Omega(n \log T)$, even when the type distribution is uniform and the type only affects the agent's probability matrix and not the costs (i.e., the costs are fixed and shared by all types).

\citet{GuruganeshSW23} show a lower bound of $\Omega(\max\{n,\log T\})$ on the potential loss from using a single contract rather than a deterministic menu of contracts. They also present a construction with $n = O(T)$ actions in which the best deterministic menu of contracts incurs a loss of $\Omega(T)$ relative to the best randomized menu of contracts. 

Finally, \citet{GuruganeshSW21} show that the worst-case loss from a deterministic menu of contracts relative to the welfare is $\Omega(n \log T)$, while the respective worst-case for randomized menus of contracts is shown to be at least $\Omega(n)$.

Together these results show that there are significant gaps between any two consecutive levels of the hierarchy. Another important insight is that (with the possible exemption of the last comparison, between randomized menus of contracts and welfare) all gaps have to grow with the number of actions \emph{and} the number of types.

\subsection{{Single-Dimensional Types: Private Cost per Unit-of-Effort}}
\label{sub:single-dim-types}

Next we discuss a natural restriction of the general multi-dimensional types model from Section~\ref{sub:multi-dim-types} to \emph{single-dimensional} types, introduced by \citet*{AlonDT21} and \citet*{AlonDLT23}. As before, there are $n$ actions and $m$ outcomes. The matrix $\{q_{i,j}\}$ that describes how each action $i$ translates into reward $t_j$ is fixed (and known). The actions take different amounts of effort, as given by a (known) vector $(\gamma_1, \ldots, \gamma_n)$. The private type consists of the agent's cost $c\in\mathbb{R}_{\ge 0}$ for expending one unit-of-effort, where $c$ is distributed according to distribution $G$. Action $i$'s total cost is then given by $c_i=c\gamma_i$. For an example illustrating this model with private cost per unit-of-effort, see Example~\ref{ex:det-vs-rand}.

\paragraph{Simple vs.~Optimal.}

\cite{AlonDT21} and \cite{AlonDLT23} focus on deterministic IC type-soliciting contracts. They give two characterizations of implementable ``allocation rules'', i.e., mappings from types to recommended actions. Here, ``implementable'' refers to the fact that these mappings can be realized in an IC type-soliciting contract. They use these characterizations to show that optimal IC type-soliciting contracts for this single-dimensional typed contract setting exhibit several undesirable features, akin to those known from \emph{multi-dimensional mechanism design}~\cite[e.g.,][]{Daskalakis15}. For instance, optimal deterministic IC type-soliciting contracts may fail to satisfy revenue monotonicity. Namely, suppose that for two type distributions $H$ and $G$ it holds that $G(c) \geq H(c)$ for all $c$. That is, types drawn from $G$ are more likely to have lower cost than those drawn from $H$. Then one would expect that the principal's expected utility under $G$ is at least as high as under $H$. However, this is not necessarily the case. 
Another observation is that optimal deterministic IC type-soliciting contracts may have a \emph{menu complexity} (size of the image of the mapping from types to contracts and recommended actions) of at least $\Omega(n)$. 
These findings further amplify the critique of optimal contracts in pure hidden-action models that has motivated the work in  Section~\ref{sec:simple}; and it is  natural to ask for conditions under which simple contracts (such as linear contracts) are near-optimal in Bayesian settings.  

The main result of \citet{AlonDLT23} is that linear contracts provide a good approximation to the optimal welfare whenever the setting is \emph{not} point-mass like and there is enough uncertainty abut the setting.\footnote{We already know from Theorem~\ref{thm:wc-approx} that linear contracts can be far from optimal for degenerate Bayesian settings, without any uncertainty about the setting.} The result is driven by a parameterization of the tail of the induced welfare distribution. 

To formally state the condition on the tail of the welfare distribution (see Definition~\ref{def:small-tail}), we need the following notation. For a fixed principal-agent instance with rewards $\{\rew_j\}$, probability matrix $\Prob = \{\prob_{ij}\}$, units of effort $\{\gamma_i\}$, and type distribution $G$ with range $[\underbar{c},\bar{c}]$ we define the \emph{welfare contribution from types in $[a,b] \subseteq [\underbar{c},\bar{c}]$} as 
\[
\text{Wel}(a,b) := \int_{a}^{b} R_{i^\dagger(c)} - \gamma_{i^\dagger(c)} \cdot c \;d G(c),
\]
where 
\[
i^\dagger(c) \in \argmax_{i \in [n]} \left(\Rew_{i} - \gamma_i \cdot c \right) 
\]
is an action that maximizes the expected welfare for type $c$.

\begin{definition}[\citet*{AlonDLT23}]\label{def:small-tail} 
Let $\eta \in (0,1]$ and $\kappa \in [\underbar{c},\bar{c}]$. 
A principal-agent instance has \emph{$(\kappa,\eta)$-thin-tail}%
\footnote{Notice that when the types represent costs rather than values, a low cost corresponds to a strong type. Consequently, the tail of the distribution is on the left, reversing the usual situation with value distributions, where the tail is on the right.} if   
\[
\emph{Wel}(\underbar{c},\kappa) \leq (1-\eta) \cdot \emph{Wel}(\underbar{c},\bar{c}).
\]
\end{definition}

Intuitively, the $(\kappa, \eta)$-thin-tail condition quantifies how much of the welfare is concentrated in
the tail around the strongest (lowest cost) types. The larger $\kappa$ is, and
the closer $\eta$ is to $1$, the thinner the tail and the further the setting is from point mass.

We remark that this condition is a property of the whole instance, and not just the type distribution. \cite{AlonDLT23} demonstrate that this is necessary: there are instances with uniform type distribution (and thus well spread out types) where the whole welfare is concentrated on the tail of the distribution and linear contracts are far from optimal.

The following theorem shows an approximation guarantee for linear contracts in terms of the parameterization of the tail, against the optimal welfare, which serves as an upper-bound on the revenue. To state the theorem, for every
quantile $q \in (0,1)$, denote by $c_q$ the cost corresponding to quantile $q$, i.e., $G(c_q) = q$. Notice that $c_q$ is increasing in $q$.

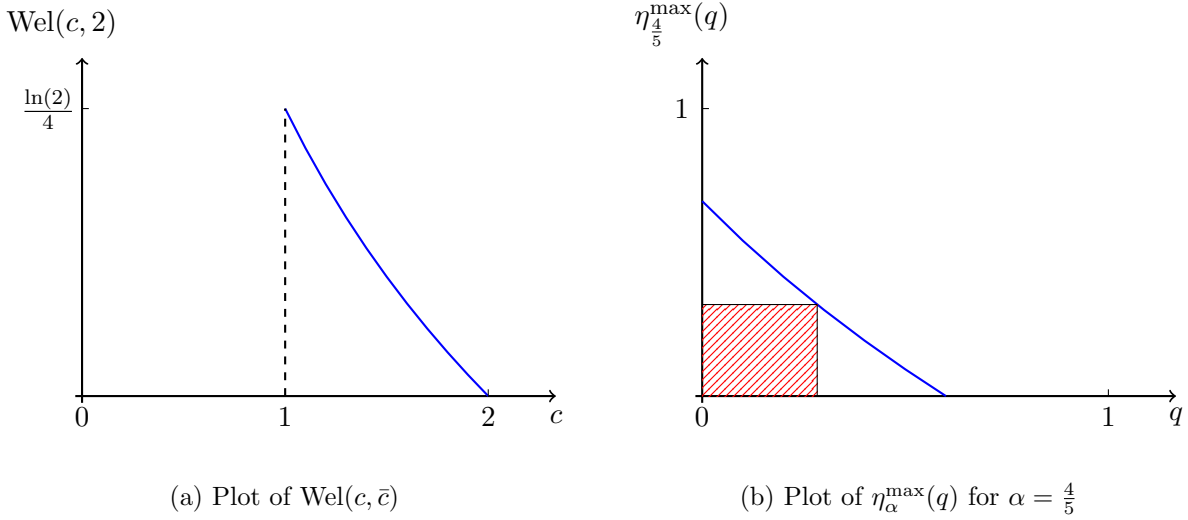
\begin{figure}[t]
\begin{subfigure}[t]{0.49\textwidth}
\begin{center}
\begin{tikzpicture}[scale = 0.9]

\draw[scale=0.5,->,thick] (-0.2,0) -- (14, 0);
\draw[scale=0.5,->,thick] (0, -0.2) -- (0, 10);

\draw [thick,blue] plot coordinates {(3,4.25) (3.3, 3.6655) (3.6,3.1320) (3.9,2.6413) (4.2, 2.1869) (4.5, 1.7639) (4.8, 1.36819) (5.1, 0.9964) (5.4, 0.64601) (5.7, 0.3145) (6,0)};

\draw [thick,dashed,black] (3,0) -- (3,4.25);

\draw[scale=0.5] (12,0) -- (12,0.2);
\draw[scale=0.5] (12, -0.6) node {$2$};
\draw[scale=0.5] (0.5*12, -0.6) node {$1$};

\draw[scale=0.5] (0,8.5) -- (0.2,8.5);
\draw[scale=0.5] (-1,8.5) node {$\frac{\ln(2)}{4}$};
\draw[scale=0.5] (0,-0.6) node {$0$};

\draw (7, -0.3) node {$c$};
\draw (-0.3, 5.5) node {$\text{Wel}(c,2)$};

\end{tikzpicture}
\end{center}
\caption{Plot of $\text{Wel}(c,\bar{c})$}
\end{subfigure}
\begin{subfigure}[t]{0.5\textwidth}
\begin{center}
\begin{tikzpicture}[scale = 0.9]

\draw[scale=0.5,->,thick] (-0.2,0) -- (14, 0);
\draw[scale=0.5,->,thick] (0, -0.2) -- (0, 10);

\def\x{12*0.5}
\def\y{8.5*0.5}
\draw [thick,blue] plot coordinates {(0*\x, 0.678072*\y) (0.1*\x, 0.540568*\y) (0.2*\x, 0.415037*\y) (0.3*\x, 0.29956*\y) (0.4*\x, 0.192645*\y) (0.5*\x,0.0931094*\y) (0.6*\x,0)};

\fill[pattern={north east lines},pattern color=red]
    (0,0) rectangle +(0.28316*\x,0.318371*\y);
\draw[-] (0,0.318371*\y) -- (0.28316*\x,0.318371*\y);
\draw[-] (0.28316*\x,0) -- (0.28316*\x,0.318371*\y);


\draw[scale=0.5] (12,0) -- (12,0.2);
\draw[scale=0.5] (12, -0.6) node {$1$};

\draw[scale=0.5] (0,8.5) -- (0.2,8.5);
\draw[scale=0.5] (-0.6,8.5) node {$1$};
\draw[scale=0.5] (0,-0.6) node {$0$};

\draw (7, -0.3) node {$q$};
\draw (-0.3, 5.5) node {$\eta_{\frac{4}{5}}^{\max}(q)$};

\end{tikzpicture}
\end{center}
\caption{Plot of $\eta_{\alpha}^\text{max}(q)$ for $\alpha = \frac{4}{5}$}\label{fig:thin-tail}
\end{subfigure}
\caption{
Visualization of the key quantities involved in applying Theorem~\ref{thm:approx-bayesian} to the contracting problem described in Example~\ref{ex:continuum-of-actions}. The left tableau gives the welfare contribution $\text{Wel}(c,2)$ from types above $c$, as a function of $c$. The right tableau gives the quantity $\eta_{\alpha}^{\text{max}}(q)$ for $\alpha = \frac{4}{5}$, as a function of $q$. The best-possible approximation guarantee that can be shown via the theorem for a fixed choice of $\alpha$ (here $\alpha = \frac{4}{5}$) is proportional to the largest-area rectangle that can fit under this curve (red, striped box).
}\label{fig:small-tail}
\end{figure}

\begin{theorem}[\citet*{AlonDLT23}]\label{thm:approx-bayesian} 
Let $q, \alpha, \eta \in (0,1)$. For
any principal-agent instance with $(\frac{c_q}{\alpha},\eta)$-thin-tail, a linear contract with parameter $\alpha$
provides expected revenue that is an $\frac{1}{(1-\alpha)\eta q}$-approximation of the optimal welfare.
\end{theorem}

The proof of this theorem exploits that if the thin-tail condition is satisfied, then the type distribution $G$ cannot grow too quickly. Moreover, in this case, the contribution of lower-cost (thus stronger) types to a linear contract’s
revenue is sufficient to cover the welfare from higher-cost ones. This leaves the welfare from lower-cost types uncovered, but the thin-tail condition ensures that this contribution is limited.

Let's carefully parse Theorem~\ref{thm:approx-bayesian}, and see how it connects approximation guarantees offered by linear contracts to properties of the tail. First note that for a fixed $\alpha$ and a fixed $q \in [0,1]$, the largest $\eta$ that satisfies the $(\frac{c_q}{\alpha},\eta)$-thin-tail condition is
\begin{align}
\eta^{\text{max}}_{\alpha}(q) := \frac{\text{Wel}(\frac{c_q}{\alpha},\bar{c})}{\text{Wel}(\underbar{c},\bar{c})},
\label{eq:eta-max}
\end{align}
which is a non-increasing function of $q$. For a fixed $\alpha$ and a fixed $q$, the best approximation guarantee that can be shown via Theorem~\ref{thm:approx-bayesian} is thus proportional to $\eta_\alpha^{\text{max}}(q) \cdot q$. Optimizing this quantity over $q \in [0,1]$  
amounts to finding the area-wise largest rectangle that can fit  under the curve defined by Equation~\eqref{eq:eta-max} 
(see Figure~\ref{fig:thin-tail}).

Intuitively, for distributions that are point-mass like, for all possible choices of $\alpha$, this area is small and the approximation guarantee is poor. In contrast, when the welfare is sufficiently well-spread out over types, there will be $\alpha$ such that this area is large, and hence the approximation ratio will be good.
The following example illustrates how Theorem~\ref{thm:approx-bayesian} enables the derivation of approximation guarantees for linear contracts.

\begin{example}[Example with a continuum of actions]\label{ex:continuum-of-actions}
We illustrate the guarantee provided by Theorem~\ref{thm:approx-bayesian} by considering a setting with a continuum of actions and an arbitrary number of outcomes.\footnote{We consider a continuum of actions to simplify the calculations. Analogous results can be obtained for a setting with a finite number of actions, by discretizing the action space.} Suppose that the agent's cost $c$ is drawn from $U[1,2]$ and that for each $c$ the agent can choose action $\gamma \in [0,1]$, with an expected reward of $\Rew_\gamma = \sqrt{\gamma}$ yielding an expected welfare of $\Wel_\gamma = \sqrt{\gamma} - \gamma \cdot c$. Since $\frac{d}{d\gamma} \Wel_\gamma = \frac{1}{2\sqrt{\gamma}} - c$ and $\frac{d^2}{d\gamma^2} \Wel_\gamma = - \frac{1}{4\gamma^{3/2}}$, the welfare maximizing action for an agent with cost $c$ is $\gamma^\dagger(c) = \frac{1}{4c^2}.$ We thus have $R_{\gamma^\dagger(c)} - \gamma^\dagger(c) \cdot c = \sqrt{\frac{1}{4c^2}} - \frac{1}{4c^2}\cdot c 
= \frac{1}{4c}$ and
\[
\emph{Wel}(c,2) = \int_{x=c}^{2} \left(\frac{1}{4x}\right) \;dx = \frac{1}{4} \cdot \big(\ln(2) - \ln(c) \big). 
\]
Note that $\emph{Wel}(1,2) = \ln(2)/4 \approx 0.1734$. Next observe that, in this case, $c_q = q+1$. Let's choose $\alpha = \frac{4}{5}$. Then, for a given $q$, the largest $\eta$ that satisfies Definition~\ref{def:small-tail} is 
\[
\eta^{\text{max}}_{\frac{4}{5}}(q) 
= \frac{\emph{Wel}\left(\frac{5}{4}(q+1),2\right)}{\emph{Wel}(1,2)} = \begin{cases}  1 - \frac{\ln\left(\frac{5}{4}(q+1)\right)}{\ln(2)} & \text{for $q \leq 3/5$}\\ 
0 & \text{for $q > 3/4$}
\end{cases}.
\]
The best approximation guarantee that can be obtained via  Theorem~\ref{thm:approx-bayesian} for this $\alpha$ is then obtained by maximizing $q \cdot \eta_{\max}(q)$ over $q \in [0,\frac{3}{5}]$. This yields $\max_{q} \left(q \cdot \eta_{\max}(q) \right) \approx 0.09015 \geq \frac{1}{12}$ at $q \approx 0.28316$, for an approximation guarantee of $(\frac{1}{4}\cdot \frac{1}{12})^{-1} = 48$. So linear contracts are near-optimal in this setting despite there being arbitrarily many actions, irrespective of the details that govern how the actions translate into outcomes.
\end{example}

\cite{AlonDLT23} also offer a version of Definition~\ref{def:small-tail} and Theorem~\ref{thm:approx-bayesian} which benchmarks against the optimal revenue rather than the optimal social welfare. This benchmark is weaker hence the guarantees that can be obtained are better.
The approximation guarantees can be further improved by utilizing additional properties of the type distributions. 

As shown in \cite{AlonDLT23}, two corollaries of these general results are that for any principal-agent setting and any type distribution $G$ with non-increasing density on $[0,\infty)$, linear contracts obtain a $4$-approximation to the optimal welfare, and a $2$-approximation to the optimal revenue. 

Importantly, as demonstrated in the full version of \cite{AlonDLT23}, guarantees similar to those shown for linear contracts cannot  be obtained for other simple classes of contracts (such as single-outcome payment contracts or debt contracts).

\paragraph{Computational Complexity.} Another direction that has been studied for single-dimensional types is the computational complexity of finding optimal deterministic menus of contracts. \cite{AlonDT21}, for example, show that---in contrast to the multi-dimensional case---in the single-dimensional case the problem of computing an optimal deterministic menu of contracts is tractable for a constant number of actions. The more general case, beyond constantly-many actions, was recently addressed by \cite{CastiglioniCL+25}, whose results we discuss in more detail below.

\subsection{A Reduction from Multi-Dimensional to Single-Dimensional Types} 
\label{sec:typed-reduction}

The work of \cite*{CastiglioniCL+25} establishes a fundamental algorithmic connection between the two models of Section~\ref{sub:multi-dim-types} and Section~\ref{sub:single-dim-types}. In the former, the agent's type determines the distribution matrix and costs. In the latter, the agent's type is simplified to a single value---his cost per unit-of-effort. It thus appears that the former model is significantly more complex than the latter model. This is strengthened by the separation result of~\cite{AlonDT21} for a constant number of actions. However, \cite{CastiglioniCL+25} show that, in general, there is an (almost) approximation-preserving polynomial-time reduction from the setting with general multi-dimensional types to the single-dimensional setting (as in Section~\ref{sub:single-dim-types}).  This rules out the hope to generalize the positive results of~\cite{AlonDT21}, and is surprising in light of the separation between single- and multi-dimensional settings in mechanism design (which are generalized by Bayesian contract design). 

\begin{theorem}[\cite*{CastiglioniCL+25}]
    \label{thm:multi-to-single-reduction}
    Fix $\epsilon>0$. For any multi-dimensional instance of Bayesian contract design $I^M$ with $n$ actions, $m$ outcomes and $T$ types, there is a poly-time (in $n$, $m$, $T$, and $\log(\nicefrac{1}{\varepsilon})$)
    construction of a single-dimensional instance $I^S$ with $Tn+1$ actions, $m+1$ outcomes and $T+1$ types, such that:\footnote{The construction/reduction of \cite{CastiglioniCL+25} relies on linear programming techniques, and is thus only weakly-polynomial time.}
    \begin{itemize}
        \item Any $\beta$-approximate single contract (joint for all agent types) 
        for $I^S$ can be converted into a $(\beta+\epsilon)$-approximate single contract for $I^M$.
        \item Any $\beta$-approximate deterministic  menu of contracts for $I^S$ can be converted to a $(\beta+\epsilon)$-approximate deterministic  menu of contracts for $I^M$.
        \item If $\beta=1$, then the dependence on $\varepsilon$ is removed and both reductions are exact. 
    \end{itemize}
\end{theorem}

While the reduction is technically involved, it is useful to mention that for each (action, type) pair in $I^M$, 
there will be a corresponding action in $I^S$. 
The multi-dimensional types are reduced to a single-dimensional type by packing them into a single dimension, using exponentially-decaying/increasing parameters. 
This compression is done in a way that ensures that an agent with a single-dimensional type corresponding to the multi-dimensional type $\theta$ will only choose actions from among (action, type) pairs with the type set to $\theta$.
This also explains why the separation result of~\cite{AlonDT21} for a constant number of actions holds despite the reduction: the polynomial-time reduction from multi-dimensional settings to single-dimensional ones blows up the number of actions by a factor of $T$, i.e., the number of agent types.%
\footnote{Theorem~\ref{thm:multi-to-single-reduction} together with the results in \cite{AlonDT21} implies that an optimal deterministic menu of contracts for multi-dimensional settings can be found in polynomial time, when both the number of actions and types is constant. The complexity of the more general case with with a general number of actions and constantly-many types is, to our knowledge, open.}

A take-away from Theorem~\ref{thm:multi-to-single-reduction} is that it is sufficient to focus on the single-dimensional setting to develop positive computational or learning algorithms, and likewise, it is sufficient to focus on the multi-dimensional setting when developing hardness results. 

\begin{corollary}[\cite*{CastiglioniCL+25}]
    Consider single-dimensional Bayesian contract design settings. For settings with $T$ types, for any $\delta\in(0,1]$ it is \textsf{NP}-hard to compute a $T^{(1-\delta)}$-approximation to the optimal single contract. Moreover, for any constant $\rho\ge 1$ it is \textsf{NP}-hard to compute a $\rho$-approximation to the optimal deterministic menu of contracts. 
\end{corollary}

As a technical tool, \cite{CastiglioniCL+25} also establish a result regarding the \emph{power of menus}; namely, a (tight) $\Omega(n)$-separation between the principal's utility via the optimal deterministic menu of contracts and the optimal single contract. In particular, this bound is independent of the number of types $T$, which presents an interesting contrast to general multi-dimensional settings (where the gap depends on both $n$ and the number of types $T$).

\subsection{Agent-Designed Contracts with Typed Principals} 
\label{sub:agent-designed}

\citet*{BernasconiEtAl24} introduce agent-designed contracts, reversing the role of the principal and the agent (see also Footnote \ref{ftnt:fourth-rubric}). They study a hidden-action setting in which the party who is more informed about the action---namely the agent---moves first and designs the contract. The friction arises from the fact that the principal has a private type, namely her rewards for different outcomes.
A deterministic menu of contracts consists of pairs
$(i,\mathbf{t}_i)$, each specifying the action and payment vector. For example, a service provider (agent) can offer a user (principal) several levels of service quality that require increasing effort. The payment vector ensures that the chosen effort level is incentive compatible for the agent.

The principal knows her private type and uses this knowledge to choose among the menu options. \cite{BernasconiEtAl24} show there is no polynomial-time algorithm that can approximate the optimal deterministic menu of contracts within any additive factor, i.e., the problem is not in \textsf{APX}. However, they find that the problem becomes tractable if the agent is restricted to menus of constant size. The model is then extended to handle randomized menus of contracts. They show that optimal menus of randomized contracts can be computed in polynomial time---and provide at least $\Omega(T)$ times more utility than optimal deterministic menus (where $T$ is the number of principal types).

\paragraph{Additional Directions and Open Questions.} 

Several interesting open questions remain. For example, for many of the worst-case comparisons between different classes of contracts and benchmarks, there are rather significant gaps between the best-known lower bounds and upper bounds. We note that, while our exposition (like most of the existing work) has focused on typed contract settings with a single agent, it is natural to extend this study to settings with multiple agents, and more generally combinatorial contracts with types. We refer the interested reader to \cite{CastiglioniEtAl23} and \cite{CacciamaniEtAl24} for results on multi-agent settings with private types. 

\section{Machine Learning for Contracts: Data-Driven Contracts}
\label{sec:data-driven}
In this and the following section, we explore interactions between contracts and machine learning. We start by considering the problem of learning (near-)optimal contracts.
The learning angle helps bridge the gap between theory and practice by making more realistic informational assumptions, and as we shall see, it also sheds additional light on the tradeoff between simple and optimal contracts.

A pioneering work in this direction is the work of \citet*{HoSV16}, who formulate the problem of learning optimal contracts as an \emph{online} learning problem, and give algorithms that achieve sublinear regret. 
In their model, the agent is drawn from an unknown distribution, and the principal repeatedly posts a contract and observes an outcome sampled from the agent's best-response action. 

In our exposition, we focus on the recent results by \citet*{ZhuEtAl22} (see Section~\ref{sub:samples-general}), who give nearly tight bounds on the regret achievable in this model, for both linear contracts and general (bounded) contracts.
The work of \citet{ZhuEtAl22} shows that while linear contracts can be learned with only
polynomial regret, general (bounded) contracts necessarily entail regret that is exponential in the number of outcomes. The hardness for general (bounded) contracts applies even when the principal repeatedly interacts with the \emph{same} agent, but requires the agent to have (exponential in the number of outcomes) many actions.

We then discuss subsequent works by \cite*{BacchiocchiC0024} (see Section~\ref{sub:samples-small}), and \cite*{ChenCDH24} (see Section~\ref{sub:samples-regular}). Motivated by the impossibility for general (bounded) contracts, these works demonstrate that---in settings where the principal repeatedly interacts with the same agent---polynomial regret bounds are possible when either the agent has few (i.e., constantly many) actions, or the setting satisfies regularity assumptions. 

We conclude with a discussion of \cite*{GuruganeshSW23} (see Section~\ref{sub:samples-feedback}). Their work implies improved regret bounds for the problem of learning linear contracts, for a setting where the principal repeatedly interacts with the same agent, and the feedback consists of the principal's expected utility under the agent's best-response action. 

\subsection{Tight Regret Bounds for General Instances}
\label{sub:samples-general}

We first discuss the results of \citet*{ZhuEtAl22}---the state-of-the-art results in the model introduced by \citet*{HoSV16}.\footnote{Also see \citet*{CohenDK22}, who study the problem of learning bounded contracts in this model, even with possibly risk-averse agents, under the additional assumption that the instances satisfy FOSD (see Section~\ref{sub:regularity}) and the contracts are \emph{monotone smooth}. Assuming rewards are sorted from low to high, a contract $\con$ is monotone smooth if $0 \leq t_{j+1}-t_{j} \leq r_{j+1} - r_j$ for all $j \in [m]$.} 
In this problem, a single principal repeatedly interacts with an agent. The interaction takes place over $\numrounds$ rounds. In each round $\round$, the agent's type $\theta^\round$ is drawn from a type distribution $\mathcal{D}$. This distribution is not known to the principal. The principal has fixed rewards $\rew_j \geq 0$ for $m$ possible outcomes $j \in [m]$. The agent can choose from $n$ actions $i \in [n]$. The agent's type $\theta^\round$ determines the cost $c^\theta_i \geq 0$ of each action $i \in [n]$, as well as the probability distribution $\Prob^\theta_i$ over outcomes $j \in [m]$. It is assumed that both rewards and costs are bounded in $[0,1]$.\footnote{Since regret is an additive metric, we need to specify the range of the key quantities involved. 
Normalization to $[0,1]$ can always be achieved through appropriate scaling, but also scales the regret with respect to the original unscaled instances accordingly.}

In each round $\round$, 
the principal posts a contract $\con^\round = (\pay^\round_1, \ldots, \pay^\round_m)$ (a non-negative payment for each outcome). The choice of contract may depend on what the principal has observed so far, and the choice of contract may be randomized. We consider two classes of contracts. In a \emph{bounded} contract we have $\con^\round \in [0,1]^m$,
while a \emph{linear} contract is defined by $\alpha \in [0,1]$ and has $t^\round_j = \alpha \cdot \rew_j$ for all $j \in [m]$. 
After the principal has posted contract $\con^\round$, a type $\theta^\round$ is drawn from $\mathcal{D}$, the agent takes a best response action $i^{\star}(\theta^\round,\con^\round)$, and an outcome $j^\round$ is sampled from $\Prob_{i^{\star}(\theta^\round,\con^\round)}$. The principal learns about the outcome $j^\round$, receives the corresponding reward $r_{j^\round}$, and pays the agent the amount specified by contract $\con^\round$ for outcome $j^\round$.

The principal's goal is to minimize regret with respect to the best single contract in hindsight. To formally define this, let $\mathcal{T}$ denote a class of contracts (e.g., linear or bounded). Let $U_P(\con \mid \theta)$ denote the expected principal utility for contract $\con$ when the agent's type is $\theta$, let $\pi$ be a policy which maps each history $\mathcal{H}^{\round-1}$ to a distribution over contracts. We then have
\[
\mathsf{regret}(\pi,\mathcal{T}) := \sup_{\bar{\con} \in \mathcal{T}} \sum_{\round=1}^{\numrounds} \mathbb{E}_{\con^\round \sim \pi(\mathcal{H}^{\round-1})} \left(\mathbb{E}_{\theta^\round}[U_P(\bar{\con} \mid \theta^\round)] - 
\mathbb{E}_{\theta^t}[U_P(\con^\round \mid \theta^\round)] \right).
\]

The main result of \cite{ZhuEtAl22} is a pair of nearly-tight upper and lower bounds on the regret achievable when the goal is to learn bounded contracts.
This problem is challenging for two reasons. First, the contract space is a continuous high-dimensional cube (namely $[0,1]^m$). Second, the expected principal utility (as a function of the contract) is not Lipschitz continuous, meaning that even a slight change in the contract can cause the expected utility to jump. 
The key insight behind the upper bound is that the problem admits a weaker form of continuity,
establishing that there is a direction (a cone) in which the utility doesn't drop off by too much. With this, the problem can be reduced to a well-understood covering problem. The lower bound is obtained through a meticulous explicit construction. 
In what follows, we use $\tilde{O}(\cdot)$ to denote $O(\cdot)$ omitting logarithmic factors.

\begin{theorem}[\citet*{ZhuEtAl22}]\label{thm:zhu1}
There is an online learning algorithm for 
bounded contracts that incurs a regret of at most $\tilde{O}(\sqrt{m} \cdot \numrounds^{1-1/(2m+1)})$, and no online learning algorithm can incur a regret better than $\Omega(\numrounds^{1-1/(m+2)})$.
\end{theorem}

This result is mostly a negative one, as it establishes that the achievable regret grows (and has to grow) exponentially in the number of outcomes $m$. 
Notably, the lower bound applies even if the principal interacts with the \emph{same} agent over all $\numrounds$ rounds. Another important feature of the lower bound construction is that it requires exponential in $m$ many actions. So it does not rule out polynomial regret bounds when the number of actions is small.

The impossibility result for general (bounded) contracts becomes particularly interesting, when contrasted with the following positive result for linear contracts, which shows that this problem admits polynomial regret bounds.

\begin{theorem}[\citet*{ZhuEtAl22}]\label{thm:zhu2}
There is an online learning algorithm for 
linear contracts that incurs a regret of at most $\tilde{O}(\numrounds^{2/3})$, and no online learning algorithm can incur a regret better than $\Omega(\numrounds^{2/3})$.
\end{theorem}

There are two main differences between linear contracts and bounded contracts that drive the difference in the asymptotic regret. First, the space of linear contracts is just the unit interval $[0,1]$ as opposed to the $m$-dimensional cube $[0,1]^m$. Second, whereas bounded contracts only admit a rather weak directional notion of continuity, linear contracts are one-sided Lipschitz continuous. Intuitively, this says that the principal's expected utility cannot drop by too much when we slightly overshoot the parameter, provided that we don't overshoot by too much. (It's ``one-sided'' because the utility can still drop a lot if we undershoot the parameter.) The upshot is that for linear contracts a uniform discretization of the unit interval with carefully chosen discretization width, together with standard regret-minimization algorithms imply the desired bound.

Note that the bound for linear contracts is polynomial, and neither depends on the number of actions nor the number of outcomes. Together the two results for bounded 
and linear contracts thus highlight another desirable feature of linear contracts, namely ``learnability.'' 
An intriguing general open problem is whether there are other ``simple'' contracts that can be learned efficiently, while allowing the principal to achieve higher expected utility.

\subsection{Improved Regret Bounds with a Small Number of Actions}
\label{sub:samples-small}

In follow-up work, \citet*{BacchiocchiC0024} consider the same online learning problem as \cite{ZhuEtAl22}, except that they assume that the principal interacts with the \emph{same} agent over all $\numrounds$ rounds. Their main contribution is a polynomial regret bound for bounded contracts for settings with a \emph{constant} number of actions. Recall that the lower bound of Theorem~\ref{thm:zhu1} for bounded contracts already applies to this setting, but requires instances where the number of actions $n$ is exponential in the number of outcomes $m$. 

More formally, \cite{BacchiocchiC0024} assume that the principal interacts with the agent over $\numrounds$ rounds. There are $m$ outcomes with rewards $\rew_j \in [0,1]$ for $j \in [m]$. The agent can take one of $n$ actions. Each action $i \in [n]$ is associated with a cost $c_i \in [0,1]$ and a probability distribution $\Prob_i$ over outcomes,
both of which are unknown to the principal. 
In each round $\round$, the principal posts a bounded contract $\con^\round = (t^\round_1, \ldots, t^\round_m) \in [0,1]^m$, the agent takes a best response action $i^\star(\con^\round)$, and then an outcome $j$ is sampled from $\Prob_{i^\star(\con^\round)}$. The principal gets to observe the sampled outcome $j$, but not the agent's action. As before the principal's goal is to minimize regret. Since the principal interacts with the same agent over all rounds, writing $U_P(\con)$ for the principal's expected utility given contract $\con$, the regret is now:
\[
\mathsf{regret}(\pi,\mathcal{T}) := \sup_{\bar{\con} \in [0,1]^m} \sum_{\round=1}^{\numrounds} \mathbb{E}_{\con^\round \sim \pi(\mathcal{H}^{\round-1})} \left(U_P(\bar{\con}) - 
U_P(\con^\round) \right).
\]

The approach of \cite{BacchiocchiC0024} relies on a (standard) reduction of the online learning problem to the following offline sample complexity question. Formally, a \emph{contract query} is given a contract $\con$, and returns an outcome $j$ sampled from the distribution over outcomes $\Prob_{i^\star(\con)}$ induced by the agent's best response action $i^\star(\con)$ to contract $\con$.\footnote{The term contract query is due to the work of \citet{ChenCDH24}, which we discuss below.} The question is, given parameters $\delta > 0$ and $\epsilon > 0$, how many contract queries are needed to identify a bounded contract $\con$ such that with probability at least $1-\delta$ it holds that
\[
U_p(\con) \geq \max_{\bar{\con} \in [0,1]^m} U_P(\bar{\con}) - \varepsilon.
\]

The offline-to-online reduction is (see \citet[][Theorem 3]{BacchiocchiC0024}): 

\begin{proposition}[Offline to online reduction]\label{prop:sample-reduction} 
Let $a, b, c > 1$ be constants.
Suppose that for any $\delta \in (0,1)$ and $\varepsilon > 0$, there is an offline learning algorithm that computes, with probability $1-\delta$, an $\varepsilon$-approximate bounded contract, with at most $\tilde{O}(m^n \cdot m^a \cdot n^b \cdot \nicefrac{1}{\varepsilon^c} \cdot \log(\nicefrac{1}{\delta}))$ contract queries. Then, for any $\delta \in (0,1)$, there exists an online learning algorithm for bounded contracts that, with probability $1-\delta$, incurs a regret of at most $\tilde{O}(m^n \cdot m^{\nicefrac{a}{(c+1)}} \cdot n^{\nicefrac{b}{(c+1)}} \cdot \numrounds^{\nicefrac{c}{(c+1)}} \cdot \log(\nicefrac{1}{\delta}))$. 
\end{proposition}

\begin{proof}[Proof sketch.] Fix $\delta > 0$ and run the offline learning algorithm with parameter $\varepsilon$, to be determined later, to learn a contract. 
Then use this contract for the remaining rounds. Let's call these two phases the \emph{exploration phase} and the \emph{exploitation phase}. Note that the length of the exploration phase is $S_1(\varepsilon) = \tilde{O}(m^n \cdot m^a \cdot n^b \cdot \nicefrac{1}{\varepsilon^c} \cdot \log(\nicefrac{1}{\delta}))$, while the length of the exploitation phase is $S_2(\varepsilon) = \max\{S - S_1(\varepsilon),0\} \leq S$.
The per-round regret in the exploration phase is at most $1$. Moreover, with probability $1-\delta$, the exploration phase succeeds in identifying an $\varepsilon$-approximate contract. In this case, the per-round regret in the exploitation phase is at most $\varepsilon$. Thus, with probability $1-\delta$, the regret is at most 
\[
\tilde{O}(\underbrace{S_1(\varepsilon) \cdot 1}_{\text{from exploration phase}} + \underbrace{\numrounds \cdot \textcolor{white}{(}\varepsilon\textcolor{white}{)}}_{\text{from exploitation phase}}). 
\]
Now, we choose $\varepsilon$ to equate the left and right terms, to get $\varepsilon = (m^n \cdot m^a \cdot n^b \cdot \nicefrac{1}{\numrounds})^{\nicefrac{1}{(c+1)}}$.
So,  with probability $1-\delta$, the regret is bounded by $\tilde{O}(2\cdot \varepsilon \cdot \numrounds \cdot \log(\nicefrac{1}{\delta}))$, yielding the claimed bound.
\end{proof}

For the offline sample complexity, \citet{BacchiocchiC0024} show the following guarantee.  The idea behind the algorithm is to approximately identify a covering of contracts
into best-response regions, each one representing a set of contracts in which a given agent’s action
is a best response. 

\begin{theorem}[\citet*{BacchiocchiC0024}]
For any $\delta \in (0,1)$ and $\varepsilon > 0$, there is an offline learning algorithm that, with probability at least $1-\delta$, computes an $\varepsilon$-approximate bounded contract, with at most $\tilde{O}(m^n \cdot \mathsf{poly}(n,m) \cdot \nicefrac{1}{\varepsilon^4} \cdot \log(\nicefrac{1}{\delta}))$ many contract queries. 
\end{theorem}

Using Proposition~\ref{prop:sample-reduction} they thus obtain: 

\begin{theorem}[\citet*{BacchiocchiC0024}]
For any $\delta \in(0,1)$, there is an online learning algorithm for bounded contracts that, with probability at least $1-\delta$, incurs a regret of at most 
$\tilde{O}(m^n \cdot \mathsf{poly}(n,m) \cdot \numrounds^{4/5} \cdot \log(\nicefrac{1}{\delta}) )$.
\end{theorem}

This shows that the regret is polynomial, when the number of actions is constant. It remains an open question to prove (or disprove) that the problem admits a polynomial regret bound when the number of actions is polynomial in $m$. It is also open what the corresponding regret bounds are when the agent is sampled afresh in each round.
 
\subsection{Improved Regret Bounds under Regularity Assumptions}
\label{sub:samples-regular}

Next we turn to recent work by \citet*{ChenCDH24}. Motivated by the results of \cite{ZhuEtAl22}, this work asks whether the learning problem becomes more tractable, when we are willing to impose regularity assumptions. They also investigate the gap incurred that results from restricting attention to bounded contracts. 

For the learning results \cite{ChenCDH24} again focus on the special case, where the principal interacts with the \emph{same} agent over all $\numrounds$ rounds. (Recall that the impossibility of \cite{ZhuEtAl22} in Theorem~\ref{thm:zhu1} applies under this restriction.) 
That is, the agent's costs $c_i \ge 0$ for actions $i \in [n]$ and the probability distributions $\Prob_{i}$ over outcomes $j \in [m]$ are fixed, but unknown to the principal. As before, costs and rewards are  assumed to be normalized so that  $c_i, r_j \in [0,1]$ for all $i$ and $j$. 
The focus of  \cite{ChenCDH24} is on the 
offline sample complexity problem, assuming that the principal has access to contract queries. Recall that in a contract query, the principal posts a bounded contract $\con \in [0,1]^m$, and receives an outcome $j$ sampled from the distribution over outcomes $\Prob_{i^\star(\con)}$ induced by the agent's best response action $i^\star(\con)$ to contract $\con$.\footnote{We remark that \cite{ChenCDH24} also consider a different form of feedback, which the authors refer to as \emph{action query}. Here the principal can specify an action $i \in [n]$, and receive a sample $j \sim \Prob_i$. We refer the interested reader to the paper of details.} This then implies a regret bound for the regret minimization problem in an online learning setup.

The main result of \cite{ChenCDH24} is the following polynomial sample complexity bound, for instances that satisfy first-order stochastic dominance (FOSD) and the concavity of distribution function property (CDFP) (see Section~\ref{sub:regularity}).

\begin{theorem}[\citet*{ChenCDH24}] 
For instances that satisfy FOSD and CDFP, for any $\delta \in (0,1)$ and any $\varepsilon > 0$ there is an offline algorithm, that, with 
probability at least $1-\delta$, computes an $\varepsilon$-approximate bounded contract with at most 
$
\tilde{O}\left(m{^{11}} \cdot \nicefrac{1}{\varepsilon^{20}} \cdot \log(\nicefrac{1}{\delta})\right)
$
many contract queries.
\end{theorem}

The proof roughly proceeds in two steps: The first ingredient is an approach for learning an empirical instance through piece-wise linear approximation of the concave distributions over outcomes.
This part leverages \emph{step} contracts (a.k.a.~\emph{threshold} contracts), to implement
(approximate) subgradient oracles.\footnote{An $\varepsilon$-approximate subgradient oracle for a non-decreasing convex function $G$ takes a positive $p$ as input and returns a point $y$ such that $p$ is a subgradient of $G$ at some point $z$ such that $y-\varepsilon \leq z \leq y$ \cite[][Definition 5]{ChenCDH24}.} 
Assuming rewards are sorted from low to high, a step contract $\con$ is such that $t_j = 0$ for all $j \leq j'$, and $t_j = t$ for all $j > j'$.
The second step  shows how to deal with the insufficiencies of the thus obtained empirical instance with respect to low-cost actions and their distributions.

Applying a similar reduction to that in Proposition~\ref{prop:sample-reduction} (with the $n^m$ term omitted), yields the following implication for the online learning version of the problem.

\begin{corollary}[\citet*{ChenCDH24}]
For any $\delta > 0$ there is an online learning algorithm for bounded contracts, that, with probability at least $1-\delta$, incurs a regret of at most $\tilde{O}(m^{\nicefrac{11}{21}} \cdot \numrounds^{\nicefrac{20}{21}} \cdot \log(\nicefrac{1}{\delta}))$.
\end{corollary}

This shows that the learning problem indeed becomes more tractable under suitable regularity assumptions. If both FOSD and CDFP are imposed, then optimal bounded contracts can be learned while incurring polynomial (rather than exponential in $m$) regret.

For the bounded vs.~unbounded contracts question, \cite{ChenCDH24} consider $H$-bounded contracts, with the requirement that $\con \in [0,H]^m$ (where the rewards are still normalized to be in $[0,1]$). They then prove that, even when restricting attention to instances that satisfy FOSD and CDFP, for any $H \geq 1$ and any $\alpha > 1$, there is an instance such that $\OPT_H < \nicefrac{1}{\alpha} \cdot \OPT$, where $\OPT_H$ and $\OPT$ denote the optimal principal utility from an $H$-bounded contract and a contract that can have arbitrarily high payments, respectively. This shows that bounded contracts can be arbitrarily worse than unbounded ones. 

There are a couple of interesting questions stemming from the work of \cite{ChenCDH24}. One such question is whether analogous sample complexity results can be obtained under weaker regularity assumptions (e.g., only one of FOSD or CDFP). Moreover, just like in the ``few actions'' case, it is unclear whether the positive results for the online learning problem carry over to a setting, where in each round the agent is sampled afresh.

\subsection{Improved Regret Bounds for Linear Contracts with Stronger Feedback}
\label{sub:samples-feedback}

We next discuss the work of \cite*{DuettingGSW23}, which shows tight regret bounds for general one-sided Lipschitz functions with ``function-value feedback.'' These bounds can be instantiated for the online learning problem of finding a linear contract when the principal interacts with the same agent over all rounds, and the principal's feedback to a contract $\alpha$ is her expected utility $U_P(\alpha)$ under this contract. 

More formally, suppose a principal interacts with the same (a priori unknown) agent over $\numrounds$ rounds. As before, suppose that the contracting problem is normalized with rewards (and hence expected rewards) as well as costs normalized to lie in $[0,1]$. 
In each round $\round \in [\numrounds]$, the principal posts a linear contract $\alpha^\round \in [0,1]$ and observes $U_P(\alpha^\round)$ --- her expected utility from contract $\alpha^\round \in [0,1]$. 
The principal's goal is a learning algorithm for finding a linear contract that incurs low regret with respect to the best linear contract in hindsight. 
Namely, denote by $\alpha^\star \in [0,1]$ the linear contract that maximizes 
the principal's total utility among all linear contracts; so $\alpha^\star \in \arg\max_{\alpha \in [0,1]}U_P(\alpha)$.
Using this notation, the principal aims to minimize the regret incurred by the algorithm, given by
$\sum_{\round=1}^{\numrounds} (U_P(\alpha^\star) - U_P(\alpha^t))$.    

The result of \cite{DuettingGSW23} applies, to any objective function that is one-sided Lipschitz, according to the following definition. Consider a single-dimensional function $f$, with domain $\mathsf{dom}(f)$. Then, $f$ is \emph{left-Lipschitz continuous} if for all $x,y \in \mathsf{dom}(f)$ with $x \leq y$ it holds that $f(x)-f(y) \leq y-x.$ Similarly, $f$ is \emph{right-Lipschitz continuous} if
for all $x,y \in \mathsf{dom}(f)$ with $x \leq y$ it holds that $f(y)-f(x) \leq y-x.$ 
Intuitively, left-Lipschitz-continuous functions cannot increase too quickly as you move to the left from a given point. Similarly, for right-Lipschitz-continuous functions, this property must hold as you move to the right.

Leveraging the perspective in Figure~\ref{fig:upper-envelope-principal}, let us convince ourselves that the principal's expected utility $U_P(\alpha)$ as a function of $\alpha$ is a left-Lipschitz continuous function. 
To see this, first recall that the principal's expected utility $U_P(\alpha)$ of a linear contract with parameter $\alpha$ is equal to $(1-\alpha) \cdot R_{i^\star(\alpha)}$, where $i^\star(\alpha)$ is the action chosen by the agent under this contract and $\Rew_{i^\star(\alpha)}$ is the expected reward of that action. Now as we vary $\alpha$ the agent's best response may change, and this may cause the principal's utility to change in a discontinuous way. Specifically, if we consider decreasing $\alpha$ to $\alpha' \leq \alpha$, the agent may switch to an action with potentially much smaller expected reward, causing $U_P(\alpha) - U_P(\alpha')$ to be much larger than $\alpha - \alpha'$ (in violation of right-Lipschitz continuity).  
However, if we consider increasing $\alpha$ to $\alpha' \geq \alpha$ we can only move to an action with higher expected reward, and the principal's expected utility drops at a negative slope of at most $\max_{i \in [n]} \Rew_i \leq 1.$ Thus, in the case where we move from $\alpha$ to $\alpha' \geq \alpha$, we have $U_P(\alpha) - U_P(\alpha') \leq \alpha' - \alpha$, showing that the principal's expected utility is indeed left-Lipschitz continuous.

The result of \citet{DuettingGSW23} is an online learning algorithm for general one-sided Lipschitz functions that achieves an $O(\log \log \numrounds)$ regret bound. 
This result generalizes the seminal work and bounds established in \citet*{KleinbergL03}, while also matching the lower bound of $\Omega(\log \log \numrounds)$ proven in that earlier work.
The intuitive idea behind the algorithm that obtains the optimal regret bound for general one-sided Lipschitz functions 
is as follows:
Given a set of historical queries and the function value at those queries, the possible one-sided Lipschitz functions that are consistent with that history trace out a sequence of parallelograms. The algorithm keeps track of these parallelograms and decides how to carve up a particular parallelogram with additional queries based on the relative height and width of these parallelograms.

Applying this result to the problem of learning linear contracts yields:

\begin{theorem}[\citet*{DuettingGSW23}] There is an online learning algorithm for linear contracts (with function-value feedback) that incurs a regret of at most $O(\log \log \numrounds)$. 
\end{theorem}

We remark that the stark improvement over the regret bound in Theorem~\ref{thm:zhu2} is possible because of two differences: First, unlike the earlier bound, this bound is for a setting where the principal interacts with a single agent, rather than an agent that is drawn afresh each round. Second, the principal receives stronger feedback, namely her expected utility $U_P(\alpha)$ for a given contract $\alpha$, rather than just an outcome sampled from the best-response action.  

An important feature of this result is that the incurred regret is again independent of the number of peaks/discontinuities of the principal's expected utility function $U_P(\cdot)$, which also makes it applicable in combinatorial extensions of the vanilla contracting problem (e.g., the extension discussed in Section~\ref{sec:multiple-actions} where $n$ can be exponential in the number of actions).

\paragraph{Additional Directions and Open Questions.} The learning perspective on contracts has already yielded some deep insights, but we expect there to be a significant amount of additional work going forward. 
First of all, there remain several 
important gaps in our understanding of the online learning direction. 
For example, there is a notable gap between the frameworks studied by 
\cite{HoSV16} and \cite{ZhuEtAl22}, where a new agent is drawn afresh in each round, and the work of \cite{BacchiocchiC0024} and \cite{ChenCDH24}, which considers a fixed agent. One could also aim to examine whether the positive result of \cite{BacchiocchiC0024} for settings with few actions can be extended to a polynomial (in $m$) number of actions. Similarly, it would be interesting to explore whether positive results akin to those established by \cite{ChenCDH24} hold under  weaker regularity assumptions. Additionally, it is worth exploring other forms of simple contracts through the lens of online learning, thereby deepening our understanding of the tradeoff between simple and optimal contracts.

Second, most of the existing work on learning in contracts 
has focused on the \emph{online learning} setup, with rather strong impossibilities stemming from the hardness of learning an agent's type with restricted (bandit-type) feedback. 
An intriguing direction for future work is to develop a more comprehensive theory of the {\em offline} sample complexity of learning contracts, with different forms of feedback (e.g., bandit- and expert-type feedback).
This direction could draw inspiration from analogous work in mechanism design (e.g., \cite{MorgensternR15}), 
potentially shedding more light on how to trade off learnability and approximation.
For a preliminary study in this direction see~\cite{sample-complexity}.

A related approach to learning optimal mechanisms from samples is known as \emph{differentiable economics} \citep{Dutting0NPR24}. The idea here is to cast the learning problem as an end-to-end differentiable neural network, enabling the automated design of mechanisms using  standard machine learning pipelines.
This approach was recently adopted to contracts by \citet*{WangEtAl2023}. This work proposes neural network architectures that are suitable for capturing piece-wise affine, discontinuous objective functions (e.g., the principal's utility in contract design); and demonstrates that these neural network architectures can be used for the end-to-end design of contracts.

\section{Contracts for Machine Learning: Incentive-Aware Classification}
\label{sec:incentive-aware}
This section complements Section~\ref{sec:data-driven} by exploring additional interactions between contracts and machine learning (ML), in which contract theory helps steer strategic behavior in ML.
Machine learning tasks often involve effort by strategic players; using contracts to incentivize and optimize this effort can be the key to successful learning. 
Our main focus in this section is on effort exerted by the \emph{subjects} of the learning process.
We outline the connection between contracts and a thriving line of research known as \emph{strategic classification} (a.k.a.~\emph{incentive-aware} ML or \emph{performative prediction}).%
\footnote{For performative prediction see, e.g.,~\citep{PerdomoZMH20,Mendler-DunnerP20,PiliourasY23}. There is also a recent related literature in economics, which studies optimal design problems where the agents have the ability to privately manipulate or fabricate the signals; see, e.g.,~\citep{PerezRS22,PerezRS24,LiQ24,FrankelK19}.}
We then discuss contracts for delegating ML-related tasks.
The section is organized as follows: 
Section~\ref{sub:eval-scheme} introduces the \emph{evaluation} model of~\citet{KleinbergR19} --- the first work to incorporate self-improvement in addition to gaming into strategic classification. Section~\ref{sub:connection} presents a result of~\citet*{AlonDPTT20}, who identify a formal connection between contracts and a simplified version of \cite{KleinbergR19}. Section~\ref{sub:multi-lin} returns to the fully general version of \cite{KleinbergR19} and discusses the power of \emph{multi-linear} evaluation for incentivizing both single and multiple agents. 
Section~\ref{sub:ML-optimizing} considers not just incentivizing certain effort investments through evaluation, but also optimizing over effort investments. Section~\ref{sub:ML-other} surveys additional results, focusing on contracts for ML delegation. 

\subsection{Incentive-Aware Evaluation} \label{sub:eval-scheme}

Strategic classification studies how strategic agents react in response to being classified or otherwise learned. This reaction typically involves expending effort by the agent, which ranges from socially undesirable \emph{gaming} attempts (see the seminal works of~\cite{BrucknerS11} and~\cite{HardtMPW16}), to \emph{self-improvement} efforts~\citep{KleinbergR19,KleinbergR20}. 
Strategic reactions to learning are abundant in real-life scenarios, ranging from school admission~\cite[e.g.,][]{HaghtalabILW20,LiuGB21} to credit assessment~\cite[e.g.,][]{GhalmeNETR21}. 
Because contracts are the main economic tool for shaping effort, they are ideally suited for steering the agent's effort toward self-improvement rather than gaming --- to the benefit of both the learning principal and society.

As an illustrative example, consider the following toy scenario from school admission~\citep*{HardtMPW16}: The number of books in a candidate's household is a well-studied predictor of academic success. 
Even if this feature could be accurately measured, could it reliably determine a candidate's admission to academic studies? The answer is no, in part because, with minimal effort, a candidate could acquire more books, thereby manipulating the admission decision.

Note that this form of manipulation requires neither dishonesty nor breaking any rule, but does involve wasting resources on unread books. Such gaming thus poses not only a risk of skewed decision-making, but also of a collective waste of effort. In other words, careless design of the admission classifier may induce agents to concentrate effort on superficially passing tests and assessments, rather than on creating true social value. To show the role contracts can play in mitigating these risks, we introduce the evaluation model of~\citet{KleinbergR19}.

\paragraph{The Evaluation Model.}
The model of~\citet{KleinbergR19} is best-described within the domain of student evaluation, but applies more generally to additional evaluation settings (e.g., evaluating loan applicants). 
We now describe the model, intentionally overloading some notation. 
An \emph{evaluation scheme} is a classifier mapping a student (agent) to his final grade.
The mapping is based on student \emph{features} $\Feat = (\feat_1, \ldots, \feat_m)$ such as homework grades, exam performance, class participation, etc. 
The student reacts strategically to the classifier by deciding how to allocate his (normalized) \emph{budget} $B=1$ of effort among his possible \emph{actions} --- which include, e.g., studying the material, memorizing, or even cheating. As demonstrated by this example, some actions correspond to positive self-improvement, while others correspond to gaming attempts as in the standard strategic classification paradigm. We refer to the former actions as \emph{admissible}.
The classifier only observes the student features, which are \emph{noisy} (stochastic) outcomes of the chosen actions. For example, a student might fail his midterm despite studying hard for it, since failure is a possible (if unlikely) outcome of studying. The evaluation scheme maps the student features to a single number, which is the student's final grade. The grade is treated as the agent's utility and determines the agent's strategic reaction. 

The agent responds to the evaluation scheme strategically by choosing an \emph{effort allocation} denoted by $\Alloc=(\alloc_1,\dots,\alloc_n)$ among the $n$ actions, where $\sum_{i\in[n]} {\alloc_i} \le B$. The effort allocation leads to features $\Feat = (\feat_1, \ldots, \feat_m)$ which determine the agent's score/utility. Mathematically, each feature $\feat_j$ is a function of the efforts $\alloc_1,\dots,\alloc_n$. These functions can take one of two forms:
\begin{itemize}
    \item The \emph{simplified} or \emph{multi-linear} model: For every $j\in[m]$, feature $\feat_j=\sum_i x_i q_{ij}$, i.e., the feature is a convex combination of $\alloc_1,\dots,\alloc_n$ with coefficients $q_{1j},\dots,q_{nj}$. The coefficients $\{q_{ij}\}_{i\in[n],j\in[m]}$ are non-negative and are given as part of the setting in matrix representation or equivalently as a weighted bipartite graph. Figure~\ref{fig:contract-vs-evaluation} depicts the simplified model (and its relation to contracts---more on this below).
    
    \item The \emph{generalized} or \emph{concave} model: For every $j\in[m]$, feature $\feat_j=f_j(\sum_i x_i q_{ij})$, where $f_j$ is a concave and strictly increasing function. Both the functions $\{f_{j}\}_{j\in[m]}$ and the weights $\{q_{i,j}\}_{i\in[n],j\in[m]}$ are given as part of the setting.
\end{itemize}
In either model, each feature $\feat_j$ is assumed to strictly increase through some effort investment (otherwise, we can simply ignore feature~$\feat_j$).
An evaluation setting is summarized by the actions, the features, and the functions mapping effort allocations to features. 

As usual in game-theoretic settings, we refer to an effort allocation $\Alloc$ that maximizes the agent's utility as a \emph{best response} to the evaluation scheme. 
The allocation can be integral (a pure strategy) or fractional (a mixed strategy).
An effort allocation is \emph{implementable} (up to tie-breaking) if there exists an evaluation scheme under which this allocation is a best response for the agent. 

\paragraph{Multi-linear and Monotone Evaluation Schemes.}
\cite{KleinbergR19} consider a natural family of \emph{multi-linear} evaluation schemes, each defined by non-negative weights $(t_1,\dots,t_m)$ applied to the features. Given a multi-linear evaluation scheme $\con=(t_1,\dots,t_m)$, the student's final grade is $\sum_j t_j F_j$. 
Multi-linear schemes are a subclass of \emph{monotone} evaluation schemes, where a scheme is monotone if for every two feature vectors $\Feat\ge\Feat'$, the final grade of a student with features $\Feat$ is at least as high as the final grade of a student with features $\Feat'$ (i.e., the classifier is a monotone mapping from the features to the score).
We remark that despite their name, multi-linear evaluation schemes are more similar to general contracts than to linear contracts; this is evident when comparing a multi-linear evaluation scheme $\con=(t_1, \dots, t_m)$ to a linear contract~$\alpha$. Multi-linear evaluation schemes are related to linear \emph{classifiers}---see Section~\ref{sub:multi-lin}.%
\footnote{For this reason, what we refer to in this survey as \emph{multi-linear} evaluation schemes (to signal their distinction from linear contracts) is usually called \emph{linear} evaluation schemes in the literature~\cite[see][]{KleinbergR19,AlonDPTT20}.}

\subsection{Formal Connection Between Evaluation and Contracts} 
\label{sub:connection}

An evaluation scheme determines the agent's best response effort allocation. This means that the design of the evaluating classifier determines whether the agent engages in true self-improvement (like studying), or in gaming efforts to superficially improve his features (like short-term memorizing or cheating).
In effect, the classifier \emph{incentivizes the strategic allocation of effort under uncertainty}---that is, functions like a contract.  
The design of a classifier that incentivizes self-improvement is thus closely related to the design of a contract that incentivizes a target action. \citet*{AlonDPTT20} make this intuition explicit in the simplified evaluation model of \citep{KleinbergR19}. To describe the connection we temporarily depart from the evaluation model and analyze a class of contract settings. We then return below to the evaluation perspective.

\paragraph{The Contracts Perspective.} 

Towards connecting evaluation and contracts, the following class of contract settings introduced by~\cite{AlonDPTT20} will be useful. This class is shown below to coincide with the multi-linear evaluation model. 
In this class of contract settings, all actions have zero cost for the agent (one can imagine an agent with a ``budget of effort'' to spend ``for free'' on taking some action). 
Additionally, the agent's matrix $\{q_{ij}\}_{i\in[n],j\in[m]}$ (where action $i$ leads to outcome $j$ with probability $q_{ij}$) is allowed to have rows summing up to less than~$1$, with the convention that with the remaining probability $1-\sum_j q_{ij}$, action $i$ leads to a fictitious \emph{null} outcome. 
Each outcome except the null outcome is assumed to be reached with nonzero probability by at least one action (otherwise it can be removed from the setting). 
The null outcome can receive no payment.
Given any contract $\con=(\pay_1,\dots,\pay_m)$, the agent chooses an action maximizing his expected utility. 
Since there are no costs, this is an action that maximizes his expected payment, i.e., $\arg\max_{i\in [n]} \sum_j \prob_{ij}t_j$.

Recall from Section~\ref{sec:opt-and-linear} that an action is called implementable (up to tie-breaking) if there exists a contract under which it maximizes the agent's expected utility. We now adapt the classic characterization of implementability in Proposition~\ref{prop:implementable}, which follows from linear programming duality (see Figure~\ref{fig:dual-lp}), to contract settings with zero costs and ``partial'' distributions as described above. A subtle point is that with zero costs, the zero-payment contract $\con=(0\dots,0)$ makes the agent indifferent among all actions; we ignore such uninteresting contracts (in practice, it is arguably implausible that an agent facing zero payments will spend any effort).

\begin{proposition}[Implementability by contracts, adopted from \cite*{AlonDPTT20}]
\label{prop:implement-by-classifier}
Consider a contract setting in which all actions have zero cost, for every action $i'$ the corresponding matrix row $\Prob_{i'}$ sums up to $\le 1$, and for every outcome $j$ there is at least one action $i_j$ with $\prob_{i_j j} > 0$. Then action~$i$ with row $\Prob_i$ is implementable (up to tie-breaking) by a non-zero contract if and only if the following condition holds:
There is no linear combination of $\{\Prob_{i'}\}_{i' \in [n]}$ that coordinate-wise dominates 
$\Prob_i$, where the coefficients $\{\lambda_{i'}\}_{i'\in[n]}$ are non-negative and sum up to $\sum_{i'\in[n]} \lambda_{i'} <1$.
\end{proposition}

\begin{proof}
To show that action $i$ is implementable if and only if no linear combination with a certain property exists (call this condition $X$), we first show that if there exists such a linear combination (i.e., if $\neg X$), then $i$ is not implementable. We then show the other direction, i.e., that if condition $X$ holds then $i$ is implementable.

\medskip\noindent
{\it ``Only if'' direction.} Assume $\neg X$, i.e., there exists a linear combination of the rows with non-negative coefficients summing up to $\sum_{i'\in [n]} \lambda_{i'} <1$, such that 
\begin{equation}
    \sum_{i'\in [n]} \lambda_{i'} \Prob_{i'} \ge \Prob_i.\label{eq:comb-dominates}
\end{equation}
We will now show that action $i$ is not implementable.

We begin by establishing that in the linear combination, $\lambda_i=0$ without loss of generality: Observe that
$$
\sum_{i'\neq i} \lambda_{i'} \Prob_{i'} + \lambda_{i} \Prob_{i} = \sum_{i'\in [n]} \lambda_{i'} \Prob_{i'} \ge \Prob_i,
$$
where the inequality is by Equation~\eqref{eq:comb-dominates}. By rearranging we get
$\sum_{i' \neq i} \lambda_{i'} \Prob_{i'} \ge (1-\lambda_i) \Prob_i$, and since $\lambda_i<1$ (as the sum of \emph{all} coefficients is $<1$), we have
\begin{equation}
    \sum_{i' \neq i} \frac{\lambda_{i'}}{1-\lambda_i} \Prob_{i'} \ge \Prob_i.\label{eq:new-comb-dominates}
\end{equation}
We can now define new coefficients $\lambda'_{i'} := \lambda_{i'}/(1-\lambda_i)$ for every $i' \neq i$, and $\lambda'_i := 0$. By Equation~\eqref{eq:new-comb-dominates}, the new linear combination maintains the property $\sum_{i'\in[n]} \lambda'_{i'} \Prob_{i'} \ge \Prob_{i}$. Moreover, the new coefficients satisfy $\sum_{i'\in[n]} \lambda'_{i'} = \sum_{i' \neq i} \lambda'_{i'} = \sum_{i' \neq i} \lambda_{i'}/(1-\lambda_i)<1$, where the equalities follow from the definition of $\{\lambda'_{i'}\}$, and the final inequality holds since $\sum_{i' \neq i} \lambda_{i'} + \lambda_i<1$. Thus from now on we assume $\lambda_i=0$.

Consider a non-zero contract $\con$. We will show that $\con$ does not implement action $i$, by identifying some other action which is strictly preferred by the agent. Since this holds for any non-zero contract $\con$, we conclude that $i$ is not implementable. 

Recall that for every action $i'\in[n]$, $\Pay_{i'}$ denotes the agent's expected payment $\Prob_{i'}\cdot \con$ for taking action $i'$ given contract $\con$.
We first deal with the case of $T_i=0$. Since $\con$ is non-zero, there must be an outcome~$j$ such that $t_j > 0$. Since each outcome is attained with positive probability by some action, there must be an action $i_j\ne i$ with $\prob_{i_j j} > 0$. For this action it holds that $T_{i_j}>0$, and so the agent strictly prefers action $i_j$ to action~$i$. 
Thus from now on we can focus on the complementary case in which $T_i>0$. We can also assume that $\Prob_i\ne 0$ (since otherwise $T_i=0$).  

By Equation~\eqref{eq:comb-dominates} and since $\lambda_i=0$, $(\sum_{i'\ne i} \lambda_{i'} \Prob_{i'})\cdot \con \ge \Prob_i\cdot \con=\Pay_i$. 
We can rewrite the linear combination $(\sum_{i'\ne i} \lambda_{i'} \Prob_{i'})\cdot \con$ as $\sum_{i'\ne i} \lambda_{i'} (\Prob_{i'}\cdot \con)$ to obtain 
\begin{equation}
\sum_{i'\ne i} \lambda_{i'} \Pay_{i'} = \sum_{i'\ne i} \lambda_{i'} (\Prob_{i'}\cdot \con) \ge T_i.\label{eq:linear-comb}
\end{equation}
We now normalize the coefficients: For every $i' \in [n]$, define $\Lambda_{i'} := \lambda_{i'}/\gamma$ where the normalization factor is $\gamma := \sum_{i'\in[n]} \lambda_{i'} = \sum_{i' \neq i} \lambda_{i'}$. We are guaranteed that $\gamma>0$ since Equation~\eqref{eq:comb-dominates} holds and $\Prob_i\ne 0$.
The new coefficients maintain $\Lambda_i=0$ and so $\sum_{i' \neq i} \Lambda_{i'} = 1$. 
Using that $\sum_{i' \neq i} \lambda_{i'} < 1$ and dividing Equation~\eqref{eq:linear-comb} by $\gamma$, we obtain
\[
\sum_{i' \neq i} \Lambda_{i'} T_{i'} = \sum_{i' \neq i} \Lambda_{i'} (\Prob_{i'} \cdot \con) \ge \Pay_i/\gamma > \Pay_i.
\]
Since the convex combination $\sum_{i' \neq i} \Lambda_{i'} T_{i'}$ is $>T_i$, we conclude there must be an action $i^* \neq i$ such that $T_{^*} > T_i$.
This completes the proof of the ``only if'' direction.

\begin{figure}[t]
\begin{center}
\begin{subfigure}[t]{0.4\textwidth}
\begin{align*}
\max_{t_j:j\in[m]} \quad & \sum_{j} \prob_{ij} t_j \\
\text{s.t.} \quad &
\sum_{j} \prob_{i' j} t_j \le 1 && \forall~i'\in [n]\\
&t_j \geq 0 &&\forall~j\in[m]
\end{align*}
\caption{$\textsf{MAXPAY-LP}(i)$}
\end{subfigure}
\quad\quad
\begin{subfigure}[t]{0.4\textwidth}
\begin{align*}
\min_{\lambda_{i'}:i'\in [n]} \quad & \sum_{i'\in[n]} \lambda_{i'} \\
\text{s.t.} \quad & \sum_{i'\in[n]} \lambda_{i'}\prob_{i' j}\ge \prob_{ij} && \forall~j\in[m]\\
&\lambda_{i'} \geq 0 &&\forall~i'\in[n]
\end{align*}
\caption{$\textsf{MAXPAY-DUAL}(i)$}
\end{subfigure}
\end{center}
\caption{The condition of Proposition~\ref{prop:implement-by-classifier} for implementability of action $i$ formulated as a dual LP ({\bf right}), and its corresponding primal ({\bf left}). The dual seeks a linear combination of the rows $\{\Prob_{i'}\}_{i'\in[n]}$ with non-negative coefficients, which minimizes the sum of coefficients while coordinate-wise dominating row $\Prob_i$. 
The primal seeks a contract that maximizes the expected payment for action~$i$ while upper-bounding the expected payment for \emph{any} action by $1$. 
}\label{fig:primal-dual-lp-for-evaluation}
\end{figure}

\medskip\noindent
{\it ``If'' direction.} 
Assume now that condition $X$ holds, i.e., for every linear combination of the rows with non-negative coefficients such that $\sum_{i'\in[n]} \lambda_{i'}\Prob_{i'}\ge \Prob_i$, the coefficients sum up to $\sum_{i'\in [n]} \lambda_{i'} \ge 1$. 
We can express such a linear combination as the solution to a linear program, with the objective of minimizing the sum of coefficients. This linear program appears as the dual in Figure~\ref{fig:primal-dual-lp-for-evaluation} (right), and since $X$ holds we know its optimal objective value is $\ge 1$. Observe that there is always a feasible dual solution that achieves an objective value of $1$ by placing all weight on action~$i$: $\lambda_i=1$ and $\lambda_{i'}=0$ for every $i'\ne i$. 
We conclude that the optimal dual objective value is $1$.
We now take the dual of the dual to get the primal program---see Figure~\ref{fig:primal-dual-lp-for-evaluation} (left). By strong duality, the primal's optimal objective is also $1$. 
Thus, there exists a feasible primal solution, that is, a contract $\con$ such that the agent's expected payment $\Pay_i$ for action $i$ is $\sum_{j} \prob_{ij} t_j=1$. 
Clearly, this contract must be non-zero. 
Since the solution is feasible, the constraints hold, and so the expected payment $\Pay_{i'}$ for any action $i'\in[n]$ is $\sum_{j} \prob_{i' j} t_j \le 1$. We conclude that action $i$ maximizes the agent's expected payment given the non-zero contract $\con$, and thus that $i$ is implementable (up to tie-breaking).
\end{proof}

\paragraph{The Evaluation Perspective.} 

A main achievement of \cite{KleinbergR19} is in characterizing the effort allocations $\Alloc=(\alloc_1,\dots,\alloc_n)$ that are implementable by evaluation schemes, both in the simplified model where features are multi-linear functions in $\alloc_1,\dots,\alloc_n$, and in the generalized model where features are concave functions. They give intuition for their characterization as follows: an action (or combination of actions) is \emph{not} implementable only if the effort invested in it can be substituted out and replaced by effort invested in a different combination of actions, while improving the agent’s utility. Reinterpreting the linear combination in Proposition~\ref{prop:implement-by-classifier} as a way to relocate effort shows the conceptual connection to (non-)implementability by contracts. 

\begin{figure}[t]
\vspace*{-10pt}
\begin{center}
\begin{tikzpicture}
\def\y{1.3}
\node[circle, inner sep=0pt, minimum size=16pt, color=red, fill=red, thick] (C) at (0,1*\y) {};
\node at (-1.25,1*\y) {\small{\textcolor{red}{Cheating}}};
\node[circle, inner sep=0pt, minimum size=16pt, color=green!50!black, fill=green!50!black, thick] (S) at (0,0*\y) {};
\node at (-1.25,0*\y) {\small{\textcolor{green!40!black}{Studying}}};
\node[circle, inner sep=0pt, minimum size=16pt, color=blue, fill=blue, thick] (E) at (3,-1*\y) {};
\node at (3,-0.5*\y) {\small{\textcolor{blue}{Exam}}};
\node[circle, inner sep=0pt, minimum size=16pt, color=blue, fill=blue, thick] (M) at (3,0*\y) {};
\node at (3,0.5*\y) {\small{\textcolor{blue}{Midterm}}};
\node[circle, inner sep=0pt, minimum size=16pt, color=blue, fill=blue, thick] (H) at (3,1*\y) {};
\node at (3,1.5*\y) {\small{\textcolor{blue}{Homework}}};
\node[circle, inner sep=1pt, minimum size=16pt, color=blue, fill=blue, thick] (N) at (3,2*\y) {};
\node at (3,2.5*\y) {\small{\textcolor{blue}{Null}}};
\node[circle, inner sep=0pt, minimum size=16pt, color=violet, fill=violet, thick,right] (G) at (6,0.5*\y) {};
\node at (7.4,0.5*\y) {\small{\textcolor{violet}{Grade}}};
\draw[thick,->] (C.east) --node[above]{\footnotesize{.2}} (N.west);
\draw[thick,->] (C.east) --node[above]{\footnotesize{.8}} (H.west);
\draw[thick,->] (S.east) --node[above]{\footnotesize{.35}} (H.west);
\draw[thick,->] (S.east) --node[above]{\footnotesize{.25}} (M.west);
\draw[thick,->] (S.east) --node[above]{\footnotesize{.4}} (E.west);
\draw[thick,->] (H.east) --node[above]{\footnotesize{$t_1$}} (G.west);
\draw[thick,->] (M.east) --node[above]{\footnotesize{$t_2$}} (G.west);
\draw[thick,->] (E.east) --node[above]{\footnotesize{$t_3$}} (G.west);
\path[draw,decorate,decoration=brace,very thick] (2.7,-2) -- (0.25,-2) 
node[midway,below,font=\small,yshift=-0.2cm]{Normalized weights};
\path[draw,decorate,decoration=brace,very thick] (6.2,-2) -- (3.3,-2) 
node[midway,below,font=\small,yshift=-0.2cm]{Contract/};
\node at (4.6,-2.9) {\small{evaluation scheme}};
\path[draw,decorate,decoration=brace,very thick] (-2.2,-0.5) -- (-2.2,1.8) 
node[midway,left,font=\small,xshift=-0.4cm,yshift=+0.3cm]{Agent};
\node at (-3.1,0.5) {\small{actions}};
\path[draw,decorate,decoration=brace,very thick] (2.25,3.6) -- (3.75,3.6) 
node[midway,above,font=\small,yshift=+0.2cm]{Student features};
\end{tikzpicture}
\end{center}
\vspace*{-30pt}
\caption{
The simplified evaluation model and its connection to contract design, shown in the context of student evaluation. 
A student (equiv., agent) chooses among cheating and studying, 
based on the multi-linear evaluation scheme (equiv., contract) $\con=(t_1,t_2,t_3)$.
The effort put into each action translates via a multi-linear function with non-negative weights---which can be normalized (equiv., probabilities)---to student features (equiv., outcomes). 
The weights $\{\prob_{ij}\}$ are depicted on the edges. The null feature is used for normalization and does not affect the final grade. 
For example, choosing the action of cheating leads to a vector $(0.8,0,0)$ of non-null features, which can be interpreted as the student's grades on the homework, midterm and exam. 
The final grade (equiv., agent utility) is a linear combination of the features, where the coefficients are determined by the evaluation scheme (equiv., contract). 
In our example, if $\con=(t_1,t_2,t_3)=(\nicefrac{3}{4},\nicefrac{1}{8},\nicefrac{1}{8})$, the student's final grade will be $0.6=60/100$.
}
\label{fig:contract-vs-evaluation}
\end{figure}
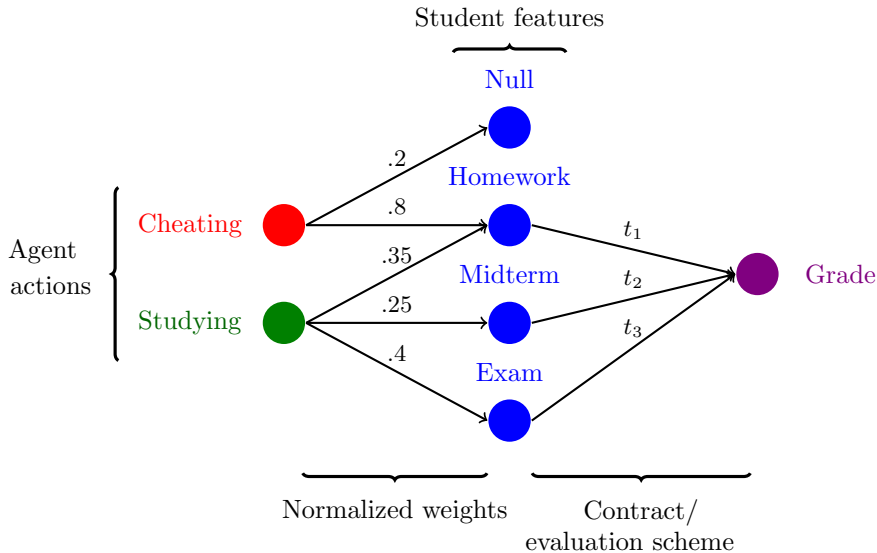

To formalize the conceptual connection, we show that Proposition~\ref{prop:implement-by-classifier} (characterizing implementability by contracts) can be obtained as a special case of \citeauthor{KleinbergR19}'s characterization of implementability by evaluation schemes. 
This is achieved by noticing that contract settings with zero-cost actions and ``partial'' distributions coincide with simplified evaluation instances. 
This unified view is depicted in Figure~\ref{fig:contract-vs-evaluation}.

The special case of \citeauthor{KleinbergR19}'s characterization that is relevant to contracts is the characterization of which actions are implementable by multi-linear evaluation schemes in the simplified evaluation model.  
When facing a multi-linear evaluation scheme $\con=(t_1,\dots,t_m)$, the student chooses an effort allocation $\Alloc$ that results in features $\Feat$ which maximize his final grade $\sum_j t_j \feat_j$. Because we are in the simplified model, 
$\feat_j=\sum_i x_i q_{ij}$ for every feature $j\in[m]$.
Notice that we can write the final grade as $\sum_j t_j \sum_i \alloc_i q_{ij}=\sum_i \alloc_i \sum_j q_{ij} t_j$.
Thus, without loss of generality, the agent chooses a pure best response to $\con$, investing his budget of effort in a single action $i$ that maximizes $\sum_j q_{ij} t_j$. In this case, the implementability characterization of~\cite{KleinbergR19} boils down to the following---which is in fact a restatement of Proposition~\ref{prop:implement-by-classifier}. 

\begin{proposition}[Implementability by evaluation schemes,~\cite{KleinbergR19}]
\label{prop:rephrased}
Consider a simplified evaluation setting in which for each feature $j$ there exists an action $i_j$ that leads to it with positive probability $\prob_{i_j j} > 0$.
For action~$i$ and every $j\in[m]$, let $\Prob_i=(q_{i1},\dots,q_{im})$ be the coefficients determining the mapping from effort invested in $i$ to feature $F_j$. Then action $i$ is implementable (up to tie-breaking) by a non-zero multi-linear evaluation scheme $\con$ if and only if the condition of Proposition~\ref{prop:implement-by-classifier} holds for $\Prob_i$.
\end{proposition}

\subsection{When are Multi-linear Evaluation Schemes Sufficient?}
\label{sub:multi-lin}

In the generalized evaluation model of \cite{KleinbergR19}, as opposed to the simplified model,
the student no longer necessarily has a \emph{pure} (single-action) best response to the evaluation scheme.
The reason why the student may strictly prefer to divide his budget of effort among multiple actions in the generalized model is the way in which effort translates into features. Recall that for every $j\in[m]$, feature $F_j$ is a concave, strictly-increasing function $f_j(\sum_i x_i q_{ij})$ of $\vec{x}$.
Due to the concavity of the $f_j$'s, the first ``unit of effort'' spent on an action is more effective than the last one towards maximizing the final grade. 
For example, investing 50\% of the budget in a certain action can result in achieving close to 100\% of a corresponding feature's maximum value, so redirecting the other 50\% of the effort budget to a different action can raise the value of additional features and result in an overall higher final grade. 

\citet{KleinbergR19} extend their implementability characterization (Proposition~\ref{prop:rephrased}) to the generalized evaluation model, and apply it to show that when evaluating a single student, the class of monotone evaluation schemes has no more implementability power than the class of multi-linear evaluation schemes. 

\begin{theorem}[\cite{KleinbergR19}]
\label{thm:linear-evaluation} 
Consider a generalized evaluation setting. An allocation of effort $\Alloc$ where $\sum_{i\in[n]} \alloc_i\le B$ is implementable (up to tie-breaking) by a non-zero \emph{monotone} evaluation scheme if and only if it is implementable (up to tie-breaking) by a non-zero \emph{multi-linear} evaluation scheme.
\end{theorem}

\paragraph{Multiple Agents.}

Many applications such as student evaluation typically involve more than one agent. In multi-agent evaluation, if the principal can apply a customized evaluation scheme to every agent, then the results stated above for single-agent evaluation continue to hold.
However, in many scenarios, the principal may need to apply a \emph{uniform} evaluation scheme to multiple agents, e.g., for fairness considerations. 

\citet{AlonDPTT20} study such uniform evaluation schemes for agents who diverge in how their effort allocation translates into features, i.e., in their weights $\{q_{ij}\}_{i\in[n],j\in[m]}$. Their model can capture strong students, who achieve high grades (features) even with small effort, along with weaker students who need to allocate more effort for similar achievements, and correspondingly have lower weights.
\cite{AlonDPTT20} demonstrate that Theorem~\ref{thm:linear-evaluation} (equivalence of multi-linear and monotone evaluation schemes) no longer holds in the multi-agent case. 
In fact, they show that general monotone evaluation schemes can implement a profile of effort allocations that is \emph{arbitrarily better} than the best profile implementable by a multi-linear evaluation scheme: Example 2.2 in \citep{AlonDPTT20} formulates an $n$-agent setting, in which action $1$ represents true studying and all other actions are forms of cheating. In this setting, there is a monotone evaluation scheme that can incentivize all agents to invest their entire budget of effort in action~$1$; but no multi-linear evaluation scheme can incentivize more than one agent to do so.

The intuition for this gap result is similar to the intuition behind the extra power of non-linear classifiers over linear ones: Consider the student evaluation application, where students are graded based on their $m$-dimensional feature vectors. Suppose our goal is to maximize the number of students who invest their budget of effort in truly studying. It is reasonable to assume that for different types of students (e.g., strong vs.~weak), there are different indicators of true study. Thus, to separate between students with high and low grades, the evaluation scheme must form a highly non-linear separator among the students' $m$-dimensional feature vectors. This can be achieved by a general evaluation scheme but not by a multi-linear one.

Having established an unbounded gap between the number of agents that can be incentivized to take an admissible action with monotone vs.~multi-linear schemes, \cite{AlonDPTT20}~study the computational complexity of finding the best evaluation scheme from each of these classes. 

\subsection{Optimizing Effort with Evaluation Schemes}
\label{sub:ML-optimizing}

Sections~\ref{sub:connection} and \ref{sub:multi-lin} discuss the \emph{implementability} of effort allocations through evaluation schemes. A natural extension considers cases where the evaluation scheme seeks to optimize an \emph{objective function} by incentivizing a desired effort allocation, subject to the constraint that the agent  takes admissible actions.
\cite{KleinbergR19} show that, even when the objective function $g:\mathbb{R}^m\to \mathbb{R}$ over effort allocations is concave, the optimization problem is NP-hard.
However, in the special case in which there are constantly-many admissible actions, optimizing $g$ over all admissible effort allocations (i.e., effort allocations that allocate nonzero effort only to admissible actions) is tractable.

\citet*{HaghtalabILW20} introduce a different model of effort optimization through incentive-aware evaluation. In their model there is a population of agents, each with a vector of $m$ preliminary features. The distribution $\mathcal{D}$ of feature vectors among the population is known. 
There is a \emph{true quality score} $f(\cdot)$, which is a multi-linear function mapping feature vectors to a score in $[0,1]$. 
An agent can modify his preliminary feature vector $\Feat$ by investing costly effort; to obtain a modified vector $\Feat'$, his cost is proportional to ${\lVert \Feat - \Feat' \rVert}_2$. 
The designer observes only a projection of the agent's modified feature vector $P \Feat'$, where $P_{n\times n}$ is a projection matrix. While the outcome of the agent's effort---the modified feature vector $\Feat'$---is not stochastic, due to the projection the effort is not fully observable to the designer. 

The goal in the work of \cite{HaghtalabILW20} is to design a \emph{perceived} quality score $g(\cdot)$, which is a (not necessarily  multi-linear) function mapping feature vectors to a score in $[0,1]$. The utility of each agent is his perceived score, while the designer aims to maximize \emph{welfare}, defined as the average true score of the population $\mathcal{D}$ after the agents' modify their features (note that effort costs do not count towards the welfare). The average true score is compared to the \emph{initial} average score before the effort investment; taking the difference yields the welfare \emph{gain}. 

They then show how to maximize the welfare gain over different classes of perceived scoring functions. Their main result is a polynomial-time algorithm with a $4$-approximation guarantee over the class of linear threshold functions, provided that certain assumptions on the distribution $\mathcal{D}$ hold. 
Furthermore, they relax the assumption that the designer has complete knowledge of $\mathcal{D}$, allowing for approximation guarantees based on sample access to the population.

\paragraph{Summary and Open Problems.}
In the evaluation model, strategic agents react to being classified by allocating their budget of effort among actions, some admissible and others not. The model comes in two flavors depending on whether agent features are multi-linear or concave in the effort allocation $\Alloc$. In the former version, the characterization of implementability by an evaluation scheme coincides with the characterization of implementability by a contract in a suitable contract setting (Propositions~\ref{prop:implement-by-classifier} and~\ref{prop:rephrased}). Multi-linear evaluation schemes have the same implementability power as monotone ones for a single agent (Theorem~\ref{thm:linear-evaluation}), 
but this equivalence does not extend to more general settings.

Optimizing the allocation of effort under different evaluation schemes raises new challenges and directions for future research.
One interesting open direction is to study a \emph{combination} of objectives for incentive-aware evaluation. 
For example, the designer of an evaluation scheme typically cares about \emph{accurate} classification, in addition to incentivizing desirable actions and maximizing self-improvement. This raises the question of what combined guarantees can be achieved using tools from algorithmic contract design.

\subsection{Contracts for ML: Other Results} 
\label{sub:ML-other}

To complement our discussion of the relevance of contract design to incentive-aware classification, we now turn to an additional application of contract design to machine learning: facilitating the delegation of ML-related tasks. Machine learning pipelines are becoming increasingly collaborative, with each task requiring expertise and specialized resources. Tasks are often distributed among different players, and typically involve uncertainty and stochasticity---the defining features of contract design. Contracts can thus be an important tool for the delegation of tasks among different stakeholders in the ML ecosystem. 
We focus on \emph{data collection}, a crucial component of learning, the delegation of which raises a novel informational challenge. For the delegation of other tasks like exploration, see, e.g., \cite{KremerMP14,FrazierKKK14,AzarM18}. 

\paragraph{Multiple Agents: Competition-Based Data Collection.}
\cite*{CaiDP15} give an early formulation of the problem: In their model, there is an unknown function $f$ to be learned, with the goal of achieving accuracy on a random test input $x^*\sim\mathcal{D}$, where distribution~$\mathcal{D}$ is supported over domain~$\mathcal{X}$. In the context of linear regression, for example, $f$ is a linear function that must be learned from samples. Each agent~$i$ receives a sample $x_i$ chosen from domain~$\mathcal{X}$ by the designer, and by exerting effort $e$ achieves a stochastic (continuous) outcome---an estimation ${\hat y}_i$ of $y_i=f(x_i)$. The outcome distribution has the following form: estimation ${\hat y}_i$ is drawn from a distribution with mean $y_i$, whose variance $\sigma_i^2=(\sigma_i(e))^2$ depends on the effort $e$, decreasing as $e$ grows. The main challenge is how to pay the agents for their effort~$e$ so as to incentivize less variance and more accuracy, when we do not know how accurate their estimation is since we do not have knowledge of~$f$. This knowledge gap is inherent to the delegation of learning tasks, and poses a new challenge for contract design. 

\cite{CaiDP15} observe that when there are multiple agents, an agent's payment can be made to depend on other agents' estimates instead of on knowledge of $f$. Using this idea, they design VCG-inspired payments that induce unique dominant strategies for the agents, creating among them a ``race for accuracy''. The resulting effort profile approximately minimizes the following measure of social cost: a combination of the agents' costs of effort, with the mean-square error of the estimated function when applied to the random test $x^*\sim \mathcal{D}$.  

\paragraph{Single Agent: Contract-Based Data Collection.}

The recent works of \cite*{Ananthakrishnan24} and \cite*{SaigTR23} take a different route. They study a single-principal, single-agent setting, in which relying on comparison among multiple agents is not possible, thus distilling the informational challenge arising from ML delegation. 
In their setting, the principal delegates a predictive task to the agent---to learn a classifier that achieves high accuracy on a distribution $\mathcal{D}$ of labeled data points. 
The agent chooses the number of samples to collect, and trains a classifier $h$ from a hypothesis class~$\mathcal{H}$. The number of samples chosen by the agent is his effort level: it determines how the classifier's accuracy is distributed, 
as well as the agent's total cost, which is~$c$ per sample. 

To incentivize the agent's effort, the principal offers contractual payments based on her assessment of the classifier's accuracy. 
Typically, much less data is needed to assess the accuracy of $h$ on $\mathcal{D}$ than to actually train~$h$, and so the principal is assumed to have a test dataset of moderate size (otherwise she could directly learn the model using the test data). However, even if the accuracy of $h$ is perfectly assessed, this is not yet sufficient for determining the payments due to the remaining information gap: Missing from the picture is the \emph{optimal} accuracy that can be achieved on $\mathcal{D}$ with a classifier from $\mathcal{H}$. Say the agent delivers a classifier with error $\theta$, is it due to using too few samples during training? Or is classifying samples from $\mathcal{D}$ inherently difficult? The contractual payments should reflect this.

\cite{Ananthakrishnan24} model the principal's utility as a combination of the accuracy of $h$ minus the payment to the agent. In their model, if the agent collects $n$ samples, then the classifier's accuracy on the test set (the continuous outcome) is drawn from a distribution with mean $1-\theta^*-d/(n^p)$, where $d$ is the hypothesis class dimension, and $p$ is the rate of error decay. The first error term, $\theta^*$, is of the best classifier $h^*$ in $\mathcal{H}$, while the second error term $d/(n^p)$ is due to learning a classifier $h$ that might be different than $h^*$. 
The main result of \cite{Ananthakrishnan24} stems from the robustness of linear contracts: they show that a linear contract achieves an $\frac{e}{e-1}$-approximation to the first-best principal's utility (i.e., the principal's utility if she were to train the classifier herself), provided that $\theta^*$ is known to be bounded by some $\overline\theta$, and the sample cost $c$ is sufficiently small: 

\begin{theorem}[\cite*{Ananthakrishnan24}]
    For any dimension $d>0$ and rate of error decay $p>0$, consider the linear contract $\alpha=1/(p+1)^{\nicefrac{p+1}{p}}$. 
    For any $\overline\theta\in[0,1)$, suppose that the optimum error $\theta^*$ is in $[0,\overline\theta)$, and that $c$ (the agent's cost per sample) is upper bounded by $\frac{p}{d^{1/p}}(\frac{1-\overline\theta}{(p+1)^2})^{\nicefrac{p+1}{p}}$. 
    Then $\alpha$ guarantees an $\frac{e}{e-1}$-approximation for the principal compared to her first-best utility. 
\end{theorem}

\citep{SaigTR23} focus on a different objective. They begin from a standard contract setting, in which the effort level is the number of samples in the training set, and the (discrete) outcome is the number of samples in the test set that are classified correctly. In their model, the principal has a budget $B$ to spend on training the classifier, and the goal is to incentivize the agent to train as accurate a classifier as possible subject to the budget constraint. 
The problem of finding an effort-maximizing contract in which no payment exceeds $B$ reduces to the following problem: design a contract that incentivizes the agent to exert a given level of effort (collect a given number of samples), while minimizing the budget (i.e., the highest payment). 
\cite{SaigTR23} do not impose a particular form of output distributions, but show that in binary-action settings, or alternatively under certain conditions on the distributions (MLRP and a notion of concavity), the optimal contract is a simple \emph{threshold} function, paying the full budget for every outcome above the threshold.

\cite{SaigTR23} then study empirically whether such contracts perform well when the outcome distributions are not fully known. 
For this they use that every outcome distribution captures the stochastic accuracy of learning for a certain sample size, and so coincides with the well-studied object of a \emph{learning curve}. In their experiments, the principal estimates the learning curves from small available data. Based on a recent learning curve database~\citep{MohrVLR22}, they conclude that threshold contracts generally perform well on estimated curves, despite the inherent uncertainty. 

\paragraph{Summary and Open Problems.}

Delegating ML-related tasks raises the challenge of contract design when the setting details are not entirely known (\emph{cf.}~Section~\ref{sec:robust}). In \citep{CaiDP15}, the accuracy of the outcome provided by the agents is unknown; in \citep{Ananthakrishnan24,SaigTR23}, the accuracy of the outcome is (effectively) known, but not how accuracy is distributed for different levels of effort. 
Current solutions in the literature are based on competition \citep{CaiDP15}, or on assuming that the distributions have a known functional form \citep{Ananthakrishnan24} or nice properties \citep{SaigTR23}. Developing additional solutions is arguably becoming increasingly important, as predictive, generative and/or collaborative tasks are increasingly delegated to learning agents.   

\section{Vague, Incomplete, and Ambiguous Contracts}
\label{sec:ambiguous}
In the vanilla model presented in Section~\ref{sec:model}, a contract fully specifies the payment for every single outcome that may occur. 
However, this level of detail is often absent in real-world contracts, either because it is typically impossible to foresee all contingencies, due to the complexity involved, or for other reasons.
For instance, university promotion guidelines from associate to full professor usually stipulate that a candidate should exhibit ``research independence and leadership." Yet, the interpretation of these criteria for each individual candidate is frequently articulated in terms that are somewhat vague, ambiguous or incomplete.

Indeed, the economic research community has identified  {\em incomplete} contracts as a rich area of research \citep{Hart88,HartMoore88} (also see \citep{AghionHolden11}  for a recent survey). This literature considers scenarios where some contingencies are left unspecified in a contract, and explores different ways of resolving such unspecified contingencies.

A different approach is taken in the literature on {\em vague} contracts \citep{BernheimW98}. This literature explores simultaneous or sequential move games between two or more parties, where each party's action set is partitioned into sets and the principal (or some trusted third-party) can distinguish between actions only if they belong to different sets. A contract can then restrict the actions of the parties to certain sets of the respective partitions, and is considered vague if it doesn't narrow it down to a single set for each agent.

In this section, we focus on a model of \emph{ambiguous} contracts, introduced by \citet*{DuettingFP23}, which draws on the concept of ambiguity in mechanism design and auctions introduced by \citet*{DiTillioEtAl17}. 
Section~\ref{sub:ambiguous-model} introduces the model, and Section~\ref{sub:ambiguous-struct} discusses both structural and computational insights. Section~\ref{sub:ambiguous-proof} explores classes of contracts where the principal cannot gain from ambiguity, Section~\ref{sub:ambiguous-gap} quantifies by how much a principal can gain when there is a gap relative to classic contracts, and Section~\ref{sub:ambiguous-mixing} shows that mixed strategies completely eliminate the power of ambiguity. 

\subsection{Ambiguous Contracts Model}
\label{sub:ambiguous-model}

\begin{figure}[t]
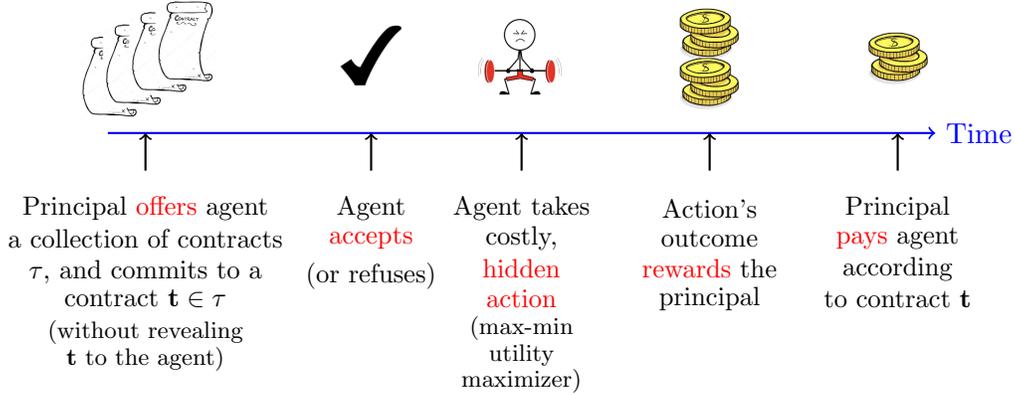

\vspace*{-10pt}
\centering
\include{figures/timing-contracts-ambiguous.tex}
\vspace*{-30pt}
\caption{Timeline of ambiguous contracts. Compared to Figure~\ref{fig:timeline}, the principal offers the agent a {\em collection} of contracts $\tau$ and commits to a single contract $\con \in \tau$ without revealing $t$ to the agent. The agent is a max-min utility maximizer, namely he maximizes his minimum utility across all contracts in $\tau$. In the final step, the principal pays the agent according to the chosen contract $\con \in \tau$ (and the revealed outcome).}
\label{fig:ambiguos-timeline}
\end{figure}

The starting point of \citet{DuettingFP23} is the vanilla contracting problem from Section~\ref{sec:model}. The principal, however, can now offer an {\em ambiguous contract}, which is defined as a collection of classic contracts $\ambpay=\{\con^1,\ldots,\con^k\}$ (each defining a payment for each outcome).
The principal commits to one of the contracts in $\tau$, without revealing the chosen contract to the agent.
The ambiguity arises from the fact that the agent observes the set of contracts but does not know which one will be applied. 
The agent is a max-min expected utility maximizer \citep{Schmeidler89,GilboaS93}, and so selects an action that maximizes their expected utility under the worst contract $\con \in \ambpay$ (i.e., the contract $\con \in \ambpay$ with the minimum expected payment).
That is, the chosen action under an ambiguous contract $\ambpay$ is 
\begin{align*}
    i^\star(\ambpay) \in \arg \max_{i\in [n]} \min_{\con \in \ambpay} (\Pay_i(\con)-\cost_i), \quad\text{where}\quad T_i(\con) = \sum_{j \in [m]} \prob_{i j} \pay_j.
\end{align*}
An outcome is then realized, based on the probability distribution of the chosen action over outcomes, and the principal pays the agent according to the contract she committed to (see Figure~\ref{fig:ambiguos-timeline}).

The expected payment of an ambiguous contract $\ambpay$ is denoted by 
$\Pay_{i^\star}(\tau)$, where, for simplicity, we denote the action incentivized by $\ambpay$ by $i^\star$. It holds that $\Pay_{i^\star}(\tau) = \min_{\con \in \tau} \Pay_{i^\star}(\con)$.

Additionally, the ambiguous contract is required to be \emph{consistent}, in that under the action chosen by the agent, denoted $i^\star$, all contracts in the support of the ambiguous contract yield the same principal utility.
Formally, an ambiguous contract $\ambpay$ is consistent if $\Pay_{i^\star}(\con) = \Pay_{i^\star}(\con')$ for any $\con,\con' \in \tau$.
This ensures that the principal is indifferent between all contracts in the support. Without this condition, the principal would strictly prefer some contracts in the support over others, and therefore it would not be believable for the agent that the principal may indeed choose any of the contracts in the support. This requirement turns out to be without loss of generality.

Ambiguity grants the principal additional power which she can use to obtain a higher utility. The following example demonstrates that the gain can be strictly positive.

\begin{example}[Strict improvement through ambiguity, \cite*{DuettingFP23}] 
Consider the following principal-agent setting with three actions:

\begin{center}
\begin{tabular}{|l|ccc|c|}
\toprule
& $r_1=0$ & $r_2 = 2$ & $r_3 = 2$  & \text{cost} \\
\midrule
action $1$: & $1$ & $0$ & $0$ & $c_1 = 0$\\
action $2$: & $\nicefrac{1}{2}$ & $\nicefrac{1}{2}$ & $0$ & $c_2 = \nicefrac{1}{4}$\\
action $3$: & $\nicefrac{1}{2}$ & $0$ &  $\nicefrac{1}{2}$ & $c_3 = \nicefrac{1}{4}$\\
action $4$: & 0 &  $\nicefrac{1}{2}$ & $\nicefrac{1}{2}$ & $c_4 = \nicefrac{3}{4}$\\
\bottomrule
\end{tabular}
\end{center}

The best classic contract in this principal-agent setting implements action $4$ with the contract $\con = (0,1,1)$, yielding the principal an expected utility of $1$. Indeed, the best classic contract implementing action $2$ is $\con = (0,1/2,0)$, for a principal's utility of $3/4$; and the same holds for action $3$, with the contract $\con = (0,0,1/2)$.
Meanwhile, the maximum principal's utility from action $1$ is $0$, as this is action $1$'s welfare.

An optimal ambiguous contract implements action $4$ by $\tau=\{\con^1,\con^2\}$, with 
$\con^1=(0,3/2,0)$ and $\con^2=(0,0,3/2)$. Both contracts in $\tau$ leave the agent with a utility of zero for action $1$.
The worst contract in $\tau$ for action $2$ is $\con^2$, giving the agent an expected payment of $0$.  Similarly, the worst contract for action $3$ is $\con^1$, for an expected payment of $0$.  Thus, both actions $2$ and $3$ give the agent negative utilities.  In contrast, the expected payment for action $4$ is $3/4$ under both $\con^1$ and $\con^2$, giving the agent an expected utility of $0$.  The ambiguous contract $\tau$ thus implements action $4$, with an expected payment of $3/4$, and an expected utility for the principal of $5/4$ strictly higher than her optimal utility under a classic contract.
\label{ex:amiguity}
\end{example}

Motivated by this example, \citet{DuettingFP23} study various aspects of ambiguous contracts, including the structure and computation of optimal ambiguous contracts, a characterization of ``ambiguity-proof'' contract classes (see Definition~\ref{def:amb-proof}), as well as upper and lower bounds on the ambiguity gap, which quantifies the principal's potential gain from employing ambiguous contracts.

\subsection{Structure and Computation of Ambiguous Contracts}
\label{sub:ambiguous-struct}

The first aspect that \citet{DuettingFP23} study is structural and computational properties of optimal ambiguous contracts. 

In particular, they show that an optimal ambiguous contract is, without loss of generality,  composed of ``simple" contracts that take the form of {\em single-outcome payment} (SOP) contracts (see Section~\ref{sec:optimal-contracts}). 
Recall that, an SOP contract is one that pays for a single outcome only. That is, a contract $\con = (\pay_1,\ldots, \pay_m)$ such that $\pay_{j} > 0$ for a single outcome $j \in [m]$ and $\pay_{j'} = 0$ for any outcome $j' \neq j$.
This is cast in the following theorem.

\begin{theorem}[\citet*{DuettingFP23}]
For every ambiguous contract $\tau$, there exists an ambiguous contract $\tau'$, consisting of at most $\min\{m,n-1\}$ contracts, such that: (i) For every  $t \in \tau'$, $t$ is an SOP contract. (ii) The same action is incentivized by $\tau$ and $\tau'$, denote it $\istar$, and (iii) $\tau$ and $\tau'$ have the same expected payment for action $\istar$. 
\label{thm:sop}
\end{theorem}

\begin{proof}
Let the ambiguous contact $\tau$ incentivize action $\istar$. 
Let $J^\star = \{j \in [m] \mid \prob_{\istar j}>0 \}$ be the outcomes that occur with positive probability under $\istar$. 
Recall that
$\Pay_{i^\star}(\tau) = \min_{\con \in \tau} \Pay_{i^\star}(\con)$, and note that $\Pay_{i^\star}(\con) = \Pay_{i^\star}(\con')$ for any $\con,\con' \in \tau$ by consistency.
For every $j \in J^\star$,  
consider the SOP contract with payment $\frac{T_{\istar}(\tau)}{\prob_{\istar j}}$ for outcome $j$. Let $\tau'$ be the ambiguous contract consisting of these SOP contracts. Clearly,  $\tau'$ satisfies property ($i$) by construction.
To see that $\tau'$ satisfies property ($iii$), note that, for every contract $t \in \tau'$, the expected payment for action $\istar$ under $t$ is 
$\prob_{\istar j} \cdot \frac{T_{\istar}(\tau)}{\prob_{\istar j}} = T_{\istar}(\tau)$ (where $j$ is the outcome corresponding to contract $t$).
We next show that $\tau'$ also satisfies property ($ii$); namely that it  incentivizes $\istar$.   Consider an action $i\neq \istar$.  Because $\tau$ incentivizes $\istar$, there exists $\con \in \tau$ with 
\[
\cost_{\istar}-\cost_i~\le~ T_{\istar}(\tau) - \sum_{j} \pay_j \prob_{ij} ~=~ \sum_{j} \pay_j \prob_{\istar j}-\sum_{j} \pay_j \prob_{ij}.
\]
To show that $\tau'$ incentivizes $\istar$, it suffices to show that 
\[
\cost_{\istar}-\cost_i~\le~ T_{\istar}(\tau) - \min_{j' \in J^\star} \prob_{ij'}\frac{T_{\istar}(\tau)}{\prob_{\istar j'}}~ =~ \sum_{j} \pay_j \prob_{i^\star j}-\min_{j'\in J^\star}\frac{\prob_{ij'}}{\prob_{\istar j'}}\sum_{j} \pay_j \prob_{\istar j}.
\]
Combining these, it suffices to show that
$\min_{j' \in J^\star}\frac{\prob_{ij'}}{\prob_{\istar j'}}\sum_{j} \pay_j \prob_{\istar j}~\le~ \sum_{j} \pay_j \prob_{ij}$,
which is equivalent to the obvious statement that
$\min_{j' \in J^\star}\frac{\prob_{ij'}}{\prob_{\istar j'}}~\le~ \frac{\sum_{j}\pay_j \prob_{ij}}{\sum_{j} \pay_j \prob_{\istar j}}$.
Notice that $\tau'$ consists of at most $m$ SOP contracts (in fact, at most $|J^\star|$ contracts).  If $m>n-1$, one can eliminate from $\tau'$ every contract that does not minimize the expected payoff to one of the alternatives $i\neq \istar$,
leaving at most $n-1$ SOP contracts.
\end{proof}

Building on the insights of Theorem~\ref{thm:sop}, \cite{DuettingFP23} give a poly-time algorithm for computing an optimal ambiguous contract.

\begin{theorem}[\citet*{DuettingFP23}]
There exists an algorithm that computes the optimal ambiguous contract in time $O(nm^2)$.
\label{thm:opt-amb-contract}
\end{theorem} 

The idea of the proof is the following. Similarly to the approach taken in Section~\ref{sec:lp-formulation} for classic contracts, here too, for every action $i$ the algorithm computes the best ambiguous contract that incentivizes action $i$, then chooses the best one among the obtained contracts.
Specifically, for any action $i$, one can, in $O(nm)$ time, decide whether action $i$ can be implemented by an ambiguous contract and if so, find the optimal ambiguous contract incentivizing it.
Notably, this algorithm is not LP-based. Instead, it extends the maximum likelihood ratio principle that underlies the optimal single contract for two actions (see Section~\ref{sec:optimal-contracts}), and combines this extended principle with a waterfilling technique, which aligns the payments of all SOP contracts in the support. 

As a byproduct of this proof,
\cite{DuettingFP23} obtain the following characterization of actions that can be implemented with an ambiguous contract.

\begin{proposition}
[\citet*{DuettingFP23}]
Action $i$ can be implemented by an ambiguous contract if and only if there is no other action $i' \neq i$ such that $\Prob_{i'} = \Prob_{i}$ and $c_{i'} < c_i$.
\end{proposition}
 
Compared with Proposition~\ref{prop:implementable}, which characterizes actions that can be implemented with a classic contract, this shows that ambiguous contracts enlarge the set of implementable actions.

Two remarks are in order. First, an SOP contract is typically non-monotone. When restricted to monotone contract, one can show that the optimal monotone ambiguous contract is also composed of simple contracts, termed ``step contracts.'' A step contract pays $0$ for all outcomes up to some outcome $j \in [m]$, and pays a fixed payment to outcomes $j+1, \ldots, m$. 
As before, using this observation, one can devise a poly-time algorithm that computes the optimal monotone ambiguous contract.
Second, the optimal ambiguous contract in instances satisfying the MLRP regularity condition (see definition in Section~\ref{sec:model}) admits an even simpler structure. 
Specifically, it is composed of only two contracts (two SOP contracts in the unrestricted case and two step contracts when restricted to monotone contracts). 
Naturally, this also leads to faster algorithms for these cases.

\subsection{Ambiguity-Proof Classes of Contracts} 
\label{sub:ambiguous-proof}

An additional natural question is whether there are classes of contracts that exhibit an inherent resistance to ambiguity.   
\citet{DuettingFP23} provide the following definition of {\em ambiguity-proofness} to capture this resistance. 

\begin{definition}[Ambiguity-proofness]\label{def:amb-proof}
    A class of contracts $\mathcal{T}$ 
is \emph{ambiguity-proof} if for any instance,  the principal cannot strictly gain from implementing any action $i$ with an ambiguous rather than a classic contract.
\end{definition}

For example, the principal-agent scenario presented in Example~\ref{ex:amiguity} demonstrates that the contract class encompassing ``all contracts" is not ambiguity-proof. 
Indeed, action $4$ can be incentivized using the ambiguous contract $\tau=\{\con^1,\con^2\}$, with expected payment of $3/4$, while the optimal classic contract that incentivizes action $4$ has expected payment of $1 > 3/4$.

We next present the condition for ambiguity-proofness given in \citet{DuettingFP23}. 
Note that here just like in Section~\ref{sec:robust} it is helpful to think of contracts as mappings from outcomes to payments. So in the following, just as we did in Section~\ref{sec:robust}, we will use $t$ rather than $\con$ to refer to a contract. 

\begin{definition}[Ordered class of contracts]
\label{definition:proper crossing}
A class of contracts $\classOfContracts$ is  \emph{ordered} if for any two contracts $t,t' \in \classOfContracts$ it holds that: 
\begin{align*}
    t(x) \geq t'(x) \quad \text{for all $x \in \reals$} \quad
    \text{or} \quad 
    t(x) \leq t'(x) \quad \text{for all $x \in \reals$.}
\end{align*} 
\end{definition}

The characterization is then given by the following theorem.
\begin{theorem}[\citet*{DuettingFP23}]
A class of payment functions $\classOfContracts$ is ambiguity-proof if and only if it is ordered.
\label{thm:amb-charac}
\end{theorem}

\begin{proof}[Proof sketch.]
We first show that ordering implies ambiguity-proofness.
Let $\classOfContracts$ be an ordered class of contracts, and let $\tau$
be an ambiguous contract composed of contracts in $\classOfContracts$. 
Ordering of $\classOfContracts$ implies that there exists a contract $\pay \in \tau$ such that $\pay(x) \leq \pay'(x)$ for any $\pay' \in \tau$ and all $x$.
It is then easy to verify that $\pay$ incentivizes the same action as $\tau$, and yields the same utility for the principal.

We next show that ambiguity-proofness implies ordering, by proving the contrapositive.
Suppose $\classOfContracts$ violates ordering. 
Then there exist $\pay,\pay' \in \classOfContracts$ and $x_1,x_2 \in \reals$ such that $\pay(x_1) > \pay'(x_1)$ and $\pay(x_2) < \pay'(x_2)$.  
Leveraging these inequalities, we derive values $q_1,q_2>0$ with $q_1+q_2=1$
satisfying
\begin{equation}
\label{spooky}
    q_1 \pay(x_1) + q_2 \pay(x_2) =  q_1 \pay'(x_1) + q_2 \pay'(x_2).\nonumber
\end{equation}
Using $q_1,q_2$, we construct an instance with two outcomes $r_1=x_1$ and $r_2=x_2$ and three actions, such that action 3 can be implemented by an ambiguous contract $\tau=\{\pay,\pay'\}$, but 
the convex combination of actions $1,2$ via vector $(q_1,q_2)$ yields the same distribution over rewards as action $3$, but at a strictly lower cost. Thus, by Proposition~\ref{prop:implementable}, action 3 cannot be implemented by a classic contract. This concludes that $\classOfContracts$ is not ambiguous-proof.
\end{proof}

As a direct corollary of Theorem~\ref{thm:amb-charac}, it follows that the class of linear contracts is ambiguity-proof.
This characteristic may provide additional insight into the widespread adoption of linear contracts and complement their max-min optimality under different notions of uncertainty, as explored in Section~\ref{sec:robust}.

\subsection{Tight Bounds on the Ambiguity Gap}
\label{sub:ambiguous-gap}

Example~\ref{ex:amiguity} demonstrates that the principal can benefit from using an ambiguous contract compared to a classic one. 
We next discuss results of \cite{DuettingFP23} that quantify how much the principal can gain by using an ambiguous contract rather than a classic one, as captured by the ambiguity gap, defined next.

The ambiguity gap of a given instance $(\mathbf{c},\mathbf{r},\Prob)$ is the ratio between the maximum principal's utility in any ambiguous contract and the maximum principal's utility in any classic contract.
The ambiguity gap of a class of instances $\mathcal{I}$ is the supremum ambiguity gap over all instances in $\mathcal{I}$. Formally, 

\begin{align*}
      \rho(\mathbf{c},\mathbf{r},\Prob) = \frac{\max_{\ambpay}\left(\Rew_{\istar(\ambpay)}-\Pay_{\istar(\ambpay)}\right)}{\max_{\con}\left(\Rew_{\istar(\con)}-\Pay_{\istar(\con)}\right)}
      \quad\quad \text{and} \quad\quad \rho(\mathcal{I}) = \sup_{(\mathbf{c},\mathbf{r},\Prob) \in \mathcal{I}} \rho(\mathbf{c},\mathbf{r},\Prob).
\end{align*}

The following is a nearly-tight bound on the ambiguity gap.\footnote{If we further assume that the zero-cost action leads to an expected reward of zero, then this bound can be strengthened to a tight bound of $n-1$.} 

\begin{proposition}[\citet*{DuettingFP23}]
\label{prop:gap-general}
Let $\mathcal{I}_n$ be all instances $(\mathbf{c},\mathbf{r},\Prob)$ with $n$ actions. It holds that $n-1 \leq \rho(\mathcal{I}_n) \leq n$.
\end{proposition}

The upper bound on the ambiguity gap is derived from the upper bound of $n$ that \citep{DuttingRT19} establish on the (possibly larger) gap between the optimal welfare and the principal's utility from a linear contract.
The lower bound of $n-1$ is established via a variant of an instance given in \citep{DuttingRT21} (see Example~\ref{ex:first-betst-second-best}).

It is worth noting that \citet{DuettingFP23} also consider cases where the rewards may be negative and show that, for this class of instances, the ambiguity gap is unbounded.

\subsection{Mixing Hedges Against Ambiguity}
\label{sub:ambiguous-mixing}

The attentive reader may notice that, in Example~\ref{ex:amiguity},  by employing a {\em mixed} strategy that mixes between the actions 2 and 3, each with probability $1/2$, the agent achieves an expected payment of $3/8$ for any of the 
two contracts in the support of the ambiguous contract (namely, $\con^1=(0,3/2,0)$ and $\con^2=(0,0,3/2)$), for an expected utility of $1/8$,  which is strictly better than the agent's utility from action $4$ (which is $0$). This is not a coincidence. Indeed, one can show that mixed strategies completely eliminate the power of ambiguity.

\begin{theorem}[\citet*{DuettingFP23}] If the agent may engage in mixed strategies, then the maximum utility the principal can achieve with an ambiguous contract is no higher than the maximum utility she can achieve with a classic contract.
\end{theorem}

Interestingly, a similar phenomenon was established by \cite*{CollinaDR24} for general Stackelberg games. 
In particular, the leader can gain utility by making an ambiguous commitment if the follower is restricted to playing a pure strategy, but no gain can be made if the follower may engage in a mixed strategy.
However, they also show that in general Stackelberg games with \emph{multiple} followers, ambiguity may be beneficial even when the followers engage in mixed strategies.

\paragraph{Open Questions and Additional Directions.} 

We believe that the algorithmic study of incomplete, vague, and ambiguous contracts has just scratched the surface of what could be a much richer theory. 
Concerning ambiguous contracts, \cite{DuettingFP23} show that for settings that satisfy MRLP, optimal ambiguous contracts are composed of two classic contracts, either two SOP contracts or two step contracts, depending on whether or not monotonicity is imposed. Given the practical appeal of such ``succinct'' ambiguous contracts, a natural direction for future work is to study ambiguous contracts of bounded size, beyond settings for which they are known to be optimal.
Another promising direction for future work is to study ambiguous contracts in settings with multiple agents. This extended setting introduces many natural structural and algorithmic challenges. A particularly intriguing open problem is whether mixed strategies still eliminate the power of ambiguity.
More generally, we see ample room for algorithmic approaches to vague and incomplete contracts, which, to the best of our knowledge, remain mostly unexplored from an algorithmic perspective.

\section{Contract Design for Social Good}
\label{sec:social-good}
The study of \emph{mechanism design for social good (MD4SG)} has grown significantly in recent years, emerging as a highly impactful area of research. 
For designers or policymakers aiming to leverage algorithms, optimization, and game theory to drive societal change, contracts represent a valuable addition to the toolbox of available techniques.
Indeed, contract design plays a crucial rule in advancing social good across a variety of domains, including environmental protection~\cite[e.g.,][]{LiAL21,LiIL21}, healthcare~\cite[e.g.,][]{BastaniEtAl16,BastaniGB19}, and education~\cite[e.g.,][see also Section~\ref{sec:incentive-aware}]{KleinbergR19,AlonDPTT20,HaghtalabILW20}.
In environmental protection and healthcare, this often takes the form of pay-for-performance programs, which align incentives with desired outcomes such as reduced emissions, afforestation or improved patient care. In education, contract design manifests in evaluation schemes that encourage meaningful learning, effectively deterring counterproductive behaviors like cheating or rote memorization. These examples illustrate the potential of contract theory to drive meaningful impact in addressing significant societal challenges.
In this section, we focus on contracts for environmental protection as a case study. 

\subsection{Contract Design and Environmental Protection}

A main driver behind the exploitation of natural resources is the classic market failure of moral hazard, where the outcomes of an agent's behavior---in this case environmental protection (or a lack thereof)---reward (or harm) different parties in different ways (e.g., the landowner, society at large, or future generations). 
Contract design is the main economic tool for combating moral hazard. Thus, optimizing contracts is highly relevant to optimizing incentives for environmental protection. 

Programs that offer contracts which reward individuals for environmental protection---called Payment for Ecosystem Services (PES)---are increasing in popularity in practice. Globally there are more than 550 PES programs, with combined annual payments of over 36 billion USD~\citep{SalzmanBC+18}. Carefully designing such programs is crucial for their success.
Indeed, studies such as~\cite{BornerBC+17} show that current PES programs vary highly in their effectiveness, highlighting the need to replace heuristics with theory-backed contract design.  
PES contracts also raise particular design challenges: They typically must be both simple and robust to be applicable, and must accommodate a rich space of possible actions as well as outcome measures to reflect reality. 

Consider for concreteness PES programs that offer contracts for afforestation (\#15 in the UN's Sustainable Development Goals \citep{undp}).
Under current PES designs, farmers often choose not to enter into the offered contracts, or opt to enter but exert a sub-optimal level of effort. The reason is, again, moral hazard: the task of growing trees to maturity or abstaining from deforestation exposes farmers to financial risks, which many of the existing contracts fail to mitigate. Pioneering works to remedy this include \cite*{LiAL21} and~\cite*{LiIL21}.
While \cite{LiAL21} consider a setting without hidden action, it illustrates the importance of incentives in encouraging desired behavior in the context of environmental protection and demonstrates the power of linear payment schemes.

\paragraph{Disincentivizing Deforestation, without Hidden Action.} 

\citet*{LiAL21} explore the use of contracts to disincentivize deforestation in a setting with no hidden actions. One of their main results is that linear contracts provide a constant-factor approximation to the optimal, more complex contract for disincentivizing deforestation. In their model, the agent (landowner) has an initial amount of forest $a_0 \in \mathbb{R}_{\geq 0}$, which is publicly observable. The agent has a private type $\theta \in [\theta_\ell,\theta_h] \subseteq [0,1]$ distributed according to a publicly known distribution $F$. The type captures the percentage of initial forest $a_0$ which the agent would preserve in the baseline case. The agent also has a convex cost function $c(a,\theta)$ that depends on both the chosen action $a$ (amount of forest actually preserved), and on the type $\theta$. Specifically, they consider the following cost function: 
\begin{itemize}
    \item for $a \leq \theta \cdot a_0$, $c(a, \theta) = 0$ (preserving less than $\theta a_0$ comes with no cost);
    \item for $a \geq \theta \cdot a_0$, $c(a,\theta) = \frac{h}{2} (a-\theta \cdot a_0)^2$ for some constant $h > 0$ 
 (the cost grows quadratically in the excess amount $a - \theta \cdot a_0$ of forest preserved).
 \end{itemize}
The principal, who does not own the land (e.g., a government or non-profit organization), has conservation value $k$ per unit of land. A contract now specifies a payment $\pay(a)$ where $a$ is the amount of land preserved (note that in~\cite{LiAL21} the action is not hidden and can be deterministically inferred from the outcome). The agent with type $\theta$ chooses to preserve an amount of forest $a^\star(\theta)=\arg\max_a \{\pay(a) - c(a,\theta)\}$. The principal's goal is to find a contract $\pay$ that maximizes her expected utility given by $\mathbb{E}_\theta[ka^\star(\theta) - \pay(a^\star(\theta))]$, where $k$ is the  conservation value. A linear contract pays the agent a fixed price $p$ per unit of forest preserved. The aforementioned main result can be formally stated as: 

\begin{theorem}[\citet*{LiAL21}] 
For every $k > 0$ and $F$, a linear contract with price $p = \nicefrac{k}{2}$ achieves at least half of the optimal contract payoff. 
\end{theorem}

The work of \cite{LiAL21} offers a variety of additional results, including results for more general convex cost functions. They also uses empirical studies to calibrate the key parameters of their model, and quantify the suboptimality of linear contracts tuned to these parameters.

\paragraph{Disincentivizing Deforestation, with Hidden Action.}

In the work of \citet*{LiIL21}, the principal has imperfect information not only about the agent's type but also about his chosen effort. 
Moreover, the agent's effort is exerted over time, affecting the tree growth process in a way that is modeled by a Markov chain.
The authors identify the structure of the optimal contract within this model, and develop a polynomial-time algorithm to calculate its payments. They also apply their approach on data from a recent afforestation program in Uganda, showcasing its  applicability.

\medskip
\noindent\emph{Model.} The model of \citet{LiIL21} is based on a Markov chain with finite state space 
$\mathcal{S}=\{0,1,\ldots,M\}$ that captures the state of the tree (i.e., the tree's growth). The state is publicly observable (e.g., through a monitoring technology). The game starts in state $s = 0$, which indicates that there is ``no tree.'' State $s > 0$ indicates the age of the tree in years, with $M$ being the number of years to get a fully mature tree. The principal and the agent derive a value of $v_s^A \geq 0$ and $v_s^P \geq 0$ for a live tree in state $s$, where $v_s^A = v_s^P = 0$ for all states $s \in \{0,\ldots,M-1\}$, so only a fully mature tree (potentially) delivers positive value. 
In every state $s \in \mathcal{S}$, the agent chooses effort/no effort and this choice is hidden from the principal. The cost of effort $c$ is also hidden and drawn from a known distribution $F$ with support $[0,\bar{c}]$ (the results also apply if the cost varies per state $c_s$, for every state $s \in S$, and under additional generalizations---see~\cite{LiIL21}). Crucially, even with effort, the tree survives only with probability $q$ at every stage. 
Thus, in every state, if the agent exerts effort, then, the tree goes to the next state (state $\min\{s+1,M\}$) with probability $q$ and to state $0$ otherwise. If the agent doesn't exert effort, then the tree goes to state $0$.

The principal defines a contract, which is a vector of payments $p_s \geq 0$ for each $s \in \mathcal{S}$, where $p_s$ is a conditional payment for a transition from state $s$ to state $\min\{s+1,M\}$ (non-negativity ensures limited-liability). 
If in state $s$ the agent does not exert effort, then his per-round utility is $u^A(s) = 0$. If the agent exerts effort then his per-round utility is $u^A(s) = q \cdot (v^A_s + p_s) - c$. The agent discounts future payoffs at a rate of $\delta$. In this setting, once the principal's payment schedule is determined, the agent operates within a Markov chain that converges to a steady-state distribution $\{p_s\}_{s \in \mathcal{S}}$. 
The principal's objective is to maximize expected revenue, 
averaged over the steady-state distribution $D$ of the Markov chain.

\medskip
\noindent{\emph{Results.}} \citet{LiIL21} observe that while the strategy space of the agent is quite large, as he can choose between two actions (effort / no effort) in every period, due to the structure of the Markov process, it suffices to consider strategies where the agent exerts effort up to a certain state. Under this observation, the principal's problem reduces to finding an optimal sub-interval $C = [c_\ell,c_h] \subseteq [0,\bar{c}]$ of the agent type to target, and for agents in this sub-interval, finding the least-costly contract such that an agent with cost $c \in C$ chooses not to drop out. 
The main result is a set of $M+1$ equalities 
that can be solved to find the optimal schedule for the subset of types corresponding to the targeted subinterval. The agent subset is found by discretization plus grid search.
The authors mention as an open direction other potential sources of heterogeneity among agents, including the agent’s discounting rate and risk preference.

\paragraph{Additional Directions and Future Work.} 

Contracts for social good present an important direction for future work. Beyond the domains mentioned above (healthcare, environmental protection, and education), we see possible applications of contract design in fostering collaborative behavior among human/AI agents in ``social dilemmas''  more broadly \citep[e.g.,][]{LeiboZLMG17,HauptCDH24}. Similarly, employing ideas from contract design to orchestrate markets of effort (e.g., through AI agents) is a very timely direction \cite[e.g.,][]{Bollini24,IvanovDP+24,WuCWWX24}, and naturally raises fairness questions.

\section{Incentivizing Effort Beyond Contracts}
\label{sec:beyond-contracts}
In this section, we discuss recent work at the intersection of economics and computation that is concerned with incentivizing effort, but either does not take a contracts approach, or combines contracts with an additional approach.
The main directions discussed are scoring rules (Section~\ref{sub:scoring-rules}), algorithmic delegation (Section~\ref{sub:delegation}), and information design (Section~\ref{sub:information}).

\subsection{Scoring Rules}
\label{sub:scoring-rules}

Scoring rules apply when one player (the \emph{forecaster}) has more information about a hidden ``state of the world'' (\emph{state of nature}) than the principal \citep{Savage71,GneitingR07}. 
Designing \emph{proper} scoring rules enables the principal to create incentives for the forecaster to reveal her \emph{true} beliefs about the unknown probabilistic state. 
Scoring rules optimization has applications to peer prediction and peer grading. 
Several recent papers formulate optimization problems that are concerned with incentivizing the forecaster to exert effort in order to refine his beliefs  
\citep{ChenYu21,NeymanNW21,LiHSW22,HartlineSLW22}, while \citep{PapireddygariW22} combines costly information acquisition with a hidden-action contracting problem.

\paragraph{Without Hidden Action.} 

We focus first on the work of \citet*{LiHSW22}, in which the prediction can be refined via costly (non-hidden) effort. 
\citet{LiHSW22} study a situation where a forecaster has a prior distribution $D \in \Delta(\Theta)$ over an unknown state $\theta \in \Theta$ from a state space $\Theta \subseteq \mathbb{R}^n$, and may exert binary effort to obtain a refined posterior distribution $G \in \Delta(\Theta)$ 
with probability $f(G)$. 
The goal is to design a proper scoring rule for eliciting the mean of the distribution that maximizes the agent's incentive for exerting effort (the difference in expected scores with and without effort) among all proper scoring rules whose score is non-negative and bounded by $B$.
 
More formally, denote by $\mu_D$ and $\mu_G$ the mean of the prior and the posterior, respectively. Let $R \subseteq \mathbb{R}^n$ be the report space, and assume it includes all possible posterior means. Let $r \in R$ be the report of the agent. A scoring rule $S: R \times \Theta \rightarrow \mathbb{R}$ takes a report $r \in R$ and a state $\theta \in \Theta$ as input, and maps it to a score $S(r,\theta) \in \mathbb{R}^n$. A scoring rule is proper for eliciting the mean of a distribution, if for any distribution $G$ and any report $r$, $\mathbb{E}_{\theta\sim G}[S(\mu_G,\theta)] \geq \mathbb{E}_{\theta\sim G}[S(r,\theta)]$. A scoring rule $S: R \times \Theta \rightarrow \mathbb{R}$ is bounded in space by $B$ if $S(r,\theta) \in [0,B]$ for all $r \in R, \theta \in \Theta$. The objective is to design a proper scoring rule for eliciting the mean, whose score is bounded in space by $B$, that maximizes 
$\mathbb{E}_{G \sim f,\theta\sim G}[S(\mu_G,\theta) - S(\mu_D,\theta)]$
among all such rules.

The authors identify optimal scoring rules for this problem and give (prior free) approximation results for both the single- and the multi-dimensional case.

\paragraph{With Hidden Action.}

\citet{PapireddygariW22} consider a problem that combines costly information acquisition with a hidden-action contracting problem. The basic scenario is that of a principal (e.g., a television company) that seeks to hire an agent (e.g., a show producer) to take a costly action (e.g., to produce a TV show). Before taking the action, the agent can acquire a costly signal (e.g., by running a market research study) to refine their beliefs about the action-to-outcome mapping (e.g., to find out which types of shows are more likely to become a hit).  In their model, nature draws a state $\sigma$ taking values in $\Sigma$ from a known prior distribution $q \in \Delta(\Sigma)$. The agent can acquire signal $\sigma$ at cost $\kappa \geq 0$. The agent takes an action $a \in A$, which incurs a cost $c_a \geq 0$. 
The chosen action $a$ induces a distribution over outcomes $\omega \in \Omega$, which depends on the state $\sigma$. 
Denote it by $p_{a,\sigma} \in \Delta(\Omega)$.
A menu of contracts $T \subseteq \mathbb{R}^\Omega$ is now a set of functions $t: \Omega \rightarrow \mathbb{R}$ (or a set of $|\Omega|$-dimensional vectors) that specify the amount of money transferred from the principal to the agent, when outcome $\omega \in \Omega$ is realized. 
A menu of contracts satisfies limited liability if the corresponding payments are all non-negative.

The timing of the problem is as follows: First, the principal posts a menu of contracts $T$. Then nature draws signal $\sigma \sim q$, and the agent decides whether to acquire signal $\sigma$ at cost $\kappa \geq 0$ or not. Afterwards, the agent chooses a contract $t \in T$ from the menu of contracts, and an action $a^\star$, which incurs a cost of $c_{a^\star}$. Finally, an outcome $\omega$ is realized from $p_{a^\star,\sigma}$, and the agent is paid $t(\omega)$. 
The question is: Given a plan that consists of always acquiring the costly signal and a mapping from signals to actions, is it implementable (i.e., is there a menu of contracts that incentivizes the agent to follow this plan)? Moreover, if it is implementable, is it possible to find a menu of contracts that incentivizes the agent to follow the plan as cheaply as possible? 

The key insight of \cite{PapireddygariW22} is that there is a close connection between menus of contracts and 
scoring rules: Namely, a scoring rule is a mapping $S: \Delta(\Omega) \times \Omega \rightarrow \mathbb{R}$. 
Observe that for a fixed $p \in \Delta(\Omega)$, $S(p,\cdot)$ is analogous to a contract $t(\cdot)$. So we can think of scoring rules as menus of contracts, and vice versa. 
Using this connection, the authors obtain a characterization of implementable plans and the limited liability condition (which requires that transfers be non-negative). They use this to gain additional structural insights into the pure information acquisition version of the problem (without the hidden-action part) and the pure hidden-action contract problem (without information acquisition). For the general case that combines both aspects they give a poly-time (linear programming based) algorithm for finding the menu of contract that incentivizes a given plan with minimum expected payment.

\subsection{Algorithmic Delegation}
\label{sub:delegation}

Another closely related direction is algorithmic delegation \citep{KleinbergK18,BechtelD21,BechtelDP22,BraunHHS23,TaggartEtAl2022}, based on the classic economic delegation model of~\citet{Holmstrom84}. 
The general problem addressed here is similar in spirit to the contract design problem in that there is an uninformed principal who consults an informed agent to make a decision.
Often the problem involves choosing an alternative from a set of alternatives, where the preferences of the parties over the alternatives are misaligned.
The agent's task is to investigate those alternatives, and propose one to the principal.
The principal cannot resort to payments. 
Instead, the principal incentivizes the agent by committing to a policy that specifies which alternatives would be accepted.

The delegation model is thus fundamentally about information: It centers on the agent's role in collecting information and communicating it back to the principal, and on the principal's strategic choice on how to act upon that information.

For example, in the delegated search problem of \citet{KleinbergK18}, which is essentially the problem considered by \citet{ArmstrongV10}, there is a publicly known distribution $F$ over outcomes $\Omega$ (e.g., candidates for a faculty position). Let $\bot$ be an outside option (not hiring), and let $\Omega_+ = \Omega \cup \{\bot\}$. 
The principal and the agent have utility functions $x,y: \Omega_+ \rightarrow \mathbb{R}$ with $x(\bot) = y(\bot) = 0$, encoding their respective preferences over outcomes (candidates). The agent who performs the search draws $n$ independent outcomes $\omega_1, \ldots, \omega_n$ from $F$, and presents one of these or $\bot$ to the principal. While the principal cannot pay the agent, the principal has the power to either accept or reject the agent's proposal $\omega \in \{\omega_1, \ldots, \omega_n\} \cup \{\bot\}$. If the principal accepts $\omega$, then the principal's and agent's utilities are $x(\omega)$ and $y(\omega)$, respectively. Otherwise, the principal's utility is $0$ and the agent's utility is $-1$ (reflecting a penalty imposed on the agent if the principal rejects the agent's proposal). 

Through a connection to prophet inequalities, \citet{KleinbergK18} show that the principal has a simple strategy that guarantees her half of what she could have achieved by performing the search on her own (drawing $n$ samples from $F$, and choosing the best option according to $y(\cdot)$); namely, an expected value of
$$
\frac{1}{2}\mathbb{E}[\max_{\omega \in \{\omega_1, \ldots, \omega_n, \bot\}} y(\omega)].
$$
This is achieved by accepting only outcomes within an {\em eligible} set of choices $R \subseteq \Omega_+$, where $R$ takes 
one of the two forms: $R = (0,\infty)$ or $R= [\theta,\infty)$. Remarkably, the choice of $R$ (whether it's of the former or the latter form, and which value $\theta$ should take in the latter case) does not depend on the agent's preferences $y(\cdot)$.

\subsection{Information Design}
\label{sub:information}

In information design, an informed party strategically reveals information about the state of the world to a decision maker, in order to incentivize the latter to make favorable choices. The model includes a hidden state of nature (as in scoring rules), and the informed party commits to a \emph{signaling scheme}---mapping every possible state of the world to a distribution over signals. The realized state is then revealed to the informed party, who draws a signal according to the signaling scheme and sends it to the decision maker. The decision maker chooses his best action in response to the signal.
If the setting is Bayesian, the state of nature is drawn from a known prior distribution, and the decision maker performs a Bayesian update of his belief about the state upon receiving the signal and before choosing his action.  

Perhaps the most natural application of information design to a contract setting relates to the matrix of distributions, which determines how the agent's actions translate into observable outcomes that signal his actions. 
Two recent works explore this application.

\cite{CastiglioniC25} study a contract setting, in which the precise nature of the task is better known to the principal than to the agent. They model this by assuming that the outcome of the agent's actions depends on a hidden state of nature, and that this state is revealed only to the principal.  
The principal simultaneously signals information about this state to the agent, while committing to contractual payments. The goal is to design the combination of signaling scheme and contract. 

In more detail, the model of \cite{CastiglioniC25} is a principal-agent setting in which the costs $\costs$ and rewards $\rvec$ are known. There is a state of nature $\theta$ drawn from a known prior $\mu$, which determines the mapping $\Prob^\theta$ from agent actions to task outcomes. The principal commits to a signaling scheme $\pi$ and a contract $\con$, knowing only the distribution $\mu$ over states of nature. The principal then observes the realized state $\theta$, and sends a signal $s$ to the agent according the signaling scheme. The agent chooses his action based on the signal $s$, signaling scheme $\pi$ and contract $\con$. 

\cite{CastiglioniC25} study several classes of contracts: They allow $\con$ to depend on both the state $\theta$ and signal $s$, only on the signal $s$, or on neither. They also consider general or linear contracts.
They show that if the contract is allowed to depend on the state, then the joint design problem does not necessarily have an optimal signaling scheme and contract pair, but the optimum can be approached within an arbitrarily small approximation factor in polynomial time. If the contract is not allowed to depend on the state, finding the optimal contract or menu of contracts turns out to be APX-hard. On the other hand, strong positive results exist for finding the optimal \emph{linear} contract and corresponding signaling scheme---an FPTAS is established. 

\cite*{BabichenkoTXZ23} combine contract design with information design (non-Bayesian). 
In the classic contract model, the outcomes have a dual role, simultaneously specifying the principal’s rewards and providing the principal with information about the agent’s action. In other words, the ``production'' technology mapping actions to outcomes also serves as a ``monitoring'' technology of the principal over the agent. This dual role makes it difficult to study the power of different monitoring technologies.

\cite{BabichenkoTXZ23} introduce a version of the principal-agent problem in which the information that the principal obtains about the agent's action is specified by an \emph{information structure}, designed by a third party (e.g., an online platform). The information structure $\Prob$ is a mapping from every agent's action to a distribution over signals. The signals have nothing to do with the principal's reward; the agent's choice of action $i$ immediately rewards the principal $R_i$. In addition, the principal receives a signal $j$ drawn from $\Prob_i$, which she can use to determine the agent's contractual payment. 

\cite{BabichenkoTXZ23} follow \cite{BergemannBM15} in studying which utility profiles for the principal and agent are \emph{implementable} through design of an appropriate information structure (\cite{BergemannBM15} study this question in the setting of monopoly pricing rather than contracting).
Here, implementability means that there exists an information structure $\Prob$ and a contract $\con$ optimal for the principal such that assuming the agent best-responds to the contract, the expected utilities of the principal and agent match the given profile. The paper provides a characterization of implementable principal-agent expected utility profiles. In particular, it turns out that a set of simple inequality conditions that are trivially necessary for implementability is also sufficient.

\section{Discussion and Future Work}
\label{sec:discussion}
This survey highlights the significance of algorithmic, learning, and general computational approaches for addressing the challenges posed by contract design in complex environments. 
By providing an overview of the current state of research in this field, we aim to inspire and inform future research directions. We believe that current research has only scratched the surface of a comprehensive algorithmic theory of contracts, and we envision a much richer area developing over the next decade.
The corresponding sections of this survey already discuss many open problems and directions for future work.

There are also additional directions not covered in previous sections. 
An important such direction is \emph{dynamic contracting} \cite[e.g.][]{milgrom-holmstrom87}, where the contractual relationship has a temporal component---with the interaction between principal and agent evolving over time. 
Such contracting problems are naturally combinatorial.  
For example, \cite{ezra2024sequentialcontracts} consider a problem where the agent takes costly actions over time, and can decide on the order of actions.
It is also natural to study dynamic contracting problems from a learning perspective. 
For example, \cite{GuruganeshKS24} consider a repeated interaction between a principal and an agent, where the agent is a no-regret learner, and study how the fact that the agent is a no-regret learner impacts the
welfare and the way it is split between the parties through an optimal contract.
Motivated by emergent marketplaces for delegating tasks to AI agents, recent work of
\cite{Bollini24,IvanovDP+24,WuCWWX24} considers situations where a principal incentivizes an agent to make sequential decisions in a Markov Decision Process (MDP).

Another important direction in economics considers \emph{contracts with inspections} \citep[e.g.][]{Dye1986,GeorgiadisS20,HalacEtAl24}, where the principal can acquire some additional information about the agent's action at extra cost.
Recent work by \citep{BallK23,EzraLR24,FallahJ24} explores contracts with inspections from a computational perspective. 
For example, \citep{EzraLR24} considers a model where the principal can inspect different sets of actions, with a cost function that assigns an inspection cost for every set, and the problem is to find the optimal inspection scheme.

More generally, we view algorithmic contract design as part of a broader theme of ``optimizing the effort of others". 
The primary objective is to design an incentive scheme, monetary or otherwise, that motivates agents to engage in desired behavior.
This wider research theme is a natural frontier for computer science research;
and we believe 
the computational perspective will play a
vital role in shaping a broad variety of applications that involve strategic effort---both those we are already aware of and those yet to emerge.


\bibliographystyle{plainnat}
\bibliography{survey-bib}

\begin{thebibliography}{209}
\providecommand{\natexlab}[1]{#1}
\providecommand{\url}[1]{\texttt{#1}}
\expandafter\ifx\csname urlstyle\endcsname\relax
  \providecommand{\doi}[1]{doi: #1}\else
  \providecommand{\doi}{doi: \begingroup \urlstyle{rm}\Url}\fi

\bibitem[Aghion and Holden(2011)]{AghionHolden11}
Phillipe Aghion and Richard Holden.
\newblock Incomplete contracts and the theory of the firm: {W}hat have we
  learned over the past 25 years?
\newblock \emph{J. Econ. Perspect.}, 25\penalty0 (2):\penalty0 181–197, 2011.

\bibitem[Alon et~al.(1995)Alon, Feige, Wigderson, and Zuckerman]{AlonFWZ95}
Noga Alon, Uriel Feige, Avi Wigderson, and David Zuckerman.
\newblock Derandomized graph products.
\newblock \emph{Comput. Complex.}, 5\penalty0 (1):\penalty0 60--75, 1995.

\bibitem[Alon et~al.(2020)Alon, Dobson, Procaccia, Talgam{-}Cohen, and
  Tucker{-}Foltz]{AlonDPTT20}
Tal Alon, Magdalen Dobson, Ariel~D. Procaccia, Inbal Talgam{-}Cohen, and Jamie
  Tucker{-}Foltz.
\newblock Multiagent evaluation mechanisms.
\newblock In \emph{Proc.~of AAAI 2020}, pages 1774--1781, 2020.

\bibitem[Alon et~al.(2021)Alon, D\"utting, and Talgam{-}Cohen]{AlonDT21}
Tal Alon, Paul D\"utting, and Inbal Talgam{-}Cohen.
\newblock Contracts with private cost per unit-of-effort.
\newblock In \emph{Proc.~of EC 2021}, pages 52--69, 2021.

\bibitem[Alon et~al.(2023)Alon, D\"utting, Li, and Talgam-Cohen]{AlonDLT23}
Tal Alon, Paul D\"utting, Yingkai Li, and Inbal Talgam-Cohen.
\newblock Bayesian analysis of linear contracts.
\newblock In \emph{Proc.~of EC 2023}, 2023.
\newblock Full version available at: \url{https://arxiv.org/abs/2211.06850}.

\bibitem[Alon et~al.(2024)Alon, Lavi, Shamash, and Talgam{-}Cohen]{AlonLST23}
Tal Alon, Ron Lavi, Elisheva~S. Shamash, and Inbal Talgam{-}Cohen.
\newblock Technical note--incomplete information {VCG} contracts for common
  agency.
\newblock \emph{Oper. Res.}, 72\penalty0 (1):\penalty0 288--299, 2024.
\newblock An extended abstract appeared in EC 2021.

\bibitem[Ananthakrishnan et~al.(2024)Ananthakrishnan, Bates, Jordan, and
  Haghtalab]{Ananthakrishnan24}
Nivasini Ananthakrishnan, Stephen Bates, Michael~I. Jordan, and Nika Haghtalab.
\newblock Delegating data collection in decentralized machine learning.
\newblock In \emph{Proc.~of AISTATS 2024}, pages 478--486, 2024.

\bibitem[Antic and Georgiadis(2023)]{AnticG23}
Nemanja Antic and George Georgiadis.
\newblock Robust contracts: {A} revealed preference approach.
\newblock In \emph{Proc.~of EC 2023}, page 112, 2023.

\bibitem[Armstrong and Vickers(2010)]{ArmstrongV10}
Mark Armstrong and John Vickers.
\newblock A model of delegated project choice.
\newblock \emph{Econometrica}, 78\penalty0 (1):\penalty0 213--244, 2010.

\bibitem[Aruguete et~al.(2019)Aruguete, Huynh, Browne, Jurs, Flint, and
  McCutcheon]{ArugueteHB+19}
Mara~S. Aruguete, Ho~Huynh, Blaine~L. Browne, Bethany Jurs, Emilia Flint, and
  Lynn~E. McCutcheon.
\newblock How serious is the `carelessness' problem on {Mechanical Turk}?
\newblock \emph{Int. J. Soc. Res. Methodol.}, 22\penalty0 (5):\penalty0
  441--449, 2019.

\bibitem[Azar and Micali(2018)]{AzarM18}
Pablo~D. Azar and Silvio Micali.
\newblock Computational principal-agent problems.
\newblock \emph{Theor. Econ.}, 13:\penalty0 553--578, 2018.

\bibitem[Azar et~al.(2013)Azar, Daskalakis, Micali, and Weinberg]{AzarDM13}
Pablo~D. Azar, Constantinos Daskalakis, Silvio Micali, and S.~Matthew Weinberg.
\newblock Optimal and efficient parametric auctions.
\newblock In \emph{Proc.~of SODA 2013}, pages 596--604, 2013.

\bibitem[Babaioff et~al.(2006)Babaioff, Feldman, and Nisan]{BabaioffFN06}
Moshe Babaioff, Michal Feldman, and Noam Nisan.
\newblock Combinatorial agency.
\newblock In \emph{Proc.~of EC 2006}, pages 18--28, 2006.

\bibitem[Babaioff et~al.(2009)Babaioff, Feldman, and Nisan]{BabaioffFN09}
Moshe Babaioff, Michal Feldman, and Noam Nisan.
\newblock Free-riding and free-labor in combinatorial agency.
\newblock In \emph{Proc.~of SAGT 2009}, pages 109--121, 2009.

\bibitem[Babaioff et~al.(2010)Babaioff, Feldman, and Nisan]{BabaioffFN10}
Moshe Babaioff, Michal Feldman, and Noam Nisan.
\newblock Mixed strategies in combinatorial agency.
\newblock \emph{J. Artif. Intell. Res.}, 38:\penalty0 339--369, 2010.
\newblock An extended abstract appeared in WINE 2006.

\bibitem[Babaioff et~al.(2012)Babaioff, Feldman, Nisan, and
  Winter]{BabaioffFNW12}
Moshe Babaioff, Michal Feldman, Noam Nisan, and Eyal Winter.
\newblock Combinatorial agency.
\newblock \emph{J. Econ. Theory}, 147\penalty0 (3):\penalty0 999--1034, 2012.

\bibitem[Babaioff et~al.(2020)Babaioff, Feldman, Gonczarowski, Lucier, and
  Talgam{-}Cohen]{BabaioffFGLT20}
Moshe Babaioff, Michal Feldman, Yannai~A. Gonczarowski, Brendan Lucier, and
  Inbal Talgam{-}Cohen.
\newblock Escaping cannibalization? {C}orrelation-robust pricing for a
  unit-demand buyer.
\newblock In \emph{Proc.~of EC 2020}, page 191, 2020.

\bibitem[Babichenko et~al.(2024)Babichenko, Talgam{-}Cohen, Xu, and
  Zabarnyi]{BabichenkoTXZ23}
Yakov Babichenko, Inbal Talgam{-}Cohen, Haifeng Xu, and Konstantin Zabarnyi.
\newblock Information design in the principal-agent problem.
\newblock In \emph{Proc.~of EC 2024}, 2024.

\bibitem[Bacchiocchi et~al.(2024)Bacchiocchi, Castiglioni, Marchesi, and
  Gatti]{BacchiocchiC0024}
Francesco Bacchiocchi, Matteo Castiglioni, Alberto Marchesi, and Nicola Gatti.
\newblock Learning optimal contracts: How to exploit small action spaces.
\newblock In \emph{Proc.~of ICLR 2024}, 2024.

\bibitem[Bachrach and Talgam{-}Cohen(2022)]{BachrachT22}
Nir Bachrach and Inbal Talgam{-}Cohen.
\newblock Distributional robustness: {F}rom pricing to auctions.
\newblock In \emph{Proc.~of EC 2022}, page 150, 2022.

\bibitem[Balamceda et~al.(2016)Balamceda, Balseiro, Correa, and
  Stier-Moses]{BalamcedaEtAl16}
Felipe Balamceda, Santiago Balseiro, Jos\'e Correa, and Nicolas~E. Stier-Moses.
\newblock Bounds on the welfare loss from moral hazard with limited liability.
\newblock \emph{Games Econ. Behav.}, 95:\penalty0 137--155, 2016.

\bibitem[Ball and Knoepfle(2023)]{BallK23}
Ian Ball and Jan Knoepfle.
\newblock Should the timing of inspections be predictable?
\newblock In \emph{Proc.~of EC 2023}, page 206, 2023.

\bibitem[Bandi and Bertsimas(2014)]{BandiB14}
Chaithanya Bandi and Dimitris Bertsimas.
\newblock Optimal design for multi-item auctions: A robust optimization
  approach.
\newblock \emph{Math. Oper. Res.}, 39:\penalty0 1012--1038, 2014.

\bibitem[Bastani et~al.(2017)Bastani, Bayati, Braverman, Gummadi, and
  Johari]{BastaniEtAl16}
Hamsa Bastani, Mohsen Bayati, Mark Braverman, Ramki Gummadi, and Ramesh Johari.
\newblock Analysis of medicare pay-for-performance contracts.
\newblock Available at
  \url{https://papers.ssrn.com/sol3/papers.cfm?abstract_id=2839143}, 2017.

\bibitem[Bastani et~al.(2019)Bastani, Goh, and Bayati]{BastaniGB19}
Hamsa Bastani, Joel Goh, and Mohsen Bayati.
\newblock Evidence of upcoding in pay-for-performance programs.
\newblock \emph{Manag. Sci.}, 65\penalty0 (3):\penalty0 1042--1060, 2019.

\bibitem[Bechtel and Dughmi(2021)]{BechtelD21}
Curtis Bechtel and Shaddin Dughmi.
\newblock Delegated stochastic probing.
\newblock In \emph{Proc.~of ITCS 2021}, pages 37:1--37:19, 2021.

\bibitem[Bechtel et~al.(2022)Bechtel, Dughmi, and Patel]{BechtelDP22}
Curtis Bechtel, Shaddin Dughmi, and Neel Patel.
\newblock Delegated {P}andora's {B}ox.
\newblock In \emph{Proc.~of EC 2022}, pages 666--693, 2022.

\bibitem[Bergemann et~al.(2015)Bergemann, Brooks, and Morris]{BergemannBM15}
Dirk Bergemann, Benjamin Brooks, and Stephen Morris.
\newblock The limits of price discrimination.
\newblock \emph{Am. Econ. Rev.}, 105\penalty0 (3):\penalty0 921--957, 2015.

\bibitem[Bernasconi et~al.(2024)Bernasconi, Castiglioni, and
  Celli]{BernasconiEtAl24}
Martino Bernasconi, Matteo Castiglioni, and Andrea Celli.
\newblock Agent-designed contracts: How to sell hidden actions.
\newblock In \emph{Proc.~of EC 2024}, 2024.
\newblock Forthcoming.

\bibitem[Bernheim and Whinston(1986)]{BernheimW86a}
B.~Douglas Bernheim and Michael Whinston.
\newblock Common agency.
\newblock \emph{Econometrica}, 54\penalty0 (4):\penalty0 923--42, 1986.

\bibitem[Bernheim and Whinston(1998)]{BernheimW98}
B.~Douglas Bernheim and Michael Whinston.
\newblock Incomplete contracts and strategic ambiguity.
\newblock \emph{Am. Econ. Rev.}, 88\penalty0 (4):\penalty0 902--32, 1998.

\bibitem[Bertelsen(2005)]{Bertelsen05}
Alejandro Bertelsen.
\newblock Substitutes valuations and m-concavity.
\newblock Master's thesis, The Hebrew University, 2005.

\bibitem[Blumrosen and Nisan(2006)]{BlumrosenN06}
Liad Blumrosen and Noam Nisan.
\newblock Combinatorial auctions.
\newblock In Noam Nisan, Tim Roughgarden, \'Eva Tardos, and Vijay~V. Vazirani,
  editors, \emph{Algorithmic Game Theory}, chapter~11. Cambridge University
  Press, 2006.

\bibitem[Blumrosen and Nisan(2009)]{BlumrosenN09}
Liad Blumrosen and Noam Nisan.
\newblock On the computational power of demand queries.
\newblock \emph{{SIAM} J. Comput.}, 39\penalty0 (4):\penalty0 1372--1391, 2009.

\bibitem[Bollini et~al.(2024)Bollini, Bacchiocchi, Castiglioni, Marchesi, and
  Gatti]{Bollini24}
Matteo Bollini, Francesco Bacchiocchi, Matteo Castiglioni, Alberto Marchesi,
  and Nicola Gatti.
\newblock Contracting with a reinforcement learning agent by playing trick or
  treat.
\newblock Available at \url{https://arxiv.org/abs/2410.13520}, 2024.

\bibitem[B\"orner et~al.(2017)B\"orner, Baylis, Corbera, de~Blas,
  Honey-Ros\'es, Persson, and Wunderg]{BornerBC+17}
Jan B\"orner, Kathy Baylis, Esteve Corbera, Driss~Ezzine de~Blas, Jordi
  Honey-Ros\'es, U.~Martin Persson, and Sven Wunderg.
\newblock The effectiveness of payments for environmental services.
\newblock \emph{World Dev.}, 96:\penalty0 359--374, 2017.

\bibitem[Braun et~al.(2023)Braun, Hahn, Hoefer, and Schecker]{BraunHHS23}
Pirmin Braun, Niklas Hahn, Martin Hoefer, and Conrad Schecker.
\newblock Delegated online search.
\newblock In \emph{Proc.~of IJCAI 2023}, pages 2528--2536, 2023.

\bibitem[Briest et~al.(2015)Briest, Chawla, Kleinberg, and
  Weinberg]{BriestEtAl2015}
Patrick Briest, Chuchi Chawla, Robert Kleinberg, and S.~Matthew Weinberg.
\newblock Pricing lotteries.
\newblock \emph{J. Econ. Theory}, 156:\penalty0 144--174, 2015.
\newblock An extended abstract appeared in SODA 2010.

\bibitem[Br{\"{u}}ckner and Scheffer(2011)]{BrucknerS11}
Michael Br{\"{u}}ckner and Tobias Scheffer.
\newblock Stackelberg games for adversarial prediction problems.
\newblock In \emph{Proc.~of KDD 2011}, pages 547--555, 2011.

\bibitem[Cacciamani et~al.(2024)Cacciamani, Bernasconi, Castiglioni, and
  Gatti]{CacciamaniEtAl24}
Federico Cacciamani, Martino Bernasconi, Matteo Castiglioni, and Nicola Gatti.
\newblock Multi-agent contract design beyond binary actions.
\newblock In \emph{Proc.~of EC 2024}, 2024.
\newblock Forthcoming.

\bibitem[Cai et~al.(2015)Cai, Daskalakis, and Papadimitriou]{CaiDP15}
Yang Cai, Constantinos Daskalakis, and Christos~H. Papadimitriou.
\newblock Optimum statistical estimation with strategic data sources.
\newblock In \emph{Proc.~of COLT 2015}, pages 280--296, 2015.

\bibitem[C{\u{a}}linescu et~al.(2011)C{\u{a}}linescu, Chekuri, P{\'{a}}l, and
  Vondr{\'{a}}k]{CalinescuCPV11}
Gruia C{\u{a}}linescu, Chandra Chekuri, Martin P{\'{a}}l, and Jan
  Vondr{\'{a}}k.
\newblock Maximizing a monotone submodular function subject to a matroid
  constraint.
\newblock \emph{{SIAM} J. Comput.}, 40\penalty0 (6):\penalty0 1740--1766, 2011.

\bibitem[Calvano et~al.(2020)Calvano, Calzolari, Denicolo, and
  Pastorello]{CalvanoCDP20}
Emilio Calvano, Giacomo Calzolari, Vincenzo Denicolo, and Sergio Pastorello.
\newblock Artificial intelligence, algorithmic pricing, and collusion.
\newblock \emph{Am. Econ. Rev.}, 110\penalty0 (10):\penalty0 3267--3297, 2020.

\bibitem[Carrasco et~al.(2017)Carrasco, Luz, Kos, Messner, Monteiro, and
  Moreira]{CarrascoEtAl17}
Vinicius Carrasco, Vitor~F. Luz, Nenad Kos, Matthias Messner, Paulo Monteiro,
  and Humberto Moreira.
\newblock Optimal selling mechanisms under moment conditions.
\newblock \emph{J. Econ. Theory}, 177:\penalty0 245--279, 2017.

\bibitem[Carroll(2015)]{Carroll15}
Gabriel Carroll.
\newblock Robustness and linear contracts.
\newblock \emph{Am. Econ. Rev.}, 105\penalty0 (2):\penalty0 536--63, 2015.

\bibitem[Castiglioni and Chen(2025)]{CastiglioniC25}
Matteo Castiglioni and Junjie Chen.
\newblock Hiring for an uncertain task: Joint design of information and
  contracts.
\newblock In \emph{Proc.~of SODA 2025}, 2025.
\newblock To appear.

\bibitem[Castiglioni et~al.(2021)Castiglioni, Marchesi, and
  Gatti]{CastiglioniM021}
Matteo Castiglioni, Alberto Marchesi, and Nicola Gatti.
\newblock Bayesian agency: {L}inear versus tractable contracts.
\newblock In \emph{Proc.~EC 2021}, pages 285--286, 2021.

\bibitem[Castiglioni et~al.(2023{\natexlab{a}})Castiglioni, Marchesi, and
  Gatti]{CastiglioniEtAl23}
Matteo Castiglioni, Alberto Marchesi, and Nicola Gatti.
\newblock Multi-agent contract design: How to commission multiple agents with
  individual outcome.
\newblock In \emph{Proc.~of EC 2023}, pages 412--448, 2023{\natexlab{a}}.

\bibitem[Castiglioni et~al.(2023{\natexlab{b}})Castiglioni, Marchesi, and
  Gatti]{CastiglioniM022}
Matteo Castiglioni, Alberto Marchesi, and Nicola Gatti.
\newblock Designing menus of contracts efficiently: The power of randomization.
\newblock \emph{Artif. Intell.}, 318:\penalty0 103881, 2023{\natexlab{b}}.
\newblock An extended abstract appeared in EC 2022.

\bibitem[Castiglioni et~al.(2025)Castiglioni, Chen, Li, Xu, and
  Zuo]{CastiglioniCL+25}
Matteo Castiglioni, Junjie Chen, Minming Li, Haifeng Xu, and Song Zuo.
\newblock A reduction from multi-parameter to single-parameter {B}ayesian
  contract design.
\newblock In \emph{Proc.~of SODA 2025}, 2025.
\newblock Forthcoming.

\bibitem[Castro-Pires et~al.(2024)Castro-Pires, Chade, and Swinkels]{ChadeS21}
Henrique Castro-Pires, Hector Chade, and Jeroen Swinkels.
\newblock Disentangling moral hazard and adverse selection.
\newblock \emph{Am. Econ. Rev.}, 114:\penalty0 1--37, 2024.

\bibitem[Chen and Yu(2021)]{ChenYu21}
Yiling Chen and Fang{-}Yi Yu.
\newblock Optimal scoring rule design.
\newblock Available at \url{https://arxiv.org/abs/2107.07420}, 2021.

\bibitem[Chen et~al.(2024)Chen, Chen, Deng, and Huang]{ChenCDH24}
Yurong Chen, Zhaohua Chen, Xiaotie Deng, and Zhiyi Huang.
\newblock Are bounded contracts learnable and approximately optimal?
\newblock In \emph{Proc.~of {EC} 2024}, 2024.

\bibitem[Clarke(1971)]{Clarke71}
Edward~H. Clarke.
\newblock Multipart pricing of public goods.
\newblock \emph{Public Choice}, 11\penalty0 (1):\penalty0 17--33, 1971.

\bibitem[Cohen et~al.(2022)Cohen, Deligkas, and Koren]{CohenDK22}
Alon Cohen, Argyrios Deligkas, and Moran Koren.
\newblock Learning approximately optimal contracts.
\newblock In \emph{Proc.~of SAGT 2022}, pages 331--346, 2022.

\bibitem[Collina et~al.(2024)Collina, Derr, and Roth]{CollinaDR24}
Natalie Collina, Rabanus Derr, and Aaron Roth.
\newblock The value of ambiguous commitments in multi-follower games.
\newblock Available at \url{https://arxiv.org/abs/2409.05608}, 2024.

\bibitem[Cong and He(2019)]{CongHe19}
Lin~W. Cong and Zhiguo He.
\newblock Blockchain disruption and smart contracts.
\newblock \emph{Rev. Financ. Stud.}, 32:\penalty0 1754–1797, 2019.

\bibitem[Conitzer and Sandholm(2003)]{ConitzerS03}
Vincent Conitzer and Tuomas Sandholm.
\newblock Automated mechanism design: complexity results stemming from the
  single-agent setting.
\newblock In \emph{Proc.~of ICEC 2003}, pages 17--24, 2003.

\bibitem[Dai and Toikka(2022)]{DaiT22}
Tianjiao Dai and Juuso Toikka.
\newblock Robust incentives for teams.
\newblock \emph{Econometrica}, 90:\penalty0 1583--1613, 2022.

\bibitem[Daskalakis(2015)]{Daskalakis15}
Constantinos Daskalakis.
\newblock Multi-item auctions defying intuition?
\newblock \emph{SIGecom Exch.}, 14\penalty0 (1):\penalty0 41--75, 2015.

\bibitem[DellaVigna and Pope(2017)]{DellaVignaPope17}
Stefano DellaVigna and Devin Pope.
\newblock {What Motivates Effort? Evidence and Expert Forecasts}.
\newblock \emph{Rev. Econ. Stud.}, 85\penalty0 (2):\penalty0 1029--1069, 2017.

\bibitem[Deo{-}Campo~Vuong et~al.(2024)Deo{-}Campo~Vuong, Dughmi, Patel, and
  Prasad]{VuongDPP23}
Ramiro Deo{-}Campo~Vuong, Shaddin Dughmi, Neel Patel, and Aditya Prasad.
\newblock On supermodular contracts and dense subgraphs.
\newblock In \emph{Proc.~of SODA 2024}, pages 109--132, 2024.

\bibitem[{Di Tillio} et~al.(2017){Di Tillio}, Kos, and Messner]{DiTillioEtAl17}
Alfredo {Di Tillio}, Nenad Kos, and Matthias Messner.
\newblock The design of ambiguous mechanisms.
\newblock \emph{Rev. Econ. Stud.}, 84:\penalty0 237--276, 2017.

\bibitem[Diamond(1998)]{Diamond98}
Peter Diamond.
\newblock Managerial incentives: On the near-linearity of optimal compensation.
\newblock \emph{J. Political Econ.}, 106:\penalty0 931--957, 1998.

\bibitem[Dunn(2024)]{Dunn24}
Erin Dunn.
\newblock How much do {YouTubers} make?
\newblock Available at
  \url{https://www.creditkarma.com/income/i/how-much-do-youtubers-make}, 2024.

\bibitem[D\"utting and Talgam-Cohen(2019)]{DuettingT19}
Paul D\"utting and Inbal Talgam-Cohen.
\newblock Tutorial ``{C}ontract theory: A new frontier for {AGT}''.
\newblock Available at \url{https://www.youtube.com/watch?v=JQsqPXCehOU} and
  \url{https://www.youtube.com/watch?v=k_c-aZdpeBo}, 2019.

\bibitem[D\"utting and Talgam-Cohen(2022)]{DuettingT22}
Paul D\"utting and Inbal Talgam-Cohen.
\newblock Tutorial ``{A}lgorithmic contract theory'''.
\newblock Available at \url{https://www.youtube.com/watch?v=duQTQFVU3sg}, 2022.

\bibitem[D\"utting et~al.(2019)D\"utting, Roughgarden, and
  Talgam{-}Cohen]{DuttingRT19}
Paul D\"utting, Tim Roughgarden, and Inbal Talgam{-}Cohen.
\newblock Simple versus optimal contracts.
\newblock In \emph{Proc.~of EC 2019}, pages 369--387, 2019.
\newblock Full version available at \url{https://arxiv.org/pdf/1808.03713}.

\bibitem[D\"utting et~al.(2021{\natexlab{a}})D\"utting, Ezra, Feldman, and
  Kesselheim]{DuettingEFK21}
Paul D\"utting, Tomer Ezra, Michal Feldman, and Thomas Kesselheim.
\newblock Combinatorial contracts.
\newblock In \emph{Proc.~of FOCS 2021}, pages 815--826, 2021{\natexlab{a}}.

\bibitem[D\"utting et~al.(2021{\natexlab{b}})D\"utting, Roughgarden, and
  Talgam{-}Cohen]{DuttingRT21}
Paul D\"utting, Tim Roughgarden, and Inbal Talgam{-}Cohen.
\newblock The complexity of contracts.
\newblock \emph{{SIAM} J. Comput.}, 50\penalty0 (1):\penalty0 211--254,
  2021{\natexlab{b}}.
\newblock An extended abstract appeared in SODA 2020.

\bibitem[D\"utting et~al.(2023{\natexlab{a}})D\"utting, Ezra, Feldman, and
  Kesselheim]{DuettingEFK23}
Paul D\"utting, Tomer Ezra, Michal Feldman, and Thomas Kesselheim.
\newblock Multi-agent contracts.
\newblock In \emph{Proc.~of STOC 2023}, pages 1311--1324, 2023{\natexlab{a}}.

\bibitem[D\"utting et~al.(2023{\natexlab{b}})D\"utting, Guruganesh, Schneider,
  and Wang]{DuettingGSW23}
Paul D\"utting, Guru Guruganesh, Jon Schneider, and Joshua Wang.
\newblock Optimal no-regret learning of one-sided {L}ipschitz functions.
\newblock In \emph{Proc.~of ICML 2023}, 2023{\natexlab{b}}.

\bibitem[D\"utting et~al.(2024{\natexlab{a}})D\"utting, Feldman, and
  Gal-Tzur]{DuettingFG23}
Paul D\"utting, Michal Feldman, and Yoav Gal-Tzur.
\newblock Combinatorial contracts beyond gross substitutes.
\newblock In \emph{Proc.~of SODA 2024}, pages 92--108, 2024{\natexlab{a}}.

\bibitem[D\"utting et~al.(2024{\natexlab{b}})D\"utting, Feldman, Gal-Tzur, and
  Rubinstein]{DuettingFGR24}
Paul D\"utting, Michal Feldman, Yoav Gal-Tzur, and Aviad Rubinstein.
\newblock The query complexity of contracts.
\newblock Available at \url{https://arxiv.org/abs/2403.09794},
  2024{\natexlab{b}}.

\bibitem[D\"utting et~al.(2024{\natexlab{c}})D\"utting, Feldman, Peretz, and
  Samuelson]{DuettingFP23}
Paul D\"utting, Michal Feldman, Daniel Peretz, and Larry Samuelson.
\newblock Ambiguous contracts.
\newblock \emph{Econometrica}, 92\penalty0 (6):\penalty0 1967–1992,
  2024{\natexlab{c}}.
\newblock An extended abstract appeared in EC 2023.

\bibitem[D\"utting et~al.(2024{\natexlab{d}})D\"utting, Feldman, Ponitka, and
  Soumalias]{sample-complexity}
Paul D\"utting, Michal Feldman, Tomasz Ponitka, and Ermis Soumalias.
\newblock The pseudo-dimension of contracts.
\newblock Working paper (available on request), 2024{\natexlab{d}}.

\bibitem[D{\"{u}}tting et~al.(2024)D{\"{u}}tting, Feng, Narasimhan, Parkes, and
  Ravindranath]{Dutting0NPR24}
Paul D{\"{u}}tting, Zhe Feng, Harikrishna Narasimhan, David~C. Parkes, and
  Sai~Srivatsa Ravindranath.
\newblock Optimal auctions through deep learning: Advances in differentiable
  economics.
\newblock \emph{J. {ACM}}, 71\penalty0 (1):\penalty0 5:1--5:53, 2024.
\newblock An extended abstract appeared in ICML 2019.

\bibitem[D\"utting et~al.(2025)D\"utting, Ezra, Feldman, and
  Kesselheim]{DuettingEFK24}
Paul D\"utting, Tomer Ezra, Michal Feldman, and Thomas Kesselheim.
\newblock Multi-agent combinatorial contracts.
\newblock In \emph{Proc.~of SODA 2025}, 2025.
\newblock Forthcoming.

\bibitem[Dye(1986)]{Dye1986}
Ronald~A. Dye.
\newblock Optimal monitoring policies in agencies.
\newblock \emph{RAND J. Econ.}, 17\penalty0 (3):\penalty0 339--350, 1986.

\bibitem[Eisner and Severance(1976)]{eisner1976mathematical}
Mark~J Eisner and Dennis~G Severance.
\newblock Mathematical techniques for efficient record segmentation in large
  shared databases.
\newblock \emph{J. ACM}, 23\penalty0 (4):\penalty0 619--635, 1976.

\bibitem[Emek and Feldman(2012)]{EmekF12}
Yuval Emek and Michal Feldman.
\newblock Computing optimal contracts in combinatorial agencies.
\newblock \emph{Theor. Comput. Sci.}, 452:\penalty0 56--74, 2012.
\newblock An extended abstract appeared in WINE 2009.

\bibitem[Emek et~al.(2014)Emek, Feldman, Gamzu, Leme, and
  Tennenholtz]{EmekFGLT14}
Yuval Emek, Michal Feldman, Iftah Gamzu, Renato~Paes Leme, and Moshe
  Tennenholtz.
\newblock Signaling schemes for revenue maximization.
\newblock \emph{{ACM} Trans. Econ. Comput.}, 2\penalty0 (2):\penalty0
  5:1--5:19, 2014.

\bibitem[Epstein and Peters(1999)]{epstein1999revelation}
Larry~G Epstein and Michael Peters.
\newblock A revelation principle for competing mechanisms.
\newblock \emph{J. Econ. Theory}, 88\penalty0 (1):\penalty0 119--160, 1999.

\bibitem[Ezra et~al.(2024{\natexlab{a}})Ezra, Feldman, and
  Schlesinger]{EzraFS24}
Tomer Ezra, Michal Feldman, and Maya Schlesinger.
\newblock On the (in)approximability of combinatorial contracts.
\newblock In \emph{Proc.~of ITCS 2024}, pages 44:1--44:22, 2024{\natexlab{a}}.

\bibitem[Ezra et~al.(2024{\natexlab{b}})Ezra, Feldman, and
  Schlesinger]{ezra2024sequentialcontracts}
Tomer Ezra, Michal Feldman, and Maya Schlesinger.
\newblock Sequential contracts.
\newblock Available at \url{https://arxiv.org/abs/2403.09545},
  2024{\natexlab{b}}.

\bibitem[Ezra et~al.(2024{\natexlab{c}})Ezra, Leonardi, and Russo]{EzraLR24}
Tomer Ezra, Stefano Leonardi, and Matteo Russo.
\newblock Contracts with inspections.
\newblock Available at \url{https://doi.org/10.48550/arXiv.2402.16553},
  2024{\natexlab{c}}.

\bibitem[Fallah and Jordan(2024)]{FallahJ24}
Alireza Fallah and Michael~I. Jordan.
\newblock Contract design with safety inspections.
\newblock In \emph{Proc.~of EC 2024}, 2024.

\bibitem[Fehr et~al.(2007)Fehr, Klein, and Schmidt]{FehrKS07}
Ernst Fehr, Alexander Klein, and Klaus~M Schmidt.
\newblock Fairness and contract design.
\newblock \emph{Econometrica}, 75\penalty0 (1):\penalty0 121--154, 2007.

\bibitem[Feldman and Lucier(2022)]{FeldmanL22}
Michal Feldman and Brendan Lucier.
\newblock Workshop ``{O}ptimizing the effort of others: {F}rom algorithmic
  contracts to strategic classification''.
\newblock Workshop website:
  \url{https://sites.google.com/corp/view/delegationworkshop2022}, 2022.

\bibitem[Fest et~al.(2020)Fest, Kval\o{}y, Nieken, and Sch\"ottner]{FestKNS20}
Sebastian Fest, Ola Kval\o{}y, Petra Nieken, and Anja Sch\"ottner.
\newblock Motivation and incentives in an online labor market.
\newblock Available at
  \url{https://www.econstor.eu/bitstream/10419/224586/1/vfs-2020-pid-39843.pdf},
  2020.

\bibitem[Fleischer et~al.(2006)Fleischer, Goemans, Mirrokni, and
  Sviridenko]{FleischerGMS06}
Lisa Fleischer, Michel~X. Goemans, Vahab~S. Mirrokni, and Maxim Sviridenko.
\newblock Tight approximation algorithms for maximum general assignment
  problems.
\newblock In \emph{Proc.~of SODA 2006}, pages 611--620, 2006.

\bibitem[Frankel and Kartik(2019)]{FrankelK19}
Alex Frankel and Navin Kartik.
\newblock Muddled information.
\newblock \emph{J. Political Econ.}, 127\penalty0 (4):\penalty0 1739--1776,
  2019.

\bibitem[Frazier et~al.(2014)Frazier, Kempe, Kleinberg, and
  Kleinberg]{FrazierKKK14}
Peter~I. Frazier, David Kempe, Jon~M. Kleinberg, and Robert Kleinberg.
\newblock Incentivizing exploration.
\newblock In \emph{Proc.~of EC 2014}, pages 5--22, 2014.

\bibitem[Gale and Hellwig(1985)]{GaleH85}
Douglas Gale and Martin Hellwig.
\newblock Incentive-compatible debt contracts: The one-period problem.
\newblock \emph{Rev. Econ. Stud.}, 52\penalty0 (4):\penalty0 647--663, 1985.

\bibitem[Gan et~al.(2024)Gan, Han, Wu, and Xu]{GanHWX22}
Jiarui Gan, Minbiao Han, Jibang Wu, and Haifeng Xu.
\newblock Generalized principal-agency: Contracts, information, games and
  beyond.
\newblock In \emph{Proc.~of WINE 2024}, 2024.
\newblock Forthcoming.

\bibitem[Garg and Johari(2021)]{GargJ21}
Nikhil Garg and Ramesh Johari.
\newblock Designing informative rating systems: Evidence from an online labor
  market.
\newblock \emph{Manuf. Serv. Oper. Manag.}, 23\penalty0 (3):\penalty0 589--605,
  2021.

\bibitem[Georgiadis and Szentes(2020)]{GeorgiadisS20}
George Georgiadis and Balazs Szentes.
\newblock Optimal monitoring design.
\newblock \emph{Econometrica}, 88\penalty0 (5):\penalty0 2075--2107, 2020.

\bibitem[Georgiadis et~al.(2024)Georgiadis, Ravid, and Szentes]{GeorgiadisRS24}
George Georgiadis, Doron Ravid, and Bal\'asz Szentes.
\newblock Flexible moral hazard problems.
\newblock \emph{Econometrica}, 92:\penalty0 387--409, 2024.

\bibitem[Ghalme et~al.(2021)Ghalme, Nair, Eilat, Talgam{-}Cohen, and
  Rosenfeld]{GhalmeNETR21}
Ganesh Ghalme, Vineet Nair, Itay Eilat, Inbal Talgam{-}Cohen, and Nir
  Rosenfeld.
\newblock Strategic classification in the dark.
\newblock In \emph{Proc.~of ICML 2021}, pages 3672--3681, 2021.

\bibitem[Ghosh et~al.(2007)Ghosh, Nazerzadeh, and Sundararajan]{GhoshNS07}
Arpita Ghosh, Hamid Nazerzadeh, and Mukund Sundararajan.
\newblock Computing optimal bundles for sponsored search.
\newblock In \emph{WINE 2007}, pages 576--583, 2007.

\bibitem[Gilboa and Schmeidler(1993)]{GilboaS93}
Itzhak Gilboa and David Schmeidler.
\newblock Updating ambiguous beliefs.
\newblock \emph{J. Econ. Theory}, 59\penalty0 (1):\penalty0 33--49, 1993.

\bibitem[Gneiting and Raftery(2007)]{GneitingR07}
Tilman Gneiting and Adrian~E. Raftery.
\newblock Strictly proper scoring rules, prediction, and estimation.
\newblock \emph{J. Am. Stat. Assoc.}, 102:\penalty0 359–378, 2007.

\bibitem[Gonczarowski and Weinberg(2021)]{GonczarowskiW21}
Yannai~A. Gonczarowski and S.~Matthew Weinberg.
\newblock The sample complexity of up-to-{\(\epsilon\)} multi-dimensional
  revenue maximization.
\newblock \emph{J. {ACM}}, 68\penalty0 (3):\penalty0 15:1--15:28, 2021.

\bibitem[Gottlieb and Moreira(2015)]{GottliebM15}
Daniel Gottlieb and Humberto Moreira.
\newblock Simple contracts with adverse selection and moral hazard.
\newblock \emph{Theor. Econ.}, 17:\penalty0 1357--1401, 2015.

\bibitem[Grossman and Hart(1983)]{GrossmanH83}
Sanford~J. Grossman and Oliver~D. Hart.
\newblock An analysis of the principal-agent problem.
\newblock \emph{Econometrica}, 51\penalty0 (1):\penalty0 7--45, 1983.

\bibitem[Gr\"otschel et~al.(1981)Gr\"otschel, Lov\'asz, and
  Schrijver]{Grotschel81}
M.~Gr\"otschel, L.~Lov\'asz, and A.~Schrijver.
\newblock The ellipsoid method and its consequences in combinatorial
  optimization.
\newblock \emph{Combinatorica}, 1\penalty0 (2):\penalty0 169--197, 1981.

\bibitem[Groves(1973)]{Groves73}
Theodore Groves.
\newblock Incentives in teams.
\newblock \emph{Econometrica}, 41\penalty0 (4):\penalty0 617--631, 1973.

\bibitem[Guesnerie(1981)]{Guesnerie81}
Roger Guesnerie.
\newblock \emph{On Taxation and Incentives: Further Remarks on the Limits to
  Redistribution}.
\newblock PhD thesis, University of Bonn, 1981.

\bibitem[Gul and Stacchetti(1999)]{gul1999walrasian}
Faruk Gul and Ennio Stacchetti.
\newblock Walrasian equilibrium with gross substitutes.
\newblock \emph{J. Econ. Theory}, 87\penalty0 (1):\penalty0 95--124, 1999.

\bibitem[Guruganesh et~al.(2021)Guruganesh, Schneider, and
  Wang]{GuruganeshSW21}
Guru Guruganesh, Jon Schneider, and Joshua Wang.
\newblock Contracts under moral hazard and adverse selection.
\newblock In \emph{Proc.~of EC 2021}, pages 563--582, 2021.

\bibitem[Guruganesh et~al.(2023)Guruganesh, Schneider, Wang, and
  Zhao]{GuruganeshSW23}
Guru Guruganesh, Jon Schneider, Joshua Wang, and Junyao Zhao.
\newblock The power of menus in contract design.
\newblock In \emph{Proc.~of EC 2023}, pages 818--848, 2023.

\bibitem[Guruganesh et~al.(2024)Guruganesh, Kolumbus, Schneider, Talgam-Cohen,
  Vlatakis-Gkaragkounis, Wang, and Weinberg]{GuruganeshKS24}
Guru Guruganesh, Yoav Kolumbus, Jon Schneider, Inbal Talgam-Cohen,
  Emmanouil-Vasileios Vlatakis-Gkaragkounis, Joshua~R. Wang, and S.~Matthew
  Weinberg.
\newblock Contracting with a learning agent.
\newblock In \emph{Proc.~of NeurIPS 2024}, 2024.

\bibitem[Gusfield(1980)]{gusfield1980sensitivity}
Daniel~Mier Gusfield.
\newblock \emph{Sensitivity analysis for combinatorial optimization}.
\newblock University of California, Berkeley, 1980.

\bibitem[Hadfield{-}Menell and Hadfield(2019)]{Hadfield-Menell19a}
Dylan Hadfield{-}Menell and Gillian~K. Hadfield.
\newblock Incomplete contracting and {AI} alignment.
\newblock In \emph{Proc.~of {AIES} 2019}, pages 417--422, 2019.

\bibitem[Haghtalab et~al.(2020)Haghtalab, Immorlica, Lucier, and
  Wang]{HaghtalabILW20}
Nika Haghtalab, Nicole Immorlica, Brendan Lucier, and Jack~Z. Wang.
\newblock Maximizing welfare with incentive-aware evaluation mechanisms.
\newblock In \emph{Proc.~of IJCAI 2020}, pages 160--166, 2020.

\bibitem[Halac et~al.(2024)Halac, Kremer, and Winter]{HalacEtAl24}
Marina Halac, Ilan Kremer, and Eyal Winter.
\newblock Monitoring teams.
\newblock \emph{AEJ: Microeconomics}, 16\penalty0 (3):\penalty0 134–61,
  August 2024.

\bibitem[Hammond(1979)]{Hammond79}
Peter~J. Hammond.
\newblock Straightforward individual incentive compatibility in large
  economies.
\newblock \emph{Rev. Econ. Stud.}, 46:\penalty0 263–282, 1979.

\bibitem[Hardt et~al.(2016)Hardt, Megiddo, Papadimitriou, and
  Wootters]{HardtMPW16}
Moritz Hardt, Nimrod Megiddo, Christos~H. Papadimitriou, and Mary Wootters.
\newblock Strategic classification.
\newblock In \emph{Proc.~of ITCS 2016}, pages 111--122, 2016.

\bibitem[Hart(1988)]{Hart88}
Oliver Hart.
\newblock Incomplete contracts and the theory of the firm.
\newblock \emph{J. Law Econ. Organ.}, 4\penalty0 (1), 1988.

\bibitem[Hart and Moore(1988)]{HartMoore88}
Oliver Hart and John Moore.
\newblock Incomplete contracts and renegotiation.
\newblock \emph{Econometrica}, 56:\penalty0 755--785, 1988.

\bibitem[Hartline and Roughgarden(2009)]{HartlineR09}
Jason~D. Hartline and Tim Roughgarden.
\newblock Simple versus optimal mechanisms.
\newblock In \emph{Proc.~of EC 2009}, pages 225--234, 2009.

\bibitem[Hartline et~al.(2023)Hartline, Shan, Li, and Wu]{HartlineSLW22}
Jason~D. Hartline, Liren Shan, Yingkai Li, and Yifan Wu.
\newblock Optimal scoring rules for multi-dimensional effort.
\newblock In \emph{Proc.~of COLT 2023}, pages 2624--2650, 2023.

\bibitem[H{\aa}stad(2001)]{Hastad01}
J.~H{\aa}stad.
\newblock Some optimal inapproximability results.
\newblock \emph{J. ACM}, 48:\penalty0 798--859, 2001.

\bibitem[Haupt et~al.(2024)Haupt, Christoffersen, Damani, and
  Hadfield{-}Menell]{HauptCDH24}
Andreas~A. Haupt, Phillip J.~K. Christoffersen, Mehul Damani, and Dylan
  Hadfield{-}Menell.
\newblock Formal contracts mitigate social dilemmas in multi-agent
  reinforcement learning.
\newblock In \emph{Proc.~of AAMAS}, volume~38, page~51, 2024.

\bibitem[H\'ebert(2017)]{Hebert17}
Benjamin H\'ebert.
\newblock Moral hazard and the optimality of debt.
\newblock \emph{Rev. Econ. Stud.}, 85\penalty0 (4):\penalty0 55--73, 2017.

\bibitem[Hermalin and Katz(1991)]{HermalinK91}
Benjamin~E. Hermalin and Michael~L. Katz.
\newblock Moral hazard and verifiability: {T}he effects of renegotiation in
  agency.
\newblock \emph{Econometrica}, 59:\penalty0 1735–1753, 1991.

\bibitem[Ho et~al.(2016)Ho, Slivkins, and Vaughan]{HoSV16}
Chien{-}Ju Ho, Aleksandrs Slivkins, and Jennifer~Wortman Vaughan.
\newblock Adaptive contract design for crowdsourcing markets: Bandit algorithms
  for repeated principal-agent problems.
\newblock \emph{J. Artif. Intell. Res.}, 55:\penalty0 317--359, 2016.
\newblock An extended abstract appeared in EC 2014.

\bibitem[Holmstr\"om(1984)]{Holmstrom84}
B.~Holmstr\"om.
\newblock On the theory of delegation.
\newblock In M.~Boyer and R.~Kihlstrom, editors, \emph{Bayesian Models in
  Economic Theory}. Elsevier, 1984.

\bibitem[Holmstr\"om(1979)]{Holmstrom79}
Bengt Holmstr\"om.
\newblock Moral hazard and observability.
\newblock \emph{Bell J. Econ.}, 10:\penalty0 74--91, 1979.

\bibitem[Holmstr\"om(1982)]{Holmstrom82}
Bengt Holmstr\"om.
\newblock Moral hazard in teams.
\newblock \emph{Bell J. Econ.}, 13:\penalty0 324–340, 1982.

\bibitem[Holmstr\"om and Milgrom(1987)]{milgrom-holmstrom87}
Bengt Holmstr\"om and Paul Milgrom.
\newblock Aggregation and linearity in the provision of intertemporal
  incentives.
\newblock \emph{Econometrica}, 55\penalty0 (2):\penalty0 303--328, 1987.

\bibitem[Holmstr\"om and Milgrom(1991)]{HolmstromMilgrom91}
Bengt Holmstr\"om and Paul Milgrom.
\newblock Multitask principal-agent analyses: Incentive contracts, asset
  ownership, and job design.
\newblock \emph{J. Law Econ. Organ.}, 7:\penalty0 24–52, 1991.

\bibitem[Innes(1990)]{Innes90}
Robert~D. Innes.
\newblock Limited liability and incentive contracting with ex-ante action
  choices.
\newblock \emph{J. Econ. Theory}, 52\penalty0 (1):\penalty0 45--67, 1990.

\bibitem[Ivanov et~al.(2024)Ivanov, D\"utting, Parkes, Talgam-Cohen, and
  Wang]{IvanovDP+24}
Dimitry Ivanov, Paul D\"utting, David Parkes, Inbal Talgam-Cohen, and Tonghan
  Wang.
\newblock Principal-agent reinforcement learning: {O}rchestrating {AI} agents
  with contracts.
\newblock Available at \url{https://arxiv.org/abs/2407.18074}, 2024.

\bibitem[Iwata et~al.(2009)Iwata, Fleischer, and Fujishige]{IwataFF01}
Satoru Iwata, Lisa Fleischer, and Satoru Fujishige.
\newblock combinatorial strongly polynomial algorithm for minimizing submodular
  functions.
\newblock \emph{J. ACM}, 48\penalty0 (4):\penalty0 761--777, 2009.

\bibitem[Kambhampati(2023)]{Kambhampati23}
Ashwin Kambhampati.
\newblock Randomization is optimal in the robust principal-agent problem.
\newblock \emph{J. Econ. Theory}, 207:\penalty0 105585, 2023.

\bibitem[Kaynar and Siddiq(2023)]{KaynarS22}
Nur Kaynar and Auyon Siddiq.
\newblock Estimating effects of incentive contracts in online labor platforms.
\newblock \emph{Manag. Sci.}, 69\penalty0 (4), 2023.

\bibitem[Kelso and Crawford(1982)]{kelso1982job}
Alexander~S Kelso and Vincent~P Crawford.
\newblock Job matching, coalition formation, and gross substitutes.
\newblock \emph{Econometrica}, pages 1483--1504, 1982.

\bibitem[Khodabakhsh et~al.(2024)Khodabakhsh, Pountourakis, and
  Taggart]{TaggartEtAl2022}
Ali Khodabakhsh, Emmanouil Pountourakis, and Samuel Taggart.
\newblock Simple delegated choice.
\newblock In \emph{Proc.~of SODA 2024}, pages 569--590, 2024.

\bibitem[Kleinberg and Raghavan(2019)]{KleinbergR19}
Jon Kleinberg and Manish Raghavan.
\newblock How do classifiers induce agents to invest effort strategically?
\newblock In \emph{Proc.~of EC 2019}, pages 825--844, 2019.

\bibitem[Kleinberg and Kleinberg(2018)]{KleinbergK18}
Jon~M. Kleinberg and Robert Kleinberg.
\newblock Delegated search approximates efficient search.
\newblock In \emph{Proc.~of EC 2018}, pages 287--302, 2018.

\bibitem[Kleinberg and Raghavan(2020)]{KleinbergR20}
Jon~M. Kleinberg and Manish Raghavan.
\newblock Algorithmic classification and strategic effort.
\newblock \emph{SIGecom Exch.}, 18\penalty0 (2):\penalty0 45--52, 2020.

\bibitem[Kleinberg and Leighton(2003)]{KleinbergL03}
Robert~D. Kleinberg and Frank~Thomson Leighton.
\newblock The value of knowing a demand curve: Bounds on regret for online
  posted-price auctions.
\newblock In \emph{Proc.~of FOCS 2003}, pages 594--605, 2003.

\bibitem[Koutsoupias and Papadimitriou(2009)]{KoutsoupiasP09}
Elias Koutsoupias and Christos~H. Papadimitriou.
\newblock Worst-case equilibria.
\newblock \emph{Comput. Sci. Rev.}, 3\penalty0 (2):\penalty0 65--69, 2009.

\bibitem[Kovalyov and Pesch(2010)]{KovalyovP10}
Mikhail~T. Kovalyov and Erwin Pesch.
\newblock A generic approach to proving {NP}-hardness of partition type
  problems.
\newblock \emph{Discrete Appl. Math.}, 158:\penalty0 1908--1912, 2010.

\bibitem[Kremer et~al.(2014)Kremer, Mansour, and Perry]{KremerMP14}
Ilan Kremer, Yishay Mansour, and Motty Perry.
\newblock Implementing the ``wisdom of the crowd''.
\newblock \emph{J. Political Econ.}, 122\penalty0 (5):\penalty0 988--1012,
  2014.

\bibitem[Laffont and Martimort(2009)]{LaffontM09}
Jean-Jacques Laffont and David Martimort.
\newblock \emph{The Theory of Incentives: The Principal-Agent Model}.
\newblock Princeton University Press, 2009.

\bibitem[Lavi and Shamash(2022)]{LaviS22}
Ron Lavi and Elisheva~S. Shamash.
\newblock Principal-agent {VCG} contracts.
\newblock \emph{J. Econ. Theory}, 201:\penalty0 105443, 2022.
\newblock An extended abstract appeared in EC 2019.

\bibitem[Lehmann et~al.(2006)Lehmann, Lehmann, and Nisan]{LehmannLN06}
Benny Lehmann, Daniel Lehmann, and Noam Nisan.
\newblock Combinatorial auctions with decreasing marginal utilities.
\newblock \emph{Games Econ. Behav.}, 55\penalty0 (2):\penalty0 270--296, 2006.
\newblock An extended abstract appeared in EC 2001.

\bibitem[Leibo et~al.(2017)Leibo, Zambaldi, Lanctot, Marecki, and
  Graepel]{LeiboZLMG17}
Joel~Z. Leibo, Vin{\'{\i}}cius~Flores Zambaldi, Marc Lanctot, Janusz Marecki,
  and Thore Graepel.
\newblock Multi-agent reinforcement learning in sequential social dilemmas.
\newblock In \emph{Proc.~of AAMAS}, pages 464--473, 2017.

\bibitem[Li et~al.(2021)Li, Immorlica, and Lucier]{LiIL21}
Wanyi~Dai Li, Nicole Immorlica, and Brendan Lucier.
\newblock Contract design for afforestation programs.
\newblock In \emph{Proc.~of WINE 2021}, pages 113--130, 2021.

\bibitem[Li et~al.(2023)Li, Ashlagi, and Lo]{LiAL21}
Wanyi~Dai Li, Itai Ashlagi, and Irene Lo.
\newblock Simple and approximately optimal contracts for {Payment for Ecosystem
  Services}.
\newblock \emph{Manag. Sci.}, 69\penalty0 (12):\penalty0 7151--7882, 2023.

\bibitem[Li and Qiu(2024)]{LiQ24}
Yingkai Li and Xiaoyun Qiu.
\newblock Screening signal-manipulating agents via contests.
\newblock Available at \url{https://arxiv.org/abs/2302.09168}, 2024.

\bibitem[Li et~al.(2022)Li, Hartline, Shan, and Wu]{LiHSW22}
Yingkai Li, Jason~D. Hartline, Liren Shan, and Yifan Wu.
\newblock Optimization of scoring rules.
\newblock In \emph{Proc.~of EC 2022}, pages 988--989, 2022.

\bibitem[Liu et~al.(2022)Liu, Garg, and Borgs]{LiuGB21}
Lydia~T. Liu, Nikhil Garg, and Christian Borgs.
\newblock Strategic ranking.
\newblock In \emph{Proc.~of AISTATS 2022}, pages 2489--2518, 2022.

\bibitem[Mason and Watts(2009)]{MasonW09}
Winter~A. Mason and Duncan~J. Watts.
\newblock Financial incentives and the ``performance of crowds''.
\newblock \emph{{SIGKDD} Explor.}, 11\penalty0 (2):\penalty0 100--108, 2009.

\bibitem[Matou{\u s}ek and G{\"a}rtner(2006)]{GartnerM06}
J.~Matou{\u s}ek and B.~G{\"a}rtner.
\newblock \emph{Understanding and Using Linear Programming}.
\newblock Springer, 2006.

\bibitem[Mendler{-}D{\"{u}}nner et~al.(2020)Mendler{-}D{\"{u}}nner, Perdomo,
  Zrnic, and Hardt]{Mendler-DunnerP20}
Celestine Mendler{-}D{\"{u}}nner, Juan~C. Perdomo, Tijana Zrnic, and Moritz
  Hardt.
\newblock Stochastic optimization for performative prediction.
\newblock In \emph{Proc.~of NeurIPS 2020}, 2020.

\bibitem[Mirrlees(1975)]{Mirrlees99}
James~A. Mirrlees.
\newblock The theory of moral hazard and unobservable behaviour: Part {I}.
\newblock \emph{Mimeo, Oxford}, 1975.
\newblock (Published in 1999 in the \emph{Rev. Econ. Stud.}, 66(1), 3--21.).

\bibitem[Mohr et~al.(2022)Mohr, Viering, Loog, and van Rijn]{MohrVLR22}
Felix Mohr, Tom~J. Viering, Marco Loog, and Jan~N. van Rijn.
\newblock {LCDB} 1.0: An extensive learning curves database for classification
  tasks.
\newblock In \emph{Proc.~{ECML} {PKDD} 2022}, pages 3--19, 2022.

\bibitem[Morgenstern and Roughgarden(2015)]{MorgensternR15}
Jamie Morgenstern and Tim Roughgarden.
\newblock On the pseudo-dimension of nearly optimal auctions.
\newblock In \emph{Proc.~of NeurIPS 2015}, pages 136--144, 2015.

\bibitem[Myerson(1979)]{Myerson79}
Roger~B. Myerson.
\newblock Incentive compatibility and the bargaining problem.
\newblock \emph{Econometrica}, 47\penalty0 (1):\penalty0 61–73, 1979.

\bibitem[Myerson(1981)]{Myerson81}
Roger~B. Myerson.
\newblock Optimal auction design.
\newblock \emph{Math. Oper. Res.}, 6\penalty0 (1):\penalty0 58--73, 1981.

\bibitem[Myerson(1982)]{Myerson82}
Roger~B. Myerson.
\newblock Optimal coordination mechanisms in generalized principal-agent
  problems.
\newblock \emph{J. Math. Econ.}, 10:\penalty0 67–81, 1982.

\bibitem[Myerson and Satterthwaite(1983)]{MyersonS83}
Roger~B. Myerson and Mark~A. Satterthwaite.
\newblock Efficient mechanisms for bilateral trading.
\newblock \emph{J. Econ. Theory}, 29\penalty0 (2):\penalty0 265--281, 1983.

\bibitem[Neyman et~al.(2021)Neyman, Noarov, and Weinberg]{NeymanNW21}
Eric Neyman, Georgy Noarov, and S.~Matthew Weinberg.
\newblock Binary scoring rules that incentivize precision.
\newblock In \emph{Proc.~of EC 2021}, pages 718--733, 2021.

\bibitem[Nisan and Segal(2006)]{NisanSegal06}
Noam Nisan and Ilya Segal.
\newblock The communication requirements of efficient allocations and
  supporting prices.
\newblock \emph{J. Econ. Theory}, 129\penalty0 (1):\penalty0 192--224, 2006.

\bibitem[nobelprize.org(2016)]{Nobel}
nobelprize.org.
\newblock The 2016 {N}obel {P}rize in {E}conomics: {S}cientific background.
\newblock Available at
  \url{https://nobelprize.org/prizes/economic-sciences/2016/advanced-information/},
  2016.

\bibitem[{Paes Leme}(2017)]{PaesLeme17}
Renato {Paes Leme}.
\newblock Gross substitutability: An algorithmic survey.
\newblock \emph{Games Econ. Behav.}, 106:\penalty0 294--316, 2017.

\bibitem[Papadimitriou(2006)]{Papadimitriou06}
Christos~H. Papadimitriou.
\newblock The complexity of finding a nash equilibrium.
\newblock In Noam Nisan, Tim Roughgarden, \'Eva Tardos, and Vijay~V. Vazirani,
  editors, \emph{Algorithmic Game Theory}, chapter~2. Cambridge University
  Press, 2006.

\bibitem[Papireddygari and Waggoner(2022)]{PapireddygariW22}
Maneesha Papireddygari and Bo~Waggoner.
\newblock Contracts with information acquisition, via scoring rules.
\newblock In \emph{Proc.~of EC 2022}, pages 703--704, 2022.

\bibitem[Peng and Tang(2024)]{PengT24}
Bo~Peng and Zhihao~Gavin Tang.
\newblock Optimal robust contract design.
\newblock In \emph{Proc.~of {EC 2024}}, 2024.
\newblock Forthcoming.

\bibitem[Perdomo et~al.(2020)Perdomo, Zrnic, Mendler{-}D{\"{u}}nner, and
  Hardt]{PerdomoZMH20}
Juan~C. Perdomo, Tijana Zrnic, Celestine Mendler{-}D{\"{u}}nner, and Moritz
  Hardt.
\newblock Performative prediction.
\newblock In \emph{Proc.~of ICML 2020}, pages 7599--7609, 2020.

\bibitem[Perez‐Richet and Skreta(2022)]{PerezRS22}
Eduardo Perez‐Richet and Vasiliki Skreta.
\newblock Test design under falsification.
\newblock \emph{Econometrica}, 90\penalty0 (3):\penalty0 1109--1142, 2022.

\bibitem[Perez‐Richet and Skreta(2024)]{PerezRS24}
Eduardo Perez‐Richet and Vasiliki Skreta.
\newblock Fraud-proof non-market allocation mechanisms.
\newblock Available at \url{https://cepr.org/publications/dp18466}, 2024.

\bibitem[Peters and Szentes(2012)]{peters2012definable}
Michael Peters and Bal{\'a}zs Szentes.
\newblock Definable and contractible contracts.
\newblock \emph{Econometrica}, 80\penalty0 (1):\penalty0 363--411, 2012.

\bibitem[Piliouras and Yu(2023)]{PiliourasY23}
Georgios Piliouras and Fang{-}Yi Yu.
\newblock Multi-agent performative prediction: From global stability and
  optimality to chaos.
\newblock In \emph{Proc.~of EC 2023}, pages 1047--1074, 2023.

\bibitem[Prat and Rustichini(2003)]{PratR03}
Andrea Prat and Aldo Rustichini.
\newblock Games played through agents.
\newblock \emph{Econometrica}, 71\penalty0 (4):\penalty0 989--1026, 2003.

\bibitem[Rogerson(1985)]{Rogerson85}
William~P. Rogerson.
\newblock The first-order approach to principal-agent problems.
\newblock \emph{Econometrica}, 53:\penalty0 1357--1367, 1985.

\bibitem[Ross(1973)]{Ross73}
Stephen~A Ross.
\newblock The economic theory of agency: The principal’s problem.
\newblock \emph{Am. Econ. Rev.}, 63:\penalty0 134--139, 1973.

\bibitem[Roughgarden(2016)]{Roughgarden16}
Tim Roughgarden.
\newblock \emph{Twenty Lectures on Algorithmic Game Theory}.
\newblock Cambridge University Press, 2016.

\bibitem[Roughgarden and Talgam{-}Cohen(2015)]{RoughgardenT15}
Tim Roughgarden and Inbal Talgam{-}Cohen.
\newblock Why prices need algorithms.
\newblock In \emph{Proc.~of EC 2015}, pages 19--36, 2015.

\bibitem[Roughgarden et~al.(2017)Roughgarden, Syrgkanis, and
  Tardos]{RoughgardenST59}
Tim Roughgarden, Vasilis Syrgkanis, and \'Eva Tardos.
\newblock The price of anarchy in auctions.
\newblock \emph{J. Artif. Intell. Res.}, 59:\penalty0 59--101, 2017.

\bibitem[Rubinstein(2018)]{Rubinstein18}
Aviad Rubinstein.
\newblock Inapproximability of nash equilibrium.
\newblock \emph{{SIAM} J. Comput.}, 47\penalty0 (3):\penalty0 917--959, 2018.

\bibitem[Saig et~al.(2023)Saig, Talgam{-}Cohen, and Rosenfeld]{SaigTR23}
Eden Saig, Inbal Talgam{-}Cohen, and Nir Rosenfeld.
\newblock Delegated classification.
\newblock In \emph{Proc.~of {NeurIPS} 2023}, 2023.

\bibitem[Saig et~al.(2024)Saig, Einav, and Talgam{-}Cohen]{SaigET24}
Eden Saig, Ohad Einav, and Inbal Talgam{-}Cohen.
\newblock Incentivizing quality text generation via statistical contracts.
\newblock In \emph{Proc.~of NeurIPS 2024}, 2024.

\bibitem[Salani\'e(2017)]{Salanie17}
Bernard Salani\'e.
\newblock \emph{The Economics of Contracts: A Primer}.
\newblock MIT press, 2017.

\bibitem[Salzman et~al.(2018)Salzman, Bennett, Carroll, Goldstein, and
  Jenkins]{SalzmanBC+18}
James Salzman, Genevieve Bennett, Nathaniel Carroll, Allie Goldstein, and
  Michael Jenkins.
\newblock The global status and trends of {Payments for Ecosystem Services}.
\newblock \emph{Nat. Sustain.}, 1:\penalty0 136--144, 2018.

\bibitem[Savage(1971)]{Savage71}
Leonard~J. Savage.
\newblock Elicitation of personal probabilities and expectations.
\newblock \emph{J. Am. Stat. Assoc.}, 66:\penalty0 783--801, 1971.

\bibitem[Scarf(1958)]{Scarf58}
Herbert~E. Scarf.
\newblock A min-max solution of an inventory problem.
\newblock In K.~J. Arrow, S.~Karlin, and H.~E. Scarf, editors, \emph{Studies in
  the Mathematical Theory of Inventory and Production}, pages 201--209.
  Stanford University Press, 1958.

\bibitem[Schmeidler(1989)]{Schmeidler89}
David Schmeidler.
\newblock Subjective probability and expected utility without additivity.
\newblock \emph{Econometrica}, 57\penalty0 (3):\penalty0 571--587, 1989.

\bibitem[Schrijver(2003)]{Schrijver03}
Alexander Schrijver.
\newblock \emph{Combinatorial Optimization: Polyhedra and Efficiency}.
\newblock Springer, 2003.

\bibitem[Shavell(1979)]{Shavell79}
Steven Shavell.
\newblock Risk sharing and incentives in the principal and agent relationship.
\newblock \emph{Bell J. Econ.}, pages 55--73, 1979.

\bibitem[Shioura(2009)]{Shioura09}
Akiyoshi Shioura.
\newblock On the pipage rounding algorithm for submodular function maximization
  - a view from discrete convex analysis.
\newblock \emph{Discret. Math. Algorithms Appl.}, 1\penalty0 (1):\penalty0
  1--24, 2009.

\bibitem[Smale(1998)]{Smale98}
Steve Smale.
\newblock Mathematical problems for the next century.
\newblock \emph{Math. Intelligencer}, 20\penalty0 (2):\penalty0 7--15, 1998.

\bibitem[Szabo(1997)]{Szabo97}
Nick Szabo.
\newblock Formalizing and securing relationships on public networks.
\newblock \emph{First Monday}, 2\penalty0 (9), 1997.

\bibitem[Tadelis and Segal(2005)]{TadelisSegal05}
Steve Tadelis and Ilya Segal.
\newblock Lectures in contract theory.
\newblock Available at
  \url{http://faculty.haas.berkeley.edu/stadelis/Econ_206_notes_2006.pdf},
  2005.

\bibitem[Trevisan(2001)]{Trevisan01}
Luca Trevisan.
\newblock Non-approximability results for optimization problems on bounded
  degree instances.
\newblock In \emph{Proc.~of STOC 2001}, pages 453--461, 2001.

\bibitem[undp.org(2024)]{undp}
undp.org.
\newblock Sustainable development goals.
\newblock \url{https://www.undp.org/sustainable-development-goals/}, 2024.

\bibitem[upwork.com(2018)]{upwork}
upwork.com.
\newblock Not happy with the job of my freelancer.
\newblock Available at
  \url{https://community.upwork.com/t5/Clients/Not-happy-with-the-job-of-my-freelancer/m-p/506341},
  2018.

\bibitem[Vickrey(1961)]{Vickrey61}
William Vickrey.
\newblock Counterspeculation, auctions, and competitive sealed tenders.
\newblock \emph{J. Finance}, 16\penalty0 (1):\penalty0 8--37, 1961.

\bibitem[Vondr\'ak(2010)]{Vondrak10}
Jan Vondr\'ak.
\newblock Lecture notes ``{C}ontinuous extensions of submodular functions''.
\newblock Available at
  \url{https://theory.stanford.edu/~jvondrak/CS369P/lec17.pdf}, 2010.

\bibitem[Walton and Carroll(2022)]{CarrollW22}
Daniel Walton and Gabriel Carroll.
\newblock A general framework for robust contracting models.
\newblock \emph{Econometrica}, 90:\penalty0 2129--2159, 2022.

\bibitem[Wang et~al.(2023)Wang, D\"utting, Ivanov, Talgam{-}Cohen, and
  Parkes]{WangEtAl2023}
Tonghan Wang, Paul D\"utting, Dmitry Ivanov, Inbal Talgam{-}Cohen, and David~C.
  Parkes.
\newblock Deep contract design via discontinuous networks.
\newblock In \emph{Proc.~of NeuRIPS 2023}, 2023.

\bibitem[Wang and Huang(2022)]{WangH22}
Yifei Wang and Peng Huang.
\newblock Contract choice, moral hazard, and performance evaluation: Evidence
  from online labor markets.
\newblock Available at \url{https://ssrn.com/abstract=4047241}, 2022.

\bibitem[Wu et~al.(2024)Wu, Chen, Wang, Wang, and Xu]{WuCWWX24}
Jibang Wu, Siyu Chen, Mengdi Wang, Huazheng Wang, and Haifeng Xu.
\newblock Contractual reinforcement learning: Pulling arms with invisible
  hands.
\newblock Available at \url{https://arxiv.org/abs/2407.01458}, 2024.

\bibitem[Yu and Kong(2020)]{YuK20}
Yimin Yu and Xiangyin Kong.
\newblock Robust contract designs: Linear contracts and moral hazard.
\newblock \emph{Oper. Res.}, 68\penalty0 (5):\penalty0 1457--1473, 2020.

\bibitem[Zhu et~al.(2023)Zhu, Bates, Yang, Wang, Jiao, and Jordan]{ZhuEtAl22}
Banghua Zhu, Stephen Bates, Zhuoran Yang, Yixin Wang, Jiantao Jiao, and
  Michael~I. Jordan.
\newblock The sample complexity of online contract design.
\newblock In \emph{Proc.~of EC 2023}, page 1188, 2023.

\bibitem[Zuo(2024)]{Zuo24}
Shiliang Zuo.
\newblock New perspectives in online contract design: Heterogeneous,
  homogeneous, non-myopic agents and team production.
\newblock Available at \url{https://arxiv.org/abs/2403.07143}, 2024.

\end{thebibliography}

\end{document}